%% file: main.tex
\documentclass[11pt]{article}

\usepackage[left=1in, right=1in, top=1in, bottom=1in]{geometry}
\usepackage{enumerate}
\usepackage[latin1]{inputenc}
\usepackage{amsmath}
\usepackage{amssymb}
\usepackage{amsthm}
\usepackage{epsfig}
\usepackage{pstricks}
\usepackage{fancyhdr}

\usepackage{graphicx}
\usepackage{float}
\usepackage{subfig}
\usepackage[format=hang]{caption}
\usepackage{multicol}
\usepackage{multirow}

\usepackage{eurosym}
\usepackage[arrow,matrix,curve]{xy}
\usepackage{color}
\usepackage{xcolor}
\usepackage{bibentry}
\usepackage{algorithmic}
\usepackage{algorithm}
\usepackage{url}
\usepackage{boxhandler}
\usepackage{pdflscape}
\DeclareSymbolFontAlphabet{\Bbb}{AMSb}

\renewcommand{\a}{\alpha}

\renewcommand{\d}{\delta}

\newcommand{\e}{\varepsilon}

\renewcommand{\r}{\rho}

   \newcommand{\PH}[1]{}

\newcommand{\snorm}[1]{\Vert #1 \Vert}

\newcommand{\innerprod}[1]{\langle #1 \rangle}
\newcommand{\clip}[1]{\overset{\frown}{#1}} 

\theoremstyle{plain}

\newtheorem{Theorem}{Theorem}
\newtheorem{Lemma}[Theorem]{Lemma}

\newenvironment{proofof}[1]{\vspace*{1ex}\noindent{\bf Proof of #1.}}{\qed\medskip}

\newcounter{nr}
\allowdisplaybreaks

\begin{document}
\title{An SVM-like Approach for Expectile Regression}

\author{\textbf{M. Farooq}\footnote{Research was funded by HEC/DAAD scholarship A/12/94543}, \,\textbf{I. Steinwart}\\
Institute for Stochastics and Applications\\
	University of Stuttgart\\
	D-70569 Stuttgart\\
	\texttt{\{muhammad.farooq, ingo.steiwnart\}@mathematik.uni-stuttgart.de}
}

\maketitle

\begin{abstract}
 Expectile regression is a nice tool for  investigating conditional distributions beyond 
 the conditional mean. It is well-known that expectiles can be described 
 with the help of the  asymmetric least square loss function, and
 this link makes it possible to  estimate expectiles in a non-parametric framework
  by a  support vector machine like approach.
  In this work we develop an efficient sequential-minimal-optimization-based  solver 
 for the underlying optimization problem.
 The behavior of the solver is investigated by conducting various experiments 
 and the results are compared with the recent R-package  \textsf{ER-Boost}. 

\end{abstract}

\section{Introduction}
In standard nonparametric regression analysis, most of the methods developed so far are based on
the least square loss function for estimating conditional expectations. In many applications, however, it is required
to study conditional distributions beyond means.  A nice tool for this purpose was offered by \cite{koenker1978regression}
in the form of quantile regression, which allows both the location and the spread of the response variable
to be studied by using asymmetric least absolute deviation loss function (ALAD).
We refer the reader to \cite{koenker2005quantile, takeuchi2006nonparametric, christmann2007svms,steinwart2011estimating} 
and references therein, for details description and different estimation methods for quantile regression.
Following the spirit of quantile regression, \cite{newey1987asymmetric} proposed 
the asymmetric least square (ALS) loss function 
\begin{equation}\label{loss function}
\begin{aligned}
 L_\tau(t)=\left\{\begin{array}{ll}
	\tau t^2 & \quad \text{if} \,\, t\geqslant0\\ 
	(1-\tau )t^2  & \quad \text{if} \,\,   t<0, \\
\end{array}\right.
\end{aligned}
\end{equation}
to compute \textit{conditional expectiles}, also called \textit{regression expectiles}. 
These expectiles were found  an interesting alternative to quantiles in many applications due to  the computational 
advantages. For example, \cite{aragon2005conditional} used the expectile-order to determine the conditional ordering 
of individual values relative to other members of data sets, \cite{stahlschmidt2014expectile} developed an expectile-based 
technique to compute the distribution of treatment effects on the tail of the outcome variable in the presence of 
confounding mechanism,  and \cite{guler2014mincer} compared expectile regression  with quantile regression 
for forecast evaluation under asymmetric loss functions and showed that expectile treatment effects provide
more efficient estimates. 
There are some other areas where  expectlies have been applied successfully, 
for instance, in demography, see \cite{schnabel2009analysis} and in education, we refer to \cite{sobotka2013estimating}.  
Moreover, in finance expectiles play an important role for risk measures of financial asserts,
see for instance \cite{anderson2012study, hamidi2014dynamic, shim2013expected, wang2011measuring}. 
For example,  it has been shown recently that expectiles are the only coherent risk measures, 
see \cite{BeKlMuGi14a,StPaWiZh14a}. 
Moreover, the frequently used  expected shortfall (ES) is a conditional mean of a random variable given that it is less than 
a certain quantile. In other words, ES can be written as a function of both quantiles and expectiles, which requires to 
establish a connection between quantiles and expectiles.This leads to the expectile-based quantile estimates,
which can be more efficient than empirical quantiles \cite{waltrup2014expectile}. 
In this regard, recall that, there is one-to-one mapping of expectiles to quantiles that was explored
by \cite{efron1991regression} and further supported by \cite{abdous1995relating, yao1996asymmetric, taylor2008estimating}.
Moreover, \cite{breckling1988m} embedded both quantiles and expectiles in the general class of M-estimators 
by proposing asymmetric M-estimators.

 Some semiparametric and nonparametric  expectile estimation methods have already been proposed in literature. 
For example, \cite{schnabel2009optimal} considered 
 penalized splines  to compute smooth expectile estimates, 
 \cite{sobotka2012geoadditive} proposed 
 a couple of 
 different procedures including least asymmetrically weighted squares in combination with mixed models, 
 boosting within an empirical risk minimization framework, and a 
 restricted expectiles regression model. 
 Moreover, \cite{sobotka2013confidence} derived asymptotic properties of 
 expectile regression estimates and used them to construct corresponding confidence intervals.
 Furthermore, a  kernel method based on local linear fits was considered in \cite{yao1996asymmetric}, 
 and a boosting method using  regression trees   was proposed in \cite{yang2014nonparametric}. 
 Finally, two expectile regression packages, \textsf{ER-Boost} \cite{yang2014nonparametric} and \textsf{expectreg} \cite{sobotka2014package}, 
have recently been made available. 

Another  family of non-parametric estimation methods are the so-called 
kernel based regularized empirical risk minimizers, which include the well known  
 \textit{support vector machines (SVMs)} \cite[p. 138ff]{vapnik2000nature}.
These kernel-based methods
often enjoy  state-of-the-art empirical performance, relatively simple implementations, 
and a high flexibility. Recall that their flexibility is based on two main ingredients, 
namely the reproducing kernel Hilbert space (RKHS) $H$ and the loss function $L$. 
Namely, the RKHS can be 
used to adapt to the nature of the input domain $X$, or more precisely, enables us to use
both standard $\mathbb{R}^n$-valued data and non-standard data such as strings and graphs.
Moreover, due to the so-called kernel-trick \cite{scholkopf2002learning},
the choice of $H $ has little to no algorithmic consequences for solving SVM optimization problems. 
On the other hand, the choice of $L$ determines 
the learning goal \cite[Chapter 3]{steinwart2008support}. 
For example, the so-called hinge loss is used for classification, 
the least squares loss leads to conditional mean regression,
and the ALAD is used to estimate quantiles.  
Unfortunately, however, different $L$ lead to  different optimization problems, 
which in turn require different solvers.
For the above mentioned loss functions various solvers have been designed, see for example
\cite{chang2011libsvm, cristianini2000introduction, glasmachers2006maximum, keerthi2003smo, takeuchi2006nonparametric}
and references therein for more detail, but besides \cite{huang2014asymmetric}, 
who considered a kernelized iteratively reweighted strategy,
no solver for the ALS has been proposed.
In this paper, we derive a sequential minimal optimization (SMO) based solver, 
see \cite{cristianini2000introduction} and particularly \cite{platt1999fast},
for the ALS, which enables us to handle large data set efficiently.
In addition, we consider different initialization methods and working set strategies in detail and 
validate them empirically, to further speed up the solver.
Finally, we report some experiments that compare our solver with 
the \textsf{ER-Boost} package. 

The rest of the paper is organized as follows: Section 2 presents the formulation of the primal and the dual optimization 
problem of SVMs. Section 3 proposes an algorithm to perform one dual variable update per iteration along with
the stopping criteria and initialization methods. 
The exact two dimensional optimization problem with some working set selection strategies is discussed in Section 4. 
Some experiments and discussion on the results can be found in section 5.
Finally, the appendices contain proofs of theorems and lemmas, and detailed results from experiments.

\section{Primal and Dual Optimization Problem}

Let us consider a training set $D := \left((x_1, y_1), (x_2, y_2),\cdots (x_n, y_n) \right) \in \left( X \times \mathbb{R}\right)^n $ 
that is sampled from some unknown distribution $P$ on $X \times Y$, where $X$ is an arbitrary set and $Y \subset \mathbb{R}$. 
In addition, we assume that $f : X \rightarrow \mathbb{R}$ is a function and  $L : Y \times \mathbb{R} \rightarrow [0, \infty)$ 
is an arbitrary convex loss function defined in (\ref{loss function}). 
Then the goal of supervised statistical learning is to find a function $f$ such that the risk
\begin{equation*}
\mathcal{R}_{L,P}(f):=\int_{X \times Y} L(y,f(x))d P(x,y)\, ,
\end{equation*}
is small. This means that $\mathcal{R}_{L,P}(f)$ has to be close to the optimal risk
\begin{equation*}
\mathcal{R}_{L,P}^* := \text{inf}\{\mathcal{R}_{L,P}(f)| f: X \rightarrow \mathbb{R}\,\, \text{measurable}\}\, ,
\end{equation*}
which is called the Bayes risk with respect to $P$ and $L$. Since the data generating distribution $P$ is unknown, 
we replace $\mathcal{R}_{L,P}(f)$ by its empirical counterpart
\begin{equation}\label{empirical risk}
\mathcal{R}_{L,D}(f):= \frac{1}{n} \sum_{i=1}^{n} L(y_i, f(x_i))\,.
\end{equation}
Now, recall that the support vector machines (SVMs) solve the regularized problem
\begin{equation}\label{regularized problem}
f_{D,\lambda} = \text{arg}\,\underset{f \in H}{\text{min}}\,\, \lambda \snorm{f}_{H}^{2} +  \mathcal{R}_{L,D}(f)\,,
\end{equation}
where $\lambda > 0$ is a user specified regularization parameter and $H$ is the reproducing kernel Hilbert space (RKHS)  over $X$
with reproducing kernel $k: X \times X \rightarrow \mathbb{R}$\,,
see e.g. \cite{berlinet2004reproducing, aronszajn1950theory, steinwart2008support}. 
For example, for input domains $X \subset \mathbb{R}^d$, 
one often uses SVMs that are equipped with Gaussian radial basis (RBF) kernels. Recall that the latter are defined by  
\begin{equation}\label{Gaussian kernel}
k_\gamma(x,x'):= \exp(-\gamma^2 \lVert x-x' \rVert_{2}^{2})\, ,\hspace*{5ex} x,x' \in \mathbb{R}^d
\end{equation}
where $\gamma > 0$ is called the width parameter that is usually determined in a data-dependent way, e.g., by cross-validation. 
Note that $k_\gamma$ is normalized, that is, $k_\gamma(x,x)=1$ for all $x \in \mathbb{R}^d$, and all kernels we consider below are 
also normalized.
By \cite[Theorem 4.56]{steinwart2008support}, $k_\gamma$ is also universal on every compact subset $X \in \mathbb{R}^n$ 
and in particular strictly positive definite. 
Furthermore, the RKHS $H_\gamma$ induced by $k_\gamma$ is dense in $L_p(\mu)$ \cite[Chapter 4]{steinwart2008support},  
where $\mu$ is a finite measure of $\mathbb{R}^n$ and $p \in [1, \infty)$. 
Therefore, the following consistency result applies to Gaussian kernels.

\begin{Theorem}\label{theorem-consistency}
 Let $P$ be a distribution on $X \times \mathbb{R}$ with
 $
  \int_{X} y^2 P(dy| x) dP_{X}(x) < \infty
 $, 
$L$ be the $\tau$-asymmetric least squares loss,
and $f_{L, P}^*$ be the conditional $\tau$-expectile function.
Moreover, let   $k$ be a bounded, measurable  kernel 
whose RKHS  is 
 separable and dense in $L_2({P}_X)$. Then for all sequences $\lambda_n \rightarrow 0$ with 
$\lambda_{n}^{4}n \rightarrow 0$ and all $\varepsilon > 0$, we have
\begin{equation*}
 \underset{n \rightarrow \infty}{\lim}  {P}^n\big(D\in (X\times \mathbb{R})^n:\mathcal{R}_{L, P}(f_{D, \lambda_n})-\mathcal{R}_{L,P}^* > \varepsilon \big)=0\, ,
\end{equation*}
and
\begin{equation*}
  \underset{n \rightarrow \infty}{\lim}  {P}^n\big(D\in (X\times \mathbb{R})^n: \snorm{f_{D, \lambda_n} - f_{L, P}^*}_{0} > \varepsilon \big) =0\,,
\end{equation*}
where $\snorm{g}_{0}:= \int \min \{1, |g|\} d  {P}_X$
is a translation-invariant metric describing convergence in probability ${P}_X$.
\end{Theorem}


To deal with (\ref{regularized problem}) algorithmically, we fix a feature space $H_0$ and a 
feature map $\Phi: X \rightarrow H_0$ of $\mathbb{R}$. Then every $f \in H $ can be represented by $w \in H_0$ via
\begin{equation}\label{expectile regression function}
f(\mathbf{x}_i)  = \innerprod{\mathbf{w}, \phi(\mathbf{x}_i)}\, ,
\end{equation}
see \cite[Theorem 4.21]{steinwart2008support} for further details. Note that the latter theorem also shows that 
\begin{equation}\label{norm of w RKHS}
 \snorm{f}_H = \inf \{\snorm{\mathbf{w}}_{H_{0}} : \mathbf{w} \in H_{0} \,\text{with}\, f =\innerprod{\mathbf{w}, \phi(\cdotp)} \}\, ,
\end{equation}
 where $\phi:= X \rightarrow H$ is the canonical feature map from the input space to RKHS.
 Using (\ref{empirical risk}) and (\ref{norm of w RKHS}) in the objective function (\ref{regularized problem}), 
 we obtain the standard regularized problem for SVMs without offset 
\begin{equation}\label{SVM without offset-General}
\text{arg}\underset{w \in H_0}{\text{min}}\,\, \lambda \snorm{\mathbf{w}}_{H_0}^{2} + \frac{1}{n} \sum_{i=1}^{n} L(y_i, f(x_i))\,. 
\end{equation}
If $L$ is the hinge loss function, then it is shown by \cite{steinwart2011training} that the SVM without offset 
not only faster but also achieves accuracy that is comparable to SVM with offset. 
One reason for the faster training time was that the offset leads to an additional equality constraint for the dual problem
and as a consequence, SMO type solvers can only update certain pairs of dual variables. In addition, 
the offset makes it relatively expensive to calculate the duality gap \cite{cristianini2000introduction}, 
which may serve as a stopping criterion for these solvers.

In the following, we will adapt the ideas of \cite{steinwart2011training}  to design a solver for (\ref{SVM without offset-General}) 
in the case of $L$ being an asymmetric least squares loss. To this end, we first reformulate the 
primal optimization problem (\ref{SVM without offset-General}) such as
\begin{equation}\label{primal objective function}
	\begin{aligned}
	{\text{arg}}\underset{\textbf{w} \in H}{\underset{(\textbf{w},\xi_{+}, \xi_{-})}{\text{min}}}\; P_C(\textbf{w},\xi_{+}, \xi_{-})&:=\frac{1}{2} \snorm{\textbf{w}}^2
	+ C \tau \sum_{i=1}^{n} \xi_{i,+}^2 +  C (1-\tau) \sum_{i=1}^{n} \xi_{i,-}^2\, ,\\
	\text{such that } \hspace*{12ex}
		\xi_{i,+} &\geq \, y_i - \innerprod{\mathbf{w},\phi(\mathbf{x}_i)}\,,  \\
		\xi_{i,-} &\geq  \, \innerprod{\mathbf{w},\phi(\mathbf{x}_i)} - y_i\,,\\
		\xi_{i,+} \; , \xi_{i,-} &\geq 0\,, \hspace*{5ex} \forall\, i = 1,\ldots,n
	\end{aligned}
\end{equation}
where $C:=\frac{1}{2n\lambda} > 0$.
Using standard Langrangian techniques, see e.g.~\cite[Chapter 6]{cristianini2000introduction}, one can easily 
obtain the 
%
dual optimization problem 
\begin{equation}\label{Dual Objective Function}
	\begin{aligned}
	\text{arg}\,\underset{(\alpha, \beta)}{\text{max}}\; W(\alpha, \beta) &:= \innerprod{\alpha-\beta, \mathbf{y}}
	- \frac{1}{2}\innerprod{ \alpha - \beta, K (\alpha - \beta)}- \frac{1}{4C\tau} \innerprod{\alpha,\alpha}
	-\frac{1}{4C(1-\tau)} \innerprod{\beta,\beta} 		
	\end{aligned}
	\end{equation}
\begin{equation*}
\alpha_i \geq 0 , \beta_i \geq 0\,.\hspace*{5ex} \forall\, i = 1, \ldots, n
\end{equation*}
Here $\mathbf{y}$ is the $n \times 1$ vector of labels and $K$ is the $n \times n$ matrix with entries 
$K_{i,j}:= k(x_i, x_j),\, i,j = 1, \ldots, n$. Note that (\ref{primal objective function}) is a 
convex function as the loss function (\ref{loss function}) is a convex suffices. 
Analogously, it is not hard to see that 
the dual optimization problem (\ref{Dual Objective Function}) is concave. This ensures the fulfillment of 
the strong duality assumptions \cite[Chapter 5]{cristianini2000introduction} and consequently, the primal optimal solution 
can be obtain from the dual optimal solution using the simple transformation, which is
\begin{equation*}
\mathbf{w}:= \sum_{i=1}^{n}(\alpha_i - \beta_i )\phi(\mathbf{x}_{i})\,.
\end{equation*}
In addition, the quadratic nature of (\ref{Dual Objective Function}) allows us to solve it using the quadratic programming (QP) techniques. 
However, many QP techniques that are implemented to solve dual optimization problems, 
for example, interior point methods \cite{wright1999numerical, scholkopf2002learning}, are impractical for large scale problems. 
Decomposition methods, such as \textit{chunking} \cite{vapnik2000nature} have been designed to handle this difficulty 
by breaking the optimization problem into smaller subproblems and solving them iteratively.
The limiting case of decomposition methods is the Sequential Minimal Optimization (SMO) methods that  optimizes two coordinates at 
each iteration \cite{platt1999fast} for SVMs with offset and hence, does not require storage of the entire kernel matrix. 
Section 4 presents this idea in more detail in view of expectile regression without offset. 
It is also worth noting that SVMs without offset allows us to develop an SMO type algorithm that performs 
one dual variable update per iteration as a starting point \cite{steinwart2011training}. 
In the following section, we introduce this algorithm in details.

\section{One Working Set Solution}
Our  goal in this section is to develop an SMO type algorithm that updates a single coordinate at each iteration. 
For this, we first compute one working set solution. Then we establish a rule to select a direction
in which update should be performed, and a criterion to stop the algorithm. 
In the end, we present the procedures to initialize the coordinates.

Let us first compute the gradients for $\alpha_i$ and $\beta_i$  from (\ref{Dual Objective Function}) that will be used
throughout this paper. For this, we take the partial derivatives of (\ref{Dual Objective Function}) w.r.t.
$\alpha_i$ and $\beta_i$ and obtain the following
\begin{equation}\label{1D-Gradients Alpha_i, Beta_i}. 
	\begin{aligned}
	\nabla W_{\alpha_i}(\alpha, \beta)& =  \innerprod{e_i,\mathbf{y}} -\innerprod{e_i, K(\alpha - \beta)}-\frac{\innerprod{e_i, \alpha}}{2C\tau}\,,\\
		\nabla W_{\beta_i}(\alpha, \beta)& =  -\innerprod{e_i,\mathbf{y}} +\innerprod{e_i, K(\alpha - \beta)}-\frac{\innerprod{e_i,\beta}}{2C(1-\tau)}\,.
	\end{aligned}
\end{equation}
We now recall \cite[p. 131ff]{cristianini2000introduction} and reformulate the dual objective function
(\ref{Dual Objective Function}). For $\alpha, \beta \in \mathbb{R}^n$ and an index $i \in \{1,\ldots,n\}$, 
we write $\alpha^{\setminus i}:= \alpha - \alpha_i e_i$ and $\beta^{\setminus i}:= \beta - \beta_i e_i$  
where $e_i$ is the $i$-th vector of standard basis of $\mathbb{R}^n$. 
Now the basic calculus together with $K_{i,i}=1$ for normalized kernels leads to the following dual objective function
for the 1D-problem
\begin{equation}\label{1D-Dual objective function}
	\begin{aligned}
	W(\alpha^{\setminus i} + \alpha_i e_i,\beta^{\setminus i} + \beta_i e_i)&:= W(\alpha^{\setminus i},\beta^{\setminus i}) + (\alpha_i - \beta_i)\innerprod{e_i, \mathbf{y}}-\frac{1}{2}(\alpha_i - \beta_i)^2 \\
		& \quad\quad- (\alpha_i - \beta_i) \innerprod{e_i, K (\alpha^{\setminus i}-\beta^{\setminus i})}-\frac{\alpha_i^2}{4C\tau} - \frac{\beta_i^2}{4C(1-\tau)}\,.
	\end{aligned}
\end{equation}
Taking partial derivative of (\ref{1D-Dual objective function}) w.r.t. $\alpha_i$ and $\beta_i$ and setting them 
to zero yields the system of equations
\begin{equation}\label{1D-globel-system of equations}
	\begin{aligned}
	b_1 \alpha_i - \beta_i &= c_i\,,\\
	\alpha_i - b_2 \beta_i &= c_i\,,
	\end{aligned}
\end{equation}
where
\begin{equation}\label{1D-coefficients-global system equations}
	\begin{aligned}
	b_1 &= \frac{2C\tau +1}{2C\tau}\,,\\
	b_2 &= \frac{2C(1-\tau)+1}{2C(1-\tau)}\,,\\
	c_i &= \innerprod{e_i, \mathbf{y}} - \innerprod{e_i, K (\alpha^{\setminus i}-\beta^{\setminus i})}=
		 \nabla W_{\alpha_i}(\alpha, \beta) + b_1 \innerprod{e_i, \alpha} -\innerprod{e_i, \beta}\,.
	\end{aligned}
\end{equation}
After solving (\ref{1D-globel-system of equations}), we obtain the global solution
\begin{equation}\label{1D-Global solution}
	\alpha_i^*  =  \frac{b_2-1}{b_1 b_2 - 1} c_i\,, \hspace*{4ex}  \beta_i^* =  \frac{1-b_1}{b_1 b_2-1} c_i\,.\\
\end{equation}
Note that $b_1,  b_2  \in (1,  \infty)$ for all $C > 0$ and $\tau \in (0,1)$. Therefore, it is not hard to see 
from (\ref{1D-Global solution}) that $\alpha_i^* = \beta_i^* = 0$ if and only if $c_i = 0$. 
On the other hand, for all $c_i \in \mathbb{R}\setminus \{0\}$, (\ref{1D-Global solution}) leads to the relation
\begin{equation}\label{1D-relation of global solution}
 \alpha_i^*  =  -\frac{\tau}{1-\tau} \beta_i^*\,, 
\end{equation}
which implies that the global solution $(\alpha_i^*,\beta_i^*)$ violates the constraints of 
the dual problem (\ref{Dual Objective Function}). 
In other words, the global maximum that is attained by (\ref{Dual Objective Function}) does not lie 
in the set of feasible vectors. The following general theorem describes the way to find the solution in this situation.

\begin{Theorem}\label{1D-optimum solution at boundary}
 Let $W:\mathbb{R}^m \rightarrow \mathbb{R}$ be a concave and twice continuous differentiable 
 function and $\mathcal{A} \subset \mathbb{R}^m$ be a closed convex set. Assume that there is 
 exactly one $\alpha^* \in \mathbb{R}^m$ with $W^\prime(\alpha^*)=0$. 
 Then the following statements hold:
\begin{itemize}
 \item[i)] For all $\alpha \neq \alpha^*$ we have $W(\alpha^*) > W(\alpha)$.
	\item[ii)] If $\alpha^* \notin \mathcal{A}$, then there exists an $\alpha^\star \in \partial \mathcal{A}$ such that $W(\alpha^\star) \geq W(\alpha)$ for all $\alpha \in \mathcal{A}$. 
\end{itemize}
 \end{Theorem}

 Theorem \ref{1D-optimum solution at boundary} says that either $\alpha^*$ is the optimal feasible solution or there is an optimal feasible solution 
 on boundary $\{(0, \beta_i):\beta_i > 0 \}\cup \{(\alpha_i, 0):\alpha_i > 0 \}$. 
 Now (\ref{1D-Global solution}) shows that we have exactly one value $(\alpha_i^*, \beta_i^*)$ at which derivative vanishes 
 and (\ref{1D-relation of global solution}) shows that $(\alpha_i^*, \beta_i^*)$ is not feasible. Consequently, we need to look at 
 the boundaries to search for an optimal feasible solution. To this end, we split the problem into two cases. 
 In the first case,  we plug $\alpha_i = 0$ in (\ref{1D-Dual objective function}) and then differentiate 
 w.r.t. $\beta_i$,  which provides
\begin{equation*}
 \frac{\partial W(\alpha^{\setminus i}, \beta^{\setminus i}+\beta_i e_i) }{\partial \beta_i} = -\innerprod{e_i, \mathbf{y}}
 +\innerprod{e_i,  K (\alpha^{\setminus i} - \beta^{\setminus i})} - b_2\innerprod{e_i,	\beta}\,.
\end{equation*}
Setting it to zero gives
\begin{equation}\label{1D-optimum Solution-case 1-alpha=0}
	\alpha_i^+= 0\,, \hspace*{4ex} \beta_i^+= -\frac{c_i}{b_2}\,.
\end{equation}
Similarly,  for the second case,  plugging $\beta_i = 0$ in (\ref{1D-Dual objective function}) 
and differentiating w.r.t. $\alpha_i$ yields
\begin{equation*}
	\frac{\partial W(\alpha^{\setminus i} + \alpha_i e_i, \beta^{\setminus i})}{\partial \alpha_i} = \innerprod{e_i, \mathbf{y}}
	- \innerprod{e_i,  K (\alpha^{\setminus i} - \beta^{\setminus i})}- b_1\innerprod{e_i, \alpha}\,.
\end{equation*}
Equating it to zero provides
\begin{equation}\label{1D-optimum Solution-case 2-beta=0}
	\beta_i^+ = 0\,, \hspace*{4ex} \alpha_i^+ = \frac{c_i}{b_1}\,.
\end{equation}
Since $b_1,  b_2 \in (1,  \infty)$ are fixed constants for certain $\tau$, therefore, 
(\ref{1D-optimum Solution-case 1-alpha=0}) and (\ref{1D-optimum Solution-case 2-beta=0}) solely 
depend on $c_i$. In particular, if $c_i \neq 0$, then we show in the following theorem that 
either (\ref{1D-optimum Solution-case 1-alpha=0}) or (\ref{1D-optimum Solution-case 2-beta=0}) gives 
the feasible optimal solution.
 
\begin{Theorem}\label{1D-theorem-feasible solution}
For $i=\{1, \ldots, n\}$, let  $c_i \in \mathbb{R}$ and $b_1, b_2 \in (1, \infty)$ be 
defined by (\ref{1D-coefficients-global system equations}). Then the following implications holds: 
	\begin{itemize}
	\item[i)] If $c_i < 0$, then (\ref{1D-optimum Solution-case 1-alpha=0}) is the feasible solution.
	\item[ii)] If $c_i = 0$, then (\ref{1D-optimum Solution-case 1-alpha=0}) and (\ref{1D-optimum Solution-case 2-beta=0}) are the same feasible solution.
	\item[iii)] If $c_i > 0$, then (\ref{1D-optimum Solution-case 2-beta=0}) is the feasible solution.
	\end{itemize}
In particular, exactly one of the two cases produces a feasible solution $(\alpha_i^+, \beta_i^+)$, and this is given by
\begin{equation*}
	\alpha_{i}^+ = \max \Big(0, \frac{c_i}{b_1}\Big)\,,\hspace*{4ex} \beta_{i}^+ = \max \Big(0, -\frac{c_i}{b_2}\Big)\,. \\
\end{equation*}
\end{Theorem}
After finding the feasible optimal solution, the next task is to determine 
the coordinate $i$ in which the update should be performed. Many approaches have been discussed 
so far for this purpose. A simple approach \cite[p. 132-133]{cristianini2000introduction} is 
to update for each coordinate $i = 1,\ldots,n$ iteratively. Another method \cite{vogt2002smo} 
is to choose the coordinate for update that violates the Karush-Kuhn-Tucker (KKT) conditions 
of optimality most. The latter approach is implemented in SVMs packages,  
SVM$^{light}$ \cite{joachims1999making} and LIBSVM \cite{chang2011libsvm}. Another idea, see 
\cite{steinwart2011training},  which is followed in this work, is to choose the coordinate $i^*$ 
whose update achieves the largest improvement for the value of dual objective function $W$. 
In other words, it performs the update in the direction
\begin{eqnarray}\label{1D-best direction search}
 i^* \in \arg\,\underset{i=1,\ldots,n} \max W(\alpha + \delta e_i, \beta + \eta e_i)-W(\alpha, \beta)\,,
\end{eqnarray}
where $\delta_i = \alpha_{i}^{+}-\alpha_i$ and  $\eta_i = \beta_{i}^{+}-\beta_i$ denote the 
difference between the new and the old values of $\alpha_i$ and $\beta_i$ respectively. 
Based on this idea, we establish a rule in the following lemma to compute the improvement in the value of dual 
objective function $W$.
\begin{Lemma}\label{1D-gain}
Let $i \in \{1,\ldots,n\}$, $\alpha, \beta \in \mathbb{R}^n$, and $\delta, \eta \in \mathbb{R}$. 
Moreover let $b_1, b_2 \in (1, \infty)$ be defined by (\ref{1D-coefficients-global system equations}), 
then we have
\begin{equation}
	\begin{aligned}
	G(\delta, \eta) &:=  W(\alpha + \delta e_i, \beta + \eta e_i)-W(\alpha, \beta)\\
			&\,\,= \delta \left( \nabla W_{\alpha_i}(\alpha, \beta)-\frac{b_1 \delta}{2}\right) 
		+\eta \left( \nabla W_{\beta_i}(\alpha, \beta)-\frac{b_2 \eta}{2}\right) +\delta \eta.
	\end{aligned}
\end{equation}
\end{Lemma}
With the above lemma, the Procedure \ref{Procedure-1D direction search} solves (\ref{1D-best direction search}) to search 
the best direction. 

\begin{algorithm}[H]
\renewcommand\thealgorithm{1}
\floatname{algorithm}{Procedure}
\renewcommand{\algorithmicrequire}{\textbf{Input:}}
\renewcommand{\algorithmicensure}{\textbf{Output:}}

\caption{\small Calculate $i^*\in \arg \max_{i=1,\ldots n} \Big(W(\alpha + \delta e_i, \beta+\eta e_i)-W(\alpha, \beta)\Big)$}
{\fontsize{10}{15}

\begin{algorithmic}\label{Procedure-1D direction search}
\STATE \textit{bestgain} $\leftarrow -1$
\FOR{ $i=1$ to $n$}
	\STATE $\delta_i\leftarrow \max \big(0, \frac{c_i}{b_1}\big)-\alpha_i$
	\STATE $\eta_i\leftarrow \max \big(0, -\frac{c_i}{b_2}\big)-\beta_i$
	\STATE $gain\leftarrow G(\delta_i, \eta_i)$
	\IF{$gain > bestgain$}
		\STATE $bestgain \leftarrow gain$
		\STATE $i^* \leftarrow i$
		\STATE $\delta_{i^*} \leftarrow \delta_i$
		\STATE $\eta_{i^*} \leftarrow \eta_i$
	\ENDIF
	\RETURN $i^*, \delta_{i^*}, \eta_{i^*}$
\ENDFOR 
\end{algorithmic}
}
\end{algorithm}

\subsection{Stopping Criteria}\label{1D-section-stopping criteria}

Solving problem (\ref{Dual Objective Function}) by some iteration method requires an appropriate stopping criteria.
Several stopping criteria have been suggested so far for SVMs \textit{with offset}.  
One method is to stop training when the KKT conditions are satisfied up to some predefined 
tolerance $\epsilon > 0$. Another method is to use the duality gap as a stopping 
criteria \cite[p. 109 and 128]{cristianini2000introduction}. This method is also adopted 
by \cite{steinwart2011training} to formulate a duality gap for SVM \textit{without offset}. 
Following this idea, we define for dual variables $\alpha \in \mathbb{R}_+$ and $\beta \in \mathbb{R}_+$
\begin{equation}\label{1D-expectile function-dual variable form}
 f_{\alpha, \beta} := \innerprod{\alpha-\beta, K e_i}\,,
\end{equation}
 which gives $\snorm{f_{\alpha, \beta}}_{H}^2 = \innerprod{\alpha- \beta , K (\alpha-\beta)}$.  
 As a result, the primal objective function (\ref{primal objective function}) is 
\begin{equation*}
P(f_{\alpha, \beta},\xi_{i,+}, \xi_{i,-})=\frac{1}{2}\innerprod{\alpha - \beta, K(\alpha - \beta)}
+ C \tau \sum_{i=1}^n \xi_{i, +}^2 + C (1-\tau)\sum_{i=1}^n \xi_{i, -}^2\,. 
\end{equation*}
Following \cite{steinwart2011training}, the duality gap of $P(f_{\alpha, \beta},\xi_{i,+}, \xi_{i,-})$ and $W(\alpha, \beta)$ is defined as
\begin{equation}\label{1D-duality gap-general}
	\begin{aligned}
	S(\alpha, \beta):= P(f_{\alpha, \beta},\xi_{i,+}, \xi_{i,-})- W(\alpha, \beta)\,,  
	\end{aligned}
\end{equation}
which tells us to stop the iteration method of solving problem (\ref{Dual Objective Function})  if $S(\alpha, \beta) < \epsilon$, 
where $\epsilon > 0$ is some predefined tolerance. 
To efficiently compute $S(\alpha, \beta)$,  we split it into 
\begin{equation}\label{1D-components of duality gap}
	\begin{aligned}
	T(\alpha, \beta) &= \frac{1}{2}\innerprod{(\alpha - \beta), K(\alpha - \beta)}- W(\alpha, \beta)\,,\\
	E(\alpha, \beta) &=    \tau \sum_{i=1}^n \xi_{i, +}^2 +  (1-\tau)\sum_{i=1}^n \xi_{i, -}^2\,,
	\end{aligned}
 \end{equation}
and as a result we have $S(\alpha, \beta) = T(\alpha, \beta)+ C\cdot E(\alpha, \beta)$. 
The value $T(\alpha, \beta)$ can be obtained at each iteration by updating it in the chosen direction $i$, such as
\begin{equation*}
 T(\alpha+\delta e_i, \beta + \eta e_i) = T(\alpha, \beta) -  U(\alpha_i,  \beta_i,  \delta, \eta)\, ,
\end{equation*}
where 
 \begin{equation}\label{1D-duality gap-update factor in T}
	\begin{aligned}
	U(\alpha_i,  \beta_i, \delta, \eta)&:= \delta \left( 2\nabla W_{\alpha_i}(\alpha, \beta) + \innerprod{y, e_i}+\frac{\innerprod{\alpha, e_i}}{2C\tau}-\frac{(b_1 + 1) \delta}{2}\right)\\
		& \quad + \eta \left( \nabla W_{\beta_i}(\alpha, \beta)+\innerprod{y, e_i}+\frac{\innerprod{\beta, e_i}}{2C(1-\tau)}-\frac{(b_2 + 1) \eta}{2} \eta \right)+ 2\delta \eta\,.
	\end{aligned}
 \end{equation}
Unlike $T(\alpha, \beta)$, the value $E(\alpha, \beta)$ can not be updated but needs to be computed 
from scratch at each iteration. To find an efficient formula, we first note that combining 
(\ref{primal objective function}) with (\ref{1D-expectile function-dual variable form}),  we have
\begin{equation*}
	\xi_{i, +}  = \max\big\{0, \innerprod{y, e_i} - \innerprod{\alpha - \beta,  K e_i}\big\}
		    = \max\Big \{0,\nabla W_{\alpha_i}(\alpha, \beta) + \frac{\innerprod{\alpha, e_i}}{2C\tau}\Big\}\, ,  
\end{equation*}
and
\begin{equation*}
	\xi_{i, -}  = \max\big\{0, \innerprod{\alpha - \beta,  K e_i}- \innerprod{y, e_i}\big\}
		    =\max\Big \{0,-\nabla W_{\alpha_i}(\alpha, \beta) - \frac{\innerprod{\alpha, e_i}}{2C\tau}\Big\}\,. 
\end{equation*}
With these formulas, the computation of $E(\alpha, \beta)$ is an $O(n)$ operation.
Let us now consider a little more involved stopping criteria based on \cite[Chapter 7]{steinwart2008support}, 
that looks for an $f_{\alpha, \beta} \in H$ with  
\begin{equation}\label{1D-clipped RERM}
 \lambda \snorm{f_{\alpha, \beta}}_{H}^{2} +  \mathcal{R}_{L,D}(\clip{f}_{\alpha, \beta}) 
 \leq
 \underset{f \in H}{\text{min}} \lambda \snorm{f}_{H}^{2} +  \mathcal{R}_{L,D}(f) + \epsilon\,,
\end{equation}
where $\clip{f}_{\alpha, \beta}$ is clipped at $\pm M \in \mathbb{R}$. Formally speaking, the clipped value
of $f_{\alpha, \beta}: X \rightarrow \mathbb{R}$ at $\pm M \in \mathbb{R}$
is defined by
\begin{equation*}
\begin{aligned}
 \clip{f}_{\alpha, \beta}=\left\{\begin{array}{ll}
	-M & \quad \text{if} \,\, f_{\alpha, \beta} < -M\,,\\ 
	f_{\alpha, \beta}  & \quad \text{if} \,\,  f_{\alpha, \beta} \in [-M, M]\,, \\
	-M & \quad \text{if} \,\, f_{\alpha, \beta} > -M\,.
\end{array}\right.
\end{aligned}
\end{equation*}
In other words, we restrict $f_{\alpha, \beta}$  to the interval $[-M, M]$, which in turns, 
reduces the risk $\mathcal{R}_{L,D}(f)$. However, clipping does not change the learning method since it is performed after 
the learning phase. Based on this idea, the clipped version of (\ref{1D-expectile function-dual variable form}) after using
(\ref{1D-expectile function-dual variable form}) is
\begin{equation}
\clip{f}_{\alpha, \beta}(x_i) = \Big[\innerprod{e_i, \mathbf{y}} -\nabla W_{\alpha_i}(\alpha, \beta) 
- \frac{\innerprod{\alpha, e_i}}{2C\tau}\Big]_{-M}^M\,,
\end{equation}
which leads to the clipped $\xi_{i, +}$ and $\xi_{i.-}$ as  
\begin{equation}\label{1D-clipped slacks}
	\begin{aligned}
	\clip{\xi}_{i, +} &= \max \left\{0, \innerprod{\mathbf{y} , e_i} - \Big[\innerprod{e_i, \mathbf{y}} -\nabla W_{\alpha_i}(\alpha, \beta) - \frac{\innerprod{\alpha, e_i}}{2C\tau}\Big]_{-M}^M\right\}\,,\\
	\clip{\xi}_{i, -} &= \max \left\{0, \Big[\innerprod{e_i, \mathbf{y}} -\nabla W_{\alpha_i}(\alpha, \beta) - \frac{\innerprod{\alpha, e_i}}{2C\tau}\Big]_{-M}^M - \innerprod{\mathbf{y}, e_i} \right\}\,. 
	\end{aligned}
\end{equation}
We further define
\begin{equation*}
 \clip{E}(\alpha,  \beta):=\tau \sum_{i=1}^n \clip{\xi}_{i, +}^2 +  (1-\tau)\sum_{i=1}^n \clip{\xi}_{i, -}^2\,.
\end{equation*}
Then we see that (\ref{1D-clipped RERM}) is satisfied if
\begin{equation}\label{1D-duality gap-clipped}
\clip{S}(\alpha,  \beta) := T(\alpha,  \beta) + C \cdot \clip{E}(\alpha,  \beta) \leq \frac{\epsilon}{2 \lambda}\,.
\end{equation}
The \textit{clipped} slack variables used in the stopping criteria (\ref{1D-duality gap-clipped}) may provide a substantial
decrease in duality gap in each iteration of learning algorithm compared to the unclipped slack variables used in
(\ref{1D-duality gap-general}), and hence the learning algorithm may require less number of iterations.
\cite{steinwart2006oracle} showed that the right hand side of the stopping criteria given in 
(\ref{1D-duality gap-general}) should be replaced by $\frac{\epsilon}{2\lambda}$  as in (\ref{1D-duality gap-clipped}), 
where $\epsilon$ has the same value for both.  Furthermore, it is argued by \cite{steinwart2011training} that unlike the 
duality gap stopping criteria for SVM \textit{with offset} given by \cite[p. 109f]{cristianini2000introduction}, 
both (\ref{1D-duality gap-general}) and (\ref{1D-duality gap-clipped}) are directly computable since they do not require 
the offset term. From this it is easy to derive an $O(n)$ procedure that updates 
$\nabla W_{\alpha}(\alpha, \beta)$, $\nabla W_{\beta}$ $(\alpha, \beta)$ and calculate $S(\alpha, \beta)$. 
The pseudocode for this is presented in Procedure \ref{procedure-duality gap comput.}. The one for $\clip{S}(\alpha, \beta)$ is an obvious
modifications and therefore omitted.

\begin{algorithm}[H]
\renewcommand\thealgorithm{2}
\floatname{algorithm}{Procedure}
\renewcommand{\algorithmicrequire}{\textbf{Input:}}
\renewcommand{\algorithmicensure}{\textbf{Output:}}

\caption{\small Update $\nabla W_{\alpha_i}(\alpha , \beta)$ and $\nabla W_{\beta_i}(\alpha , \beta) $ in direction $i^*$ and calculate $S(\alpha, \beta)$}
\begin{algorithmic}\label{procedure-duality gap comput.}
{\fontsize{10}{15}
\STATE $T(\alpha, \beta) \leftarrow  T(\alpha, \beta)-U(\alpha_i, \beta_i,  \delta,  \eta)$
\STATE 	$E(\alpha, \beta)\leftarrow 0$
	\FOR{$k=1 \,\text{to}\,\, n$}
		\STATE $\nabla W_{\alpha_k}(\alpha , \beta)\leftarrow \nabla W_{\alpha_k}(\alpha, \beta) - (\delta-\eta)K_{ik}-\frac{\delta^*}{2C\tau}\delta_{ik}$
		\STATE $\nabla W_{\beta_k}(\alpha , \beta)\leftarrow \nabla W_{\beta_k}(\alpha, \beta) + 	(\delta-\eta)K_{ik}-\frac{\eta^*}{2C(1-\tau)}\delta_{ik}$
		\STATE $\xi_{k,+}\leftarrow \max \{ 0,\nabla W_{\alpha_k}(\alpha , \beta)+ \frac{\alpha_k}{2C\tau}\}$
		\STATE $\xi_{k,-}\leftarrow \max \{ 0,-\nabla W_{\alpha_k}(\alpha , \beta)- \frac{\alpha_k}{2C\tau}\}$
		\STATE  $E(\alpha, \beta)\leftarrow E(\alpha, \beta) +\big(\tau \xi_{k,+}^{2} + (1-\tau) \xi_{k,-}^{2}\big) $
	\ENDFOR
	\STATE $S(\alpha, \beta)= T(\alpha,\beta)+C \cdot E(\alpha,\beta)$     
}
\end{algorithmic}
\end{algorithm}
With all the above computation, we now summarize the basic idea of the 1D-SVM in Algorithm 1. This tells us
to look repeatedly for the best direction $i^*$ and performs update in that direction until the predefined stopping
criteria is satisfied.

\begin{algorithm}[H]
\renewcommand\thealgorithm{1}
\floatname{algorithm}{Algorithm}
\renewcommand{\algorithmicrequire}{\textbf{Input:}}
\renewcommand{\algorithmicensure}{\textbf{Output:}}
{\fontsize{10}{15}

\caption{ 1D-SVM solver }
\begin{algorithmic}
\STATE initialize $\alpha, \beta,\nabla W_{\alpha}(\alpha , \beta),\nabla W_{\beta}(\alpha , \beta)$ and $T(\alpha, \beta)$  
	\WHILE{$S(\alpha, \beta)> \frac{\varepsilon}{ 2 \lambda}$}
		\STATE $(i^*, \delta_{i^*}, \eta_{i^*}) \leftarrow$ Procedure 1
		\STATE $ \alpha_{i^*} \leftarrow \alpha_{i^*}+\delta_{i^*}$
		\STATE $\beta_{i^*} \leftarrow \beta_{i^*}+\eta_{i^*}$ 
		\STATE use Procedure $2$ to update$ \nabla W_{\alpha}(\alpha , \beta),\nabla W_{\beta}(\alpha , \beta)$ in direction $i^*$ by $\delta_{i^*}$ and $ \eta_{i^*}$ and calculate $S(\alpha, \beta)$  
	\ENDWHILE
\end{algorithmic}
}
\end{algorithm}
A closer look of the Algorithm 1 reveals that there is still need to develop some procedures to initialize 
$\alpha$ and $\beta$, and the corresponding gradients. The following section presents some initialization methods to fulfill this requirement.

\subsection{Initialization} 

Various approaches are available to initialize $\alpha$ and $\beta$ and their corresponding gradients. 
We here briefly describe two approaches, namely, \textit{cold start} and \textit{warm start} that will be used in the 
implementation of the solver.

\textit{\textsf{I0 $\&$ W0}: Cold Start With Zeros}.  This is the most simplest initialization in which we take 
$\alpha \leftarrow 0$ and $\beta \leftarrow 0$ to initialize. After a simple calculation, it is not hard to initialize 
the corresponding gradients and the duality gap.

\textit{\textsf{W1}: Warm Start by Recycling Old Solution}. Recall that typically the hyper-parameter $\lambda$ is chosen 
by a search over a grid $\varLambda = \{\lambda_1, \ldots, \lambda_m\}$ of candidates values. If these values are ordered 
in the form  $\lambda_1> \ldots> \lambda_m$ and the SVM  is trained in this order, then the resulting $C^{(1)}, \ldots, C^{(m)}$
satisfy the property that  $C^{(j)} < C^{(j+1)}$ for all $j=1,\ldots, m-1$. For $C^{(1)}$ we initialize the solver with the
above cold start  and for $j\geq 2$, we initialize it with a warm start 
$\alpha \leftarrow \alpha^*$ and $\beta \leftarrow \beta^*$ where $\alpha^*, \beta^*$ is the approximate solution obtained 
by training with $C^{old}= C^{j-1}$. Obviously, in this case, we can also recycle parts of 
$\nabla_{\alpha}(\alpha, \beta)$, $\nabla_{\beta}(\alpha, \beta)$ and $S(\alpha, \beta)$ such as described in the 
Procedure \ref{1D-procedure-W1}.

\begin{algorithm}[H]
\renewcommand\thealgorithm{3}
\floatname{algorithm}{Procedure}
\renewcommand{\algorithmicrequire}{\textbf{Input:}}
\renewcommand{\algorithmicensure}{\textbf{Output:}}

\caption{\small Initialize by $\alpha \leftarrow \alpha^*$, $\beta \leftarrow \beta^*$, compute gradients and dual gap }
{\fontsize{10}{15}

\begin{algorithmic}\label{1D-procedure-W1}

\STATE $E(\alpha, \beta)\leftarrow 0$\\
	\FOR{ $i=1$ to $n$}
		\STATE $\alpha_i \leftarrow \alpha_i^*$
		\STATE $\beta_i \leftarrow \beta_i^*$
		\STATE $\nabla_{\alpha_i}(\alpha, \beta)\leftarrow \nabla_{\alpha_i}(\alpha^*, \beta^*) + \frac{\alpha_i^*}{2\tau}\left(\frac{1}{C^{\text{old}}}-\frac{1}{C^{\text{new}}}\right)$
		\STATE $\nabla_{\beta_i}(\alpha, \beta)\leftarrow \nabla_{\beta_i}(\alpha^*, \beta^*) + \frac{\beta_i^*}{2(1-\tau)}\left(\frac{1}{C^{\text{old}}}-\frac{1}{C^{\text{new}}}\right)$
		\STATE $\xi_{i,+} \leftarrow \max\left(0,\nabla_{\alpha_i}(\alpha, \beta) + \frac{\alpha_i}{2\tau C^{\text{new}}}\right)$
		\STATE $\xi_{i,-} \leftarrow \max\left(0,-\nabla_{\alpha_i}(\alpha, \beta) - \frac{\alpha_i}{2\tau C^{\text{new}}}\right)$
		\STATE $E(\alpha,\beta) \leftarrow E(\alpha, \beta) + \left(\tau \xi_{i,+}^2 + (1-\tau)\xi_{i,-}^2\right)$
	\ENDFOR 
	\STATE $T(\alpha, \beta) \leftarrow T(\alpha, \beta)-\frac{1}{4}\left(\frac{1}{C^{\text{old}}}-\frac{1}{C^{\text{new}}}\right)\sum_{i=1}^n \left(\frac{\alpha_i^2}{\tau}+\frac{\beta_i^2}{1-\tau}\right)$
	\STATE $S(\alpha, \beta)\leftarrow T(\alpha, \beta) + C^{\text{new}} E(\alpha, \beta)$
\end{algorithmic}
}
\end{algorithm}

\section{Working Set of Size Two}

The Algorithm 1 performs an update for one coordinate per iteration. In this section, 
we extend this idea and develop an algorithm to perform an update for \textit{two} coordinates per iteration. 
For this, we first solve the 2D- problem exactly in the following section. Then we will describe a low cost working set selection
strategy based on the 1D-SVM solver. In the end, we establish a stopping criteria for the 2D-problem.

\subsection{Exact Solution of Two Dimensional Problem}

Let us fix two coordinates $i,j \in \{1,\ldots,n\}$ with $i \neq j$. We further assume that $e_i$ and $e_j$ are the $i$-th 
and $j$-th vectors of standard basis of $\mathbb{R}^n$, and write $\alpha^{\setminus i,j}:= \alpha - \alpha_i e_i - \alpha_j e_j$ 
and $\beta^{\setminus i,j}:= \beta - \beta_i e_i - \beta_j e_j$. By this and using $K_{ii}= K_{jj}=1$ for normalized kernels,
the dual objective function for 2D-problem is
\begin{equation}\label{2D-Dual Objective Function}
	\begin{aligned}
	\tilde{W} &:= W(\alpha^{\setminus i,j} + \alpha_i e_i + \alpha_j e_j,\beta^{\setminus i,j} + \beta_i e_i + \beta_j e_j)\, \\
		 &\,\, = W(\alpha^{\setminus i,j},\beta^{\setminus i,j}) + W(\alpha_i, \beta_i) + W(\alpha_j, \beta_j) -(\alpha_i - \beta_i)(\alpha_j - \beta_j)K_{ij}\,,
	\end{aligned}
\end{equation}
where
\begin{equation*}
	\begin{aligned}
	W(\alpha_i, \beta_i) &:= (\alpha_i - \beta_i) \innerprod{e_i, \mathbf{y}} - (\alpha_i-\beta_i)\innerprod{e_i, K(\alpha^{\setminus i,j}-\beta^{\setminus i,j})}- \frac{1}{2}(\alpha_i - \beta_i)^2 \\
		& \quad- \frac{1}{4C\tau(1-\tau)}((1-\tau)\alpha_i^2 + \tau \beta_i^2)\,,\\
	W(\alpha_j, \beta_j) &:= (\alpha_j - \beta_j) \innerprod{e_j, \mathbf{y}} - (\alpha_j-\beta_j)\innerprod{e_j, K(\alpha^{\setminus i,j}-\beta^{\setminus i,j})}- \frac{1}{2}(\alpha_j - \beta_j)^2 \\
		&\quad - \frac{1}{4C\tau(1-\tau)}((1-\tau)\alpha_j^2 + \tau \beta_j^2)\,.
\end{aligned}
\end{equation*}
Taking partial derivatives of (\ref{2D-Dual Objective Function}) w.r.t.
$\alpha_i, \alpha_j, \beta_i$ and $ \beta_j$, we obtain the gradients
\begin{equation}\label{2D-gradients}
	\begin{aligned}
	\nabla \tilde{W}_{\alpha_i} &= \innerprod{e_i, \mathbf{y}} -\innerprod{e_i, K(\alpha^{\setminus i,j}-\beta^{\setminus i,j})} -b_1 \alpha_i + \beta_i - (\alpha_j - \beta_j)K_{i,j}\, ,\\ 
	\nabla \tilde{W}_{\beta_i} &= -\innerprod{e_i, \mathbf{y}} +\innerprod{e_i, K(\alpha^{\setminus i,j}-\beta^{\setminus i,j})} +\alpha_i -b_2 \beta_i + (\alpha_j - \beta_j)K_{i,j} \, ,\\ 
	\nabla \tilde{W}_{\alpha_j} &= \innerprod{e_j, \mathbf{y}} -\innerprod{e_j, K(\alpha^{\setminus i,j}-\beta^{\setminus i,j})} -b_1 \alpha_j + \beta_j - (\alpha_i - \beta_i)K_{i,j}\,  ,\\ 
	\nabla \tilde{W}_{\beta_j} &= -\innerprod{e_j, \mathbf{y}} +\innerprod{e_j, K(\alpha^{\setminus i,j}-\beta^{\setminus i,j})} +\alpha_j - b_2\beta_j + (\alpha_i - \beta_i)K_{i,j}\,  ,
	\end{aligned}
\end{equation}
where $b_1, b_2$ are defined in (\ref{1D-coefficients-global system equations}). By setting partial derivatives 
(\ref{2D-gradients}) to zero, we obtain the following system of equations
\begin{equation}\label{2D-set of equation after setting derivatives to zero}
	\begin{aligned}
	b_1 \alpha_i - \beta_i + k \alpha_j - k \beta_j &= c_i\, ,\\
	\alpha_i - b_2 \beta_i + k \alpha_j - k \beta_j &= c_i\, ,\\
	k \alpha_i - k \beta_i + b_1 \alpha_j - \beta_j &= c_j\, ,\\
	k \alpha_i - k \beta_i + \alpha_j - b_2 \beta_j &= c_j\, ,
	\end{aligned}
\end{equation}
where
\begin{equation*}
	\begin{aligned}
	k &:= K_{ij}\, ,\\
	c_i  & :=\innerprod{e_i, \mathbf{y}} - \innerprod{e_i,  K (\alpha^{\setminus i,j} - \beta^{\setminus i,j})}\, ,\\
		& \,\,=  \nabla W_{\alpha_i}(\alpha, \beta) + b_1 \innerprod{\alpha, e_i} -\innerprod{\beta, e_i}+ \innerprod{\alpha-\beta ,e_j} k \, ,\\
	c_i  & :=\innerprod{e_j, \mathbf{y}} - \innerprod{e_j,  K (\alpha^{\setminus i,j} - \beta^{\setminus i,j})} \\
		&\,\, = \nabla W_{\alpha_j}(\alpha, \beta) + b_1 \innerprod{\alpha, e_j}-\innerprod{\beta, e_j} + \innerprod{\alpha-\beta, e_i} k\,. 
	\end{aligned}
\end{equation*}
Let $\alpha_i^*, \alpha_j^*, \beta_i^*$ and $\beta_j^*$ be the solution of 
(\ref{2D-set of equation after setting derivatives to zero}). 
Then solving (\ref{2D-set of equation after setting derivatives to zero})
by matrix operations leads to the following global solution 
\begin{equation}\label{2D-global solution}
	\begin{aligned}
	\begin{vmatrix} M \end{vmatrix} \alpha_i^* & = (b_2-1)(b_1 b_2 -1) c_i + (1-b_2)(b_1+b_2-2)kc_j\, ,\\
	\begin{vmatrix} M \end{vmatrix} \beta_i^* & =(1-b_1)(b_1 b_2 -1) c_i + (b_1-1)(b_1+b_2-2)kc_j\, ,\\
	\begin{vmatrix} M \end{vmatrix} \alpha_j^* & =(b_2-1)(b_1b_2-1)c_j+(1-b_2)(b_1+ b_2 -2)k c_i\, , \\
	\begin{vmatrix} M \end{vmatrix} \beta_j^* & = (1-b_1)(b_1b_2-1)c_j + (b_1-1)(b_1+ b_2 -2) kc_i\, .
	\end{aligned}
\end{equation}
Here
\begin{eqnarray*}
 \begin{vmatrix} M \end{vmatrix} := b_1^2(b_2^2-k^2)-2 b_1 (b_2 k^2+b2-2k^2)-(b_2-2)^2 k^2 +1\, , 
\end{eqnarray*}
 is always positive. This is shown in the following lemma
\begin{Lemma}\label{2D-lemma-positivity of determinant}
 For $b_1,  b_2 \in [1, \infty)$ and $\begin{vmatrix} k \end{vmatrix} \leq 1$, we have $\begin{vmatrix} M \end{vmatrix} > 0$.  
\end{Lemma}
Note that, in the case of  $c_i = c_j = 0$, we have $\alpha_i^* = \beta_i^*=\alpha_j^* = \beta_j^*=0$. On the other hand, 
if $c_i \neq 0$ or $c_j \neq 0$,  then (\ref{2D-global solution}) together with 
Lemma \ref{2D-lemma-positivity of determinant} leads, after some calculations, to the following equations
\begin{equation*}
	\begin{aligned}
	\alpha_i^* &=  - \frac{\tau}{1-\tau} \beta_i^*\,, \hspace*{4ex}	\alpha_j^* &=  - \frac{\tau}{1-\tau} \beta_j^*\, .
	\end{aligned}
\end{equation*}
Since $\tau \in (0,1)$, the global solution (\ref{2D-global solution}) thus violates the constraints of (\ref{Dual Objective Function}) iff 
$c_i \neq 0$ or $c_j \neq 0$, that is, the solution is not feasible. To obtain the feasible solution, we know by 
the Theorem \ref{1D-optimum solution at boundary} that we need to look at the boundaries of the feasible region. 
In our case, this means that we need to set some of the dual variables to zero. Note that this is a simple extension of 
the idea that is presented in 1D-problem. Let us begin by setting one dual variable to zero, say $\alpha_i = 0$. 
Computing the gradients with the remaining variables, we get the last three expressions of (\ref{2D-gradients}) where 
we set $\alpha_i = 0$. After setting the gradients to zero, we obtain the system of equations
\begin{equation}\label{2D-alpha = 0-set of equations}
	\begin{aligned}
	-b_2 \beta_i + k \alpha_j -k \beta_j &= c_i\, ,\\
	-k \beta_i + b_1 \alpha_j - \beta_j &= c_j\, ,\\
	-k \beta_i + \alpha_j - b_2 \beta_j &= c_j\, ,
	\end{aligned}
\end{equation}
where $k, c_i, c_j, b_1$ and $b_2$ are the same as in (\ref{2D-set of equation after setting derivatives to zero}). 
Let us write $\alpha_j^+,  \beta_i^+$ and $\beta_j^+ $ be the solution of (\ref{2D-alpha = 0-set of equations}). Then,
by subtracting the last two equations of (\ref{2D-alpha = 0-set of equations}),  we obtain
\begin{equation}\label{2D-infeasible solution when alpha_i = 0}
 \alpha_j^+ = -\frac{\tau}{1-\tau}\beta_j^+\, ,
\end{equation}
and hence this solution is again not feasible. In a similar way, setting $\beta_i = 0$ provides the following 
system of equations
\begin{equation*}\label{2D-beta_i = 0-set of equations}
	\begin{aligned}
	b_1 \alpha_i + k \alpha_j -k \beta_j &= c_i\, ,\\
	k \alpha_i + b_1 \alpha_j - \beta_j &= c_j\, ,\\
	k \alpha_i + \alpha_j - b_2 \beta_j &= c_j\, ,
\end{aligned}
\end{equation*}
which again leads to (\ref{2D-infeasible solution when alpha_i = 0}) and thus the same conclusion. 
The remaining two cases where $\alpha_j = 0$ and $\beta_j = 0$ can be treated analogously. After this, we now consider 
the situation where two variables are set to zero. For this, we split the problem into six subcases. Let us consider
the first subcase where we set $\alpha_i=0$ and $\beta_i=0$ in (\ref{2D-Dual Objective Function}). Taking derivatives 
w.r.t. $\alpha_j$ and $\beta_j$ provides
\begin{equation}\label{2D-part1-case1-gradients-alpha_i=0, beta_i=0}
 \begin{aligned}
	\nabla W_{\alpha_j}(\alpha^{\setminus i}, \beta^{\setminus i}) & = \innerprod{e_j, \mathbf{y}} - \innerprod{e_j, K(\alpha^{\setminus i,j}-\beta^{\setminus i,j})}+\beta_j -b_1 \alpha_j\, ,\\
	\nabla W_{\beta_j}(\alpha^{\setminus i}, \beta^{\setminus i}) & = -\innerprod{e_j, \mathbf{y}} + \innerprod{e_j, K(\alpha^{\setminus i,j}-\beta^{\setminus i,j})}+\alpha_j -b_2 \beta_j\, .
 \end{aligned}
\end{equation}
Setting (\ref{2D-part1-case1-gradients-alpha_i=0, beta_i=0}) to zero, we obtain the system of equations
\begin{equation}\label{2D-part1-case1-system of equation-alpha_i=0 and beta_i=0}
 \begin{aligned}
b_1 \alpha_{j} - \beta_{j} &= c_j\, ,\\
\alpha_{j} - b_2 \beta_{j} &=c_j\, .
 \end{aligned}
\end{equation}
Let $\alpha_{j}^{+}$ and $\beta_{j}^{+}$ be the solution of (\ref{2D-part1-case1-system of equation-alpha_i=0 and beta_i=0}). 
Then subtracting equations of (\ref{2D-part1-case1-system of equation-alpha_i=0 and beta_i=0}) leads to
\begin{equation}
 \alpha_{j}^{+} = -\frac{\tau}{1-\tau}\beta_{j}^{+},
\end{equation}
which shows that the solution is not feasible. Analogously, the second subcase where $\alpha_j=0$ and $\beta_j = 0$ leads to
the same conclusion. In the third subcase, we set $\alpha_i = 0$ and $\alpha_j = 0$ in (\ref{2D-Dual Objective Function})
and differentiate w.r.t. $\beta_i$ and $\beta_j$  which gives
\begin{equation}\label{2D-optimum solution-derivaties- alpha_i = 0 and alpha_j = 0}
	\begin{aligned}
	\nabla W_{\beta_i}(\alpha^{\setminus i,j}, \mathbf{\beta}) & = -\innerprod{e_i, \mathbf{y}}+\innerprod{e_i, K(\alpha^{\setminus i,j}-\beta^{\setminus i,j})}-\beta_j K_{ij}  -b_2 \beta_i\, ,\\
	\nabla W_{\beta_j}(\alpha^{\setminus i,j}, \mathbf{\beta}) & = -\innerprod{e_j, \mathbf{y}}+\innerprod{e_j, K(\alpha^{\setminus i,j}-\beta^{\setminus i,j})}-\beta_i K_{ij}  -b_2 \beta_j\, .
	\end{aligned}
\end{equation}
Setting (\ref{2D-optimum solution-derivaties- alpha_i = 0 and alpha_j = 0}) to zero, we obtain a system of equations which,
after some calculations, provides the solution
\begin{equation}\label{2D_Optimum Solution-Case 1}
	\alpha_{i}^+=0\,, \hspace*{3ex} 
	\alpha_{j}^+=0\,, \hspace*{3ex} 
	\beta_{i}^+ =\begin{vmatrix}B_1 \end{vmatrix}^{-1}(kc_j-b_2c_i)\,, \hspace*{3ex} 
	\beta_{j}^+ =\begin{vmatrix}B_1 \end{vmatrix}^{-1} (kc_i-b_2c_j),
\end{equation}
where $\begin{vmatrix}B_1 \end{vmatrix}:= b_2^2 - k^2 > 0$. Considering the forth subcase, we set $\beta_i = 0$ and 
$\beta_j = 0$. Analogous to third subcase, the gradients are
\begin{equation*}
	\begin{aligned}
	\nabla W_{\alpha_i}(\alpha, \beta^{\setminus i,j}) & = \innerprod{e_i, \mathbf{y}}-\innerprod{e_i, K(\alpha^{\setminus i,j}-\beta^{\setminus i,j})} -\alpha_j K_{ij} -b_1 \alpha_i\, ,\\
	\nabla W_{\alpha_j}(\alpha, \beta^{\setminus i,j}) & = \innerprod{e_j, \mathbf{y}}-\innerprod{e_j, K(\alpha^{\setminus i,j}-\beta^{\setminus i,j})} -\alpha_i K_{ij}  -b_1 \alpha_j\, ,
	\end{aligned}
\end{equation*}
which leads to the solution
\begin{equation}\label{2D_Optimum Solution-Case 2}
	\beta_{i}^+ = 0\, ,\hspace*{3ex} 
	\beta_{j}^+ = 0\,, \hspace*{3ex} 
	\alpha_{i}^+ = \begin{vmatrix}B_2 \end{vmatrix}^{-1} (b_1 c_i-k c_j)\,, \hspace*{3ex} 
	\alpha_{j}^+ = \begin{vmatrix}B_2 \end{vmatrix}^{-1} (b_1 c_j-k c_i)\,, 
\end{equation}
where $\begin{vmatrix}B_2 \end{vmatrix}:= b_1^2 - k^2 > 0$. For fifth subcase, we set $\alpha_i =0$ and $\beta_j =0$ and
obtain the following solution
\begin{equation}\label{2D_Optimum Solution-Case 3}
	\alpha_{i}^+ = 0\,, \hspace*{3ex} 
	\beta_{j}^+ = 0\,, \hspace*{3ex} 
	\beta_{i}^+  = \begin{vmatrix}B_3  \end{vmatrix}^{-1} (b_1 c_i-k c_j)\,, \hspace*{3ex} 
	\alpha_{j}^+ = \begin{vmatrix}B_3  \end{vmatrix}^{-1}(kc_i-b_2 c_j)\,,
\end{equation}
where $\begin{vmatrix} B_3  \end{vmatrix} := k^2-b_1 b_2 < 0$. Finally, for the last subcase where $\alpha_j = 0$ and 
$\beta_i = 0$, the  solution can be obtained by interchanging $i$ with $j$ in the solution of fifth subcase, which is
\begin{equation}\label{2D_Optimum Solution-Case 4}
	\begin{aligned}
	\beta_{i}^+ = 0\,,   \hspace*{3ex} 
	\alpha_{j}^+ = 0\,,  \hspace*{3ex} 
	\alpha_{i}^+ =\begin{vmatrix}B_3  \end{vmatrix}^{-1}(kc_j-b_2 c_i)\,,  \hspace*{3ex} 
	\beta_{j}^+  = \begin{vmatrix}B_3  \end{vmatrix}^{-1} (b_1 c_j-k c_i)\,.
	\end{aligned}
\end{equation}
It is interesting to note that the solutions (\ref{2D_Optimum Solution-Case 1}), (\ref{2D_Optimum Solution-Case 2}), 
(\ref{2D_Optimum Solution-Case 3}) and (\ref{2D_Optimum Solution-Case 4}) have the following common expressions 
\begin{align}
	T_1 &:=kc_j-b_2c_i\, , \label{2D-first common equation}\\
	T_2 &:=kc_i-b_2c_j\, , \label{2D-second common equation}\\
	T_3 &:=b_1 c_i-k c_j\, , \label{2D-third common equation}\\
	T_4 &:=b_1 c_j-k c_i\, . \label{2D-forth common equation}
\end{align}
The following lemma investigates the behavior of the above four expressions. 

\begin{Lemma}\label{2D-Lemma-simaltenously positivity of 4 equations}
Assume that $c_i \neq 0$ or $c_j \neq 0$. Then the following implications hold:
	\begin{itemize}
	\item[i)] If $T_1 \geq 0$ and $T_2 \geq 0$ then we have $c_i < 0$ and $c_j < 0$.
	\item[ii)] If $T_3 \geq 0$ and $T_4 \geq 0$ then we have $c_i > 0$ and $c_j > 0$.
	\end{itemize}
In particular,  the expressions $T_1,  T_2,  T_3$ and $T_4$ are not simultaneously positive or negative.
\end{Lemma}

Using Lemma \ref{2D-Lemma-simaltenously positivity of 4 equations}, 
the following theorem shows that only one case from (\ref{2D_Optimum Solution-Case 1}), (\ref{2D_Optimum Solution-Case 2}),
(\ref{2D_Optimum Solution-Case 3}) and (\ref{2D_Optimum Solution-Case 4}) provides the feasible optimal solution.

\begin{Theorem}\label{2D-Theorem- one feasible optimum solution}
Assume that $c_i \neq 0$ or $c_j \neq 0$,  then exactly one of the four cases (\ref{2D_Optimum Solution-Case 1}),
(\ref{2D_Optimum Solution-Case 2}),  (\ref{2D_Optimum Solution-Case 3}) and (\ref{2D_Optimum Solution-Case 4}) produces
a feasible solution. Moreover,  the following implications hold:
	\begin{itemize}
	\item[i)] If $T_1 \geq 0$ and $T_2 \geq 0$,  then (\ref{2D_Optimum Solution-Case 1}) is the feasible solution.
	\item[ii)] If $T_3 \geq 0$ and $T_4 \geq 0$,  then (\ref{2D_Optimum Solution-Case 2}) is the feasible solution.
	\item[iii)] If $T_1 \leq 0$ and $T_3 \leq 0$,  then (\ref{2D_Optimum Solution-Case 3}) is the feasible solution.
	\item[iv)] If $T_2 \leq 0$ and $T_4 \leq 0$,  then (\ref{2D_Optimum Solution-Case 4}) is the feasible solution.
	\end{itemize}
\end{Theorem}

Theorem \ref{2D-Theorem- one feasible optimum solution} also suggests to impose \textit{if} conditions based on 
expressions (\ref{2D-first common equation}), (\ref{2D-second common equation}), (\ref{2D-third common equation}) and
(\ref{2D-forth common equation}) in the implementation of the algorithm for 2D-SVM solver. This helps to reach directly
to the feasible optimum solution.

\subsection{Working Set Selection Strategies}

In this section, we address the question how to choose the directions $i^*$ and $j^*$ in which 
the 2D-SVM solver performs an update. Several possibilities are available for this task. A straightforward approach is
to consider all pairs of directions $(i,j)$ and choose the one for which the 2D-gain of $W$ is maximum.
Note that the 2D-gain is simply an extension of the idea presented in Lemma \ref{1D-gain}. Formally, for 
$\alpha,\beta\in \mathbb{R}^n$ and $\delta, \eta \in \mathbb{R}$, the 2D-gain is 
\begin{equation}\label{2D-gain}
 W(\alpha + \delta e_i + \delta e_j, \beta + \eta e_i + \eta e_j)-W(\alpha, \beta)= 
 G(\delta_i, \eta_i) + G(\delta_j, \eta_j) - (\delta_i - \eta_i)(\delta_j - \eta_j) K_{i,j}\, , 
\end{equation}
where $G(\delta_k, \eta_k)$ for $k=i,j$  is the 1D-gain defined in Lemma \ref{1D-gain}.

It is worth noting that the above described working set selection strategy is not a good choice because the search is $O(n^2)$.
However it may be viewed as an "optimal" two dimensional strategy and served as a baseline to all other subset selection strategy
that can be interpreted as the low cost approximations to this approach. In the following, we describe two low cost working set selection 
strategies that we consider in this work.

\textit{\textsf{WSS 1}: Two 1D-direction With Maximal Gain From Separate Subsets.} A simple way to preserve the low cost
search from 1D-solver is to split the index set $\{1,\ldots,n\}$ into two parts $\{1,\ldots,\frac{n}{2}\}$ and 
$\{\frac{n}{2}+1,\ldots,n\}$ and search for the 1D directions with maximum gain over these two parts 
separately. In other words, we can choose the directions $i^*$ and $j^*$ by
\begin{equation}\label{2D-WSS-directions search}
	\begin{aligned}
	i^* &\in \arg\,\underset{i \leq n/2} \max W(\alpha + \delta e_i, \beta + \eta e_i)-W(\alpha, \beta)\, ,\\
	j^* &\in \arg\,\underset{i > n/2} \max W(\alpha + \delta e_i, \beta + \eta e_i)-W(\alpha, \beta)\, ,
	\end{aligned}
\end{equation}
where $\delta$ and $\eta$ are defined in 1D-SVM solver. These chosen directions are used for the first iteration. 
For the subsequent iterations, we search for the new 1D directions, $i_{\text{new}}^*$ and $j_{\text{new}}^*$, 
again by using (\ref{2D-WSS-directions search}). Then we compute the 2D-gain of $W$ for all pairs of old and new directions
of previous and current iterations respectively and look for a pair for which this gain is maximum. 

\textit{\textsf{WSS 2:} 1D-direction With Maximal Gain And A Direction Of A Nearby Sample.}
This is simply an extension of \textit{\textsf{WSS 1}}. After determining $(i^*, j^*)$ by \textit{\textsf{WSS 1}}, we fix $i^*$ and then search for another
direction $j^*$ from $k$-nearest neighbors of $x_{i^*}$ with respect to the metric
\begin{equation*}
 d(x,x^{\prime}):=\snorm{x - x^{\prime}}^2\,.
\end{equation*}

\subsection{Stopping Criteria}\label{2D-section-gain and dual gap}
To formulate the stopping criteria for 2D-problem, we follow the idea that is presented in Section 3.1.
Let us first consider the component $T(\alpha,  \beta)$ of (\ref{1D-components of duality gap}) and by using 
(\ref{2D-gain}), we find the following update of $T(\alpha,  \beta)$ in the directions of $i$ and $j$
\begin{equation*}
\begin{aligned}
 T(\alpha+\delta e_i +\delta e_j, \beta + \eta e_i + \eta e_j) &= T(\alpha, \beta)-U(\alpha_i, \beta_i, \delta_i, \eta_i)-U(\alpha_j, \beta_j, \delta_j, \eta_j)\\
	 &\quad\,+ 2(\delta_i-\eta_i)(\delta_j-\eta_j)K_{i,j}\, ,
\end{aligned}
\end{equation*}
where $U(\alpha_k, \beta_k, \delta_k, \eta_k)$ for $k=i,j$ is defined in (\ref{1D-duality gap-update factor in T}). 
To compute $E(\alpha, \beta)$, we first obtain the updated gradients in 
the directions of $i$ and $j$,
and then subsequently compute $\xi_{l,+}, \xi_{l,-}$. Moreover,  $\clip{E}(\alpha,  \beta)$ can also be computed for 
the 2D-problem similar to 1D-problem by using 
(\ref{1D-clipped slacks}). With all above computations, 
we now summarize the idea of 2D-SVM solver
in Algorithm 2.

\begin{algorithm}[H]
\renewcommand\thealgorithm{2}
\floatname{algorithm}{Algorithm}
\renewcommand{\algorithmicrequire}{\textbf{Input:}}
\renewcommand{\algorithmicensure}{\textbf{Output:}}
{\fontsize{10}{15}

\caption{ 2D-SVM Solver }
\begin{algorithmic}
\STATE initialize $\alpha, \beta,\nabla W_{\alpha}(\alpha , \beta),\nabla W_{\beta}(\alpha , \beta)$ and $T(\alpha, \beta)$  
	\WHILE{$S(\alpha, \gamma)> \frac{\varepsilon}{ 2 \alpha}$}
		\STATE select directions $i^*$ and $j^*$
		\STATE use procedure $5$ to obtain the optimum solution for direction $i^*$ and $j^*$
		\STATE update $\alpha$ and $\beta$ in the direction $i^*$ and $j^*$
		\STATE update $\nabla W_{\alpha}(\alpha , \beta),\nabla W_{\beta}(\alpha , \beta)$ in the directions $(i^*, j^*)$  and calculate $S(\alpha, \beta)$ 
	\ENDWHILE
\end{algorithmic}
}
\end{algorithm}

\section{Experiments}

To evaluate the performance of the proposed solver for expectile regression, we perform several experiments to 
address the following questions:
\begin{itemize}
	
	\item[1.] Which subset selection strategy leads to the smallest number of iterations or shortest run time?
	\item[2.] What is the  number of nearest neighbors that leads to the smallest number of iterations and shortest run time?
	\item[3.] Is there advantage of warm start initialization when the parameter search is performed over a grid? 
	\item[4.] Does the clipping provide a significant reduction in the training time and iterations?
	\item[5.] How well does the 2D-SVM-solver work as compared to \textsf{ER-Boost} that is proposed by \cite{yang2014nonparametric}?
\end{itemize}

To answer these questions, we implemented the 2D-SVM-solver in C$^{++}$. The algorithm was compiled by LINUX's \textsf{gcc}
version 4.7.2 with various software and hardware optimization enabled. All experiments were conducted on a computer 
with INTEL CORE i7 950 (3.07 GHz) and 8GB RAM under 64bit version of Debian Linux 7.8 (Debian 3.2.0-4-amd64). 
During all experiments that incorporated measurement of run time, one core was used solely for the experiments, 
and the number of other processes running on the system were minimized.

In order to perform the experiments, we have considered nine data sets that were downloaded from different sources. 
These data sets comprises various number of features and vary in sample sizes from 630 to 20639. 
The data sets \textsc{concrete-comp}, \textsc{updrs-motor}, \textsc{cycle-pp}, \textsc{airfoil-noise} and \textsc{hour} 
were downloaded from UCI repository. The two data sets   \textsc{nc-crime} and \textsc{head-circum} are available 
and documented in R packages \textsf{Ecdat} and \textsf{AGD} respectively. The remaining two data sets  \textsc{cal-housing}
and \textsc{munich-rent} were downloaded from StatLib and  the data archive of the Institute of Statistics, 
Ludwig-Maximilians-University of Munich respectively. We scaled the data sets componentwise such that all the samples
including labels lie in $[-1,1]^{d+1}$, where $d$ is the dimension of the input data. In addition to that, 
we generated a random split for all data sets that contained approximately $70\%$ training and $30\%$ test samples. 
Table \ref{Table-characteristics of datasets} describes the characteristics of the considered data sets.

In all our experiments with the SVM solver, we used Gaussian kernels (\ref{Gaussian kernel}). To determine the hyper-parameters,
we have considered a geometrically spaced 10 by 10 grid for
$\lambda$ and $\gamma$ over the interval $[c_1 n^{-1},1]$ and $[c_2n^{-1/d}, c_3]$ respectively, 
where $n$ is the number of training samples, $d$ is the input dimension, and $c_1:=0.001, c_2:=0.1$ and $c_3:=0.2$.
Here, the values of the constants were chosen with the help of our experience with least square SVMS \cite{eberts2011optimal}. 
To choose the best values of these hyper-parameters, we used $k$-fold 
cross validation with randomly generated folds. In our case, we have considered $k=5$. During the $k$-fold cross validation,
the hyper-parameter $\gamma$ was internally converted to $\tilde{\gamma}:=\frac{(k-1)n\gamma}{k}$ and $\lambda$ to
$C:=\frac{k}{2(k-1)n \lambda}$, where $(k-1)n/k$ is approximate \textit{actual} training set size for $k$-fold 
cross validation.

\begin{table}[H]
\begin{center}
 
\begin{tabular}{lcccc}
 \hline
											\\	
data			&sample sizes	&training size	&test size	&dimension	 \\ \hline \hline	
\textsc{nc-crime}	&630		&441		&189		&19\\
\textsc{concrete-comp}	&1030		&721		&309		&8\\
\textsc{airfoil-noise}	&1503		&1052		&451		&5\\
\textsc{munich-rent}	&2053		&1437		&616		&12\\
\textsc{updrs-motor}	&5875		&4112		&1763		&19\\
\textsc{head-circum}	&7020		&4914		&2106		&4\\
\textsc{cycle-pp}	&9568		&6697		&2871		&5\\
\textsc{hour}		&17379		&12165		&5214		&12\\
\textsc{cal-housing}	&20639		&14447		&6192		&8\\\hline

 \end{tabular}
\end{center}

  \caption[format=hang]{Characteristics of data sets together with the training sizes and the test sizes that refer to the 
  splits used in the run time experiments.}
	\label{Table-characteristics of datasets}
 \end{table}
Let us now explore the answers of the above stated questions one by one. To address the first question, we
performed experiments with warm start initialization method and clipped duality gap. In addition, 
we have considered $N=15$ nearest neighbors for \textsf{WSS 2}. The results are
presented in Figure \ref{figure-time and ratio-WSS vs datasets} and \ref{figure-iter and ratio-WSS vs datasets}, which depict 
that \textsf{WSS 2} needs substantially less iterations as well as training time than \textsf{WSS 1} on all data sets. 
For larger data sets such as \textsc{updrs-motor}, \textsc{head-circum}, \textsc{cycle-pp}, \textsc{hours} and 
\textsc{cal-housing}, the run time and iterations with \textsf{WSS 2} is at least $50\%$ less than
\textsf{WSS 1}.
Moreover, a closer analysis, see Figure  \ref{figure-per grid time for WSS-cal-housing} and 
\ref{figure-per grid iter for WSS-cal-housing} shows that the savings are obtained at the hyper-parameters pairs for which
training is particularly expensive, that is, for small $\lambda$ and medium to small $\gamma$.

We have fixed $N=15$ for  \textsf{WSS 2} so far to address the previous question. To investigate
how the computational requirements change with the number of nearest neighbors, 
we performed the experiments for 
$N$-nearest neighbors, where $N= 5, 10, 15, 20, 25,$ $ 30, 35, 40$ for each $\tau=0.25, 0.50, 0.75$. 
Again we used warm start initialization and clipped duality gap. 
Here, it was observed that the number of iterations tends to decrease with increasing $N$.
However, for $N\geq 25$, only a slight improvement in the number of iterations 
was found whereas the required run time tended to increase compared to smaller $N$. We 
therefore plotted the results for $N=5,10,15,20$ only. Figure \ref{figure-time and ratio-NN vs datasets} shows that 
the solver attains the minimum training time for $N=15$ on almost all data sets. Moreover, 
Figure \ref{figure-iter and ratio-NN vs datasets} shows that the number of iterations decreases with 
increasing $N$. However, this decrease becomes negligible when $N \geq 15$. All this together leads us to conclude that $N=15$
is the best choice for our \textsc{er-svm} solver. 
Finally, Figure \ref{figure-per-grid time 
for NN-cal-housing} and \ref{figure-per-grid iter for NN-cal-housing} illustrate the computational requirements for 
different hyper-parameters pairs. Again the largest savings for $N=15$ were obtained for small $\lambda$.

To answer the third question regarding the initialization methods, 
we trained with $N=15$ and clipped duality gap.
The results, which are presented in Figure \ref{figure-time and ratio-initialization vs datasets}
and \ref{figure-iter and ratio-initialization vs datasets} show that using the warm start initialization saves between
$20\%$ and $40\%$ of both training time and iterations. The detailed behavior for different hyper-parameter pairs is 
illustrated in Figure \ref{figure-per-grid time for initialization-cal-housing} and 
\ref{figure-per-grid iter for initialization-cal-housing}. Again the savings are more pronounced for smaller $\lambda$.

To answer the forth question, we considered stopping criteria with clipped duality gap and with unclipped duality gap. 
Here, we used the warm start 
initialization option and \textsf{WSS 2} with $N=15$ nearest neighbors. The corresponding results are
shown in Figures \ref{figure-time and ratio-duality gap vs datasets} 
and \ref{figure-iter and ratio-duality gap vs datasets}. In the case of hinge loss function, \cite{steinwart2011training} showed
that using the clipped duality gap yields significant reduction, both in run times and iterations. 
However, in our case, 
we get only a small reduction in iterations, that is, $1\%$ on almost all data sets.
On the other hand, this stopping criteria causes $2\%$ to $17\%$ increase in run times on different data sets. This indicates that 
the  
unclipped duality gap is the better choice in our case.
The per grid plot of hyper-parameters for data set \textsc{cal-housing}, as presented in 
Figure \ref{figure-figure-per-grid time for duality gap-cal-housing} 
and \ref{figure-figure-per-grid iter for duality gap-cal-housing}, shows that clipping reduces the run time only for few pairs of 
hyper-parameter when $\lambda$ is small and $\gamma$ is large. For rest of the pairs, unclipped duality
gap leads to smaller run time.

Finally, we compare our SVM solver with \textsf{ER-Boost} on the basis of test error and 
training time.
For this, we considered our 2D-SVM solver with unclipped duality gap (\textsc{er-svm}), our 2D-SVM solver with clipped duality gap
($\textsc{er-svm}^*$) and \textsf{ER-Boost} \cite{yang2014nonparametric}. Since the experiments using large data sets entail
long run times, we splitted the data sets into three categories, namely, small ($n < 5000$), medium 
($5000 \leq n < 10000$) and large ($n \geq 10000$). We then conducted experiments for \textsc{er-svm},
$\textsc{er-svm}^*$ and  \textsf{ER-Boost} by repeating  5-fold cross validation 25, 10 and 5 times for the small, medium
and large data sets respectively. For the 2D-SVM solvers, we used the 10 by 10 default grid of hyper-parameters described above. 
For \textsf{ER-Boost}, we used the default value of boosting steps ($M=100$) and performed 5 fold cross validation to choose 
the best value of the interaction level (L) between variables, as by the \textsf{ER-Boost} manual.  The resulting, 
average test error (standard deviation) and 
training time are shown in 
Table \ref{Table-averge test error} and Table \ref{Table-run time} respectively. It turns out that both SVMs 
solvers have better test performance than \textsf{ER-Boost} on all data sets, but all reported errors are relatively small. 
Examining the achieved training times for each data set, we observe that SVM solvers are more sensitive to the training set 
size and less sensitive to the dimensions of data set, whereas, \textsf{ER-Boost} behaves the other way around. 
In addition to that, the test performance of $\textsc{er-svm}^*$ is slightly better than \textsc{er-svm} at 
the cost of almost $10\%$ longer training times.

\begin{figure}[!t]
\vspace{-0.65cm}
\centering
 \subfloat{\includegraphics[scale=0.38]{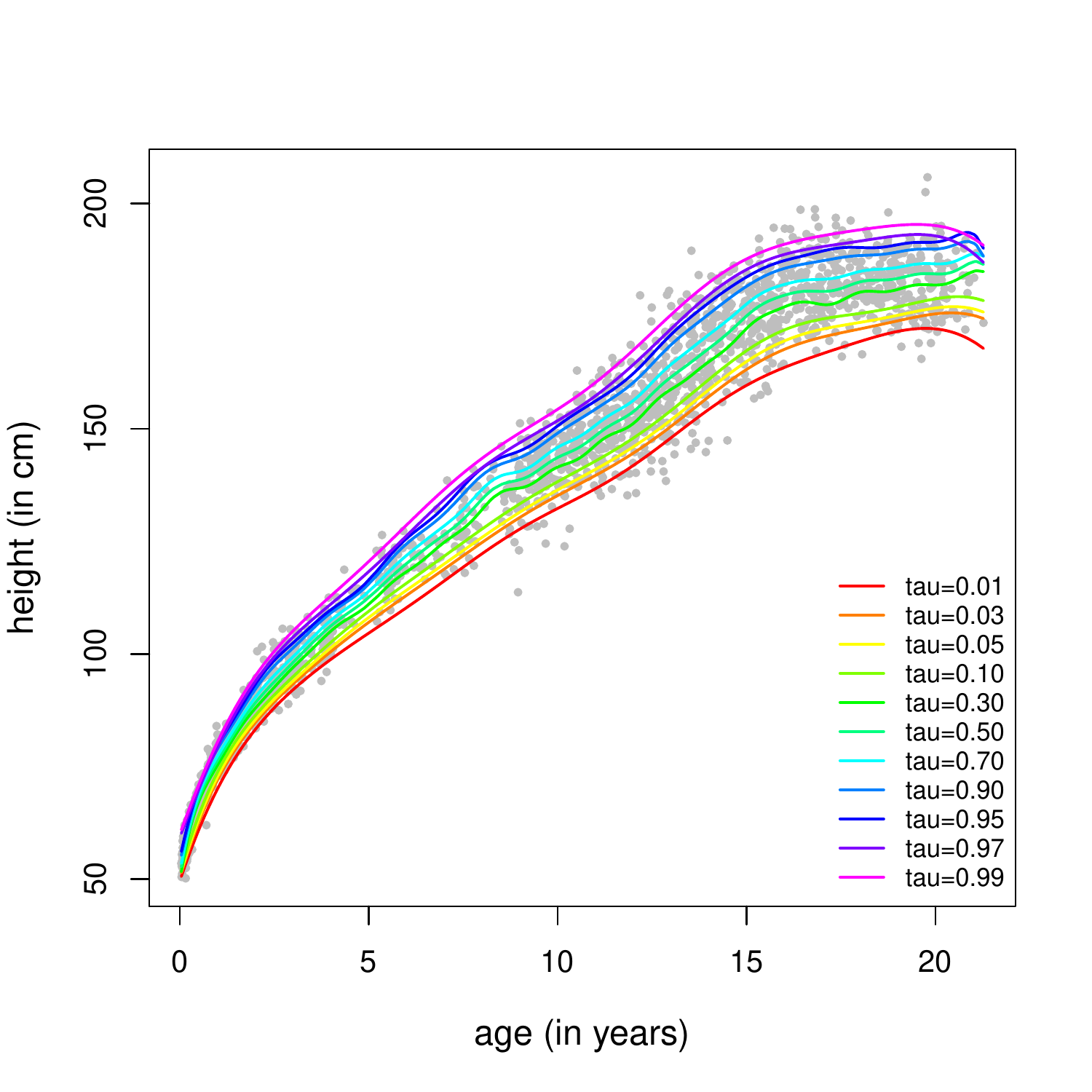}} \hspace{5ex}
 \subfloat{\includegraphics[scale=0.38]{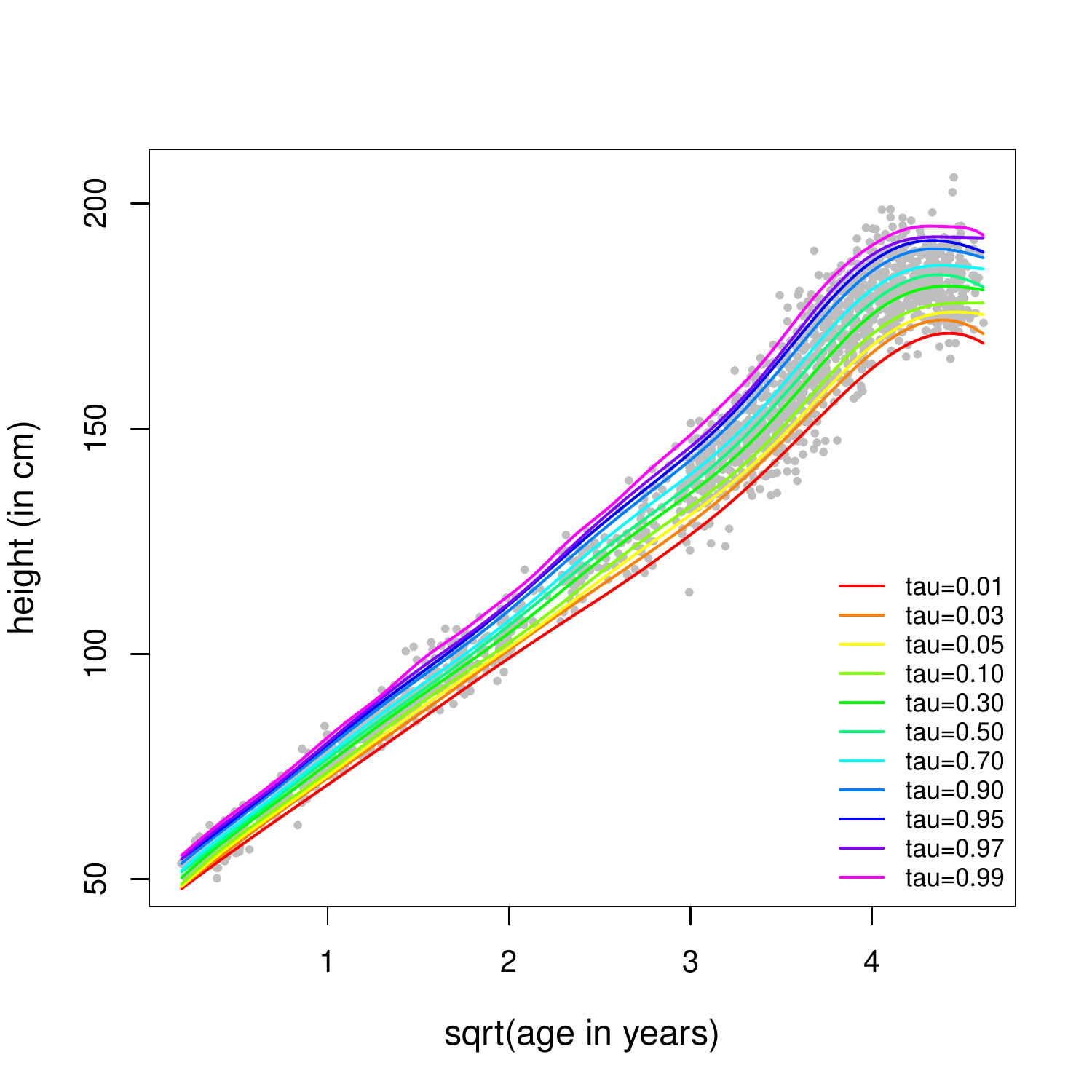}}\\

 \caption{\small{Estimated expectiles for $\tau=0.01, 0.03, 0.05, 0.10, 0.30, 0.50, 0.70, 0.90, 0.95, 0.97, 0.99$ for height
 against age of \textsc{head-circum}. The graphs comprises expectile curves for original data set (left) and 
 data set with transformed age.}
}
 \label{figure-set of expectiles}
\end{figure}

In the end, Figure \ref{figure-set of expectiles} presents the expectile curves for different $\tau$ considering height against 
age from data set \textsc{head-circum}. On the left we see some crossing and wiggling problems. 
Following \cite{schnabel2009optimal}, the use of square root transformation 
on age resolves these issues as the right figure shows.

\newpage 
\begin{landscape}
\begin{table}[H]
\footnotesize
\begin{center}
 
\begin{tabular}{|l|c|c|c|c|c|c|c|c|c|}

\hline
    \multirow{2}{*}{\textsc{data}}
    & \multicolumn{3}{|c|}{$\tau=0.25$}  & \multicolumn{3}{|c|}{$\tau=0.50$} & \multicolumn{3}{|c|}{$\tau=0.75$} \\
        \cline{2-10}
				
				&ER-SVM		&$\text{ER-SVM}^*$&ER-Boost	&ER-SVM		&$\text{ER-SVM}^*$& ER-Boost	&ER-SVM		&$\text{ER-SVM}^*$& ER-Boost\\ \hline 	
				
\textsc{nc-crime}		&0.00616	&0.00555	&0.00948	&0.00669	&0.00605	&0.01367	&0.00536	&0.00509	&0.01459		\\
				&(0.00182)	&(0.00169)	&(0.00177)	&(0.00194)	&(0.00161)	&(0.00305)	&(0.00172)	&(0.00157)	&(0.00405)		\\\hline
\textsc{concrete-comp}		&0.00901	&0.00893 	&0.03961	&0.01021	&0.01013	&0.05038	&0.00889	&0.00879	&0.04556		\\
				&(0.00130)	&(0.00128)	&(0.00365)	&(0.00122)	&(0.00117)	&(0.00417)	&(0.00112)	&(0.00101)	&(0.00339)		\\\hline
\textsc{airfoil-noise}		&0.00814	&0.00806	&0.04223	&0.00947	&0.00939	&0.04817	&0.00855	&0.00850	&0.03832		\\
				&(0.00121)	&(0.00119)	&(0.00211)	&(0.00134)	&(0.00115)	&(0.00256)	&(0.00092)	&(0.00087)	&(0.00218)		\\\hline
\textsc{munich-rent}		&0.00131	&0.00126	&0.01569	&0.00122	&0.00121	&0.01812	&0.00101	&0.00101	&0.01598		\\
				&(0.00033)	&(0.00030)	&(0.00087)	&(0.00029)	&(0.00029)	&(0.00113)	&(0.00018)	&(0.00016)	&(0.00103)		\\\hline
\textsc{updrs-motor}		&0.02518	&0.02502	&0.05345	&0.02844	&0.02828	&0.06257	&0.02585	&0.02569	&0.015229		\\
				&(0.00152)	&(0.00152)	&(0.00069)	&(0.00159)	&(0.00152)	&(0.001496)	&(0.00166)	&(0.00169)	&(0.001787)		\\\hline
\textsc{head-circum}		&0.00323	&0.00323	&0.02419	&0.00390	&0.00390	&0.02482	&0.00333	&0.00333	&0.01855		\\
				&(0.00008)	&(0.00008)	&(0.00047)	&(0.00011)	&(0.00011)	&(0.00057)	&(0.00009)	&(0.00096)	&(0.00045)		\\\hline
\textsc{cycle-pp}		&0.00420	&0.00421	&0.03588	&0.00516	&0.00516	&0.04536	&0.00479	&0.00477	&0.03930		\\
				&(0.00009)	&(0.00011)	&(0.00079)	&(0.000197)	&(0.00019)	&(0.00097)	&(0.00027)	&(0.000289)	&(0.00076)		\\\hline
\textsc{hour}			&0.01575	&0.01543	&0.02888	&0.01664	&0.01627	&0.04021	&0.01285	&0.01259	&0.03821		\\
				&(0.00029)	&(0.00034)	&(0.00077)	&(0.00046)	&(0.00043)	&(0.00110)	&(0.00031)	&(0.00035)	&(0.00103)		\\\hline
\textsc{cal-housing}		&0.02426	&0.02415	&0.05406	&0.02546	&0.02518	&0.07473	&0.01919	&0.01912	&0.07337		\\
				&(0.00126)	&(0.00117)	&(0.00135)	&(0.00123)	&(0.00119)	&(0.00158)	&(0.00071)	&(0.00064)	&(0.00144)		\\\hline

 \end{tabular}
\end{center}
\vspace*{-3ex}
  \caption[format=hang]{\scriptsize Average test error (standard deviation) for 2D-SVM with unclipped duality gap stopping criteria 
  (ER-SVM), 2D-SVM with clipped duality gap stopping criteria ($\text{ER-SVM}^*$) and \textsf{ER-Boost}. 
  The average test error (standard deviation) was computed on 25 random splits for small data sets, 10 random splits 
  for medium size data sets and 5 random split for larger size data sets.}
	\label{Table-averge test error}
 \end{table}

%
\begin{table}[H]
\footnotesize
\begin{center}
 
\begin{tabular}{|l|c|c|c|c|c|c|c|c|c|}

\hline
    \multirow{2}{*}{\textsc{data}}
    & \multicolumn{3}{|c|}{$\tau=0.25$}  & \multicolumn{3}{|c|}{$\tau=0.50$} & \multicolumn{3}{|c|}{$\tau=0.75$} \\
        \cline{2-10}
				
				&ER-SVM		&$\text{ER-SVM}^*$&ER-Boost	&ER-SVM		&$\text{ER-SVM}^*$& ER-Boost	&ER-SVM		&$\text{ER-SVM}^*$& ER-Boost\\ \hline 	
				
\textsc{nc-crime}		&0.305		&0.317		&20.954		&0.323		&0.318		&21.545		&0.298		&0.311		&21.595			\\\hline		
\textsc{concrete-comp}		&0.983		&1.028	 	&1.861		&1.027		&1.089		&1.899		&0.964		&1.018		&1.8025			\\\hline				
\textsc{airfoil-noise}		&2.078		&2.173		&0.645		&2.234		&2.342		&0.656		&2.122		&2.232		&0.649			\\\hline				
\textsc{munich-rent}		&2.413		&2.485		&9.288		&2.385		&2.476		&9.542		&2.364		&2.426		&9.460			\\\hline				
\textsc{updrs-motor}		&43.874		&46.737		&110.853	&47.819		&47.819		&114.967	&42.614		&45.537		&114.335		\\\hline				
\textsc{head-circum}		&34.352		&36.173		&1.7826		&36.928		&39.029		&1.7529		&36.744		&37.256		&1.796			\\\hline
\textsc{cycle-pp}		&83.452		&85.893		&2.7473		&91.127		&93.897		&2.758		&85.690		&87.309		&2.714			\\\hline
\textsc{hour}			&307.249	&318.357	&70.376		&315.692	&327.897	&69.576		&281.972	&288.479	&68.536			\\\hline				
\textsc{cal-housing}		&506.679	&529.945	&39.913		&535.835	&550.364	&39.974		&458.223	&479.735	&38.880			\\\hline

 \end{tabular}
\end{center}
\vspace*{-3ex}
  \caption[format=hang]{\scriptsize Training time (in seconds) for 2D-SVM with unclipped duality gap stopping criteria (ER-SVM), 
  2D-SVM with clipped duality gap stopping criteria ($\text{ER-SVM}^*$) and \textsf{ER-Boost}.}
	\label{Table-run time}
 \end{table}

\end{landscape}

\begin{appendix}
 
\section{Proofs}

The proofs of Lemma \ref{1D-gain} and Lemma \ref{2D-lemma-positivity of determinant} are trivial and therefore omitted. The rest of the proofs are given below.

\begin{proofof}{Theorem \ref{theorem-consistency}}
 The first convergence follows from \cite[Theorem 9.1]{steinwart2008support} and the second convergence is a consequence of the 
 first convergence and \cite[Corollary 3.62]{steinwart2008support}, where we note that we do not need the completeness of 
 $\mathbf{X}$ since we already know the existence and uniqueness of $f_{L,P}^*$.
\end{proofof}

\begin{proofof}{Theorem \ref{1D-optimum solution at boundary}}
 i) We first show that $W$ has a global maximum at $\alpha^*$. To do this, we proceed by contradiction, 
that is, we assume that there exists an $\alpha \in \mathbb{R}^m$ with
\begin{equation}\label{part1-contradition relation-1}
 W(\alpha^*) < W(\alpha)\,.
\end{equation}
By concavity of $W$, we conclude that for $t \in [0,1]$ 
\begin{equation}\label{part1-contradition relation 2}
 W((1-t)\alpha^* + t\alpha) \geq (1-t)W(\alpha^*)+tW(\alpha)\,.
\end{equation}
On the other hand, $h:=t(\alpha-\alpha^*) \in \mathbb{R}^m$ and Taylor's theorem in the multiple dimensional version yields
\begin{equation*}
 \begin{aligned}
	W((1-t)\alpha^* + t\alpha) &= W(\alpha^* + h)\,,\\
				   &= W(\alpha^*) + \langle W^{\prime} (\alpha^*), h \rangle + \frac{1}{2} \langle h, W^{\prime \prime} (\alpha^*) h \rangle + O(\parallel h^2 \parallel)\,,\\
				   &= W(\alpha^*) + \frac{t^2}{2} \langle \alpha-\alpha^*, W^{\prime \prime} (\alpha^*) (\alpha-\alpha^*) \rangle + O(t^2)\,.
 \end{aligned}
\end{equation*}
Using this in (\ref{part1-contradition relation 2}) we obtain
\begin{equation}\label{part1-contradition relation 3}
  W(\alpha^*) + t\big(W(\alpha)-W(\alpha^*)\big) \leq W(\alpha^*) + \frac{t^2}{2} \langle \alpha-\alpha^*, W^{\prime \prime} (\alpha^*) (\alpha-\alpha^*) \rangle + O(t^2)\,,
\end{equation}
and thus
\begin{equation*}
 c_1 t \leq \frac{c_2}{2}t^2 + O(t^2)\,,
\end{equation*}
where $c_1: = W(\alpha)-W(\alpha^*)$ and $c_2:= \langle \alpha-\alpha^*, W^{\prime \prime} (\alpha^*) (\alpha-\alpha^*)\rangle$. 
Furthermore, we have $c_2 \leq 0$ since $W$ is concave and $c_1 > 0$ by (\ref{part1-contradition relation-1}). 
For sufficiently small $t > 0$, (\ref{part1-contradition relation 3}) is therefore impossible and 
hence (\ref{part1-contradition relation-1}) can not be true. Let us now show that $W$ has no other global maximum. 
To show this, we assume the converse, that is, $W$ has a global maximum at some $\alpha^{**} \neq \alpha^*$. 
Then we obtain $W'(\alpha^{**})=0$ by usual calculus, and hence our assumptions are violated. Consequently, $W$ has its 
only global maximum at $\alpha^*$.

ii) If $\alpha^* \notin \mathcal{A}$ then we also have $\alpha^* \notin \mathring{\mathcal{A}}$, where $\mathring{\mathcal{A}}$ 
denotes the interior of $\mathcal{A}$, and for $\alpha \in \mathring{\mathcal{A}}$ we thus have $\alpha \neq \alpha^*$. 
Let us now show that for all $\alpha \in \mathring{\mathcal{A}}$ there exists an $\alpha^{\star} \in \partial \mathcal{A}$ with 
\begin{equation}\label{part2-main goal}
 W(\alpha^{\star}) > W(\alpha).
\end{equation}
To this end, we fix an $\alpha \in \mathring{\mathcal{A}}$ and consider the function
\begin{equation*}
 \begin{aligned}
	\gamma:[0,1] & \rightarrow \mathbb{R}^m\\
	t & \mapsto (1-t) \alpha^* +t\alpha\,.
 \end{aligned}
\end{equation*}
Furthermore, we set
\begin{equation*}
 h    := W \circ \gamma\,.	
\end{equation*}
Then it is easy to see that $h$ is concave. Moreover, since $\alpha \neq \alpha^*$, 
we find $\gamma(t) \neq \alpha^*$ for all $t \in (0,1]$ and thus $h(t) < h(0)$ for all $t \in (0,1]$. 
By the concavity of $h$ we conclude that $h$ is strictly decreasing.
We now show that there exists $t^{\star} \in (0,1]$ with $\gamma(t^{\star}) \in \partial \mathcal{A}$.
Let us assume the converse, that is, $\Gamma \cup \partial \mathcal{A} = \emptyset$, where $\Gamma := \gamma([0,1])$. 
Considering the partition $\mathring{\mathcal{A}}$, $\partial \mathcal{A}$, $\mathbb{R}^m \backslash \bar{\mathcal{A}}$, 
where $\bar{\mathcal{A}}$  denotes the closure of $\mathcal{A}$, we then find by the assumed 
$\mathcal{A}=\bar{\mathcal{A}}$ and $\Gamma \cup \partial \mathcal{A} = \emptyset$
that 
\begin{equation*}
 \begin{aligned}
	B_1 &:= \Gamma \cap \mathring{\mathcal{A}}\\
	B_2 &:= \Gamma \cap \mathbb{R}^m \backslash \bar{\mathcal{A}} = \Gamma \cap (\mathbb{R}^m \backslash \mathcal{A}),
 \end{aligned}
\end{equation*}
is a partition of $\Gamma$. Since $\alpha \in \mathring{\mathcal{A}}$ and $\alpha^* \notin \mathcal{A}$, 
we further find $B_1 \neq \emptyset$ and $B_2 \neq \emptyset$. Moreover, since $\mathbb{R}^m \backslash \mathcal{A}$ is open, 
the sets $B_1$ and $B_2$ are relatively open in $\Gamma$ and $\Gamma$. However, the continuous image of a connected set, 
is connected and thus $\Gamma$ is connected. This leads to a contradiction, and hence there exists a $t^{\star} \in [0,1]$ 
with $\gamma(t^{\star}) \in \partial \mathcal{A}$. Clearly, we have $t^{\star} < 1$ since $\alpha \notin \partial \mathcal{A}$.
For $\alpha^{\star}:=\gamma (t^{\star})$,  the already established strict monotonicity of $h$ then shows 
\begin{equation*}
 W(\alpha^{\star})=h(t^{\star}) > h(1)=W(\alpha)\,.
\end{equation*}
Consequently we have shown (\ref{part2-main goal}) and thus 
\begin{equation*}
 \underset{\alpha \in \partial \mathcal{A}}{\text{sup}} W(\alpha)= \underset{\alpha \in \mathcal{A}}{\text{sup}} W(\alpha).
\end{equation*}
In other words, it suffices to show that the supremum over $\partial \mathcal{A}$ is attained at some 
$\alpha^{\star} \in \partial \mathcal{A}$. To this end, we first show that $\{W \geq \rho \}$ is bounded 
for all $\rho < W^*:=W(\alpha^*)$. For $\alpha \in S$, where $S \subset \mathbb{R}^m$ denotes the Euclidean unit sphere, 
we define
\begin{eqnarray*}
   h_\a:[0,\infty) &\to & \mathbb{R}^m\\
   t&\mapsto& W(\alpha^* + t \alpha)\, .
\end{eqnarray*}

Then $h_{\alpha}$ is concave and continuously differentiable, and has a global maximum at $t=0$. 
Moreover, $h_{\alpha}$ is strictly decreasing with $\underset{t \rightarrow \infty}{\lim} h_{\alpha}(t)= -\infty$.
We define
\begin{equation*}
 \begin{aligned}
	t_{\alpha} :=\max \{t \geq 0: h_{\alpha}(t) \geq \rho\},
 \end{aligned}
\end{equation*}
where we note that the maximum is indeed attained by the continuity of $h_{\alpha}$ and $t_{\alpha} < \infty$. 
Our next intermediate goal is to show that $\alpha \mapsto t_{\alpha}$ is continuous. 
To this end, we fix an $\alpha_0 \in S$, and 
an $\varepsilon > 0$ with $\sqrt{\varepsilon} < -h'_{\alpha_0}(t_{\alpha_0})$,
where we note that  $h'_{\alpha_0}(t_{\alpha_0})<0$ since $h_{\a_0}$ is strictly decreasing and $W^*>\r$.
Since $W$ is continuous differentiable, then there exist a $\d > 0$ such that for 
all $\a\in S$ with $\snorm {\a_0-\a}\leq \d$ we have 
\begin{equation*}
 \begin{aligned}
  \lvert h'_{\alpha_0}(t_{\alpha_0}) - h'_{\alpha}(t_{\alpha_0})\rvert &\leq \varepsilon\, .
 \end{aligned}
\end{equation*}
For $t_\a \geq t_{\alpha_0}$, the concavity, or more precisely, the subdifferential inequality of $-h'_\a(t_{\a_0})$,
then gives
\begin{equation*}
 \begin{aligned}
	h_{\alpha}(t_\a) & \leq h_{\alpha}(t_{\alpha_0})+h'_{\alpha}(t_{\alpha_0}) ( t_\a - t_{\alpha_0} )\, ,\\
		      & \leq h_{\alpha_0}(t_{\alpha_0}) + \varepsilon +(h'_{\alpha_0}(t_{\alpha_0}) + \varepsilon) (t_\a-t_{\alpha_0})\, ,\\
		      & \leq h_{\alpha_0}(t_{\alpha_0}) + \varepsilon + \frac{1}{2} h'_{\alpha_0}(t_{\alpha_0})(t_\a-t_{\alpha_0})\, .
 \end{aligned}
\end{equation*}
Now recall that $h_{\alpha}(t_\a) = \rho = h_{\alpha_0}(t_{\a_0})$. Thus we obtain
\begin{equation*}
\begin{aligned}
  0 &\leq \varepsilon + \frac{1}{2} h'_{\alpha_0} (t_{\alpha_0})(t_\a-t_{\alpha_0})\, ,
\end{aligned}
\end{equation*}
and since $h'_{\alpha_0}(t_{\a_0}) < 0$, we conclude that 
\begin{equation*}
  \frac{-2 \varepsilon}{h'_{\alpha_0}(t_{\alpha_0})} \geq t_\a-t_{\alpha_0}\, , 
\end{equation*}
and thus 
\begin{equation*}
    t_{\alpha}  
    \leq 
    t_{\alpha_0} + \frac{-2 \varepsilon}{h'_{\alpha_0}(t_{\alpha_0})}
    \leq
    2 \sqrt \e
    \, .
\end{equation*}
Since an analogous bound can be established in the case $t \leq t_{\alpha_0}$, we conclude that 
$\alpha \mapsto t_{\alpha}$ is continuous.
Consequently,  there exist an $\alpha_0\in S$ with 
$t_{\alpha_0} = \underset{\alpha \in S}{\sup} \,\,t_{\alpha} $, and thus
 $\{ W \geq \rho\}$ is bounded. Now we show that there exist $\alpha^{\star} \in \mathcal{A}$ with 
\begin{equation*}
 W^{\star}:= \underset{\alpha \in \mathcal{A}}{\sup}\,\, W(\alpha)=W(\alpha^{\star})\, .
\end{equation*}
Clearly there is an $(\alpha_n) \subset \mathcal{A}$ with 
\begin{equation*}
 W(\alpha_n) \rightarrow W^{\star}\, ,
\end{equation*}
and since $\{ W  \geq \rho\}$ is bounded, the sequence $\alpha_n$ is also bounded. Then there is a subsequence $\alpha_{n_k}$ and an $\alpha^{\star}$ with $\alpha_{n_k} \rightarrow \alpha^{\star}$ and 
the continuity of $W$ yields
$W(\alpha_{n_k}) \rightarrow W(\alpha^{\star})$. Consequently, we have shown $W(\alpha^{\star}) = W(\alpha)$. Finally
$ \alpha^{\star} = \lim \alpha_{n_k} \in \mathcal{A}$ follows from $\mathcal{A}=\bar{\mathcal{A}}$.
\end{proofof}

\begin{proofof}{ Theorem \ref{1D-theorem-feasible solution}}
If $c_i=0$, there is nothing to prove. Let us assume that $c_i > 0 $. Since $b_1, b_2 \in (1,\infty)$, then only (\ref{1D-optimum Solution-case 2-beta=0}) provides a feasible solution $c_i > 0$ because $\beta_i^+ < 0$ in (\ref{1D-optimum Solution-case 1-alpha=0}). Similarly, if we assume that $c_i < 0$, then $\alpha_i^+ < 0$ in (\ref{1D-optimum Solution-case 2-beta=0}) while $\beta_i^+ > 0$ in (\ref{1D-optimum Solution-case 1-alpha=0}) which makes it feasible solution. We finally conclude that only one of two cases provides the feasible optimal solution when $c_i \neq 0$.
\end{proofof}

\begin{proofof}{Lemma \ref{2D-Lemma-simaltenously positivity of 4 equations}}
i) Since $T_1 \geq 0$ and $T_2 \geq 0$, we have $\frac{b_2}{k} c_i \leq c_j \leq \frac{k}{b_2}c_i$. 
Since we assumed that $c_i \neq 0$ or $c_j \neq 0$, we conclude from the latter and $b_2, k \geq 0$ that we actually 
have $c_i \neq 0$ and $c_j \neq 0$. Moreover, $b_2 > 1$ and $k \leq 1$ shows that $c_i < 0$ and $c_j < 0$.\\

ii) Since $T_3 \geq 0$ and $T_4 \geq 0$, we have $\frac{k}{b_1} c_i \leq c_j \leq \frac{b_1}{k}c_i$. 
Since we assumed that $c_i \neq 0$ or $c_j \neq 0$, we conclude from the latter and $b_1, k \geq 0$ that we actually
have $c_i \neq 0$ and $c_j \neq 0$. Moreover, $b_1 > 1$ and $k \leq 1$ shows that $c_i > 0$ and $c_j > 0$.

Finally, this leads to conclude that $T_1$, $T_2$, $T_3$ and $T_4$ are not simultaneously positive. 
By similar arguments, it can be shown that these expressions are not simultaneously negative.
\end{proofof}

\begin{proofof}{Theorem \ref{2D-Theorem- one feasible optimum solution}}
Our first goal is to show that at most one of the four cases  (\ref{2D_Optimum Solution-Case 1}), (\ref{2D_Optimum Solution-Case 2}), (\ref{2D_Optimum Solution-Case 3}) and (\ref{2D_Optimum Solution-Case 4}) leads to a feasible solution. To this end we note that (\ref{2D_Optimum Solution-Case 1}) is feasible if and only if $T_1$ and $T_2$ are non-negative. Similar consideration from (\ref{2D_Optimum Solution-Case 2}) to (\ref{2D_Optimum Solution-Case 4}) leads to the Table \ref{Table-expression behaviour for feasible solution}.
\begin{table}[h]
	\begin{center}
		\begin{tabular}{c|cccc}
		Optimal Solution 				& $T_1$ 	& $T_2$ 	& $T_3$ 	& $T_4$\\
		\hline
		(\ref{2D_Optimum Solution-Case 1}) feasible & $\geq 0$ 	& $\geq 0$ 	& \textendash 	& \textendash\\
		(\ref{2D_Optimum Solution-Case 2}) feasible & \textendash 	& \textendash 	& $\geq 0$ 	& $\geq 0$\\
		(\ref{2D_Optimum Solution-Case 3}) feasible & \textendash 	& $\leq 0$ 	& $\leq 0$ 	& \textendash \\
		(\ref{2D_Optimum Solution-Case 4}) feasible & $\leq 0$ 	& \textendash 	& \textendash 	& $\leq 0$ \\
		\hline
		\end{tabular}
		\caption{Behavior of expressions $T_1$, $T_2$, $T_3$ and $T_4$ when any optimal solutions is feasible}
		\label{Table-expression behaviour for feasible solution}
	\end{center}
\end{table}
Now let assume that (\ref{2D_Optimum Solution-Case 1}) is feasible. 
By Lemma \ref{2D-Lemma-simaltenously positivity of 4 equations} we then see that (\ref{2D_Optimum Solution-Case 2}) is not feasible.
Moreover, if (\ref{2D_Optimum Solution-Case 3}) was feasible, we would have
\begin{equation*}
 kc_i = b_2 c_j\,,
\end{equation*}
which implies $c_i = c_j = 0$ by $k \leq 1$ and $b_2 > 1$. Since the latter contradicts the 
assumed $c_i \neq 0$ or $c_j \neq 0$, we therefore conclude that  (\ref{2D_Optimum Solution-Case 3}) is not feasible. 
Analogously, (\ref{2D_Optimum Solution-Case 4}) is not feasible. Hence, we have shown that if (\ref{2D_Optimum Solution-Case 1}) 
is feasible, the remaining cases (\ref{2D_Optimum Solution-Case 2}) to   (\ref{2D_Optimum Solution-Case 4}) are not feasible.
Since the arguments can be repeated using Table \ref{Table-expression behaviour for feasible solution} when one of the remaining
cases  (\ref{2D_Optimum Solution-Case 2}) to   (\ref{2D_Optimum Solution-Case 4}) is considered feasible, we finally conclude 
that at most one of the four cases is feasible, that is, we have shown our intermediate result.

Let us now assume that none of the four cases yield a feasible solution. Then we obtain  
Table \ref{Table-expression behaviour for non-feasible solution},
\begin{table}[h]
	\begin{center}
		\begin{tabular}{c|cccc}
		Optimal Solution 				& $T_1$ 	& $T_2$ 	& $T_3$ 	& $T_4$\\
		\hline
		(\ref{2D_Optimum Solution-Case 1}) not feasible & $ < 0$ 	& $ < 0$ 	& \textendash 	& \textendash\\
		(\ref{2D_Optimum Solution-Case 2}) not feasible & \textendash 	& \textendash 	& $ < 0$ 	& $ < 0$\\
		(\ref{2D_Optimum Solution-Case 3}) not feasible & \textendash	& $ > 0$ 	& $ > 0$ 	& \textendash \\
		(\ref{2D_Optimum Solution-Case 4}) not feasible & $ > 0$ 	& \textendash 	& \textendash	& $ > 0$ \\
		\hline
		\end{tabular}
		\caption{Behavior of expressions $T_1$, $T_2$, $T_3$ and $T_4$ when none of the optimal solutions is feasible}
		\label{Table-expression behaviour for non-feasible solution}
	\end{center}
\end{table}
where in each row, at least one of the inequalities needs to be true. Let us assume that $T_1 < 0$, then by 
Table \ref{Table-expression behaviour for non-feasible solution}, we conclude that we have following set of inequalities
\begin{eqnarray}
  k c_j & < & b_2 c_i\label{theorem-T1 based inequality}\, ,\\
  k c_i & > & b_2 c_j\label{theorem-T2 based inequality}\, ,\\
  b_1 c_i & < & k c_j\label{theorem-T3 based inequality}\, ,\\
  b_1 c_j & > & k c_i\label{theorem-T4 based inequality}\, .
\end{eqnarray}
Combining (\ref{theorem-T1 based inequality}) and (\ref{theorem-T3 based inequality}) as well 
as (\ref{theorem-T2 based inequality}) and (\ref{theorem-T4 based inequality}), we obtain
\begin{eqnarray}
 b_1 c_i & < & b_2 c_i\label{theorem-T1, T3 based inequality}\, ,\\
 b_2 c_j & < & b_1 c_j\label{theorem-T2, T4 based inequality}\, . 
\end{eqnarray}
Now if $c_i < 0$, we find $c_j < 0$ by (\ref{theorem-T1 based inequality}). Moreover (\ref{theorem-T1, T3 based inequality}) 
together with $c_i < 0$ implies $b_2 < b_1$, while (\ref{theorem-T2, T4 based inequality}) together with $c_j < 0$ 
implies $b_1 < b_2$, that is, we have found a contradiction. Analogously, we obtain a contradiction in the case $T_1 \geq 0$. 
As a consequence exactly one of the four cases produces a feasible solution. Finally, the implications are a direct consequence 
of the form of the solutions in (\ref{2D_Optimum Solution-Case 1}) to (\ref{2D_Optimum Solution-Case 4}) and the fact that only 
one case provides a feasible solution.
\end{proofof}
\end{appendix}

\newpage
\input{figures}

\end{document}

%% file: figures.tex
\begin{appendix}

\section{Detailed Results of Experiments}


\subsection{Results for Different Working Set Selection Methods}
\begin{figure}[!ht]
\begin{scriptsize}
 \subfloat{\includegraphics[scale=0.52]{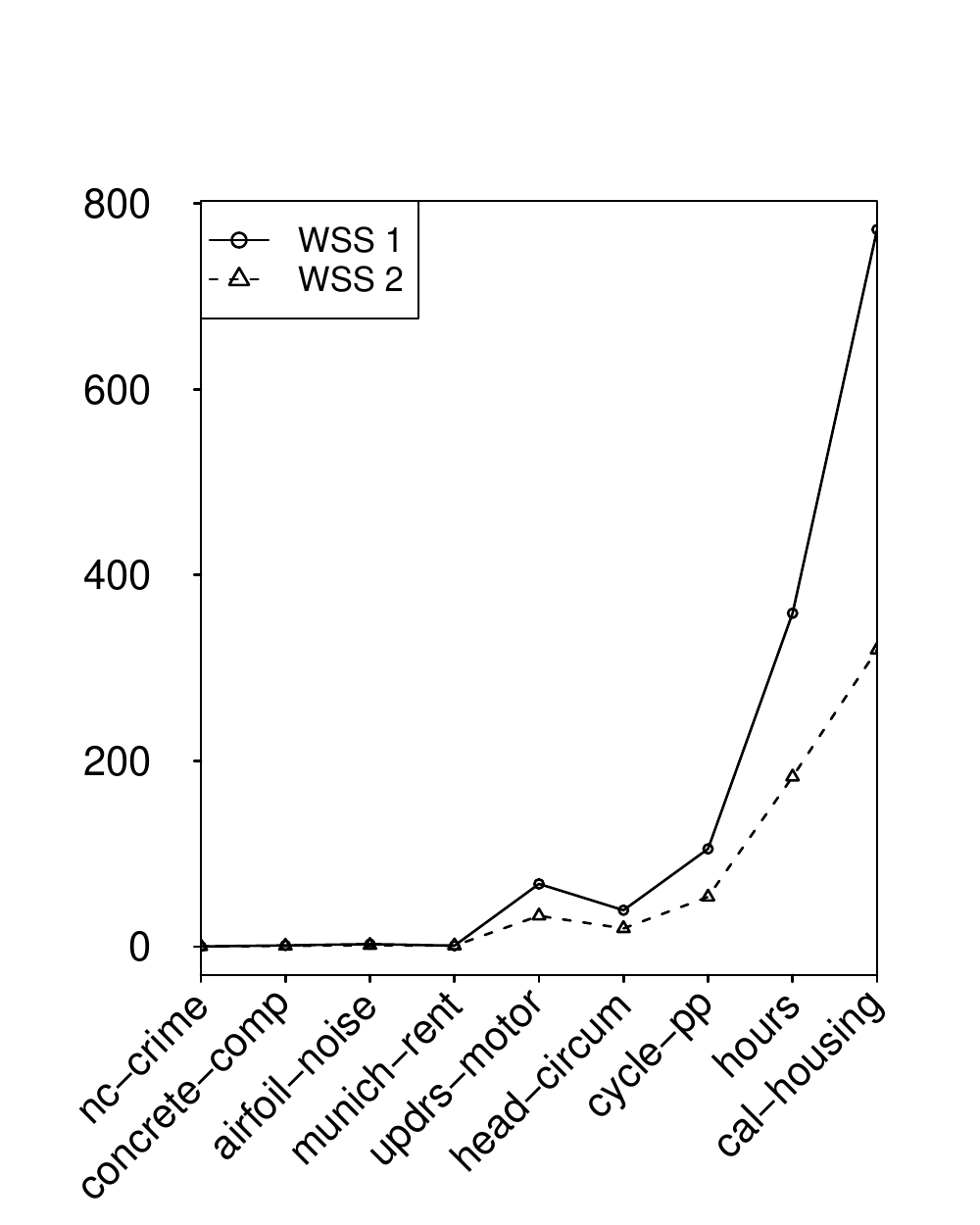}}\hspace{-0.7cm}
 \hfill
 \subfloat{\includegraphics[scale=0.52]{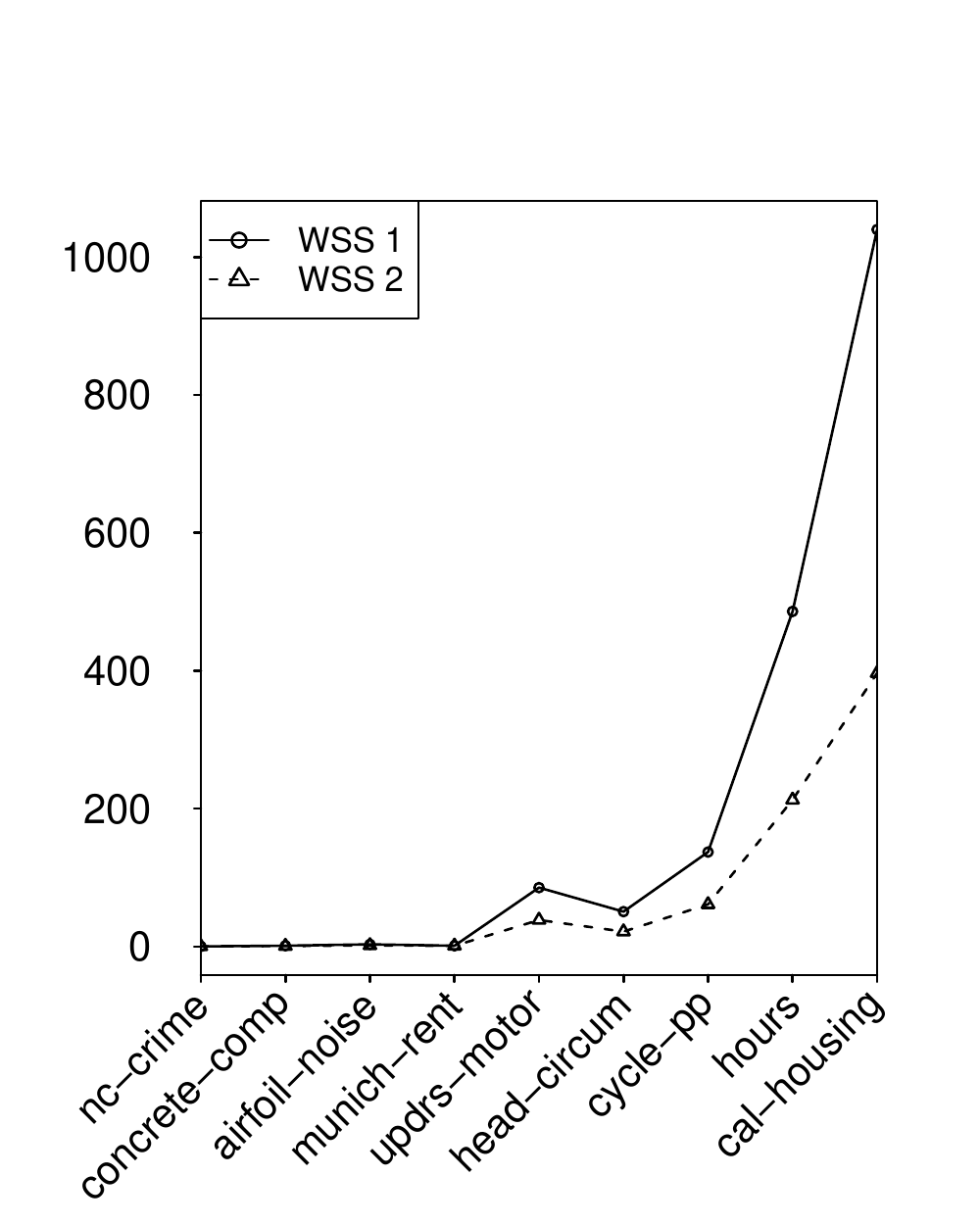}}\hspace{-0.7cm}
\hfill
 \subfloat{\includegraphics[scale=0.52]{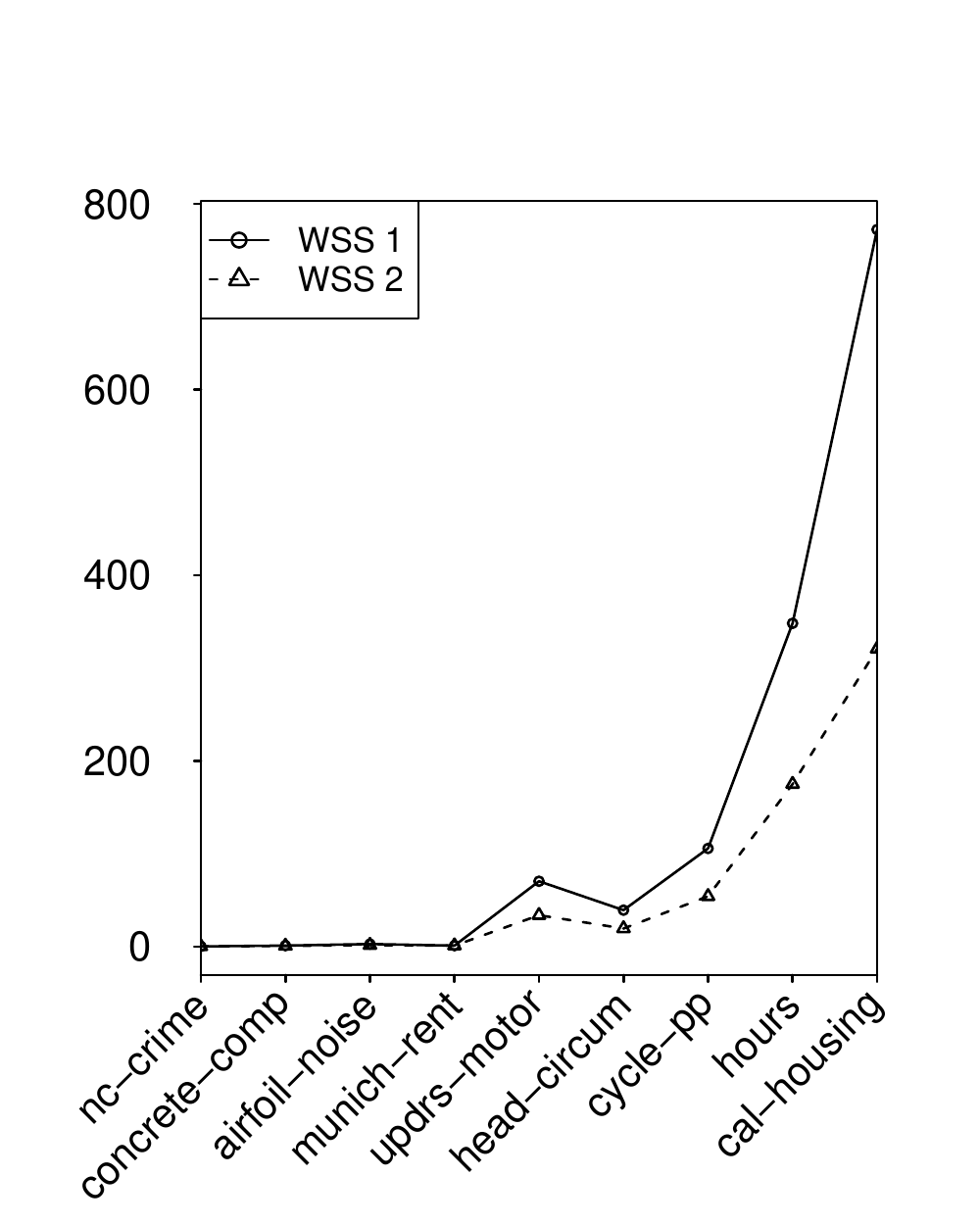}}
 
 \vspace{-1.0cm}
\subfloat{\includegraphics[scale=0.52]{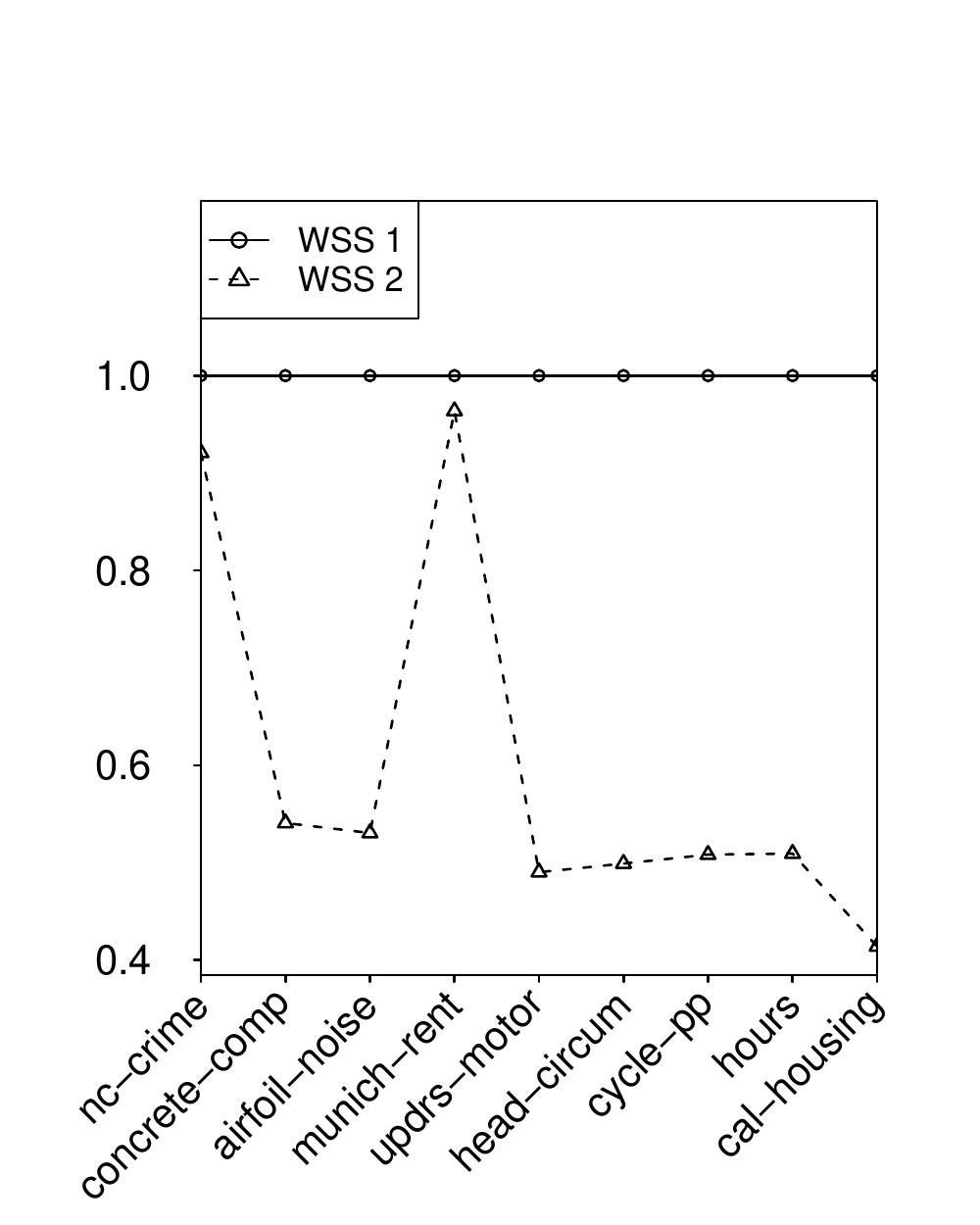}}\hspace{-0.7cm}
 \hfill
 \subfloat{\includegraphics[scale=0.52]{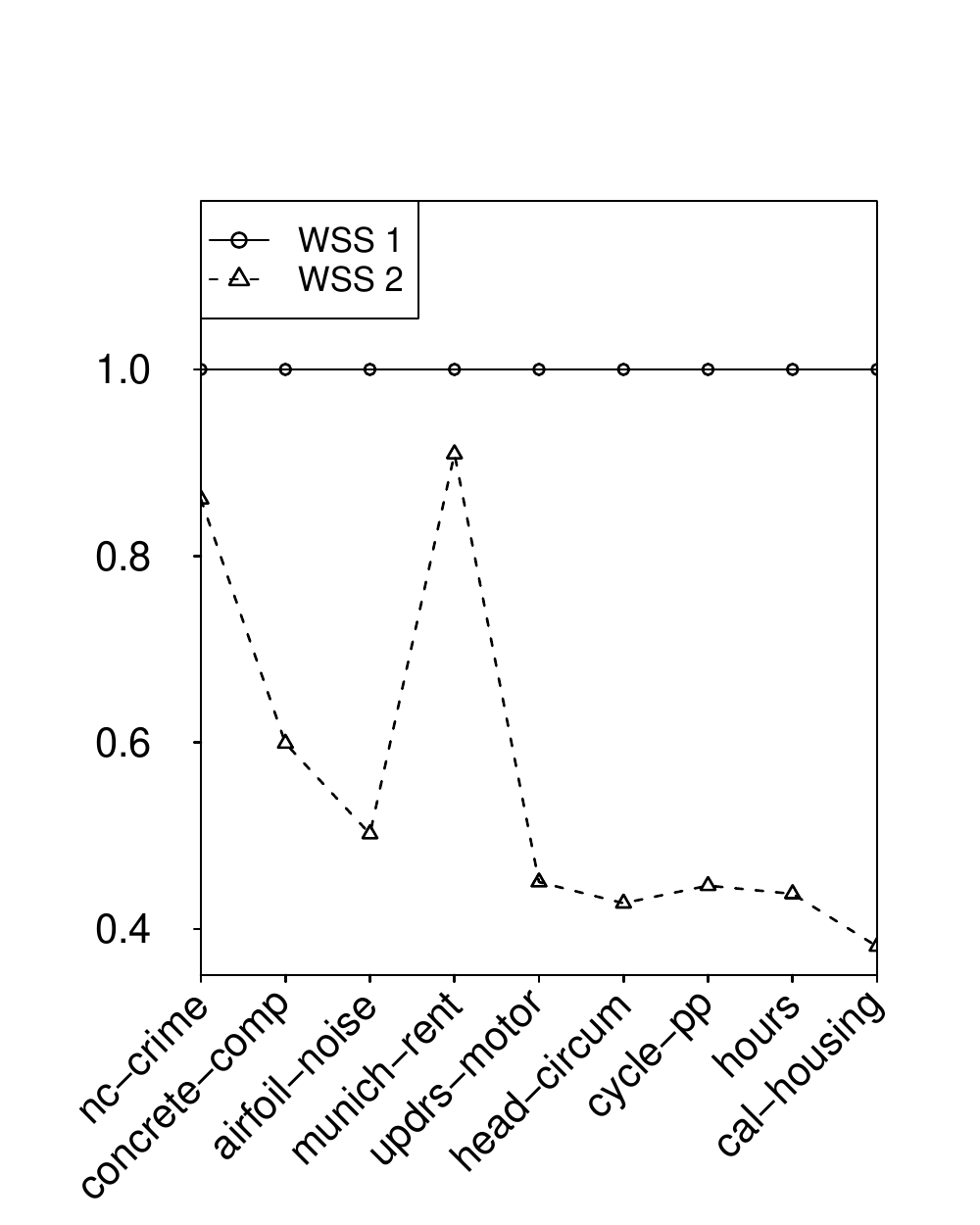}}\hspace{-0.7cm}
\hfill
 \subfloat{\includegraphics[scale=0.52]{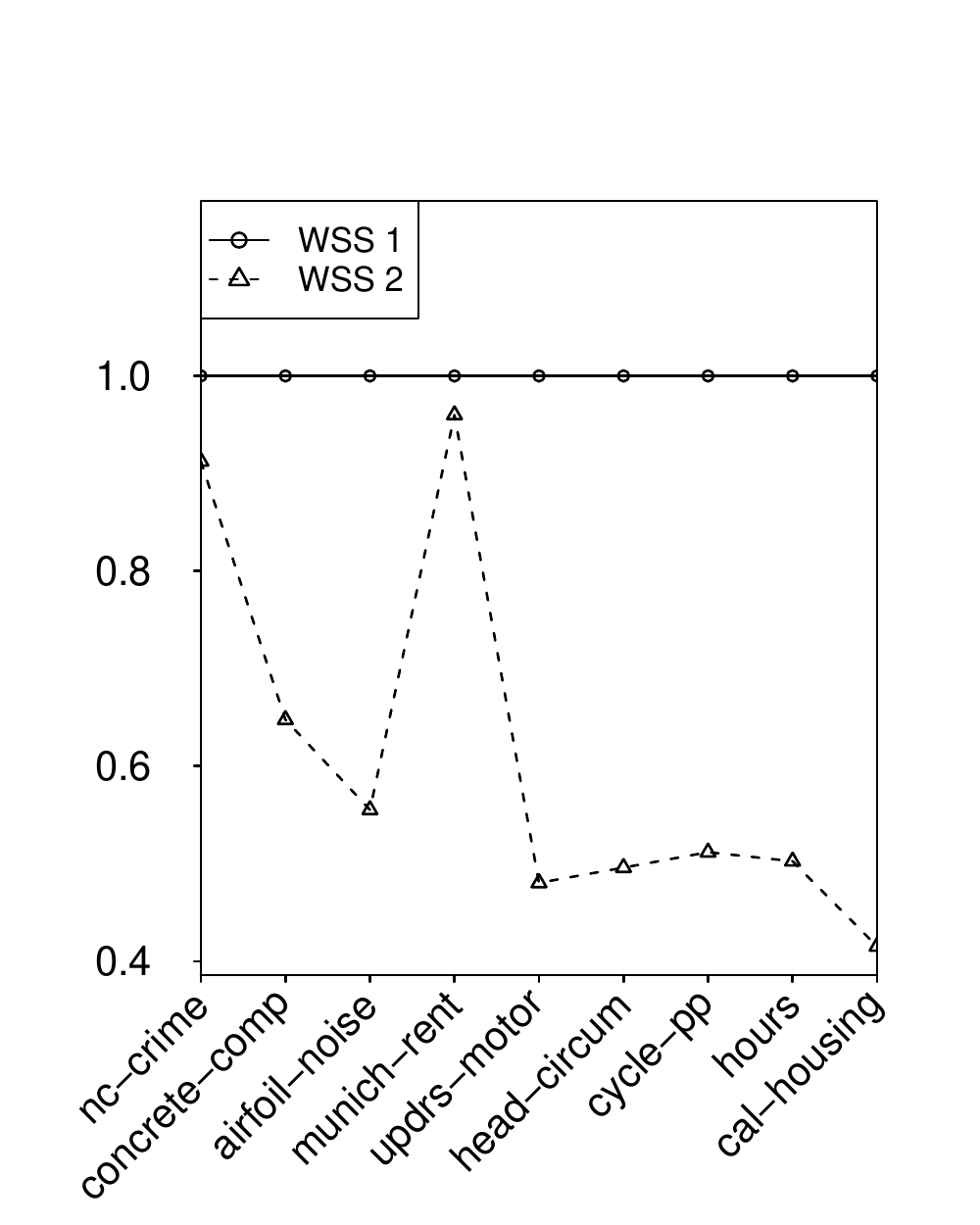}}
 \caption{Train time (top) and corresponding ratio (bottom) of different data sets for different working set selection methods
 after fixing warm start initialization and stopping criteria with clipped duality gap. The graphs comprises of 
 $\tau=0.25$ (left), $\tau=0.50$ (middle) and $\tau=0.75$ (right).}
 \label{figure-time and ratio-WSS vs datasets}
\end{scriptsize}
\end{figure}

\newpage
\begin{figure}[!ht]
\begin{scriptsize}
 \subfloat{\includegraphics[scale=0.52]{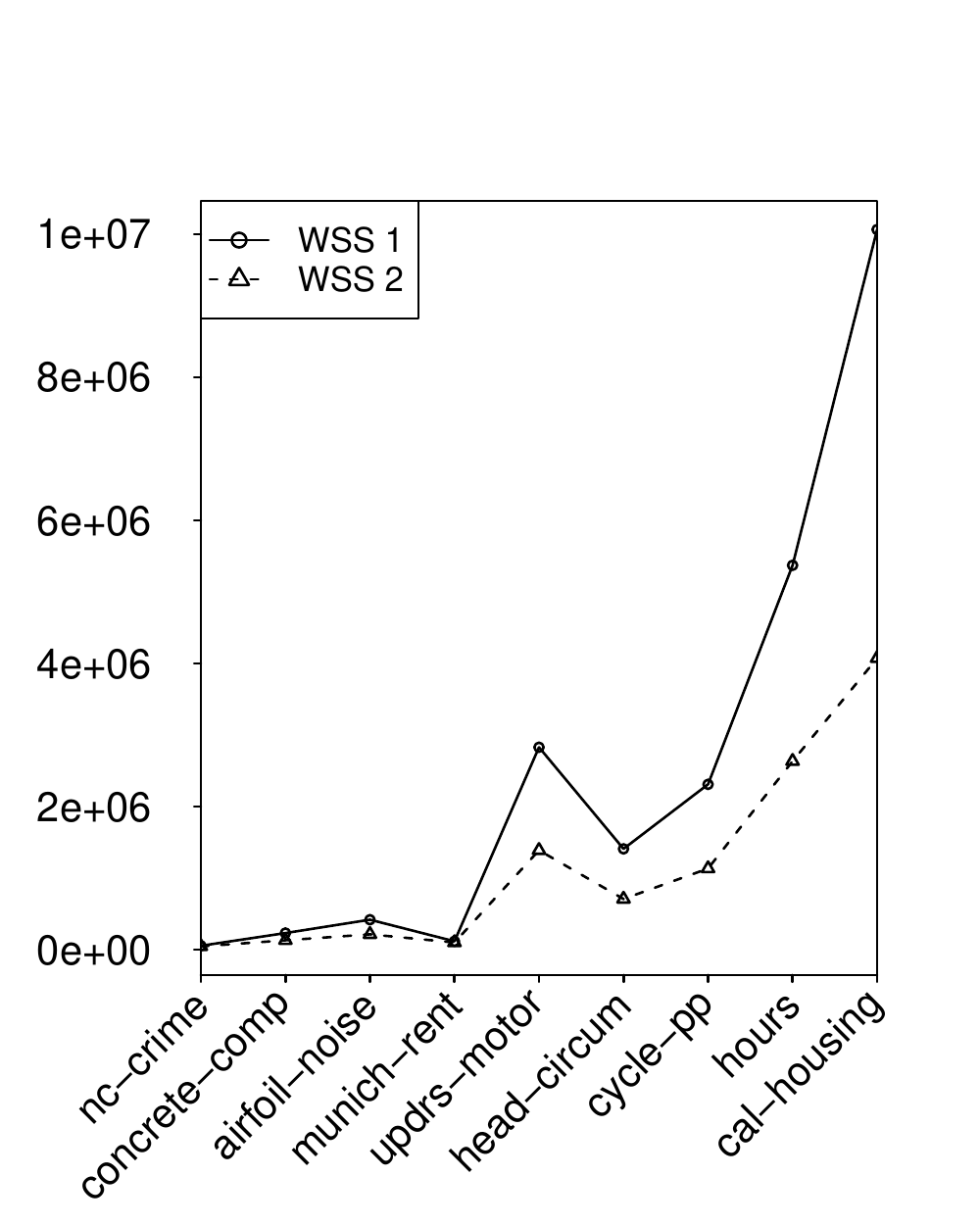}}\hspace{-0.7cm}
 \hfill
 \subfloat{\includegraphics[scale=0.52]{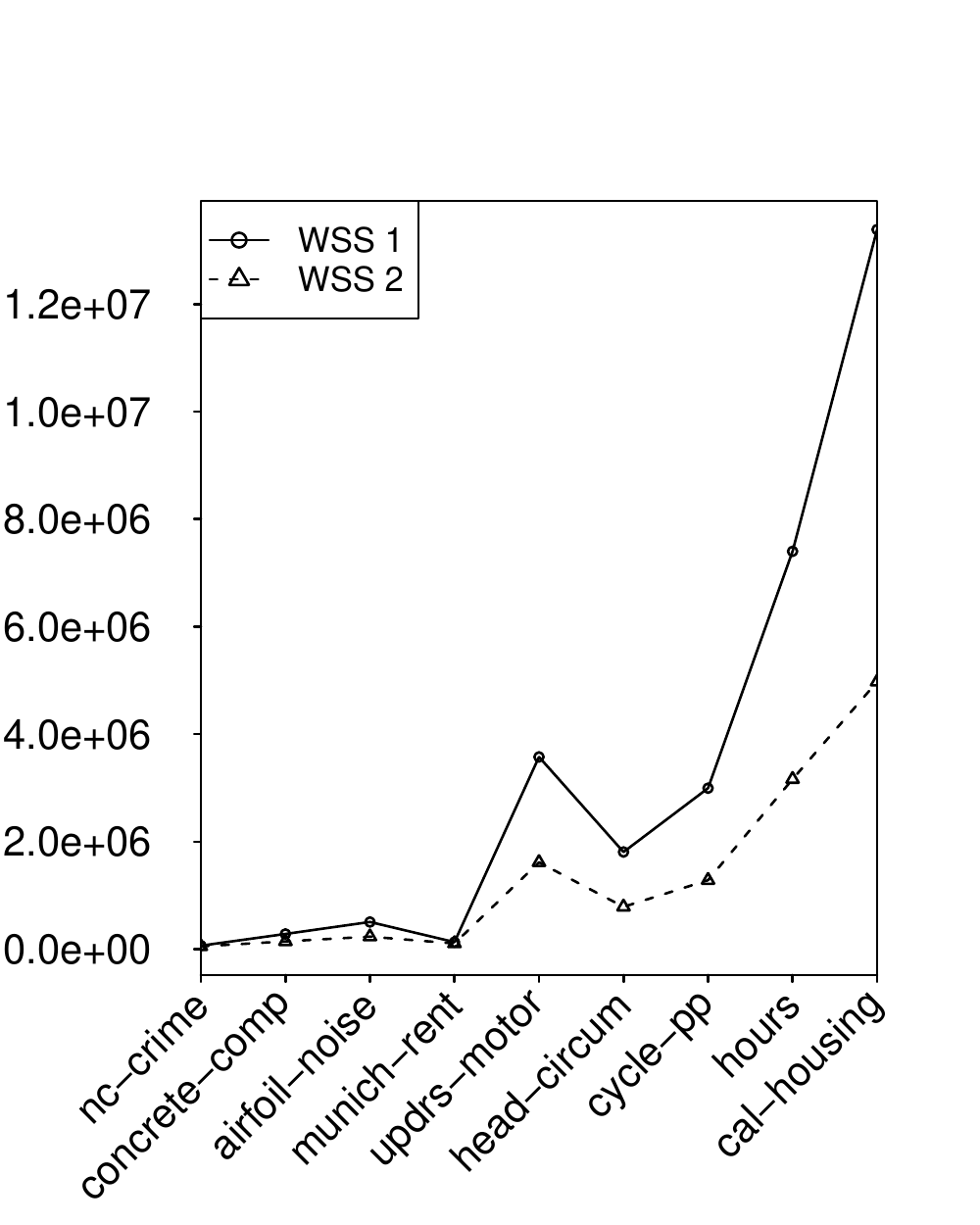}}\hspace{-0.7cm}
\hfill
 \subfloat{\includegraphics[scale=0.52]{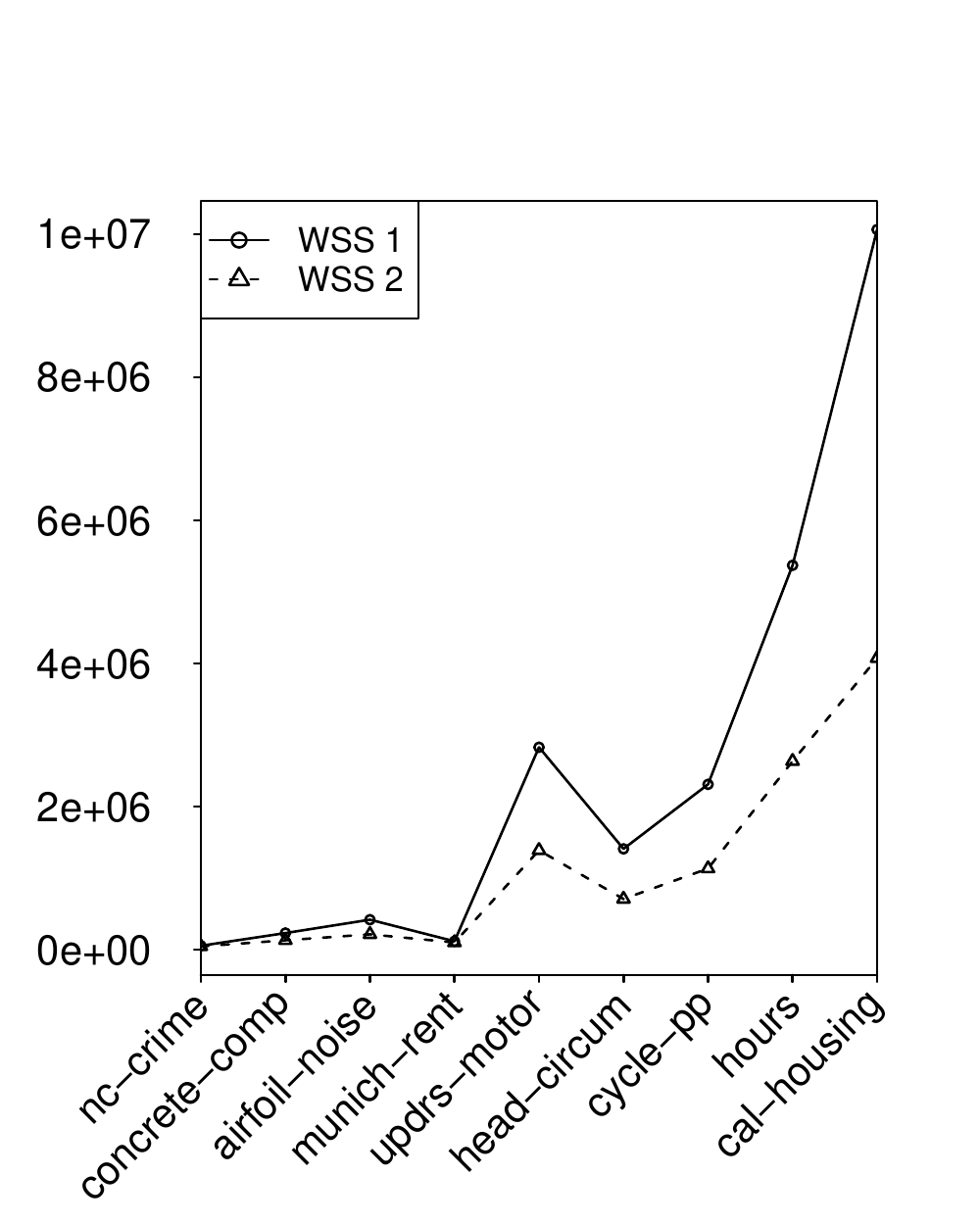}}
 
 \vspace{-1.0cm}
\subfloat{\includegraphics[scale=0.52]{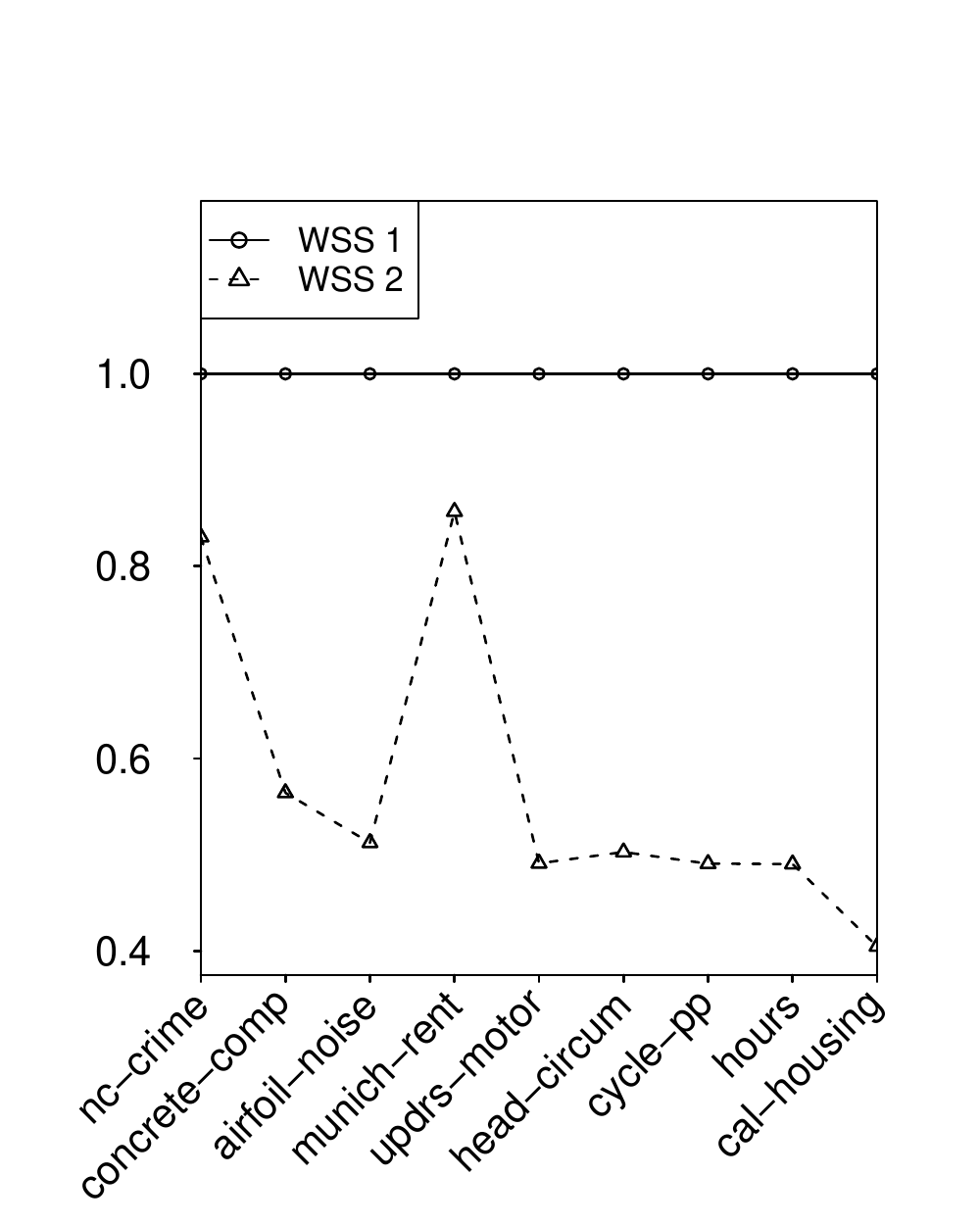}}\hspace{-0.7cm}
 \hfill
 \subfloat{\includegraphics[scale=0.52]{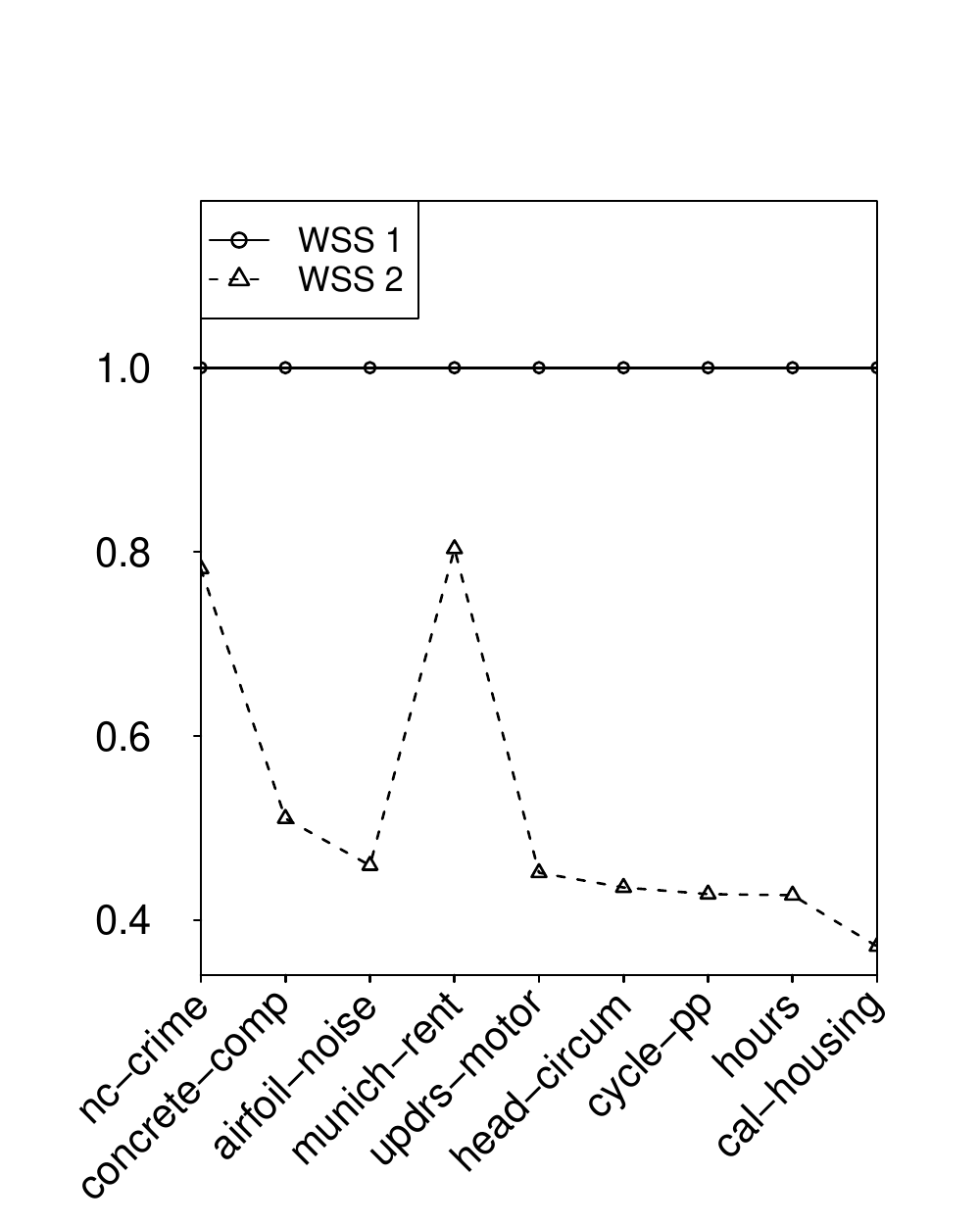}}\hspace{-0.7cm}
\hfill
 \subfloat{\includegraphics[scale=0.52]{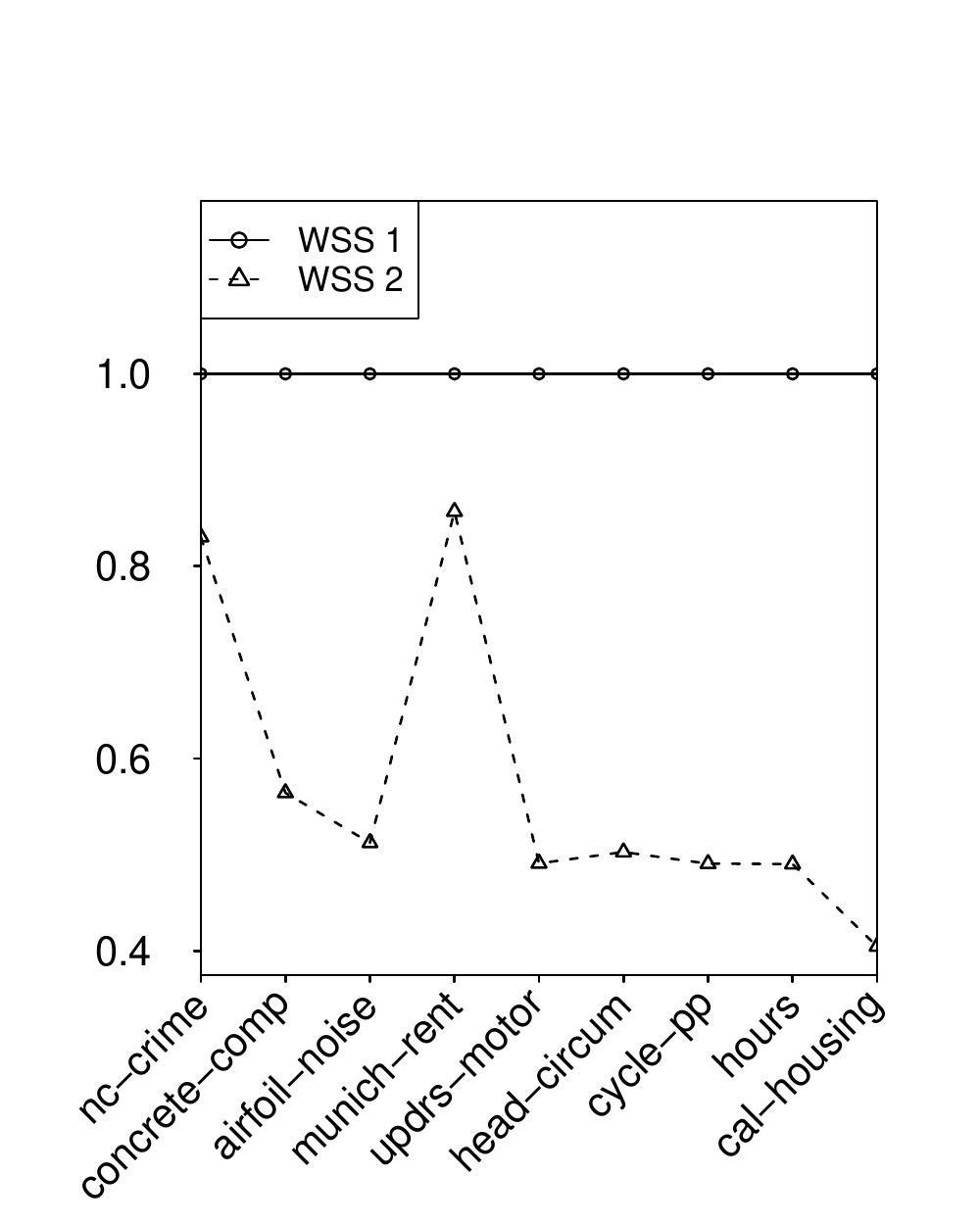}}
 \caption{Train iterations (top) and corresponding ratio (bottom) of different data sets for different working set selection 
 methods after fixing warm start initialization and stopping criteria with clipped duality gap. The graphs comprises of 
 $\tau=0.25$ (left), $\tau=0.50$ (middle) and $\tau=0.75$ (right).}
  \label{figure-iter and ratio-WSS vs datasets}
\end{scriptsize}
\end{figure}

\newpage

\begin{figure}[!ht]
\begin{scriptsize}
\vspace{-1cm}
\subfloat{\includegraphics[scale=0.49]{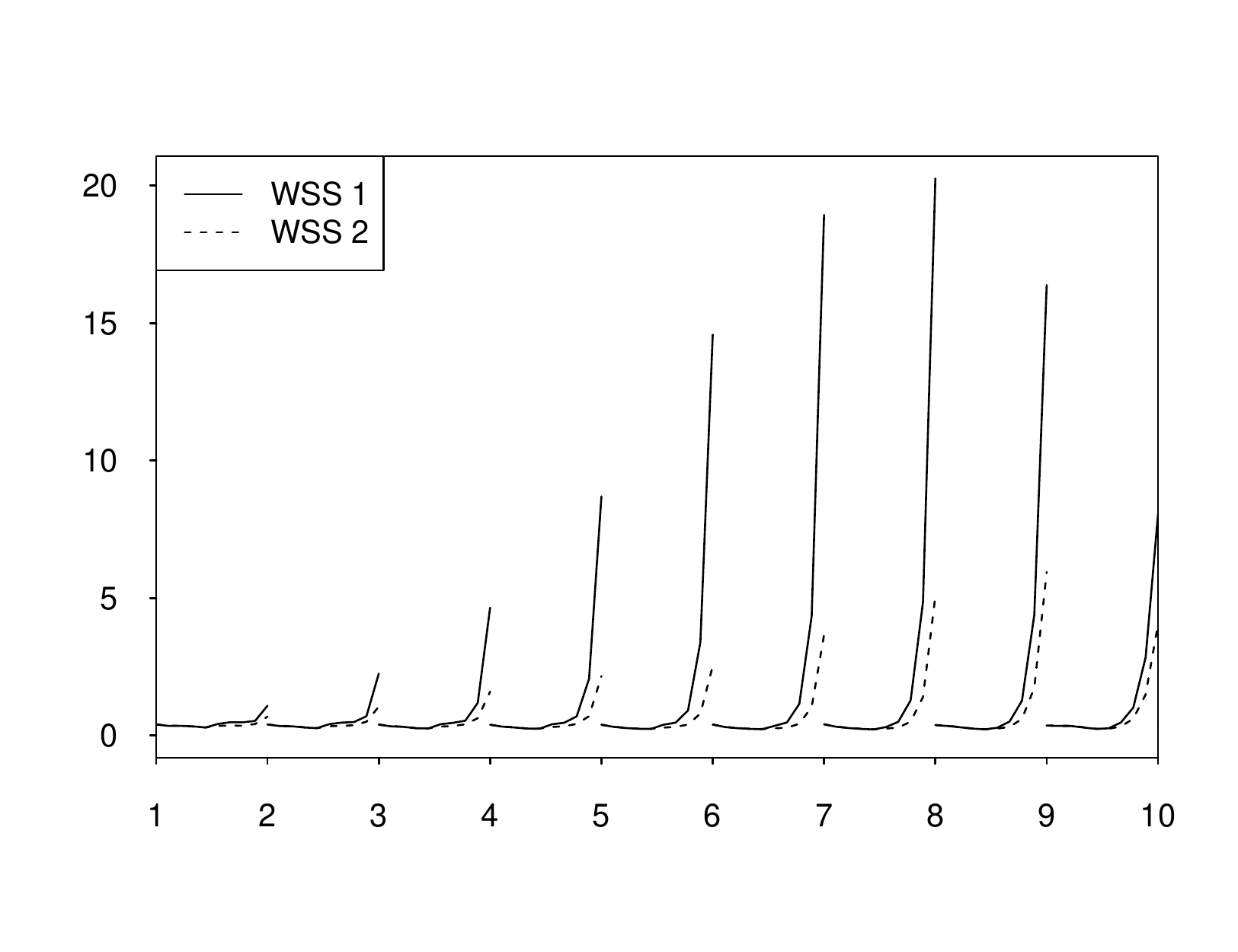}}\hspace{-0.7cm}
\hfill
\subfloat{\includegraphics[scale=0.49]{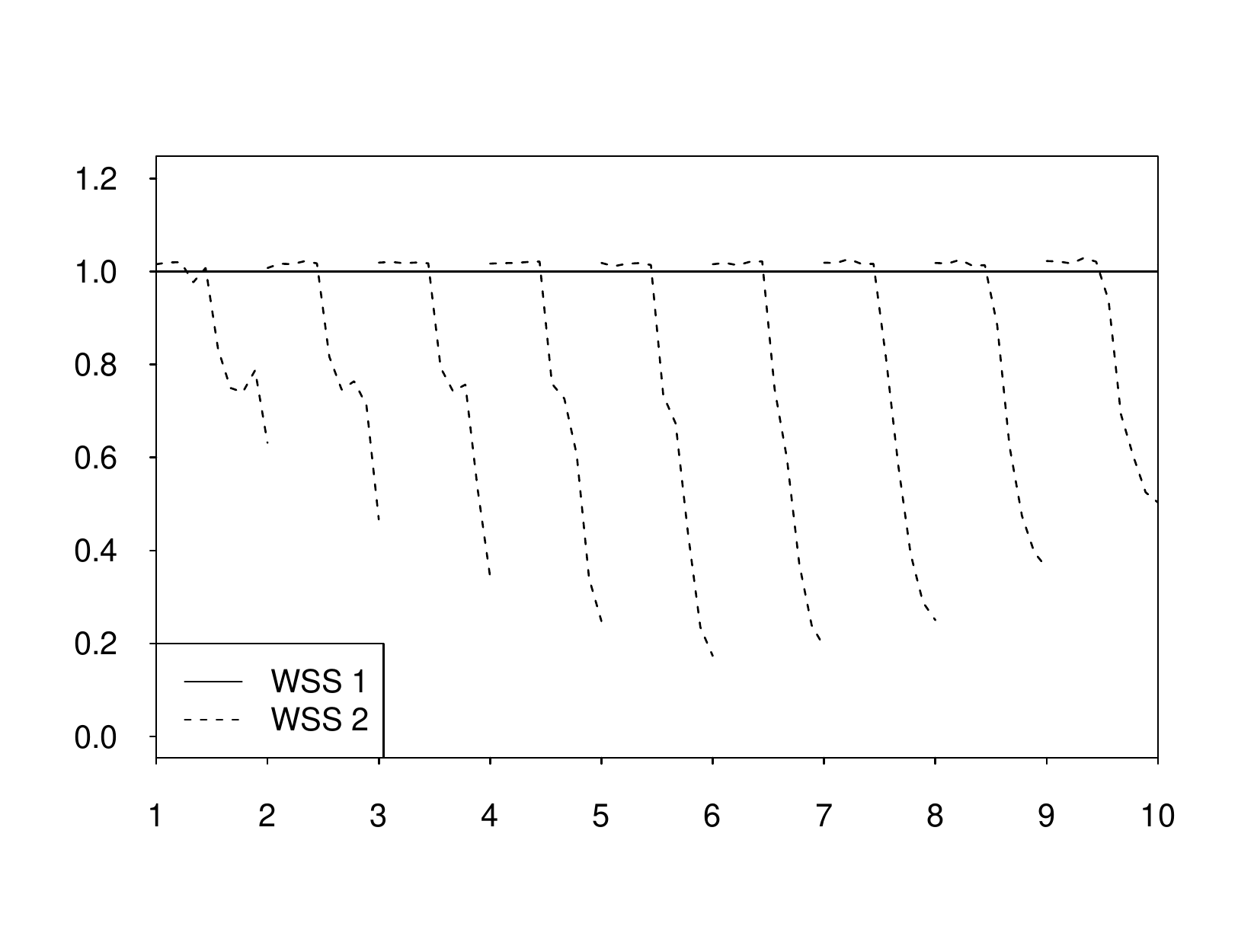}}
\vspace{-1.9cm}
\subfloat{\includegraphics[scale=0.49]{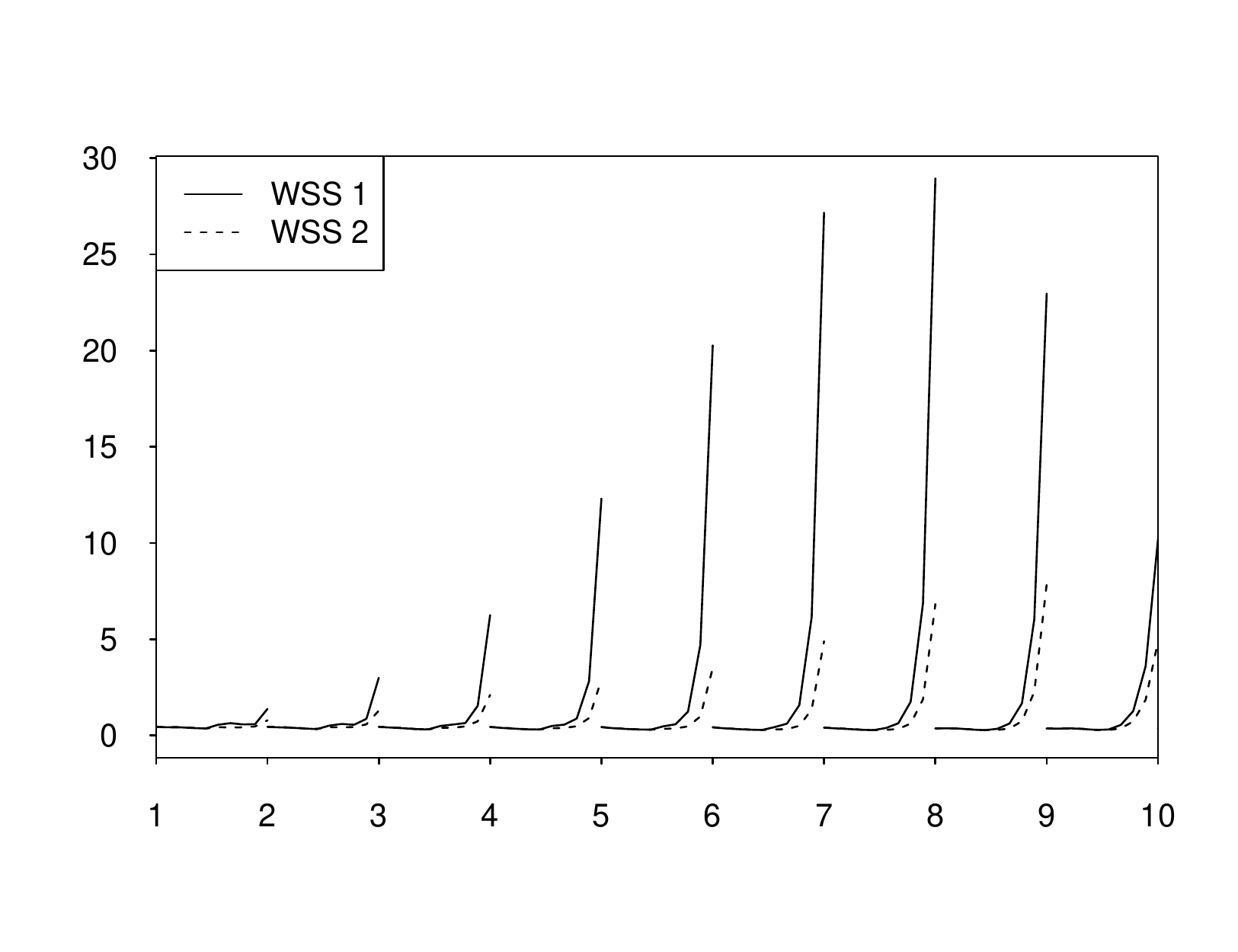}}\hspace{-0.7cm}
\hfill
\subfloat{\includegraphics[scale=0.49]{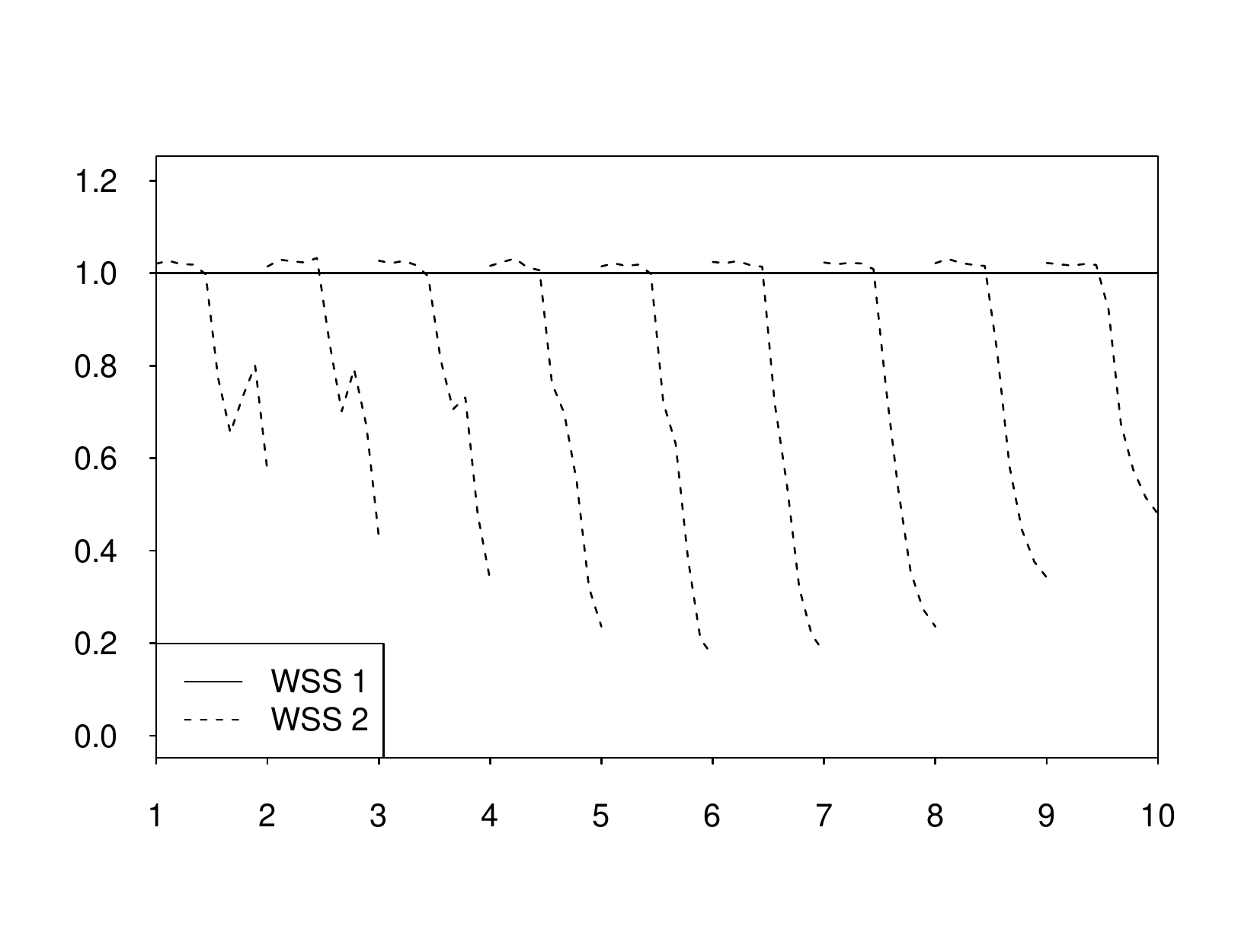}}
\vspace{-1.9cm}
\subfloat{\includegraphics[scale=0.49]{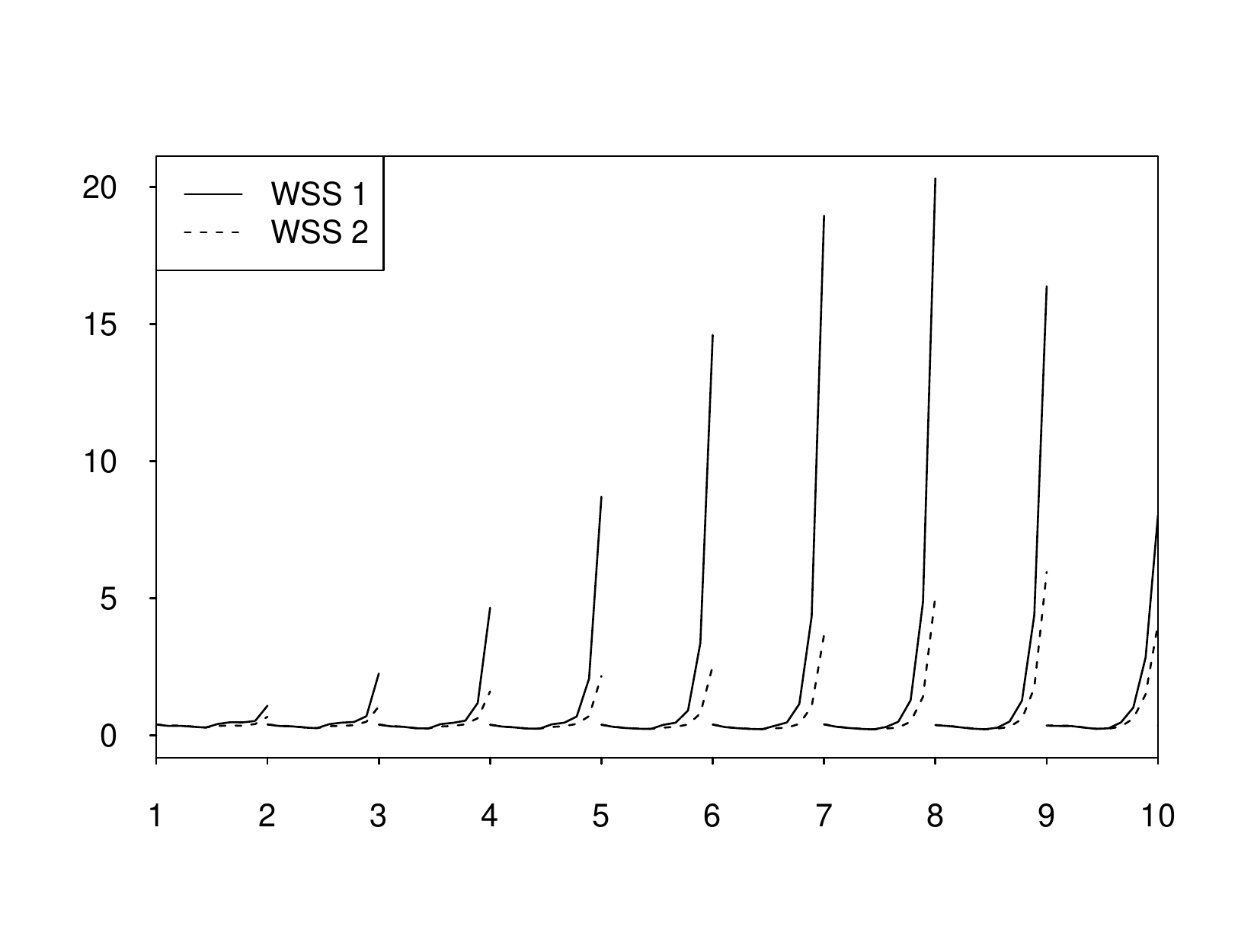}}\hspace{-0.7cm}
\hfill
\subfloat{\includegraphics[scale=0.49]{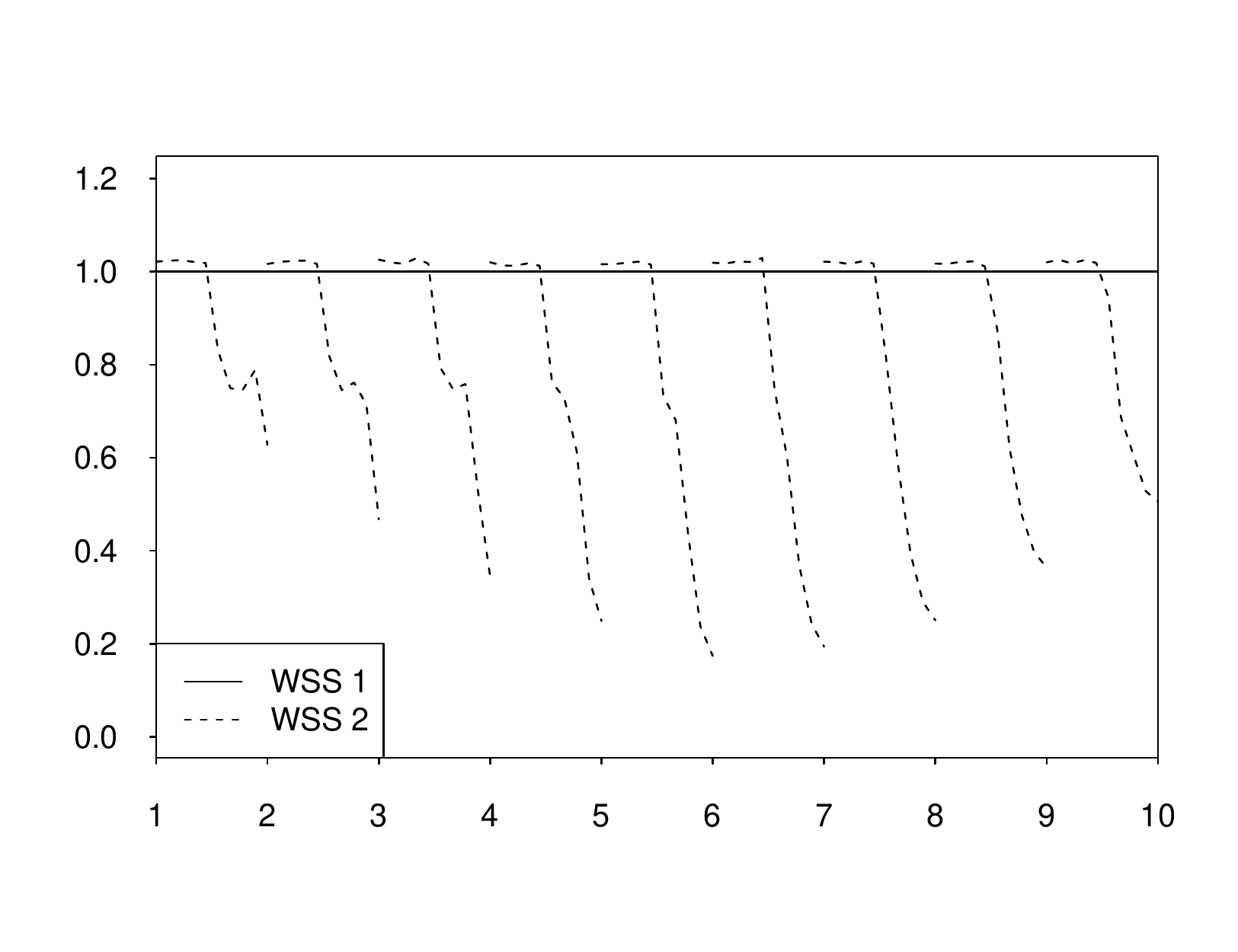}}

\caption{Average train time (left) and corresponding ratio (right) per grid point for different working set selection 
strategies using clipped duality gap criteria and initializing solver with warm start for data set \textsc{cal-housing}.
For \textsf{WSS 2}, 15 nearest neighbors are considered. The graphs comprises for $\tau=0.25$ (top), $\tau=0.50$ (middle) 
and $\tau=0.75$ (bottom).}
\label{figure-per grid time for WSS-cal-housing}
\end{scriptsize}
\end{figure}

\newpage
\vspace{0cm}
\begin{figure}[!ht]
\begin{scriptsize}
\vspace{-1cm}
\subfloat{\includegraphics[scale=0.49]{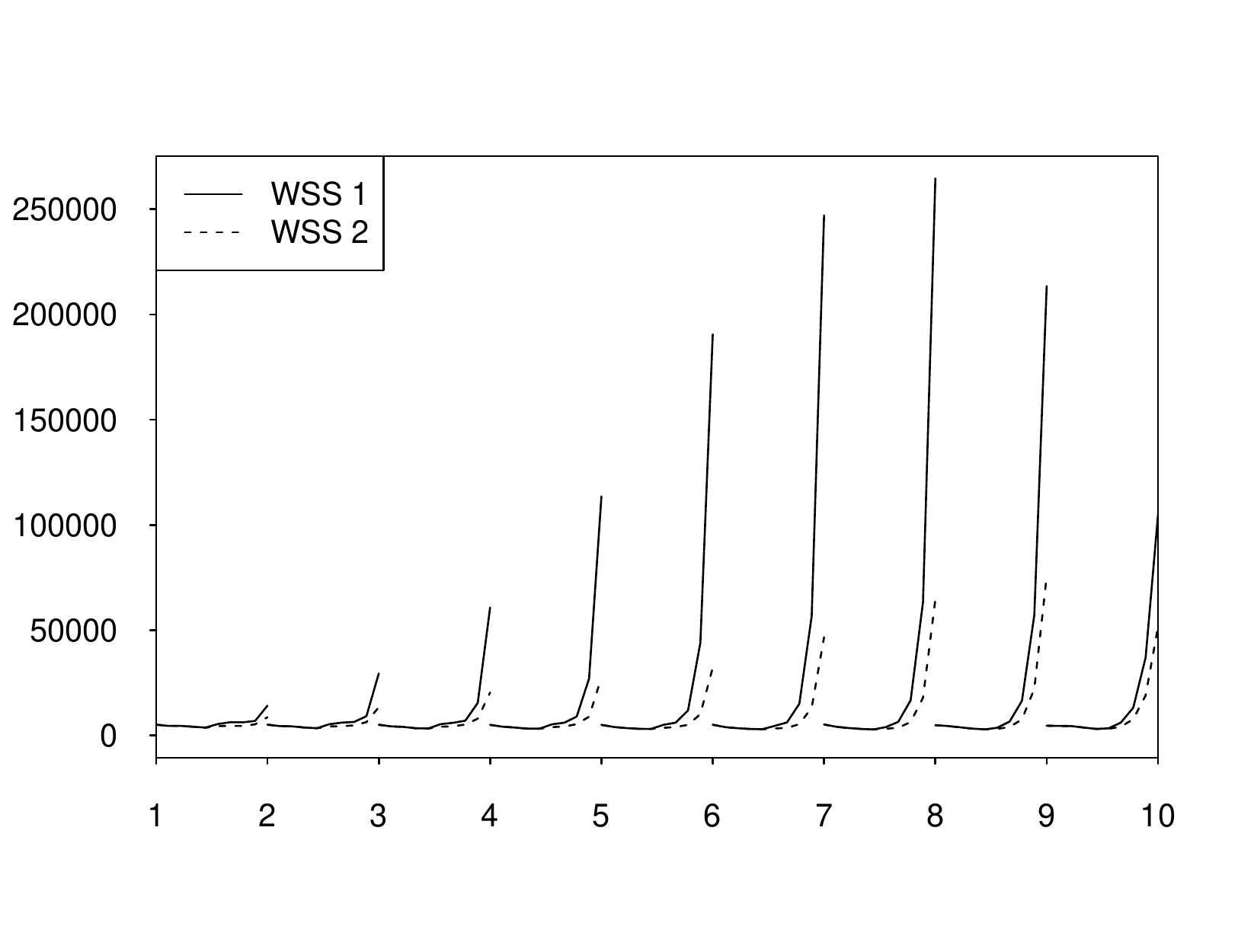}}\hspace{-0.7cm}
\hfill
\subfloat{\includegraphics[scale=0.49]{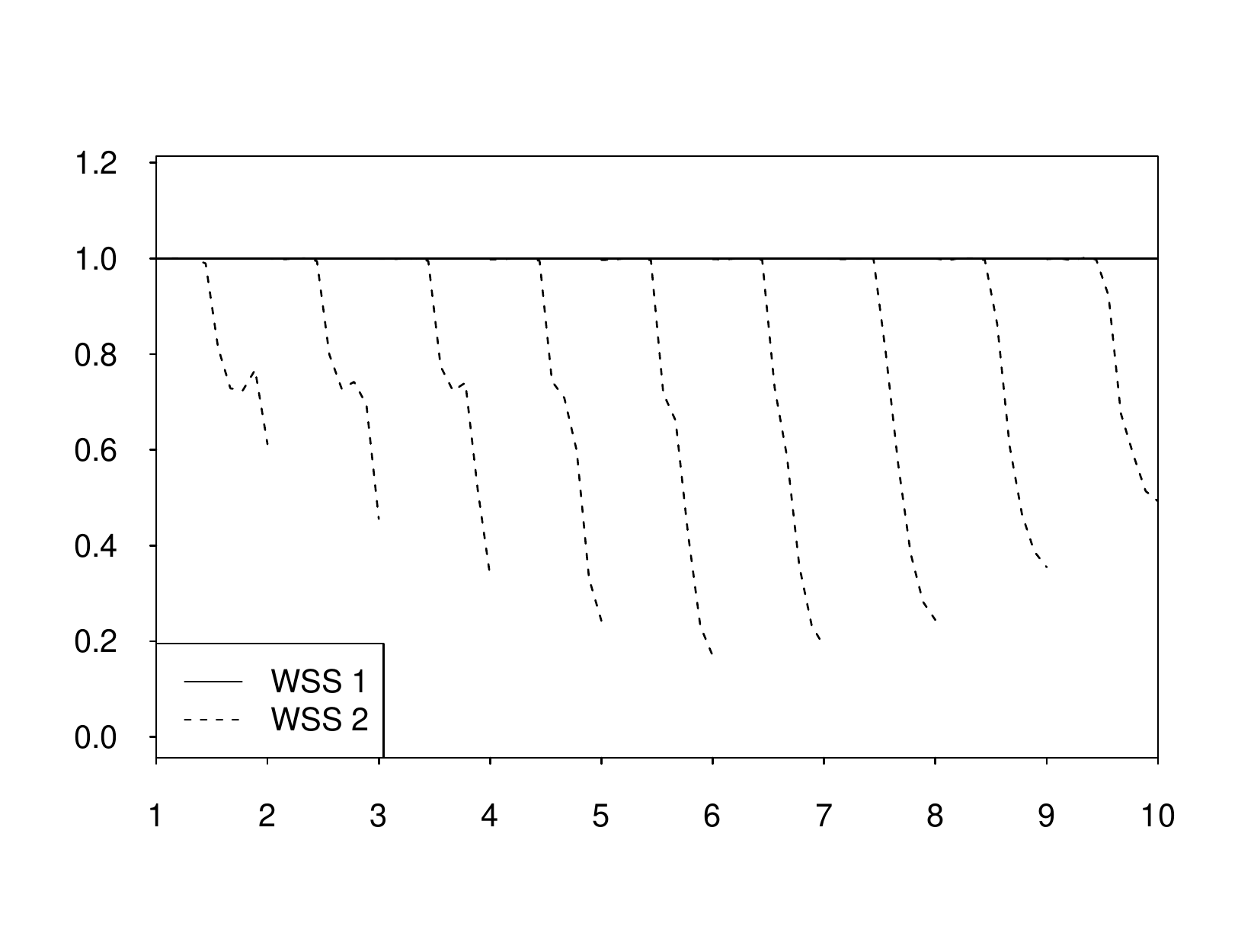}}
\vspace{-1.9cm}
\subfloat{\includegraphics[scale=0.49]{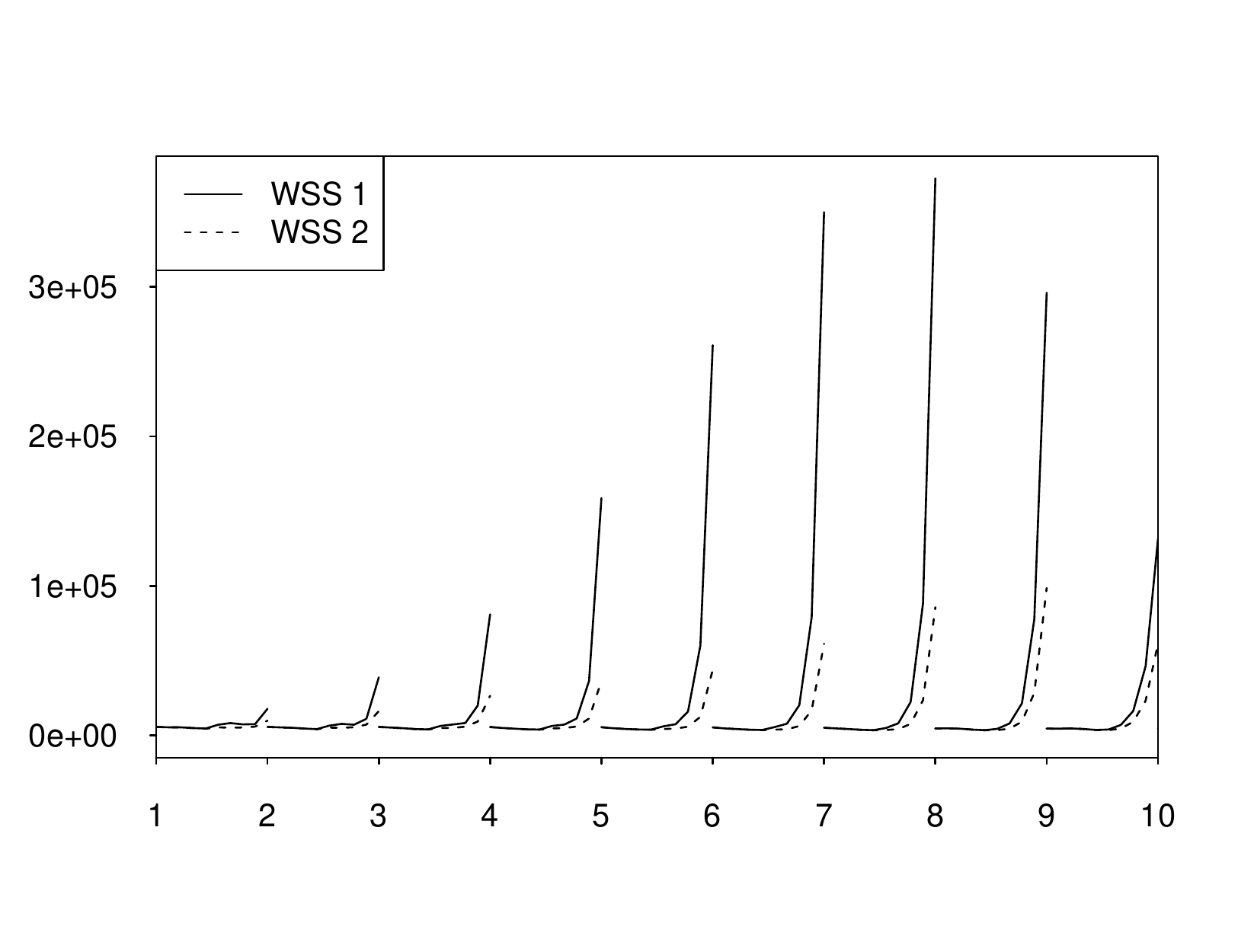}}\hspace{-0.7cm}
\hfill
\subfloat{\includegraphics[scale=0.49]{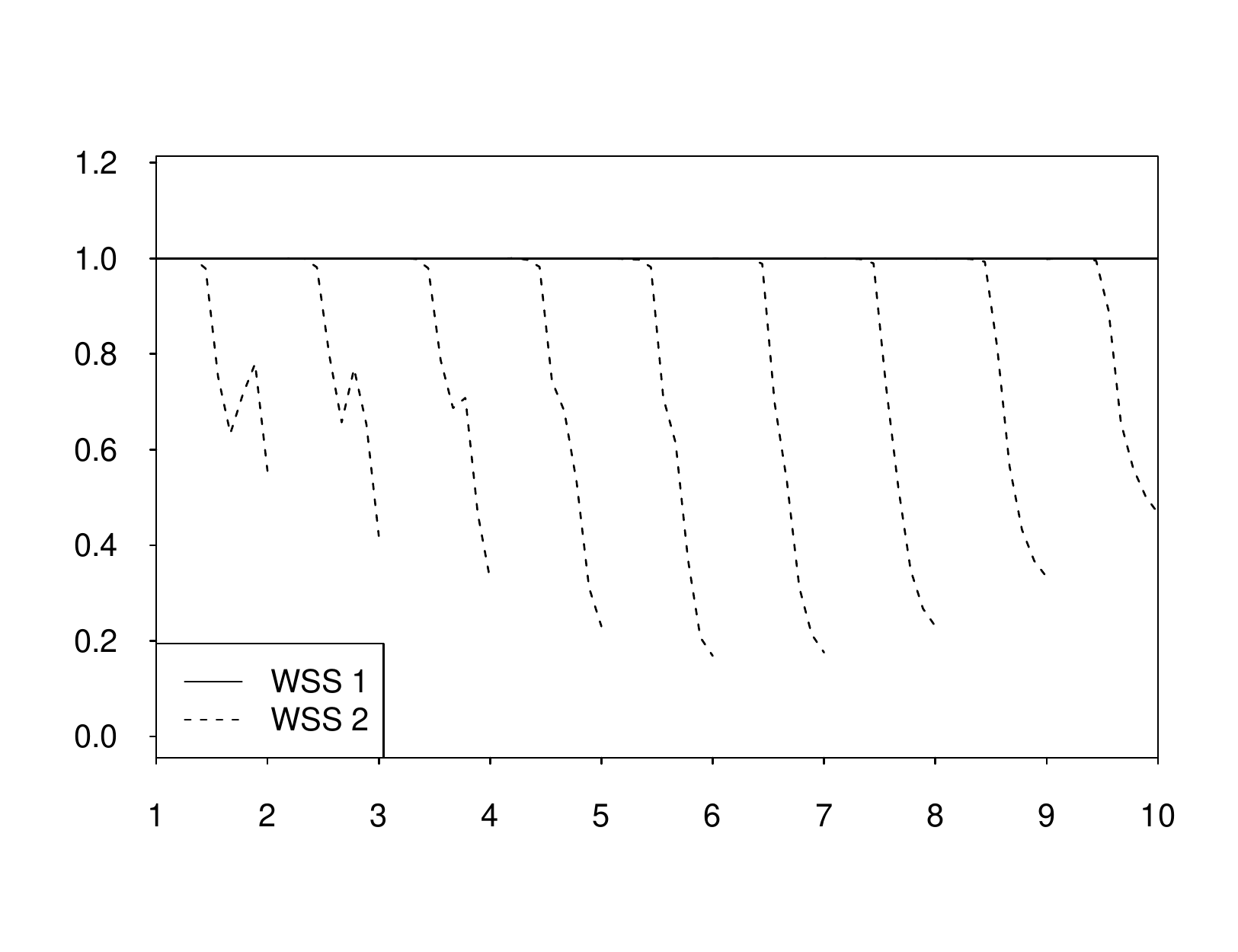}}
\vspace{-1.9cm}
\subfloat{\includegraphics[scale=0.49]{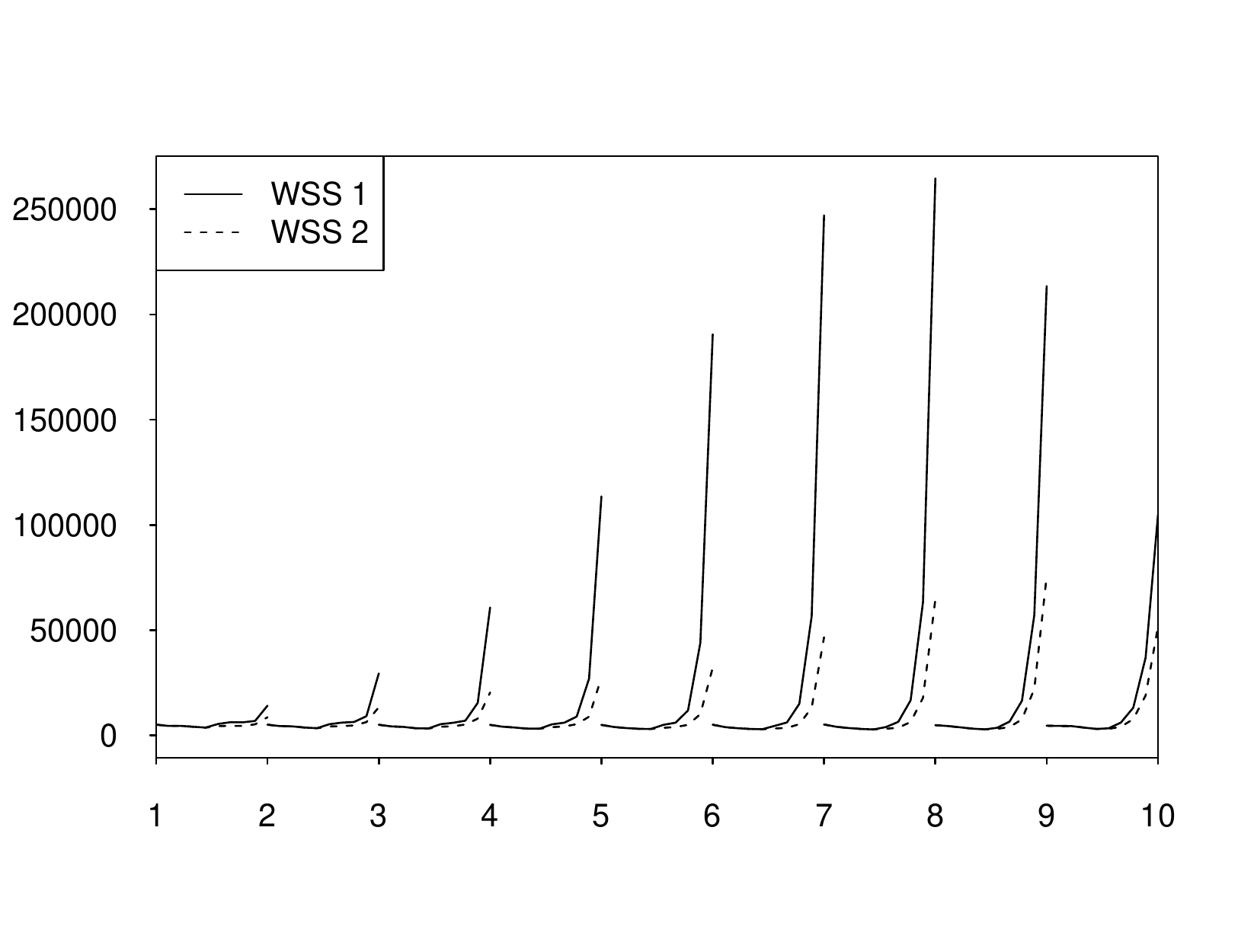}}\hspace{-0.7cm}
\hfill
\subfloat{\includegraphics[scale=0.49]{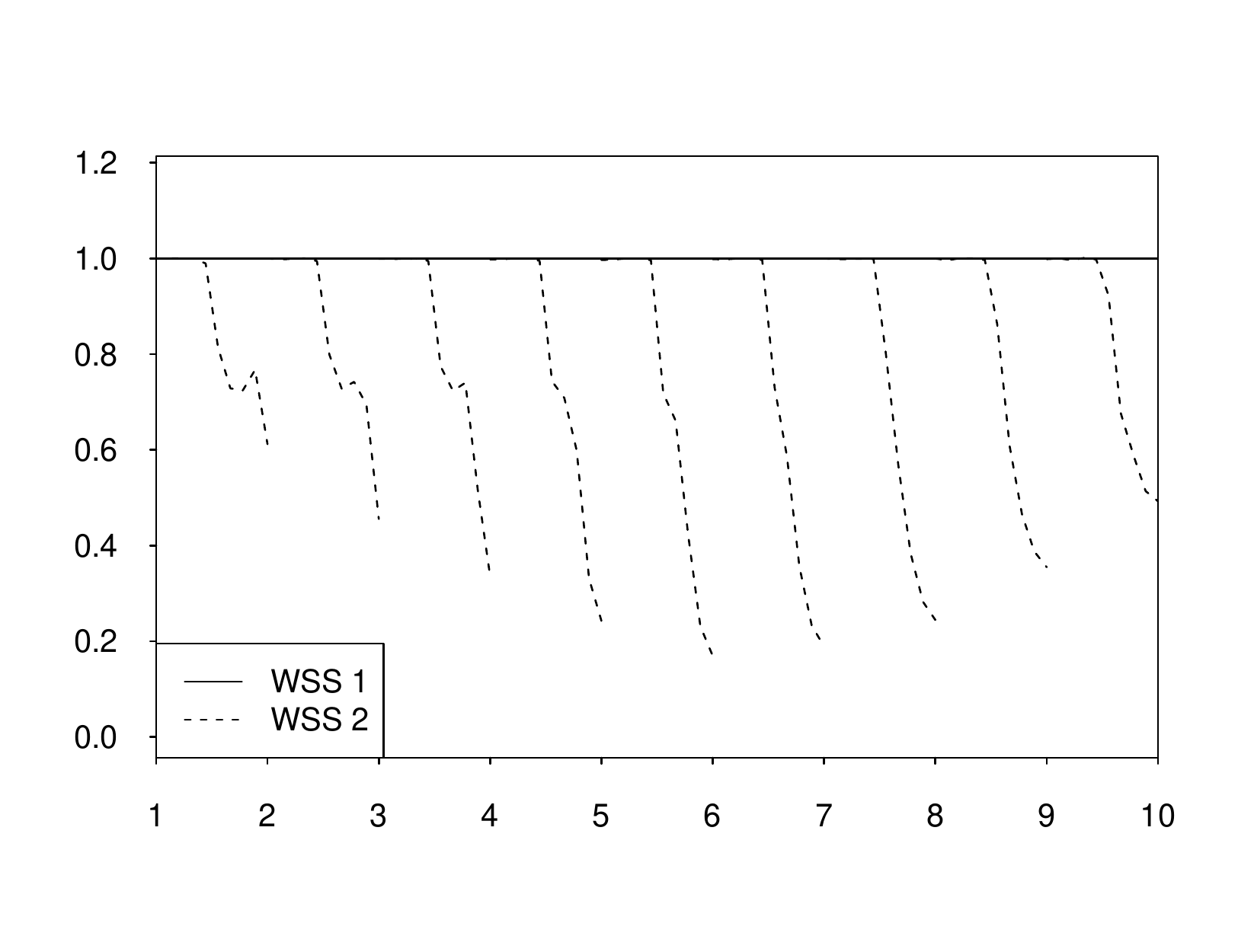}}

\caption{Average number of iterations (left) and corresponding ratio (right) per grid point for different working set 
selection strategies using clipped duality gap criteria and initializing solver with warm start for data set \textsc{cal-housing}.
For \textsf{WSS 2}, 15 nearest neighbors are considered. The graphs comprises for $\tau=0.25$ (top), $\tau=0.50$ (middle)
and $\tau=0.75$(bottom).}
\label{figure-per grid iter for WSS-cal-housing}
\end{scriptsize}
\end{figure}

\newpage
\subsection{Results for Different Number of Nearest Neighbors}

\begin{figure}[!ht]
\begin{scriptsize}
 \subfloat{\includegraphics[scale=0.52]{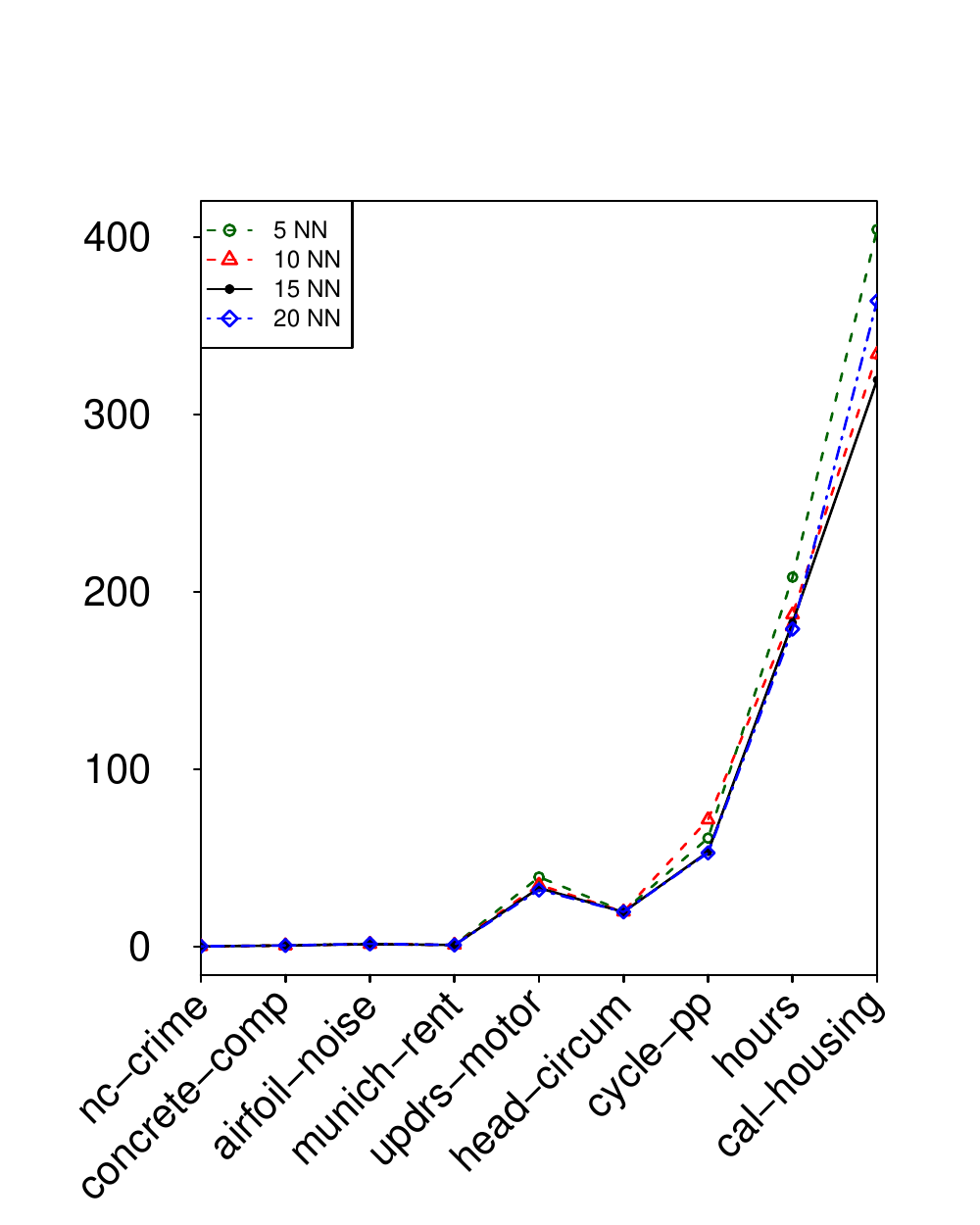}}\hspace{-0.7cm}
 \hfill
 \subfloat{\includegraphics[scale=0.52]{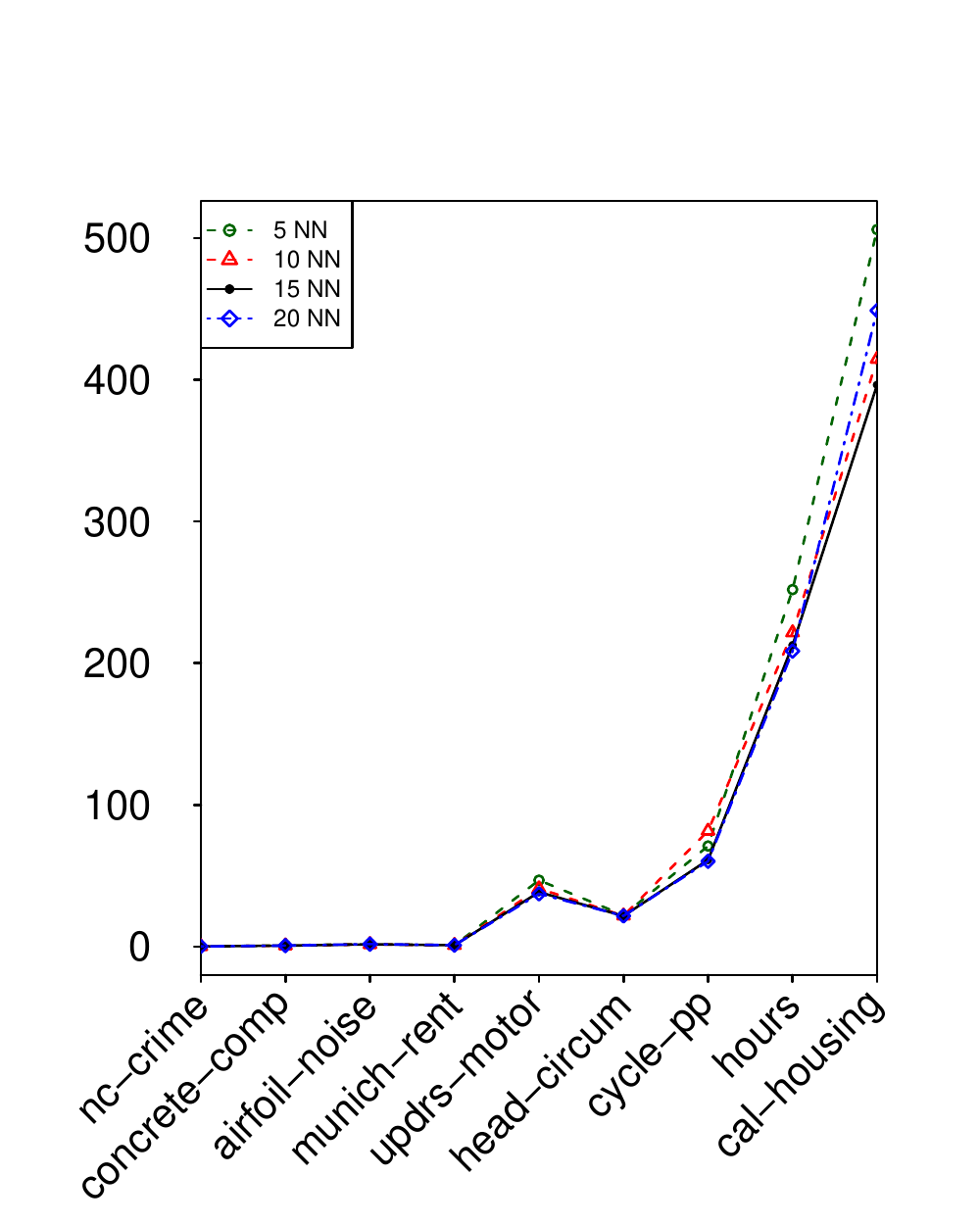}}\hspace{-0.7cm}
\hfill
 \subfloat{\includegraphics[scale=0.52]{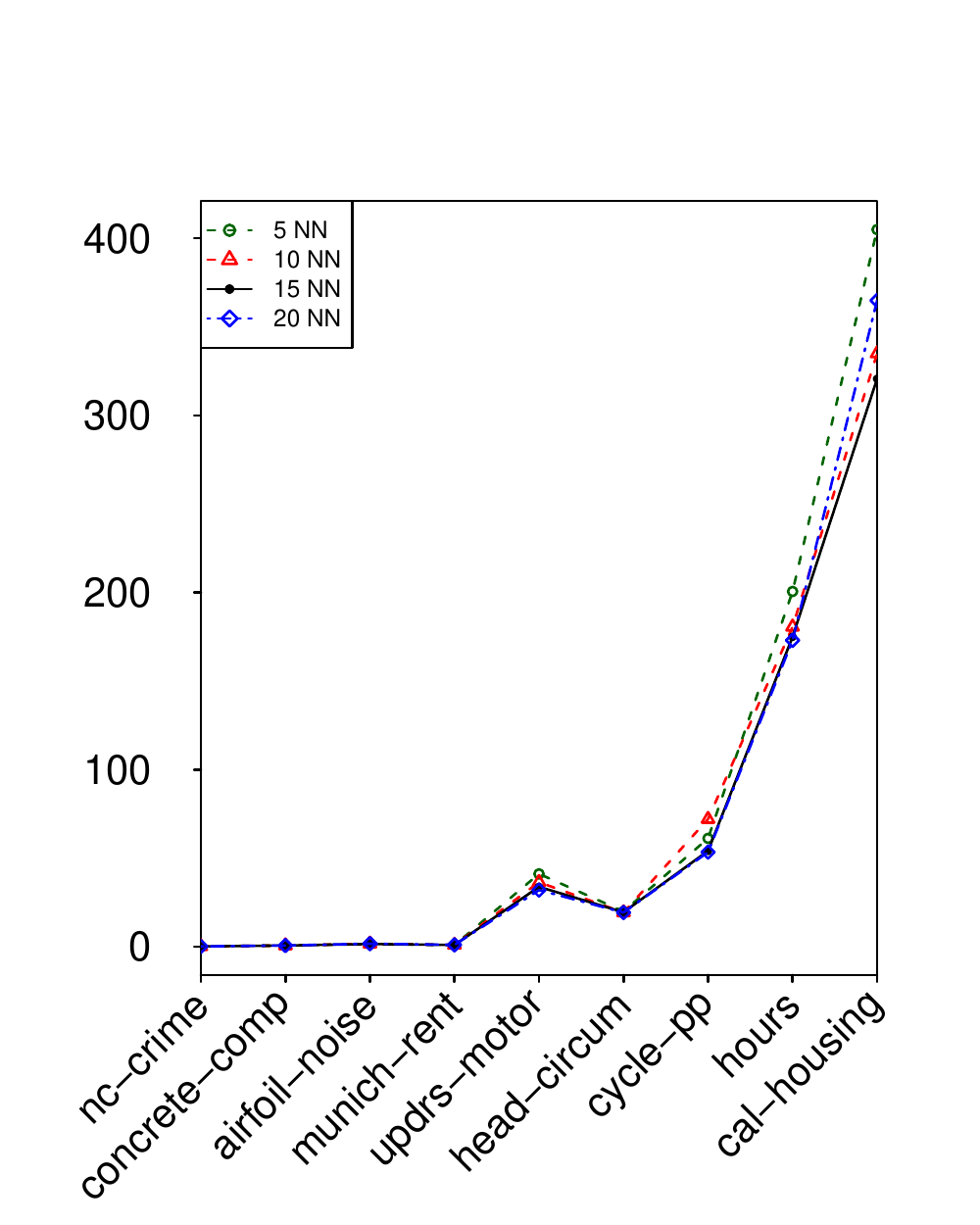}}
 
 \vspace{-1.0cm}
\subfloat{\includegraphics[scale=0.52]{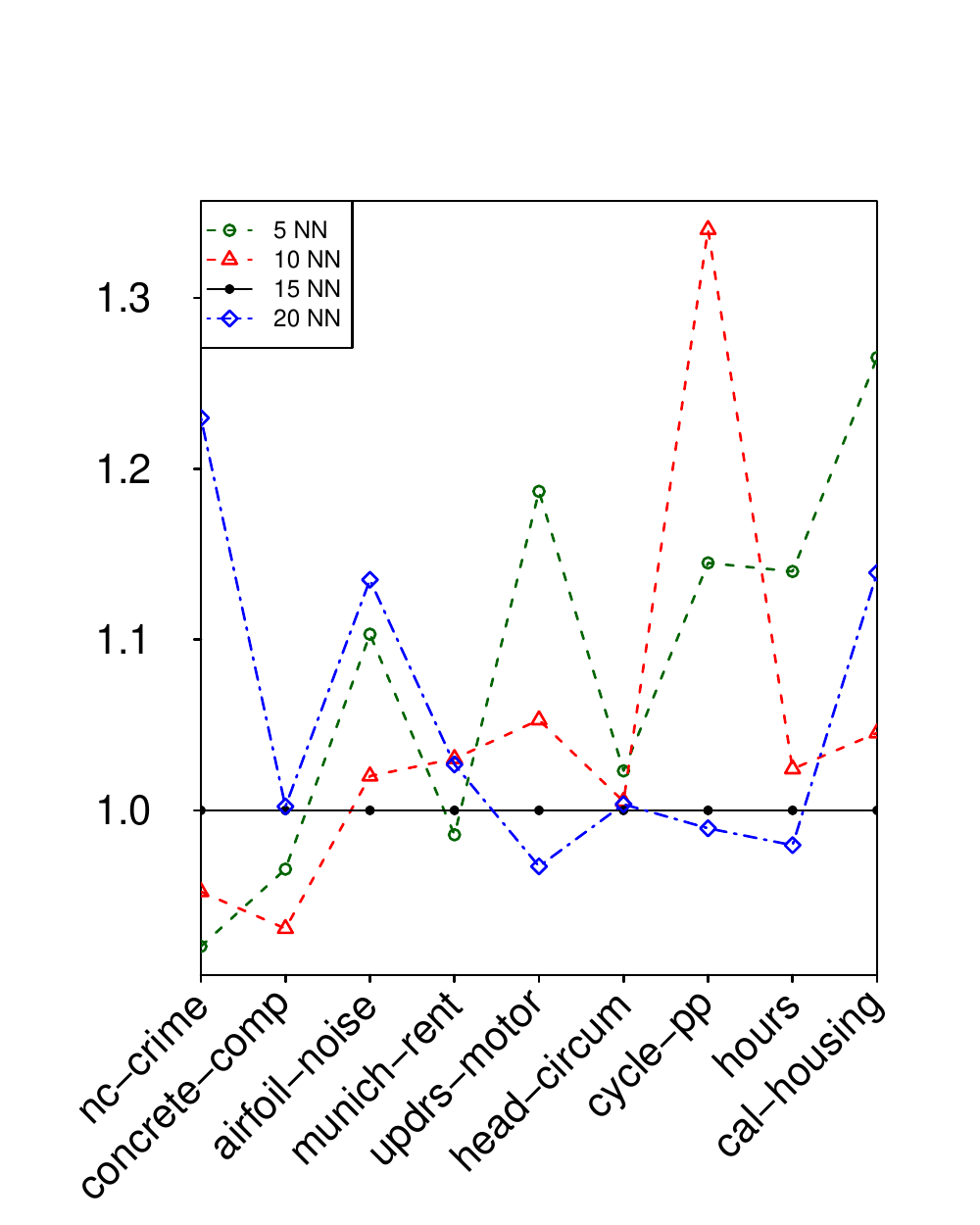}}\hspace{-0.7cm}
 \hfill
 \subfloat{\includegraphics[scale=0.52]{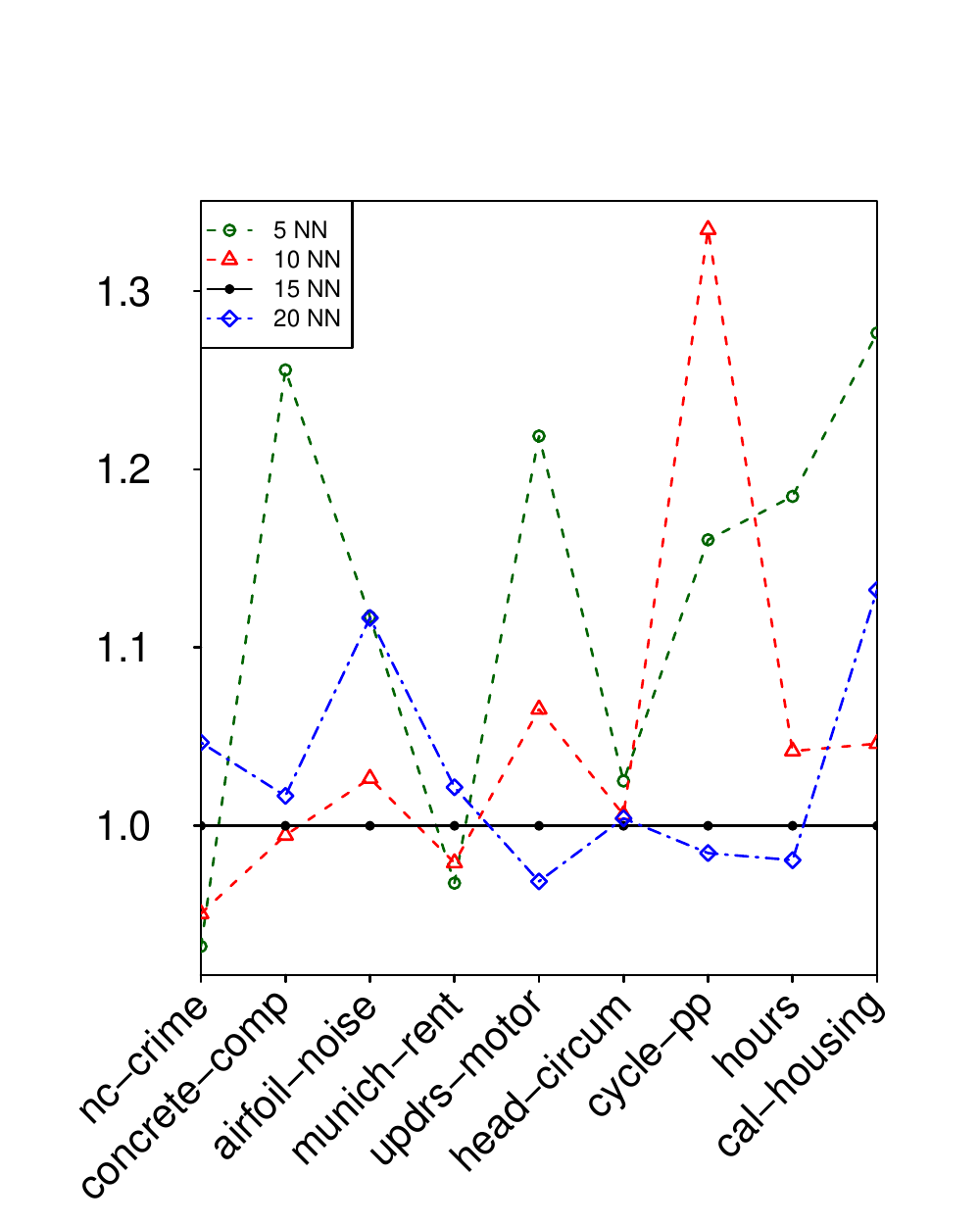}}\hspace{-0.7cm}
\hfill
 \subfloat{\includegraphics[scale=0.52]{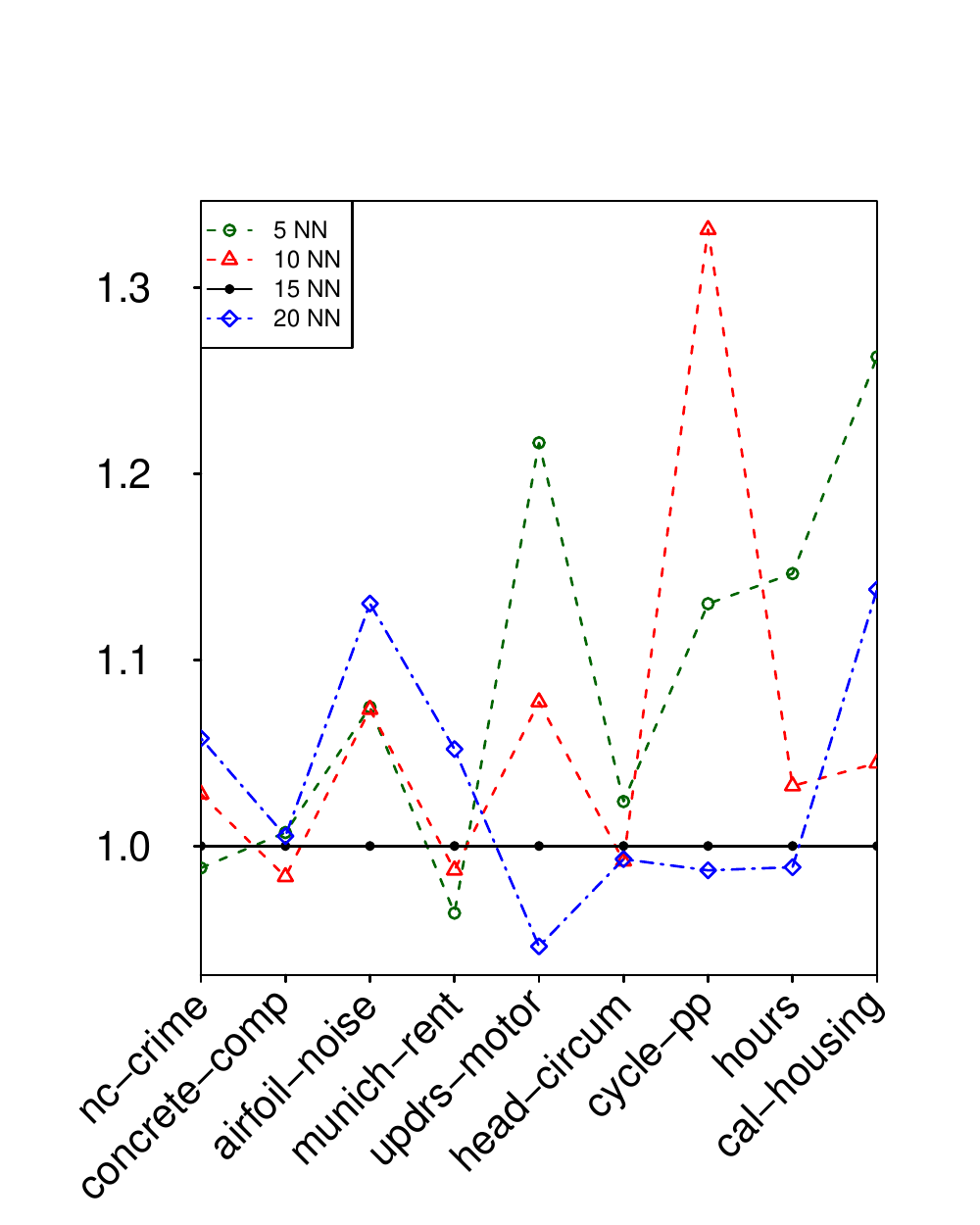}}
 \caption{Train time (top) and corresponding ratio (bottom) of different data sets for different number of nearest neighbors 
 after fixing warm start initialization and stopping criteria with clipped duality gap. The graphs comprises of $\tau=0.25$ (left),
 $\tau=0.50$ (middle) and $\tau=0.75$ (right).}
 \label{figure-time and ratio-NN vs datasets}
 \end{scriptsize}
\end{figure}

\newpage
\begin{figure}[!ht]
\begin{scriptsize}
 \subfloat{\includegraphics[scale=0.52]{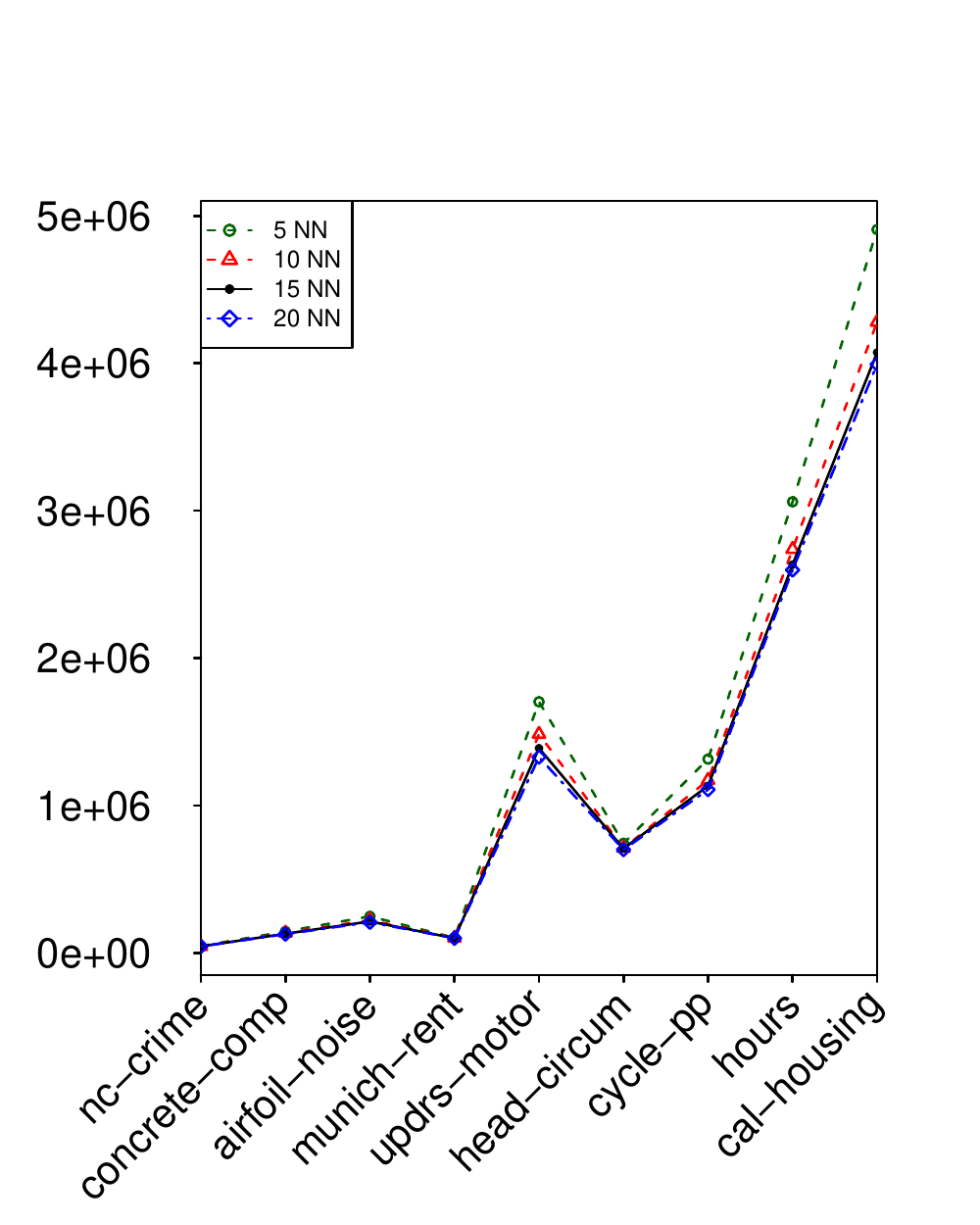}}\hspace{-0.7cm}
 \hfill
 \subfloat{\includegraphics[scale=0.52]{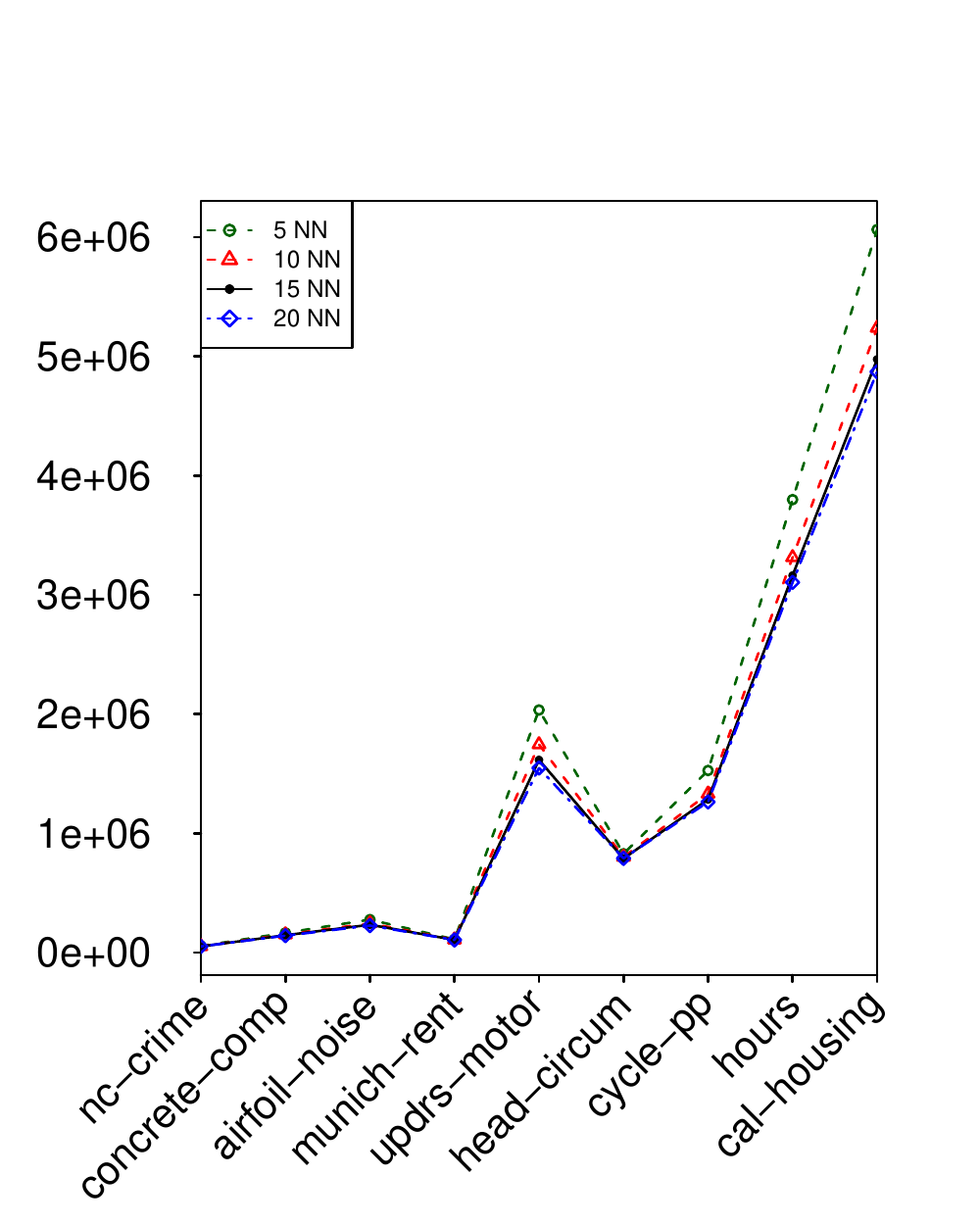}}\hspace{-0.7cm}
\hfill
 \subfloat{\includegraphics[scale=0.52]{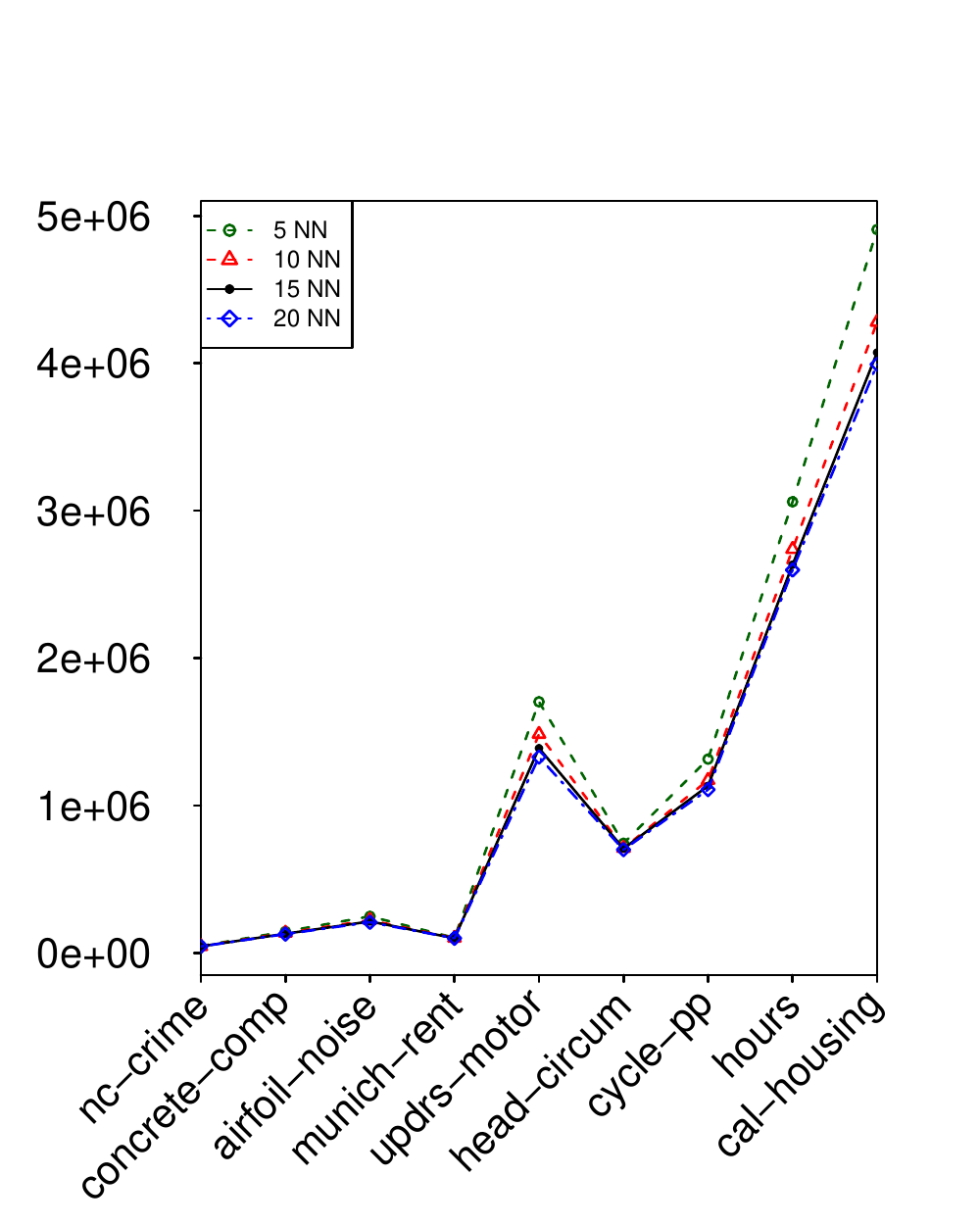}}
 
 \vspace{-1.0cm}
\subfloat{\includegraphics[scale=0.52]{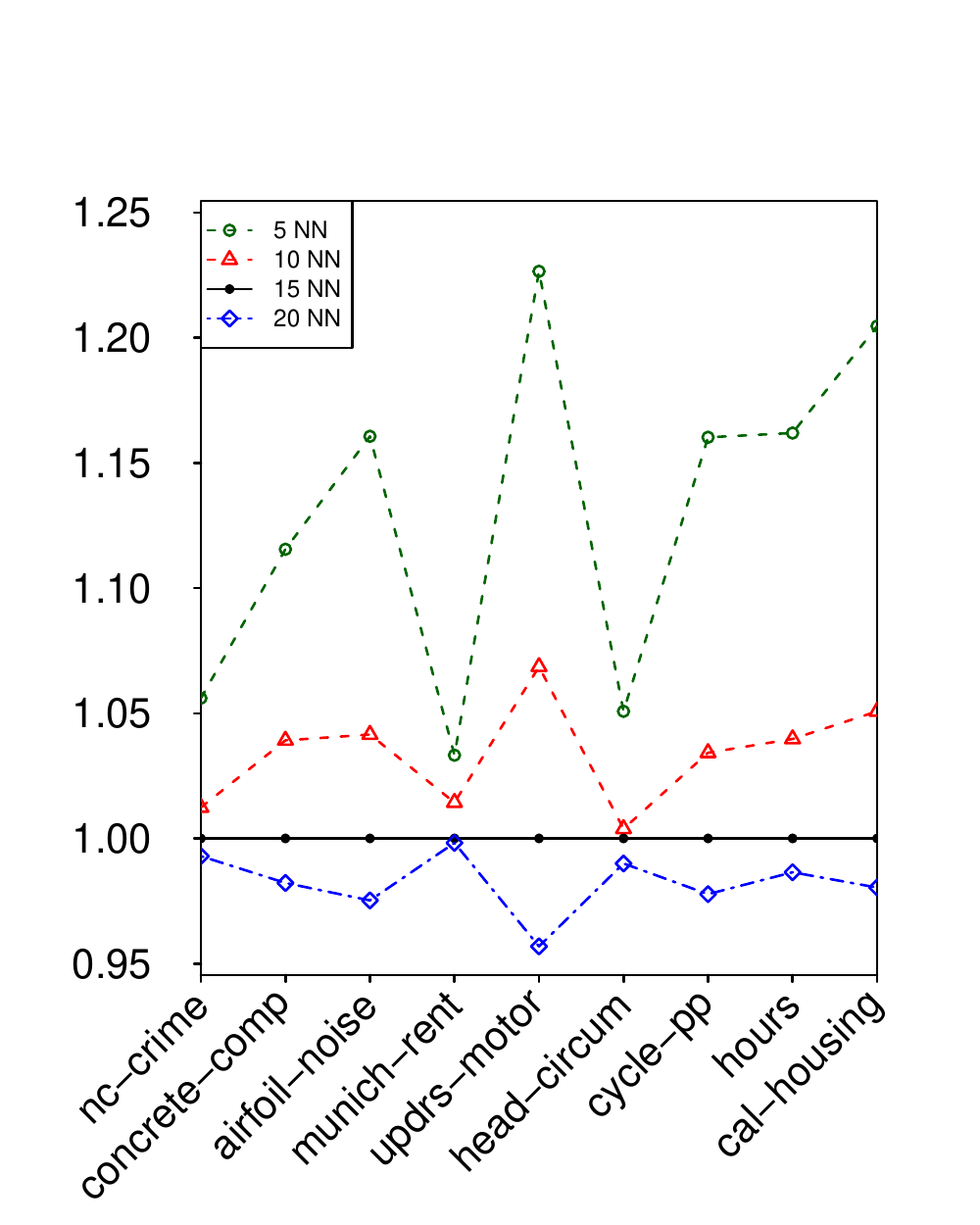}}\hspace{-0.7cm}
 \hfill
 \subfloat{\includegraphics[scale=0.52]{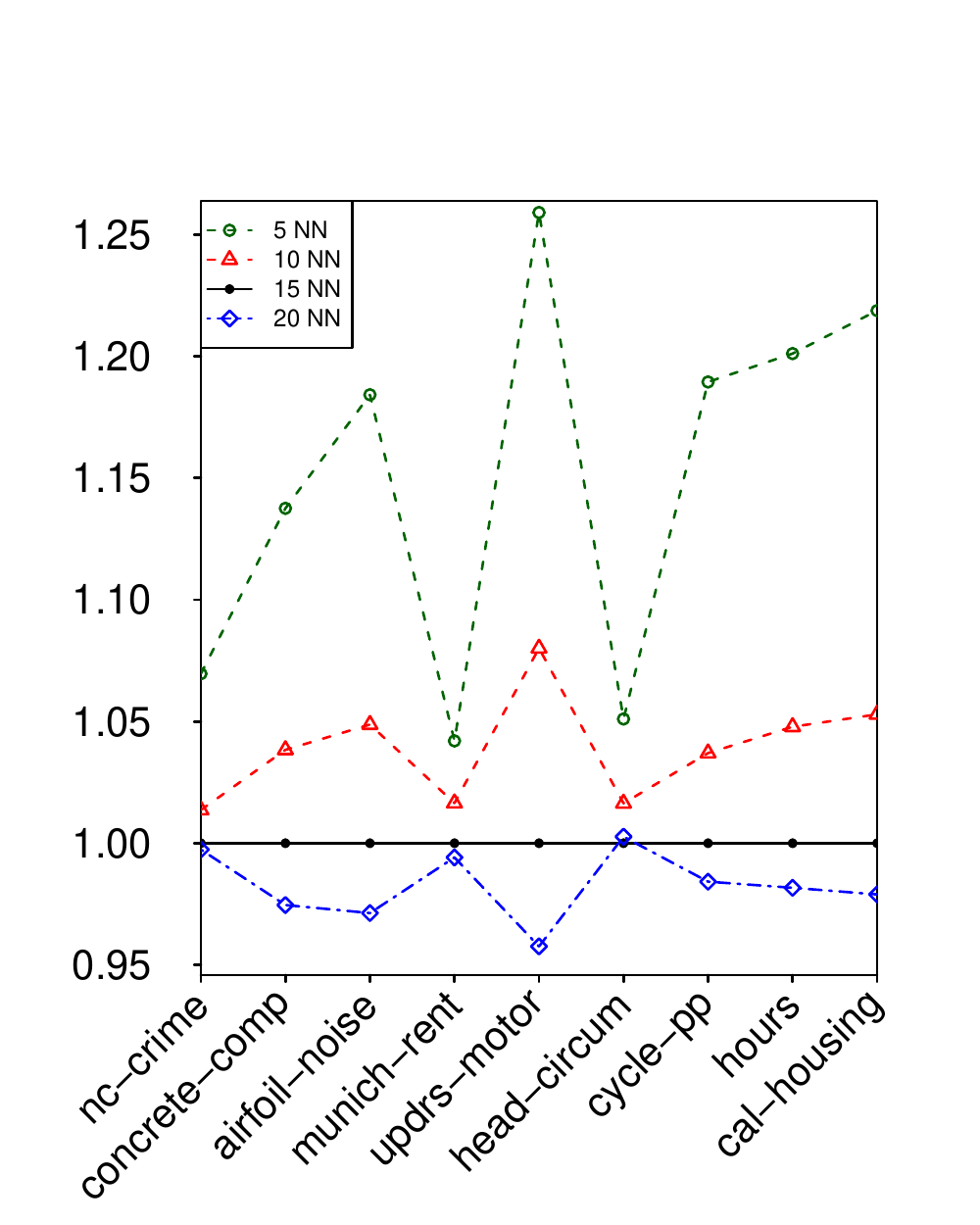}}\hspace{-0.7cm}
\hfill
 \subfloat{\includegraphics[scale=0.52]{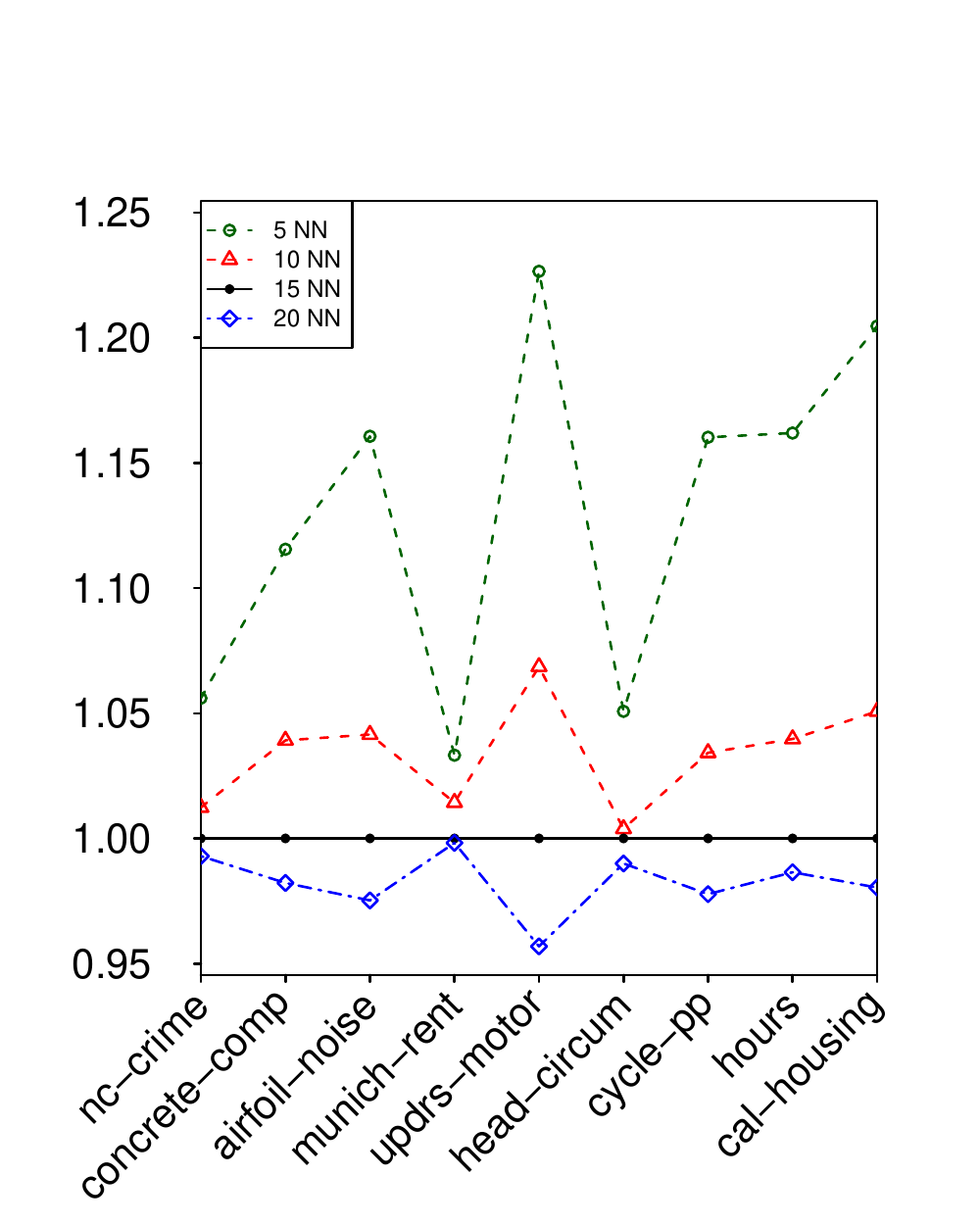}}
 \caption{Train iterations (top) and corresponding ratio (bottom) of different data sets for different number of nearest 
 neighbors after fixing with warm start initialization and stopping criteria with clipped duality gap. The graphs comprises 
 of $\tau=0.25$ (left), $\tau=0.50$ (middle) and $\tau=0.75$ (right).}
  \label{figure-iter and ratio-NN vs datasets}
\end{scriptsize}
\end{figure}

\newpage
\vspace{-1cm}

\begin{figure}[!ht]
\begin{scriptsize}
 \subfloat{\includegraphics[scale=0.49]{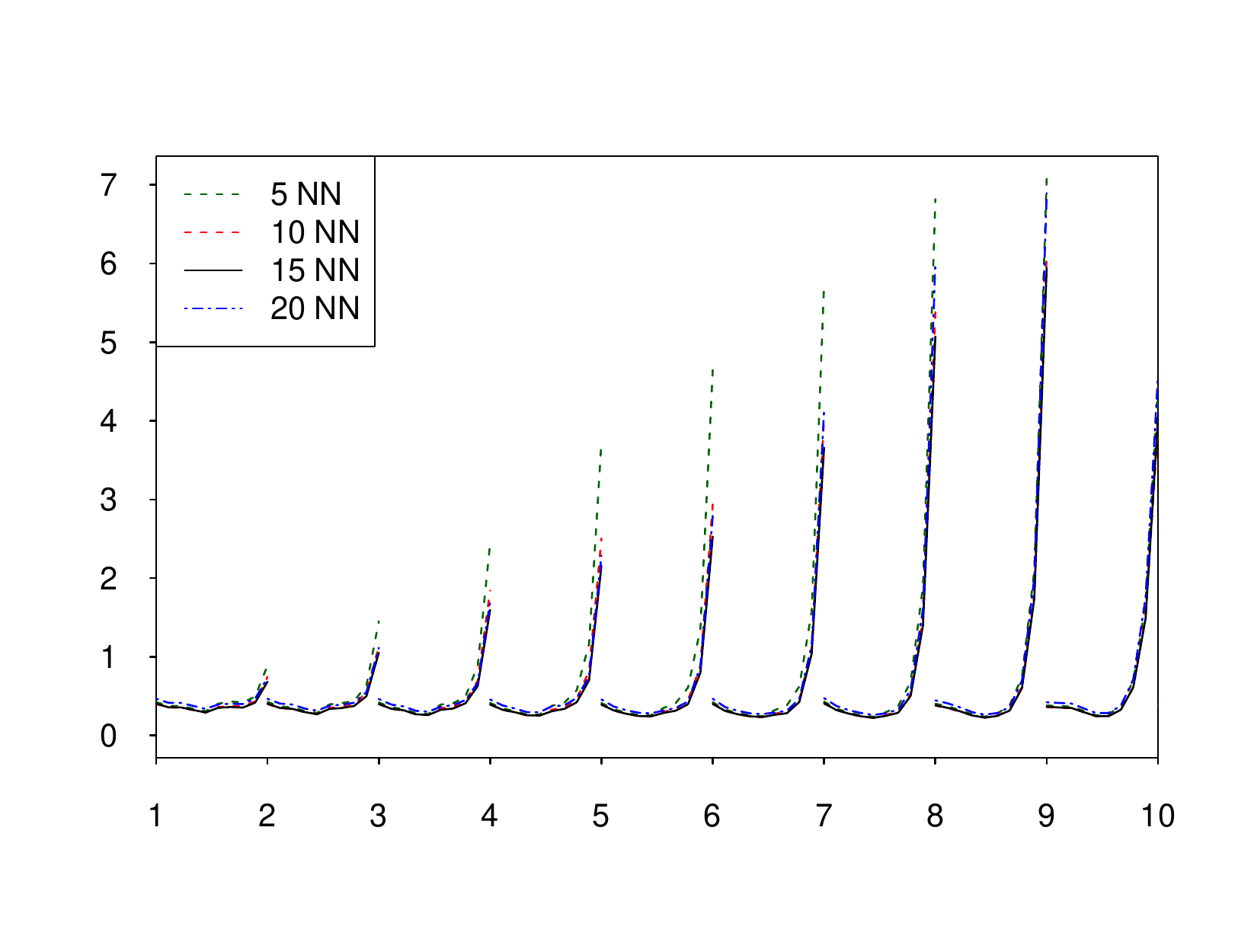}}\hspace{-0.7cm}
 \hfill
 \subfloat{\includegraphics[scale=0.49]{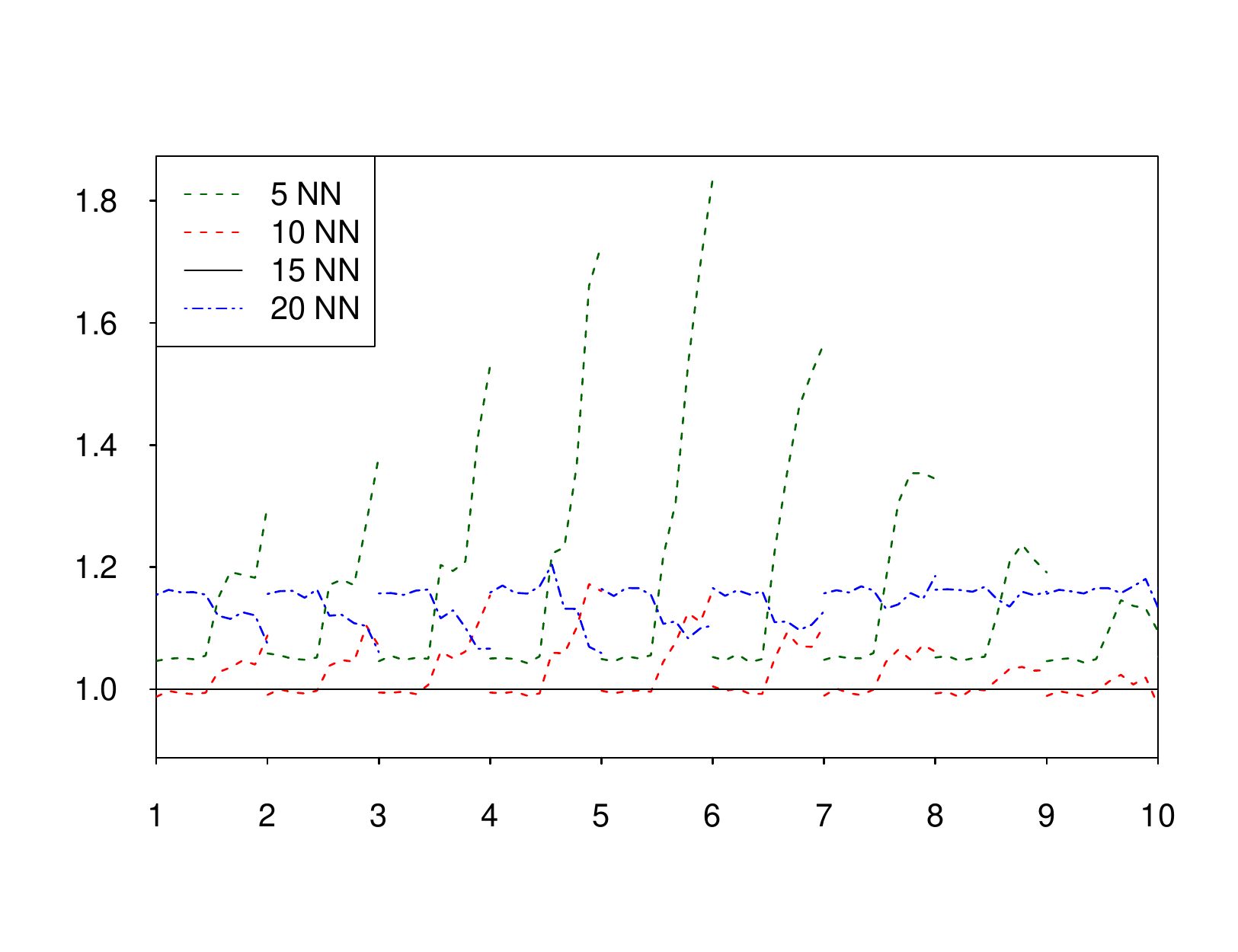}}
 \vspace{-1.9cm}
\subfloat{\includegraphics[scale=0.49]{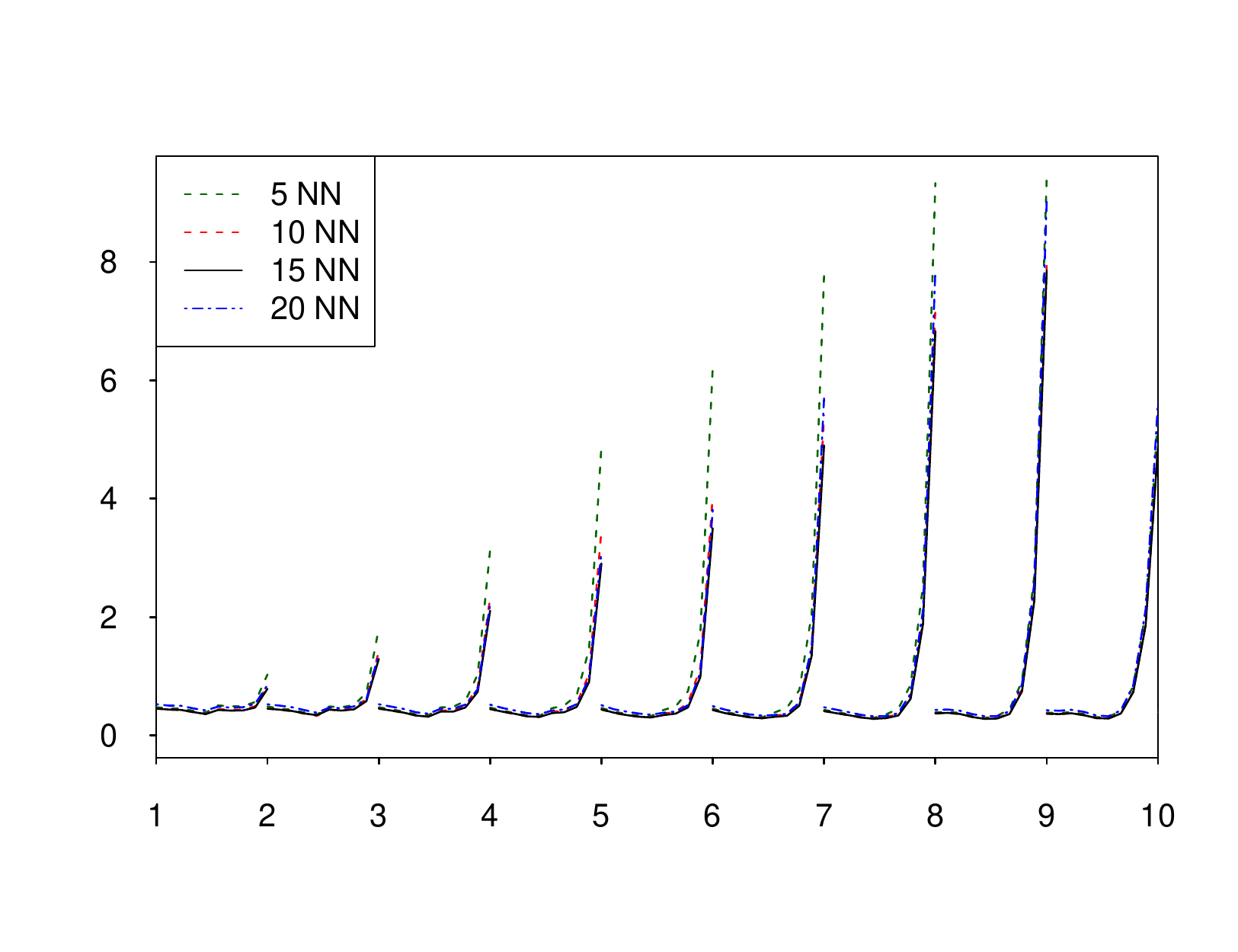}}\hspace{-0.7cm}
\hfill
\subfloat{\includegraphics[scale=0.49]{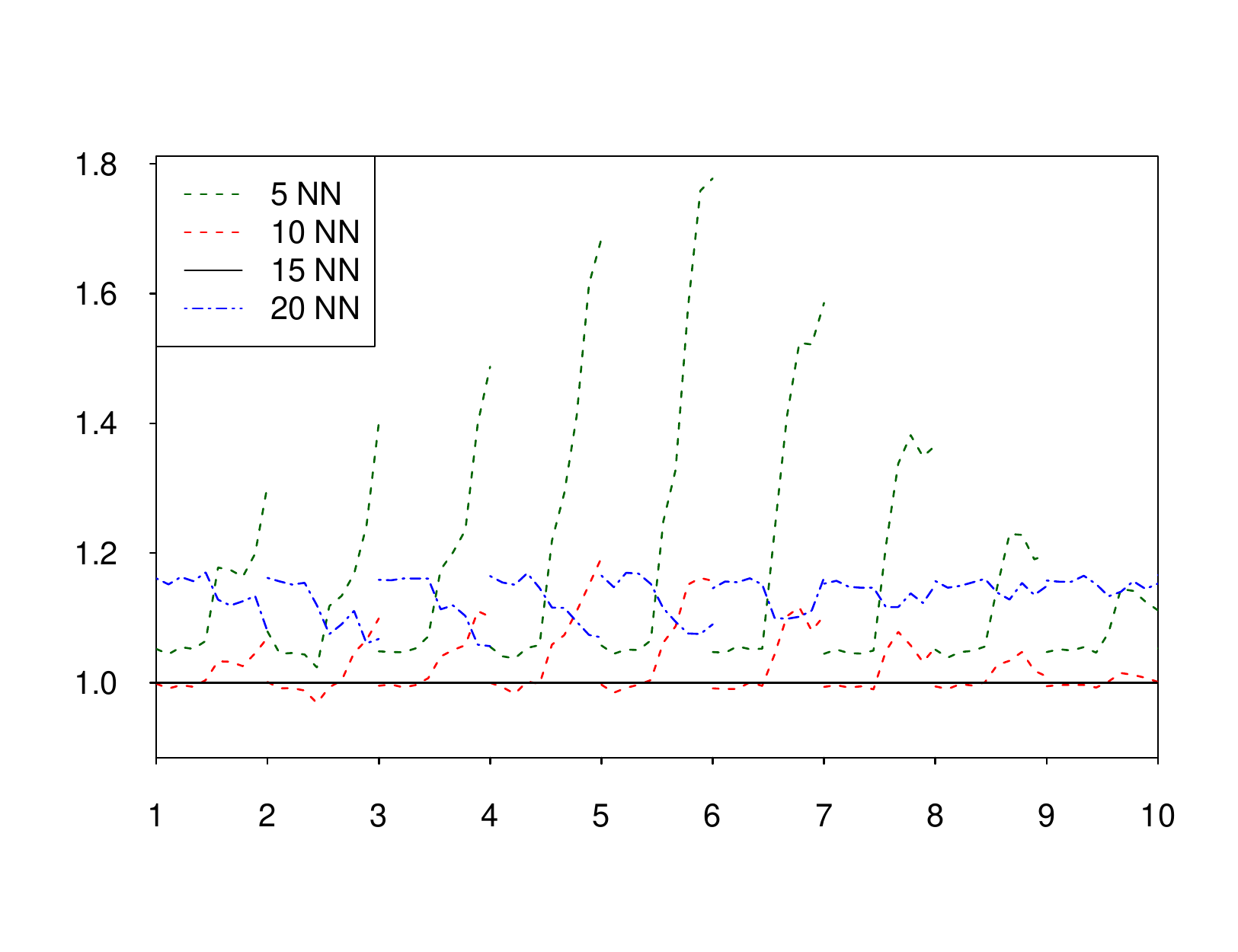}}
\vspace{-1.9cm}
\subfloat{\includegraphics[scale=0.49]{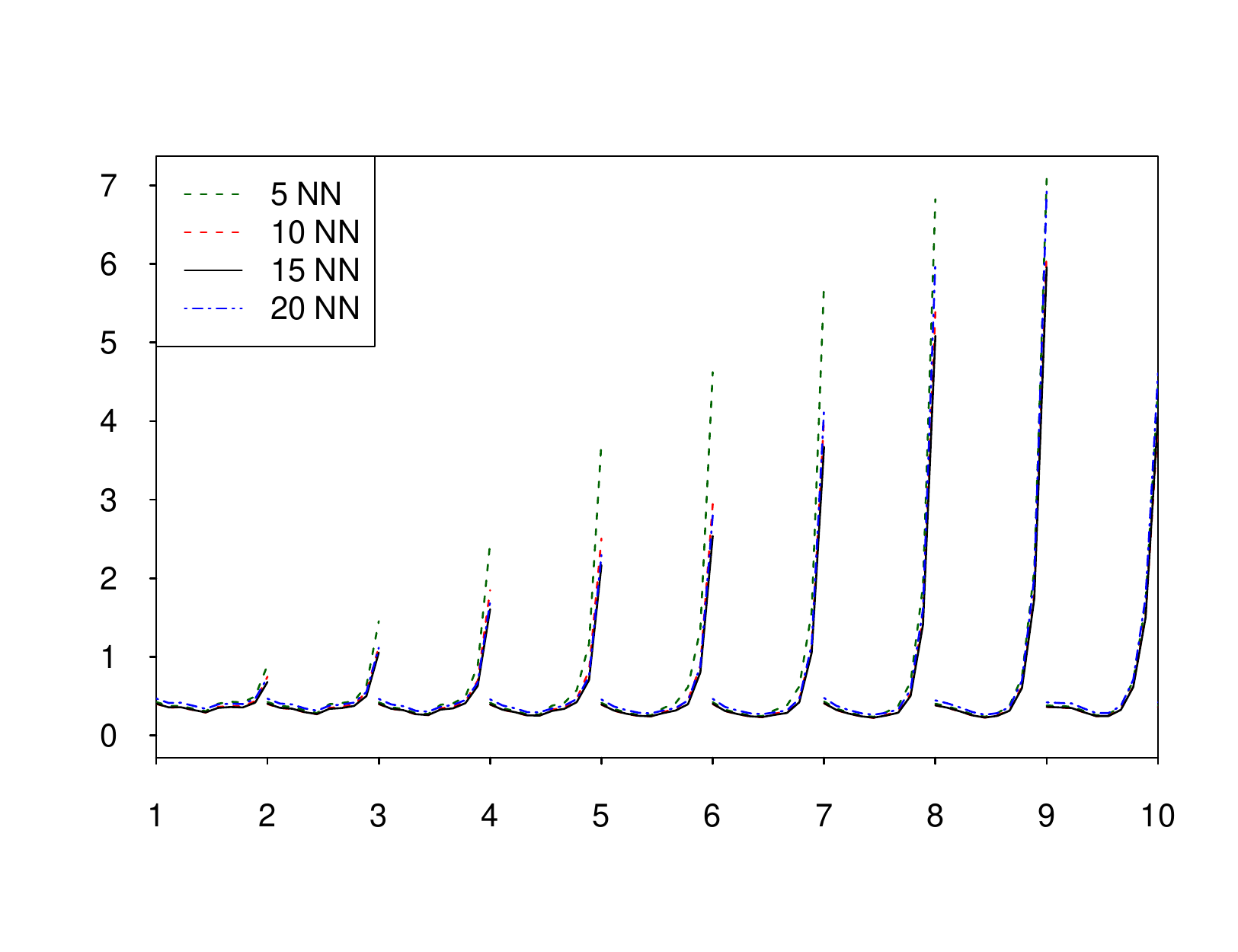}}\hspace{-0.7cm}
\hfill
\subfloat{\includegraphics[scale=0.49]{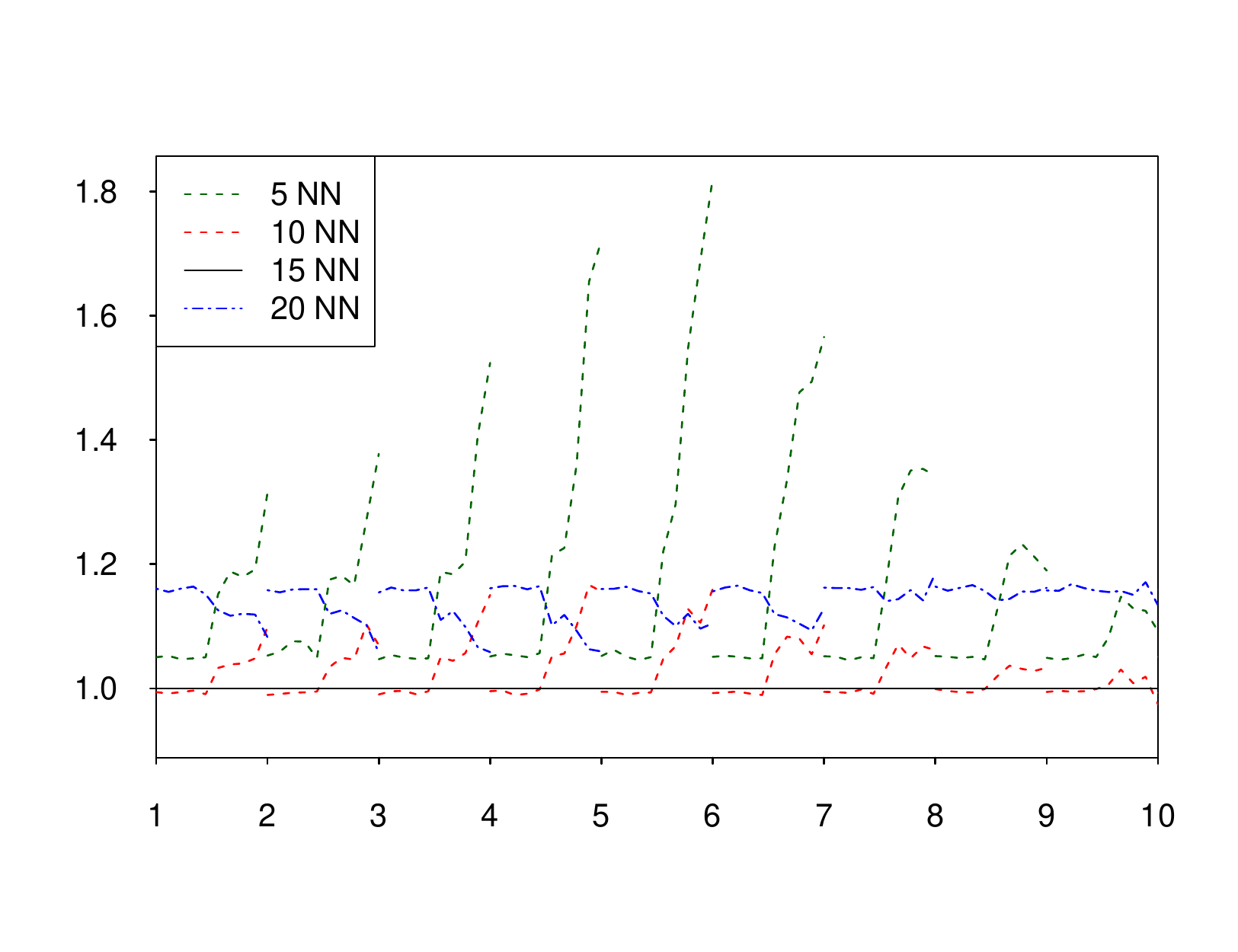}}

\caption{Average train time (left) and corresponding ratio (right) per grid point for different  different number of nearest 
neighbors considering warm start and stopping criteria with clipped duality gap for the data set \textsc{cal-housing}. 
The graphs comprises for $\tau=0.25$ (top), $\tau=0.50$ (middle) and $\tau=0.75$ (bottom).}
  \label{figure-per-grid time for NN-cal-housing}
\end{scriptsize}
\end{figure}
\newpage
\vspace{0cm}
\begin{figure}[!ht]
\begin{scriptsize}
\vspace{-1cm}
 \subfloat{\includegraphics[scale=0.49]{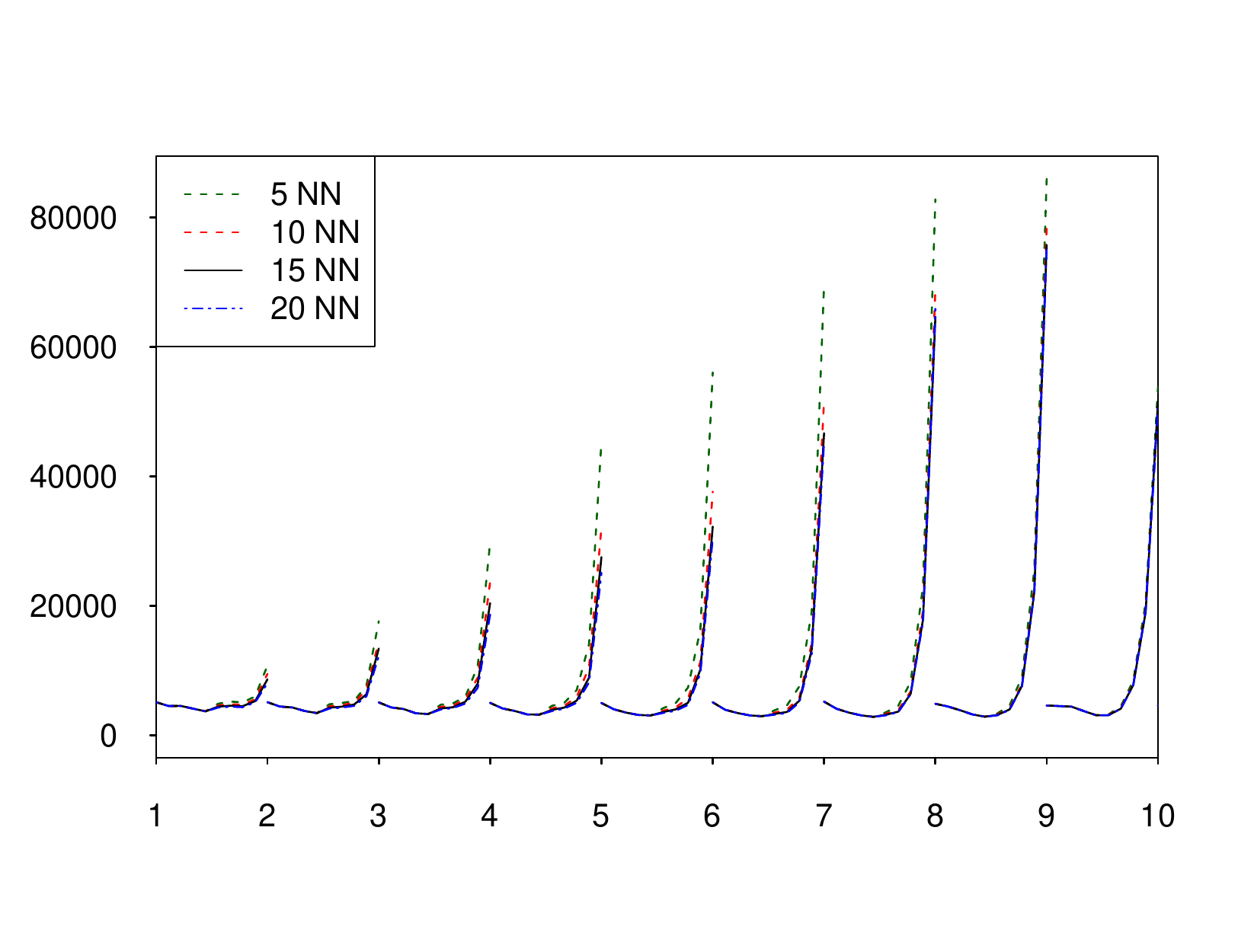}}\hspace{-0.7cm}
 \hfill
 \subfloat{\includegraphics[scale=0.49]{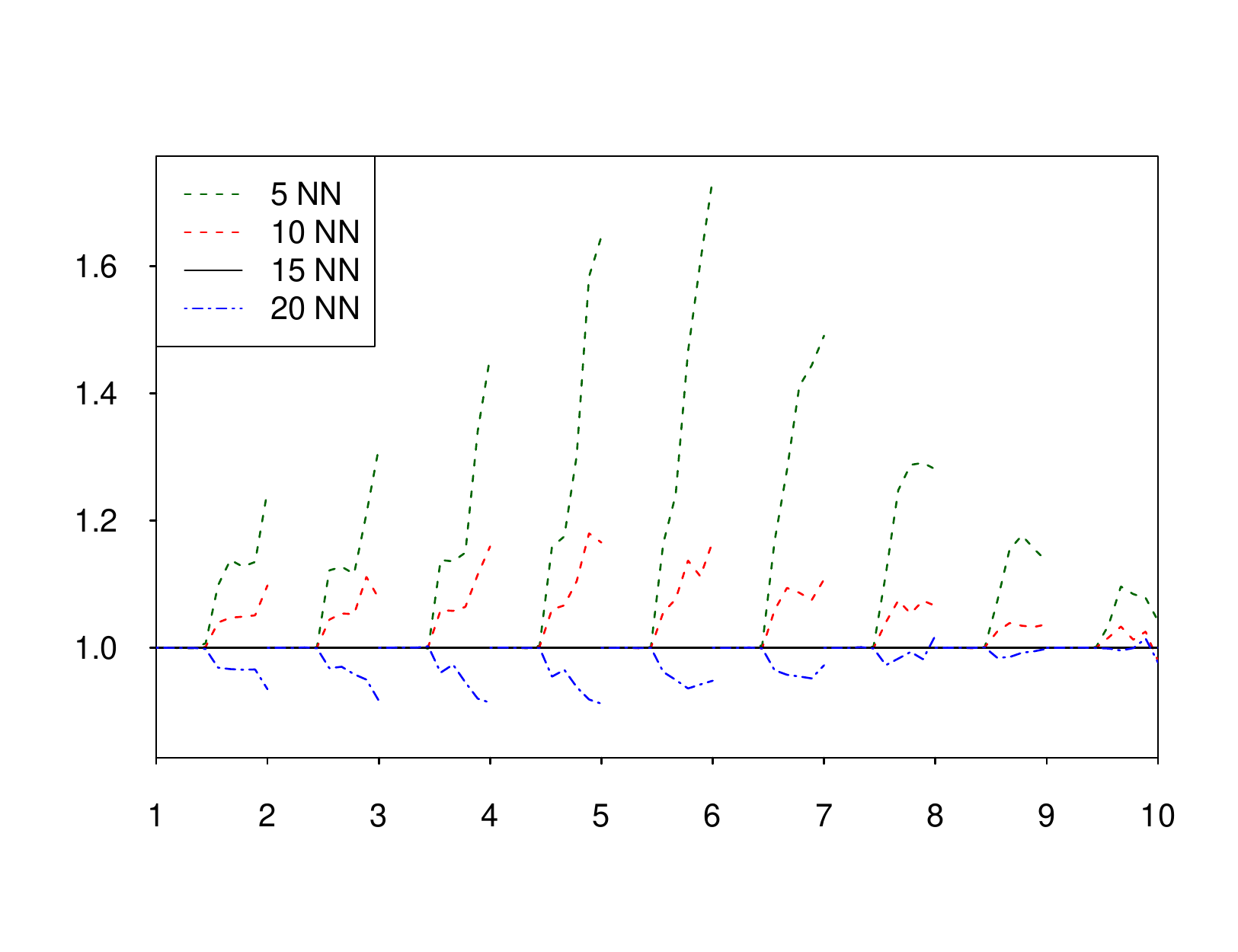}}
 \vspace{-1.9cm}
\subfloat{\includegraphics[scale=0.49]{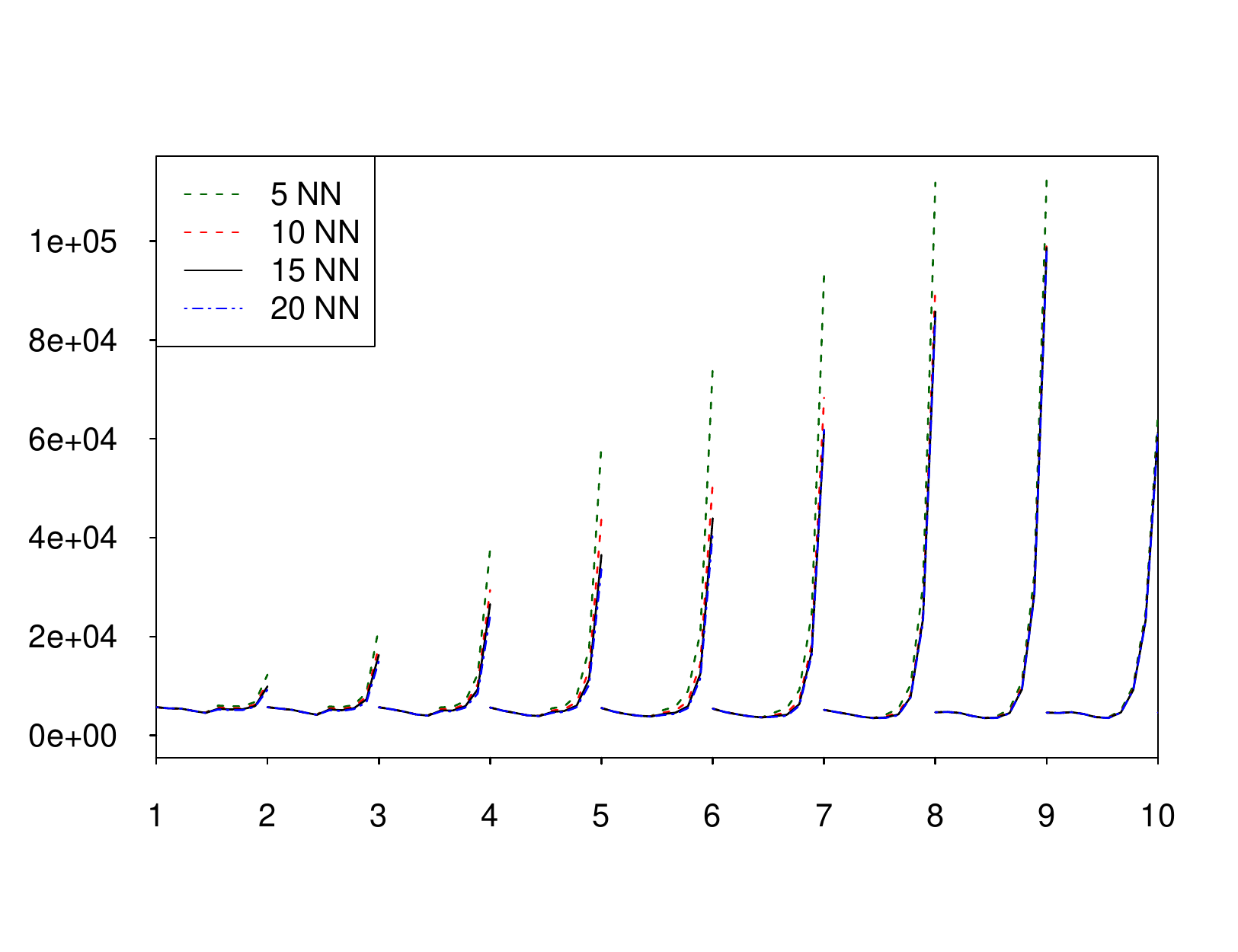}}\hspace{-0.7cm}
\hfill
\subfloat{\includegraphics[scale=0.49]{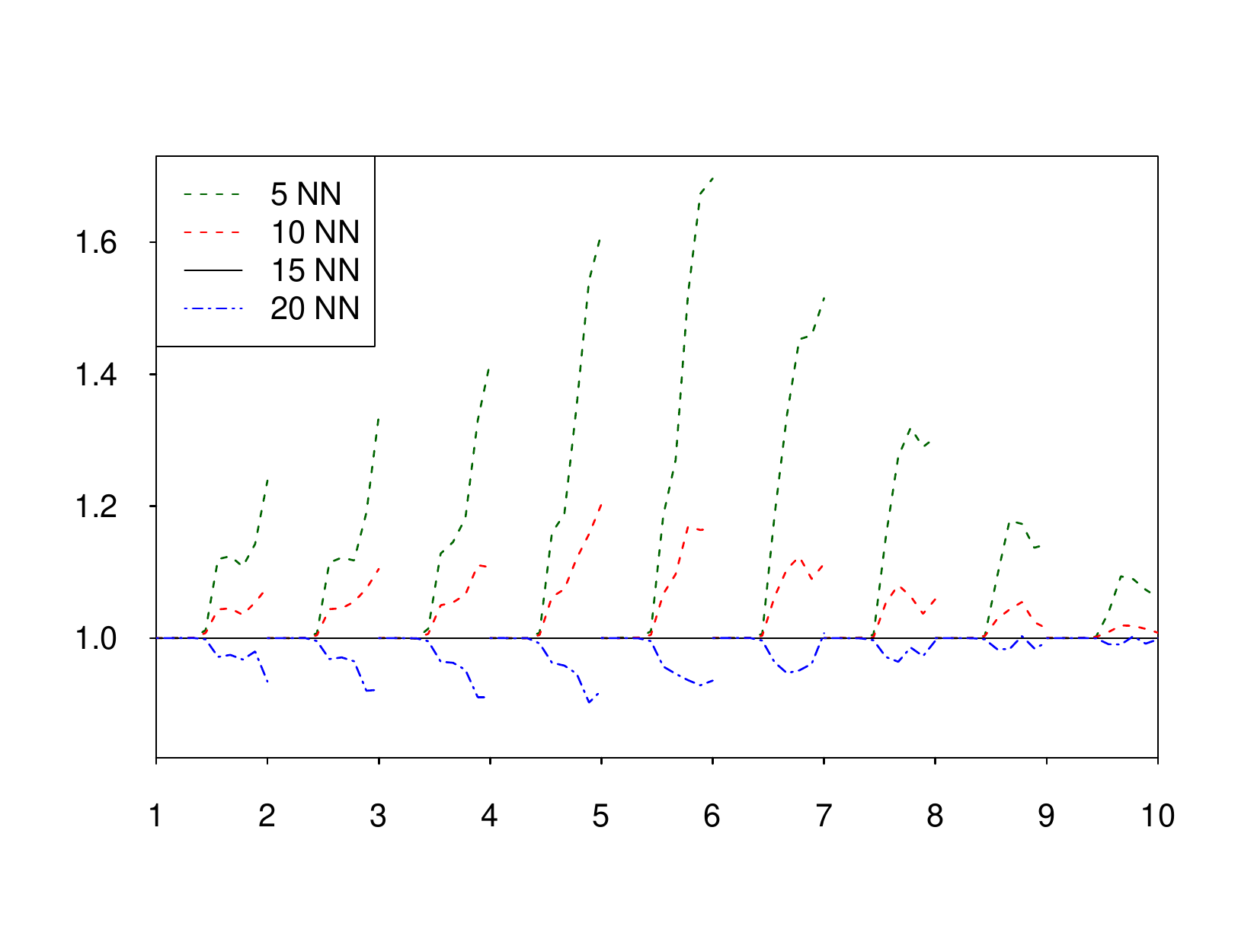}}
\vspace{-1.9cm}
\subfloat{\includegraphics[scale=0.49]{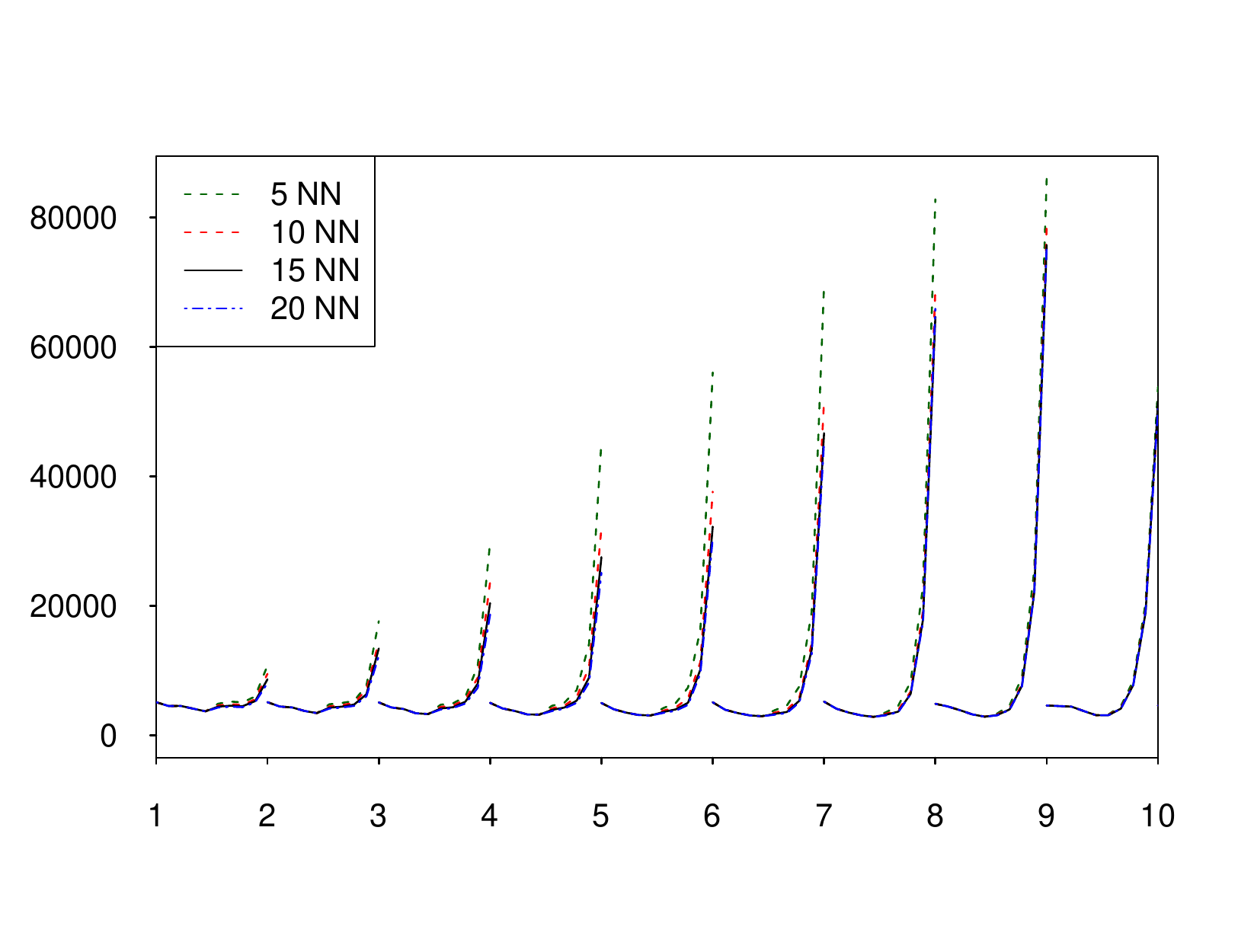}}\hspace{-0.7cm}
\hfill
\subfloat{\includegraphics[scale=0.49]{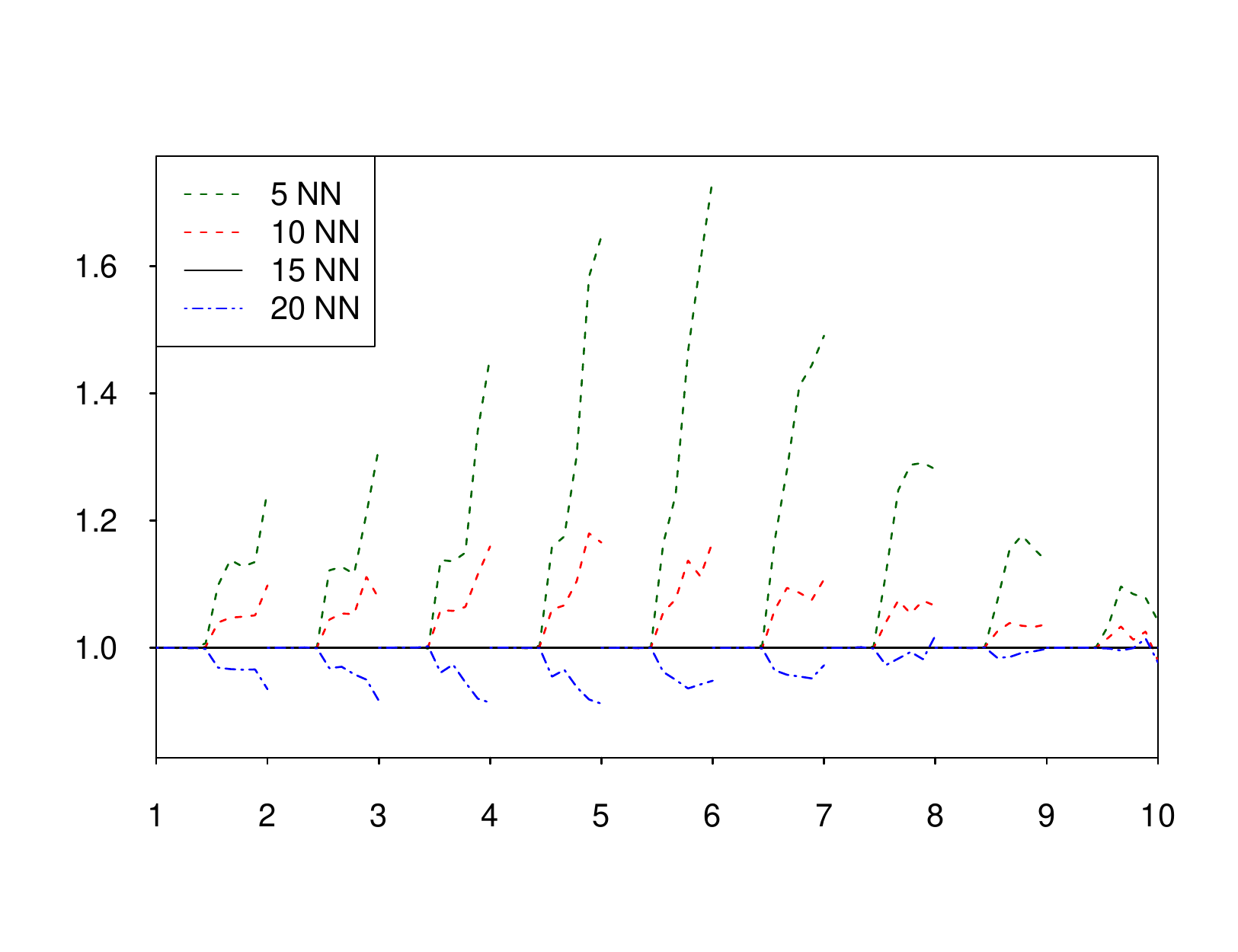}}

\caption{Average number of iterations (left) and corresponding ratio (right) per grid point for different  different number 
of nearest neighbors considering warm start and stopping criteria with clipped duality gap for the data set \textsc{cal-housing}.
The graphs comprises for $\tau=0.25$ (top), $\tau=0.50$ (middle) and $\tau=0.75$ (bottom).}
\label{figure-per-grid iter for NN-cal-housing}
\end{scriptsize}
\end{figure}


\newpage
\vspace{-3cm}
\subsection{Results for Different Initialization Methods}

\begin{figure}[!ht]
\begin{scriptsize}
 \subfloat{\includegraphics[scale=0.52]{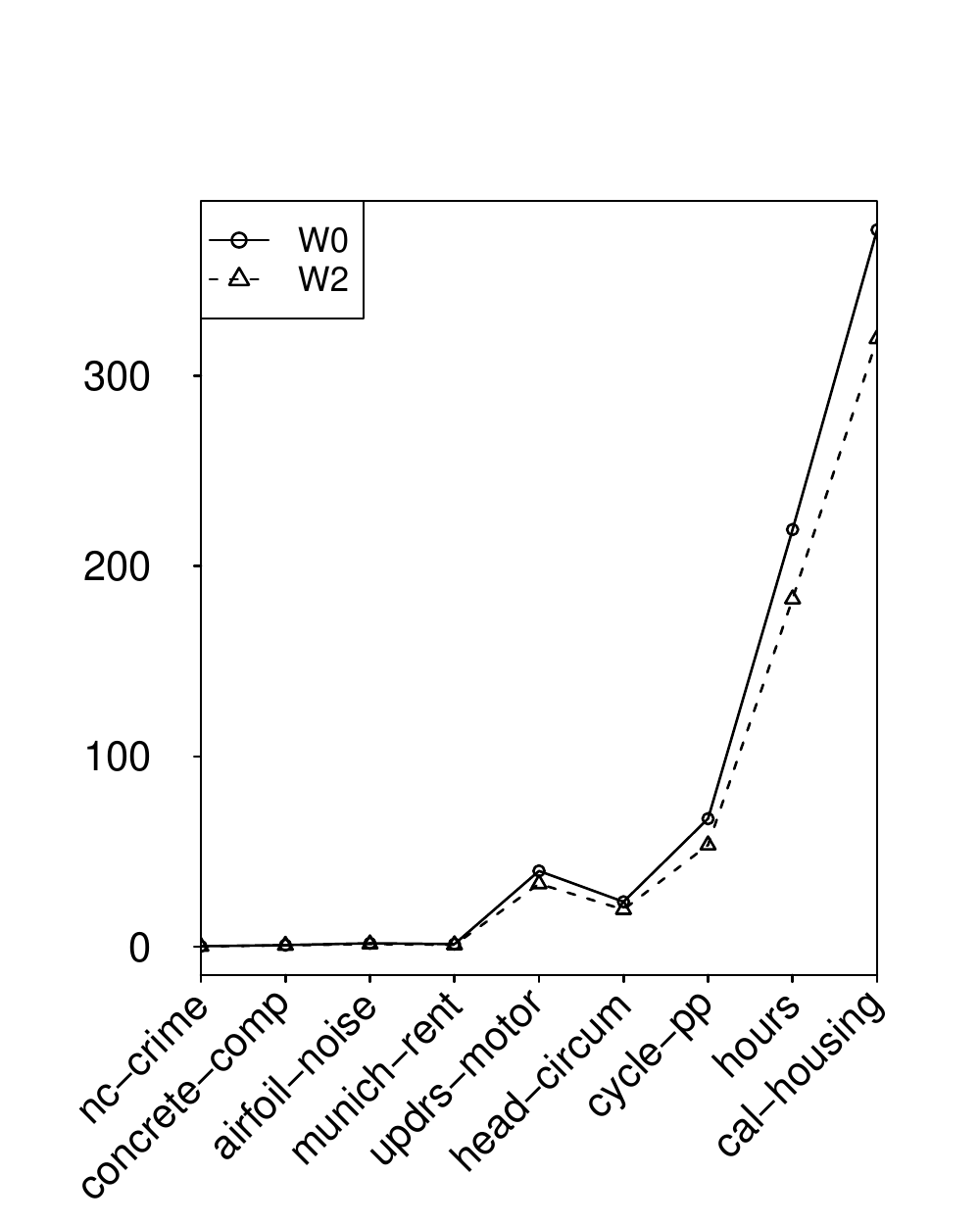}}\hspace{-0.7cm}
 \hfill
 \subfloat{\includegraphics[scale=0.52]{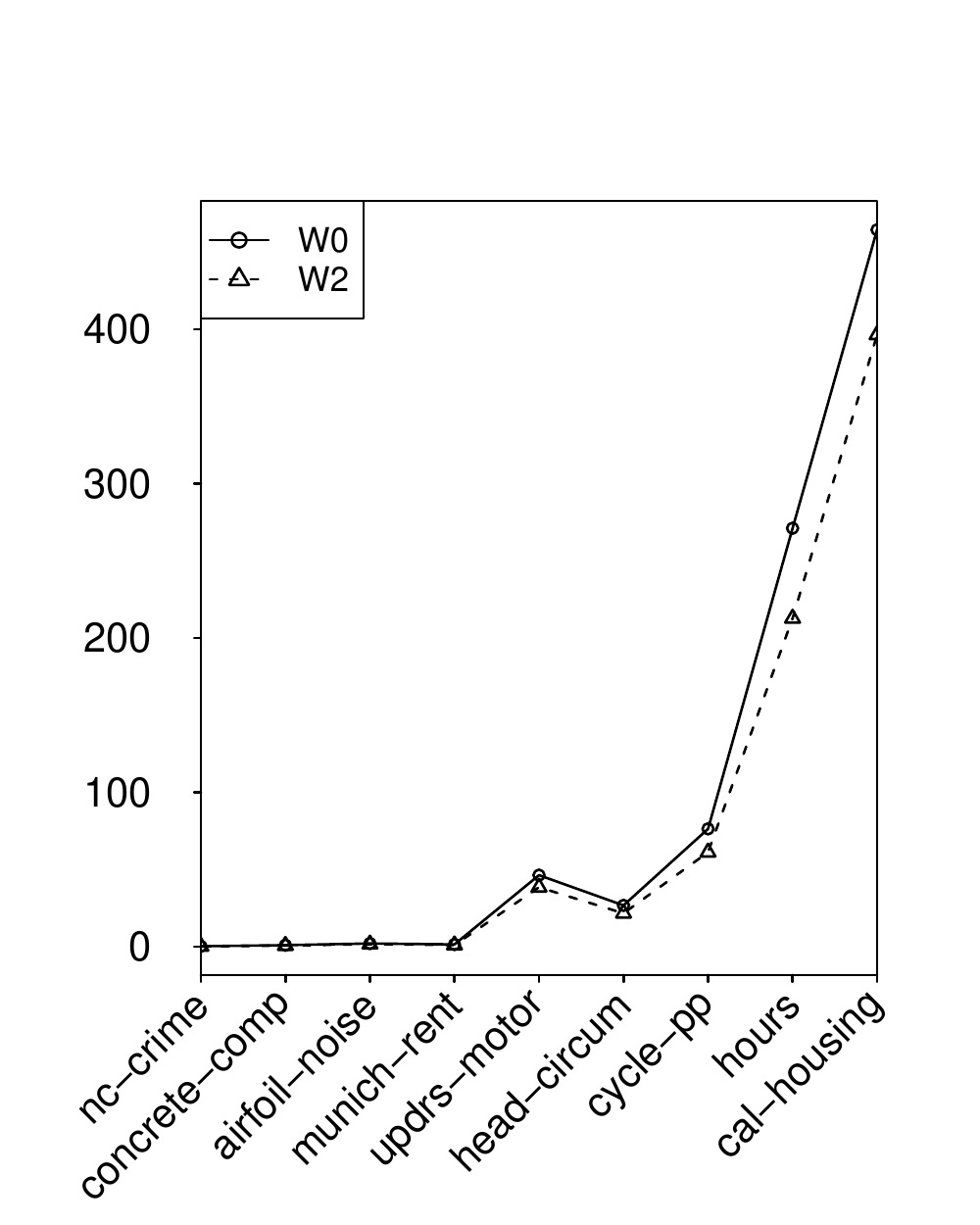}}\hspace{-0.7cm}
\hfill
 \subfloat{\includegraphics[scale=0.52]{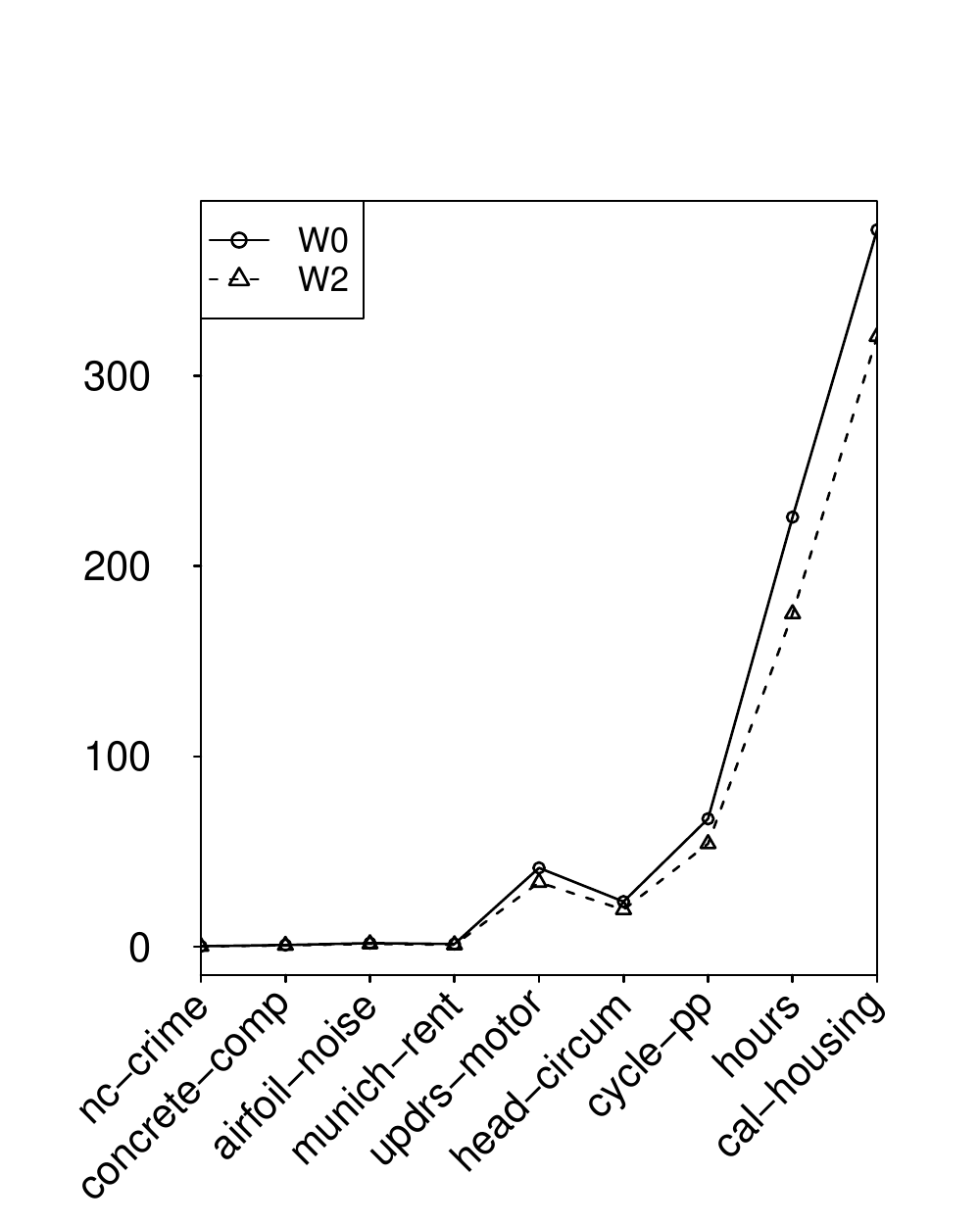}}
 
 \vspace{-1.0cm}
\subfloat{\includegraphics[scale=0.52]{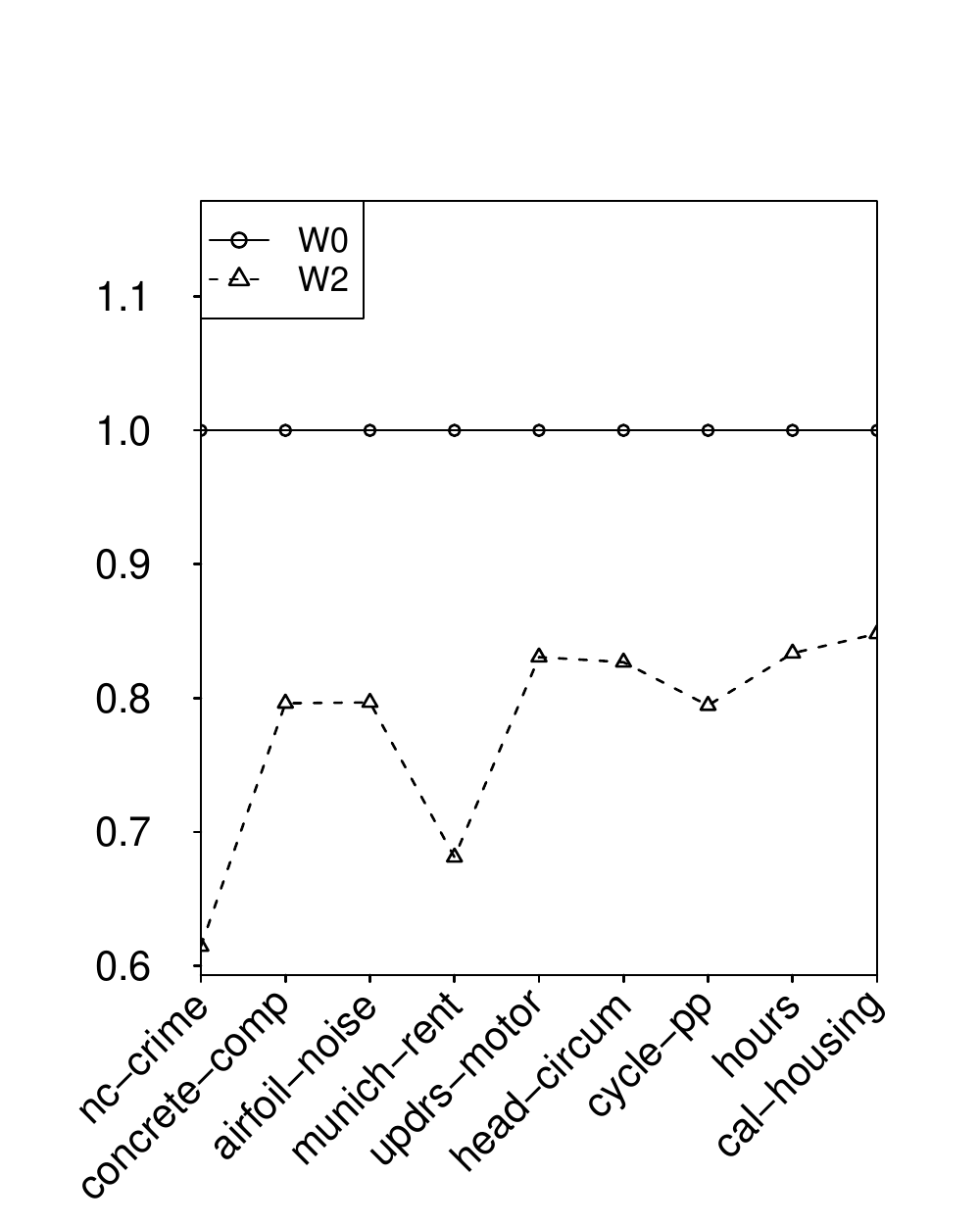}}\hspace{-0.7cm}
 \hfill
 \subfloat{\includegraphics[scale=0.52]{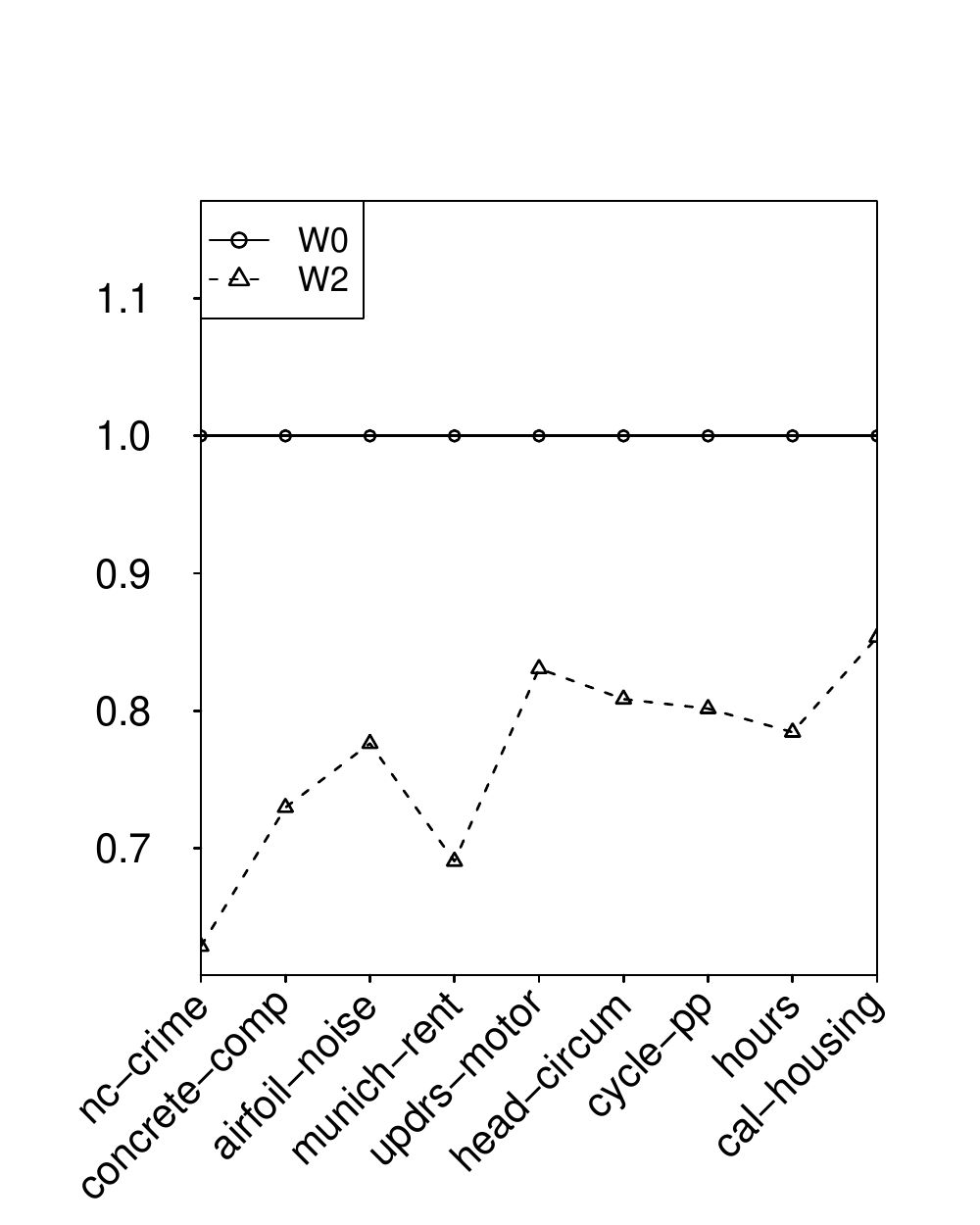}}\hspace{-0.7cm}
\hfill
 \subfloat{\includegraphics[scale=0.52]{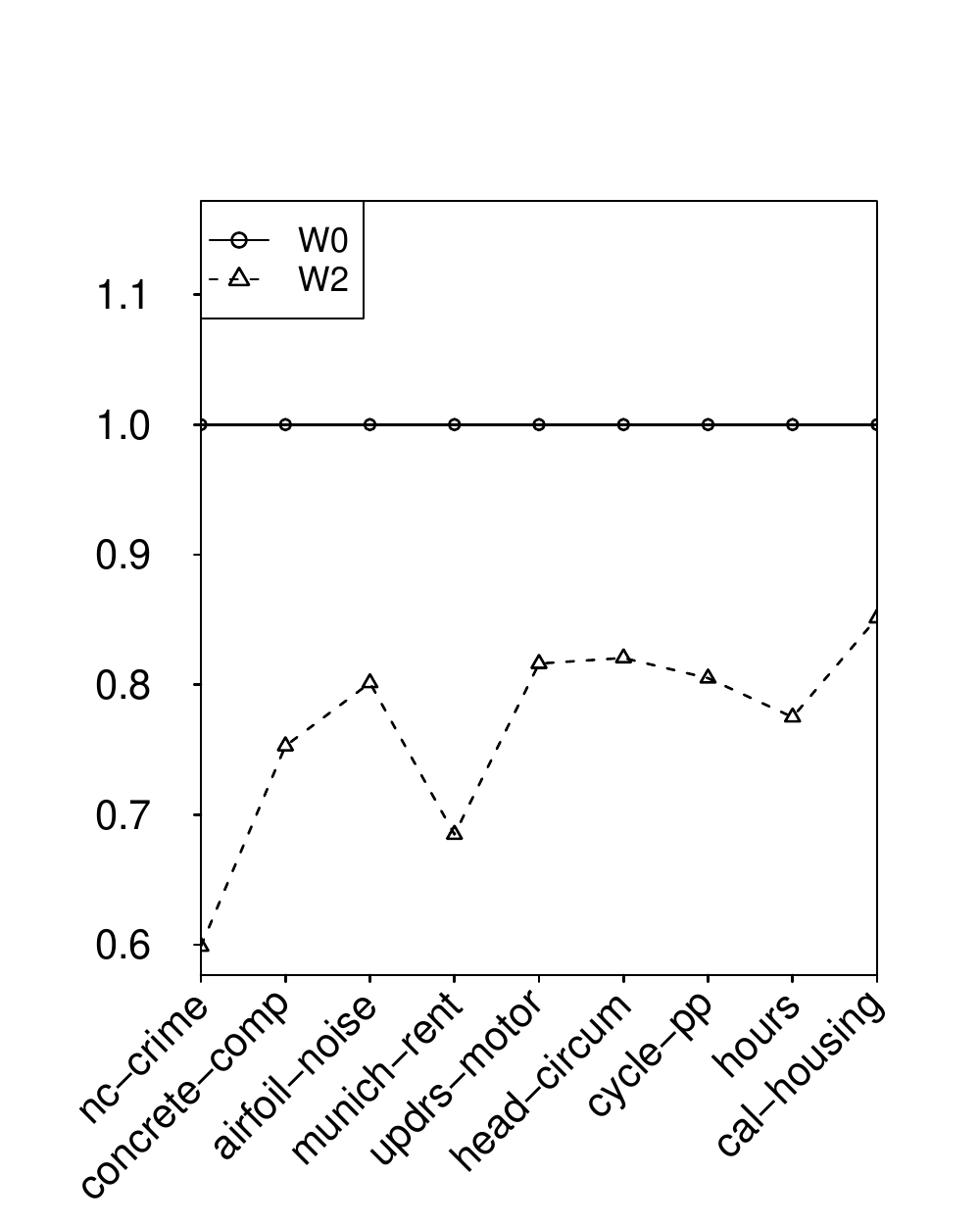}}
 \caption{Train time (top) and corresponding ratio (bottom) of different data sets for different initialization methods after
 fixing stopping criteria with clipped duality gap and $NN=15$. The graphs comprises of $\tau=0.25$ (left), $\tau=0.50$ (middle)
 and $\tau=0.75$ (right).}
  \label{figure-time and ratio-initialization vs datasets}
\end{scriptsize}
\end{figure}

\newpage
\vspace{-3cm}
\begin{figure}[!ht]
\begin{scriptsize}
 \subfloat{\includegraphics[scale=0.52]{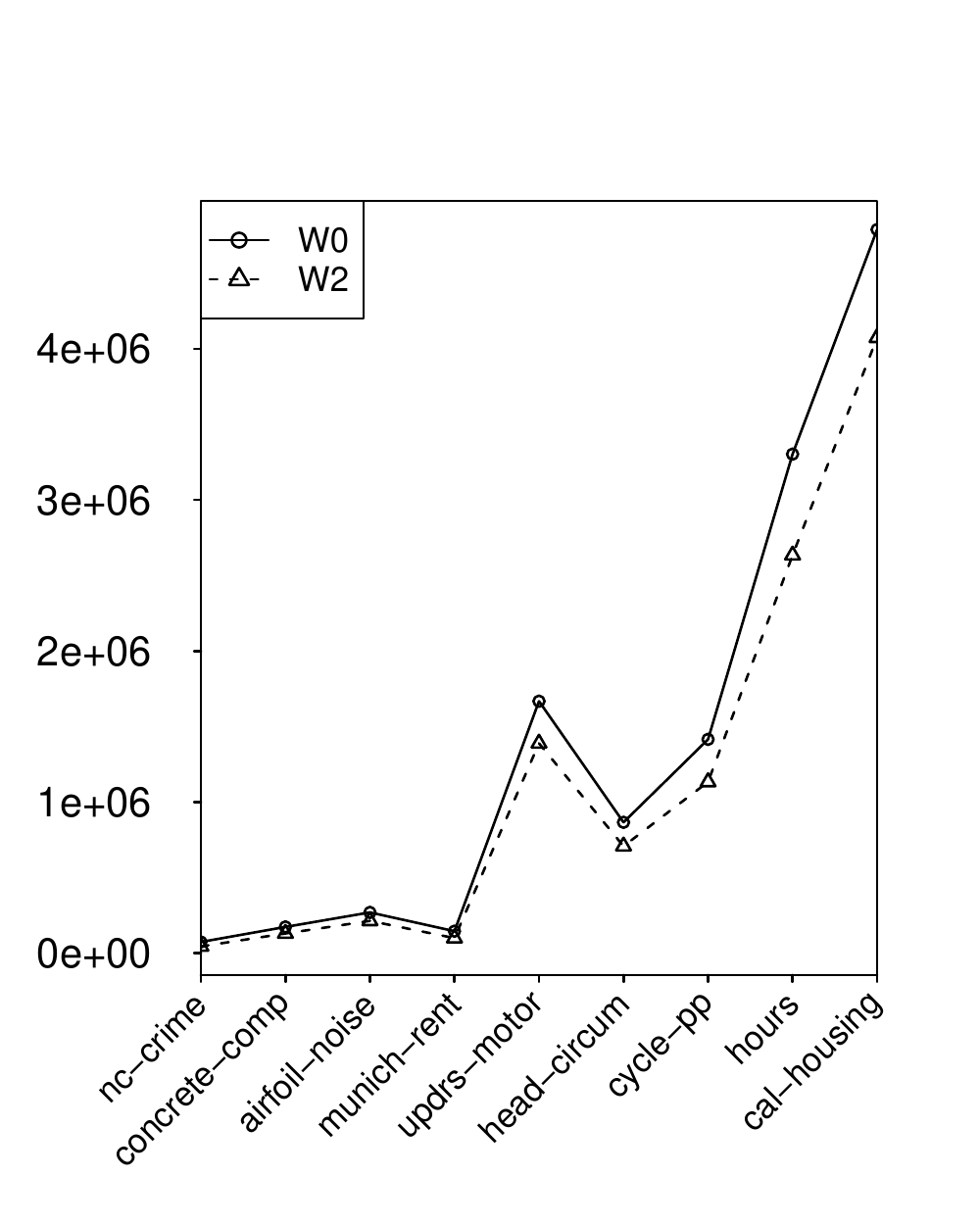}}\hspace{-0.7cm}
 \hfill
 \subfloat{\includegraphics[scale=0.52]{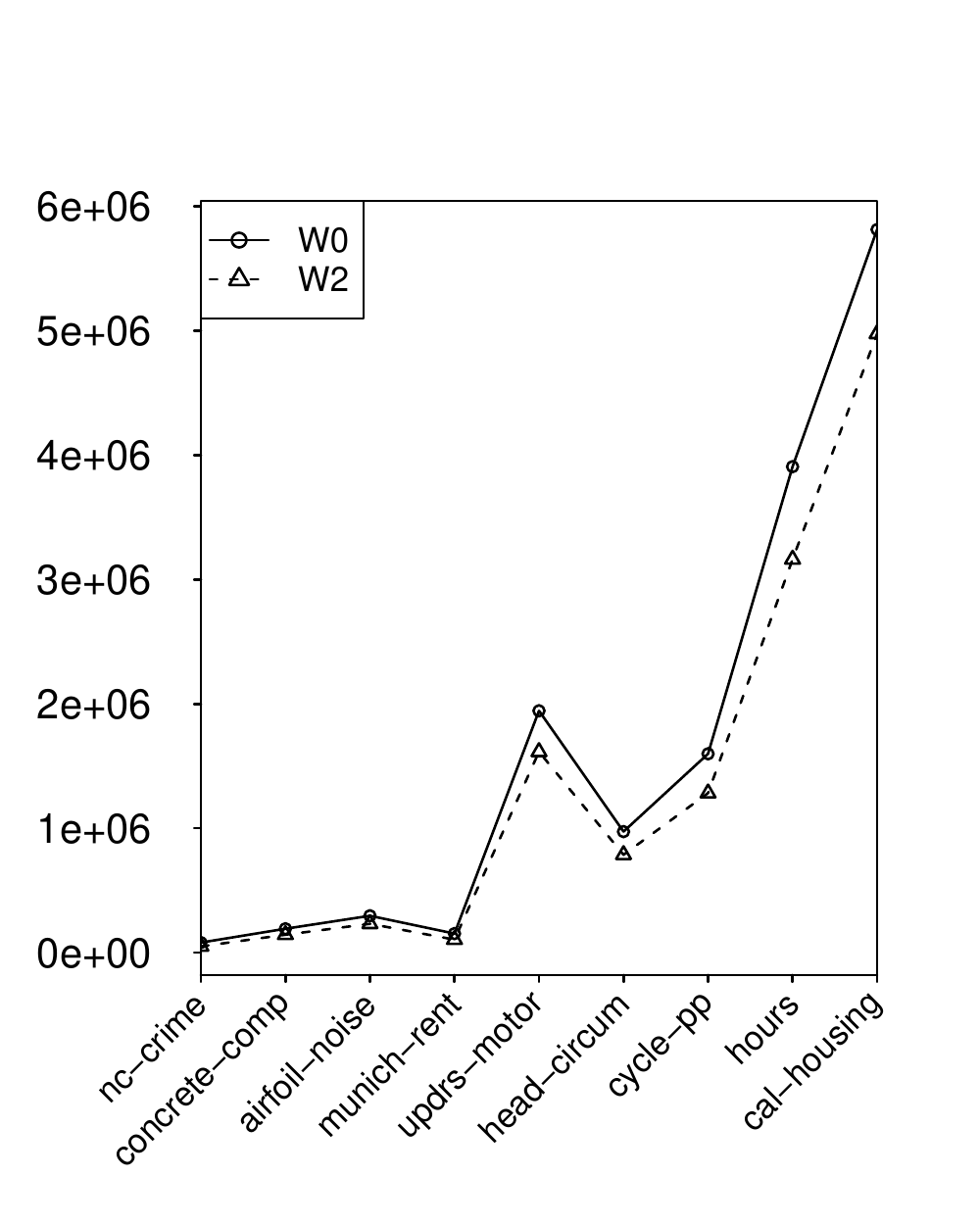}}\hspace{-0.7cm}
\hfill
 \subfloat{\includegraphics[scale=0.52]{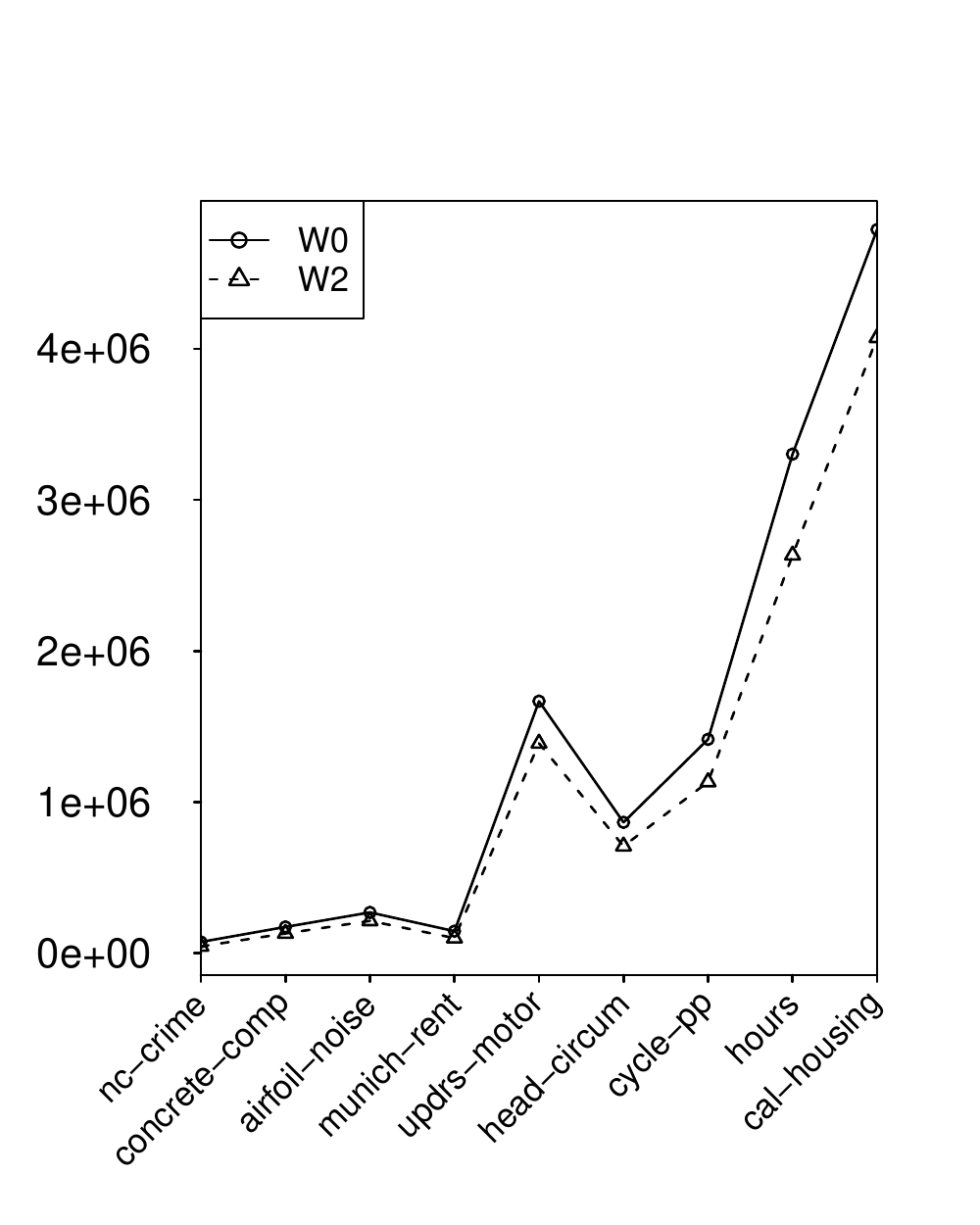}}
 
 \vspace{-1.0cm}
\subfloat{\includegraphics[scale=0.52]{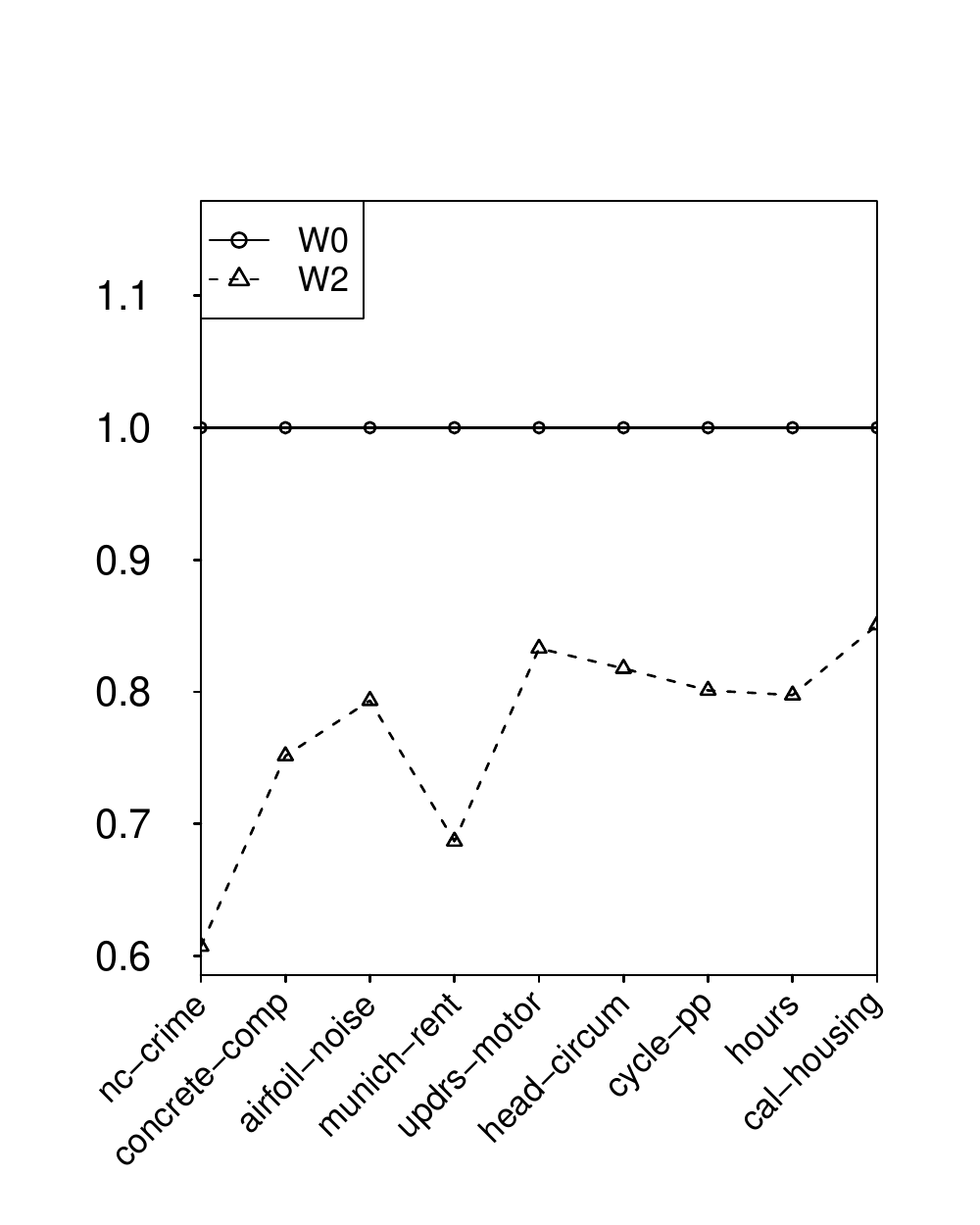}}\hspace{-0.7cm}
 \hfill
 \subfloat{\includegraphics[scale=0.52]{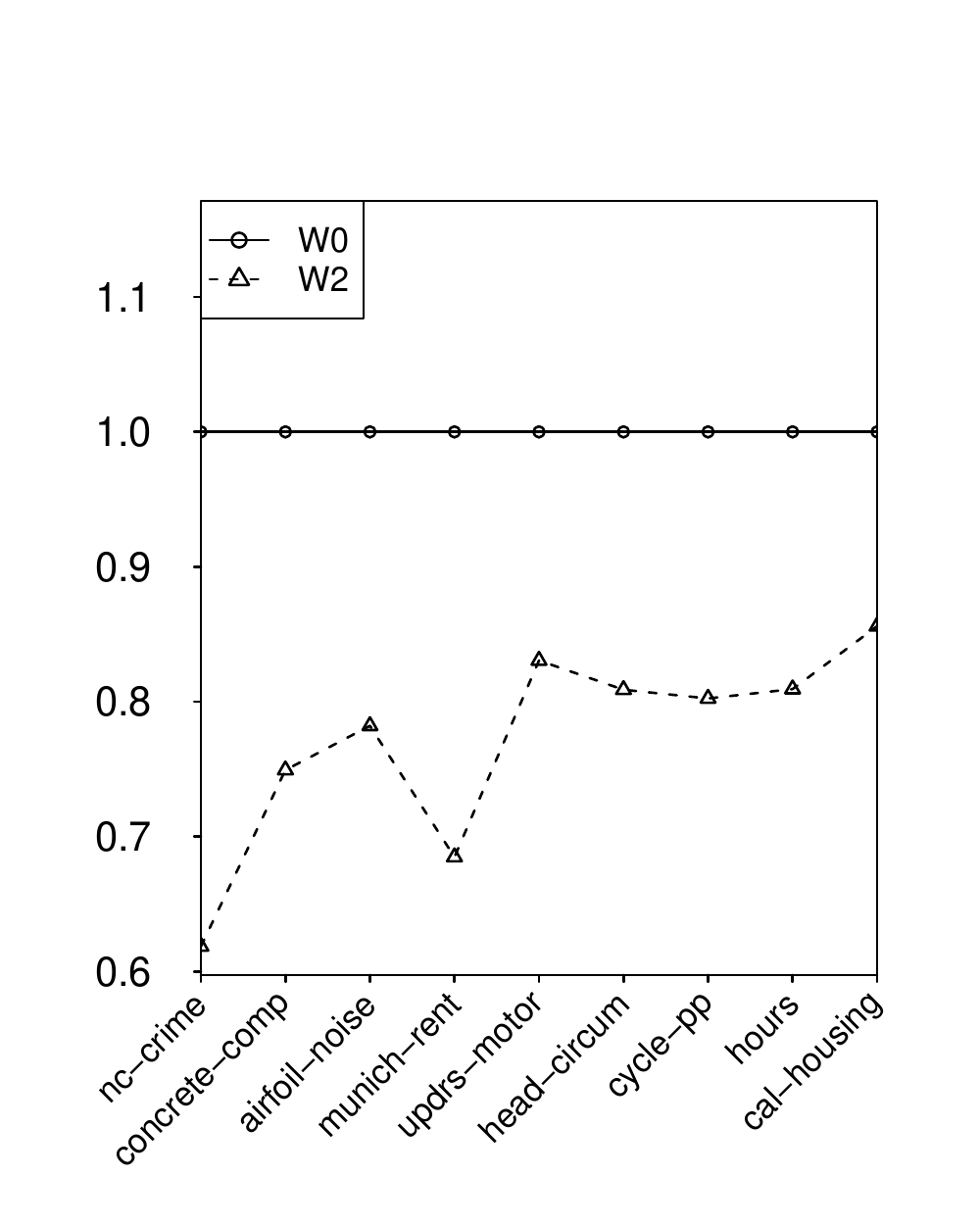}}\hspace{-0.7cm}
\hfill
 \subfloat{\includegraphics[scale=0.52]{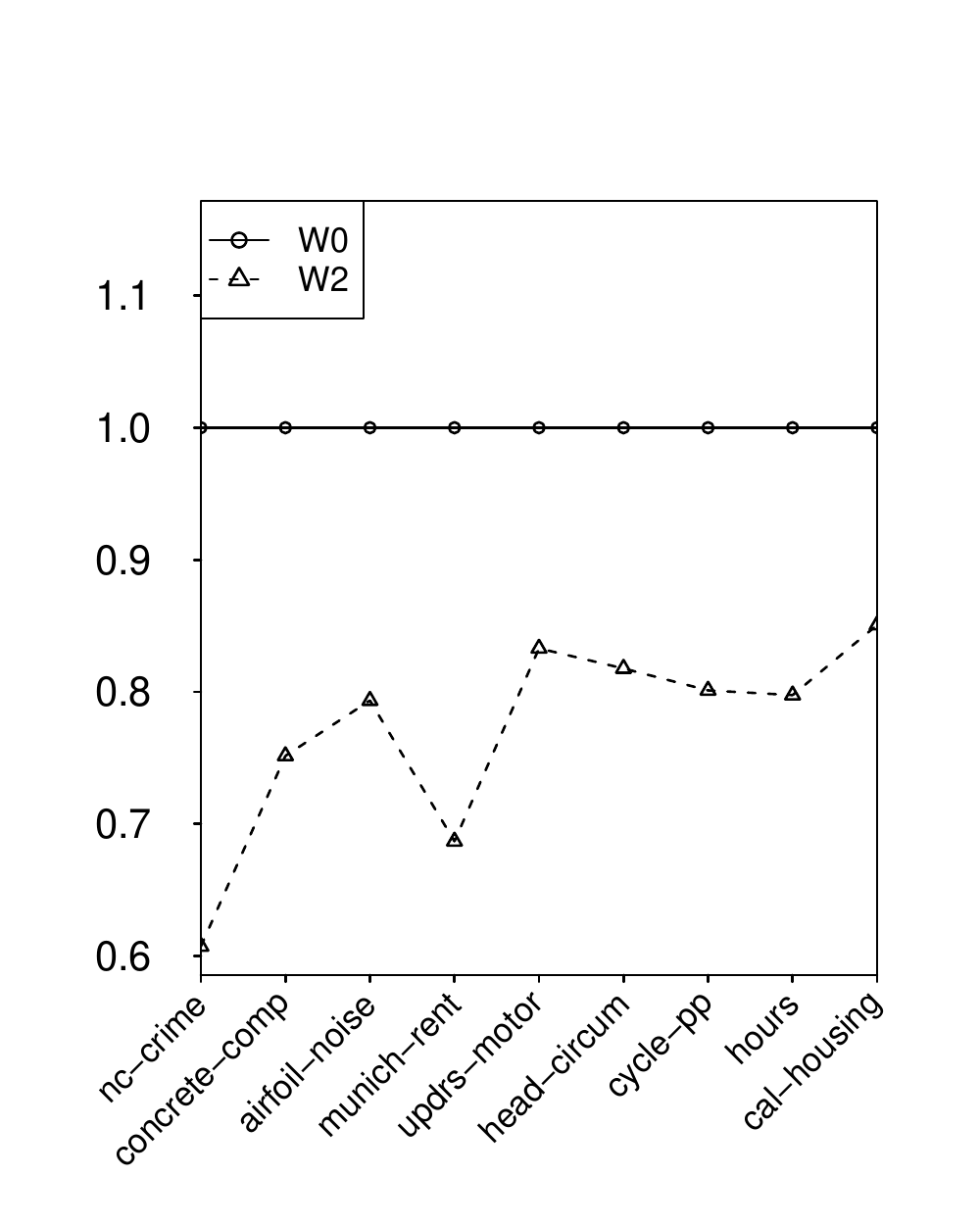}}
 \caption{Train iterations (top) and corresponding ratio (bottom) of different data sets for different initialization methods 
 after fixing stopping criteria with clipped duality gap and $NN=15$. The graphs comprises of $\tau=0.25$ (left), $\tau=0.50$ 
 (middle) and $\tau=0.75$ (right).}
 \label{figure-iter and ratio-initialization vs datasets}
\end{scriptsize}
\end{figure}

\newpage
\vspace{0cm}
\begin{figure}[!ht]
\begin{scriptsize}
\vspace{-2cm}
 \subfloat{\includegraphics[scale=0.49]{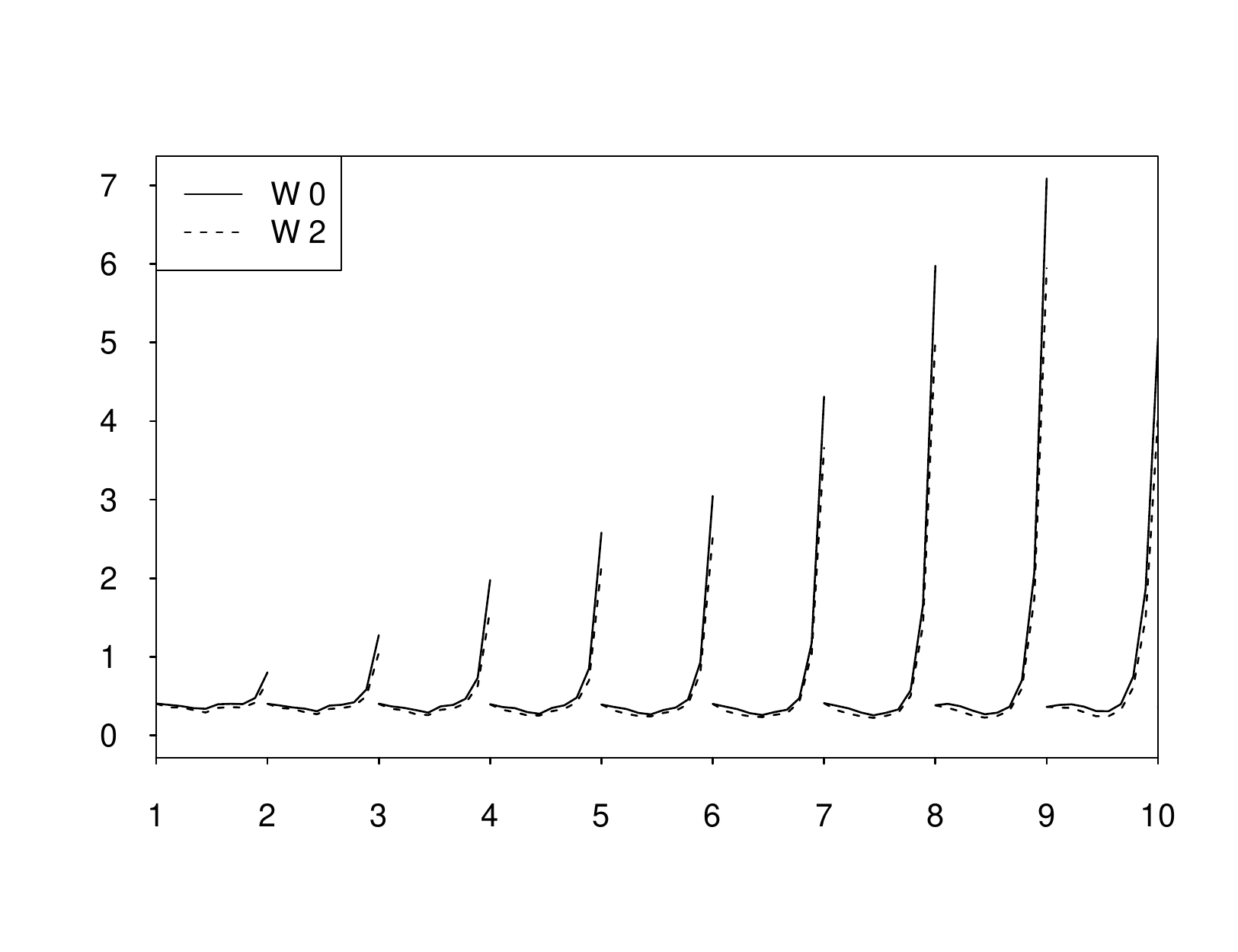}}\hspace{-0.7cm}
 \hfill
 \subfloat{\includegraphics[scale=0.49]{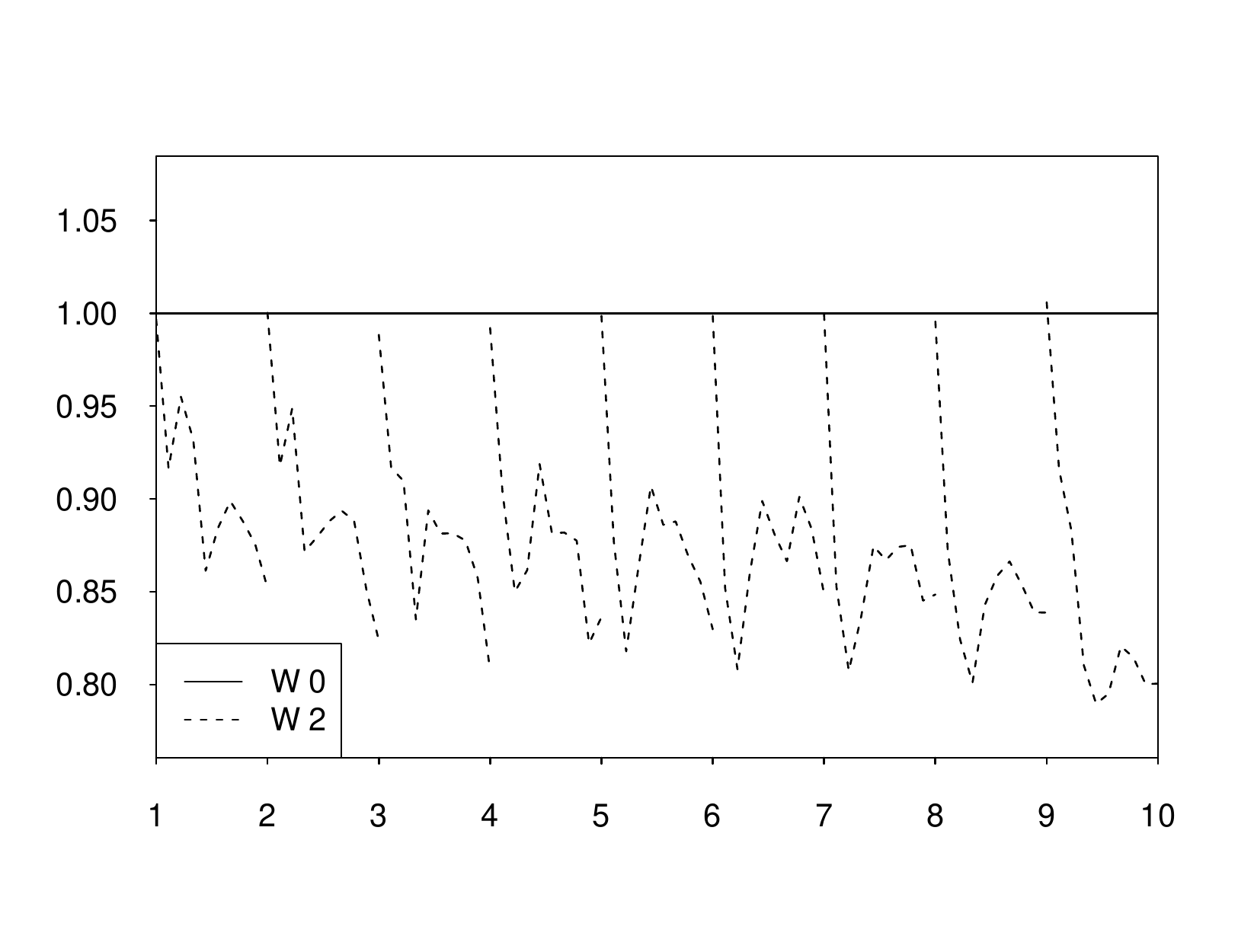}}
 \vspace{-1.2cm}
\subfloat{\includegraphics[scale=0.49]{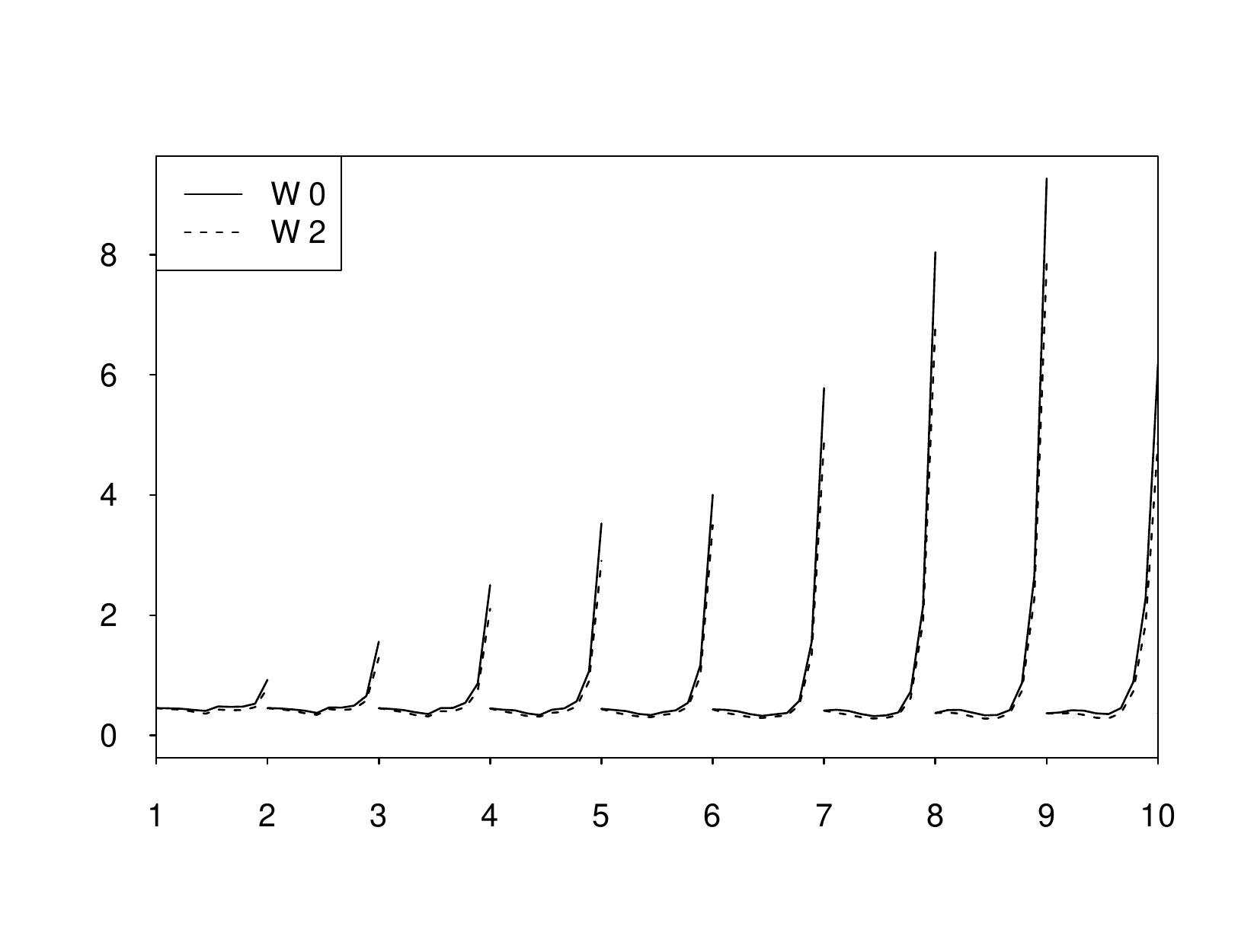}}\hspace{-0.7cm}
\hfill
\subfloat{\includegraphics[scale=0.49]{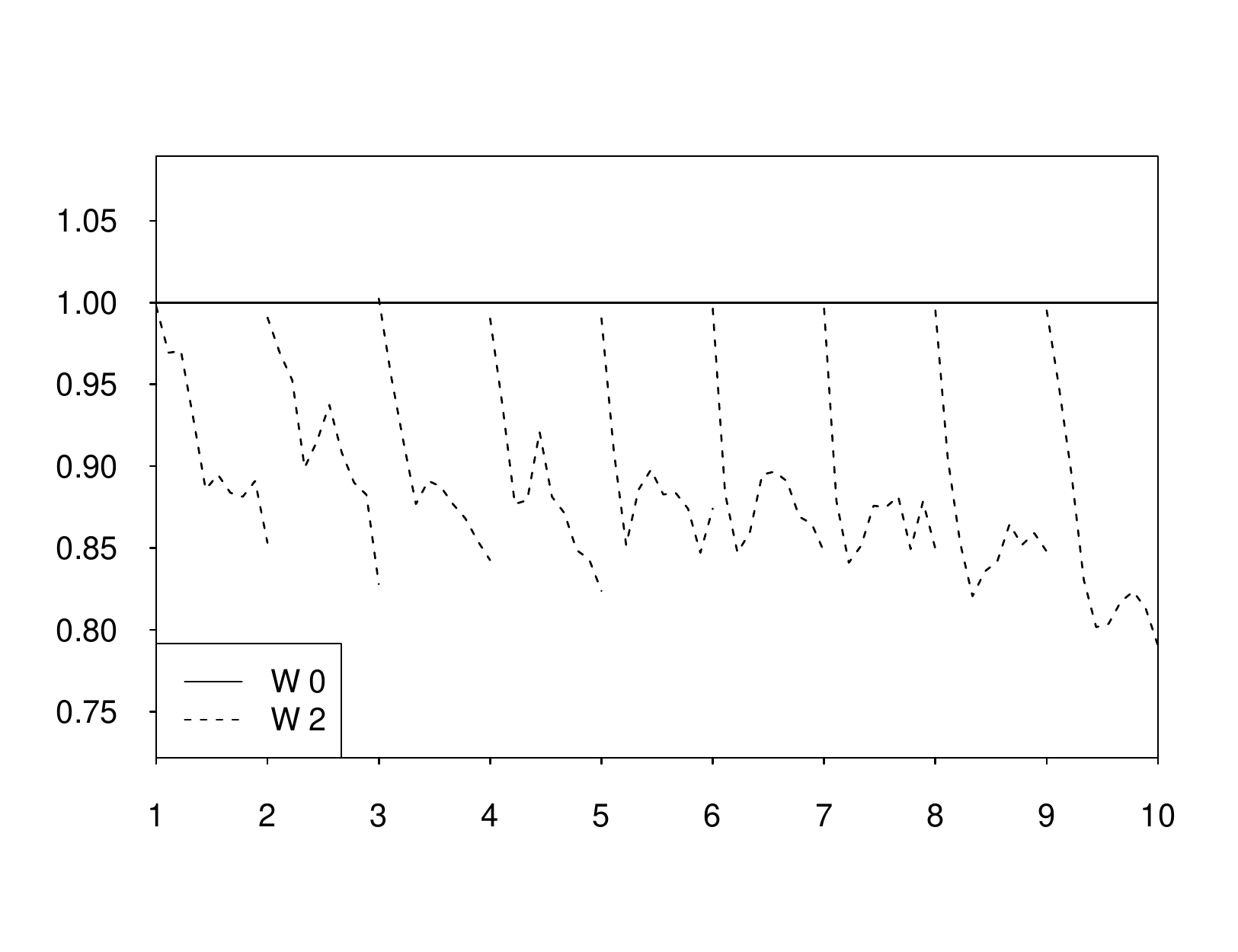}}
\vspace{-1.2cm}
\subfloat{\includegraphics[scale=0.49]{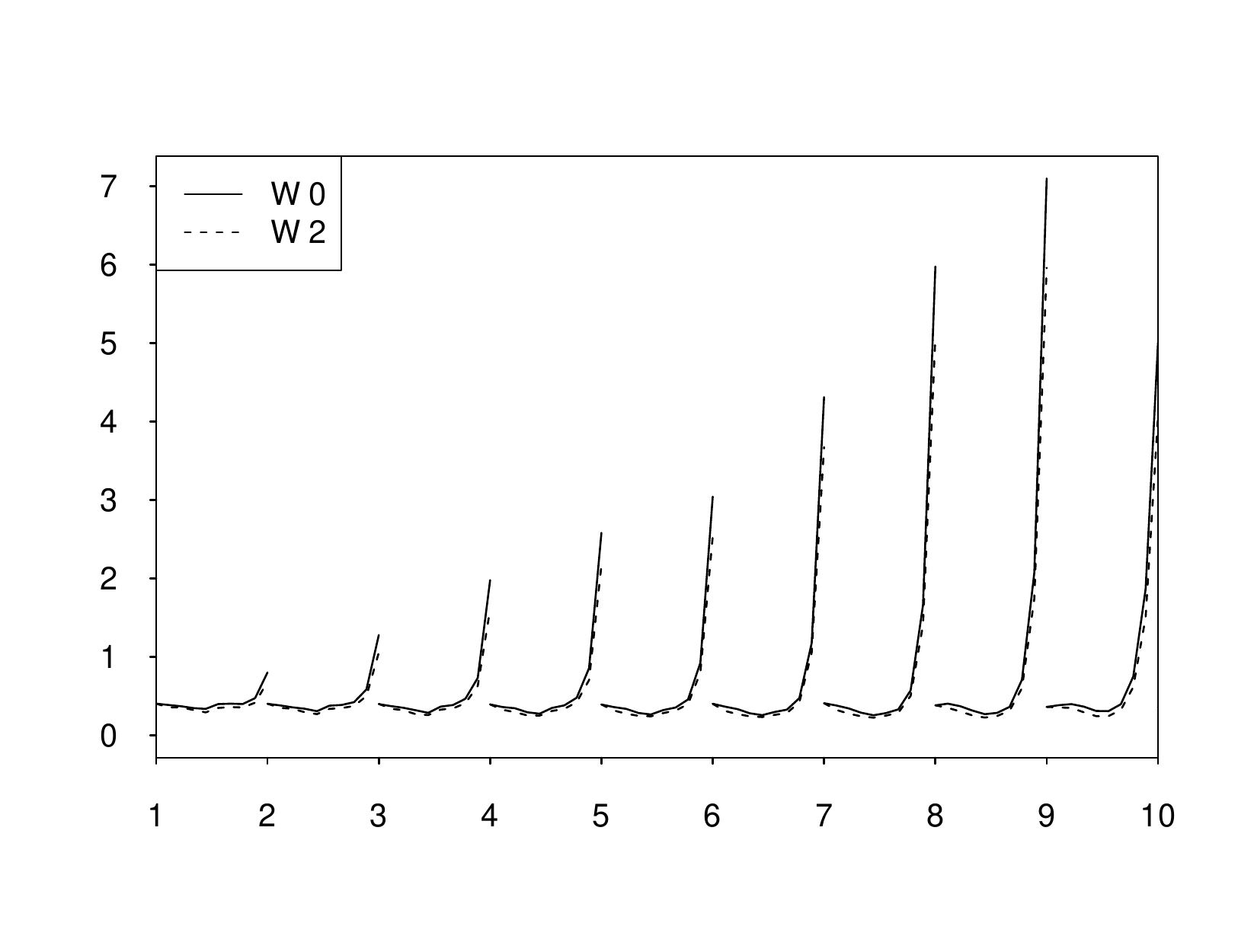}}\hspace{-0.7cm}
\hfill
\subfloat{\includegraphics[scale=0.49]{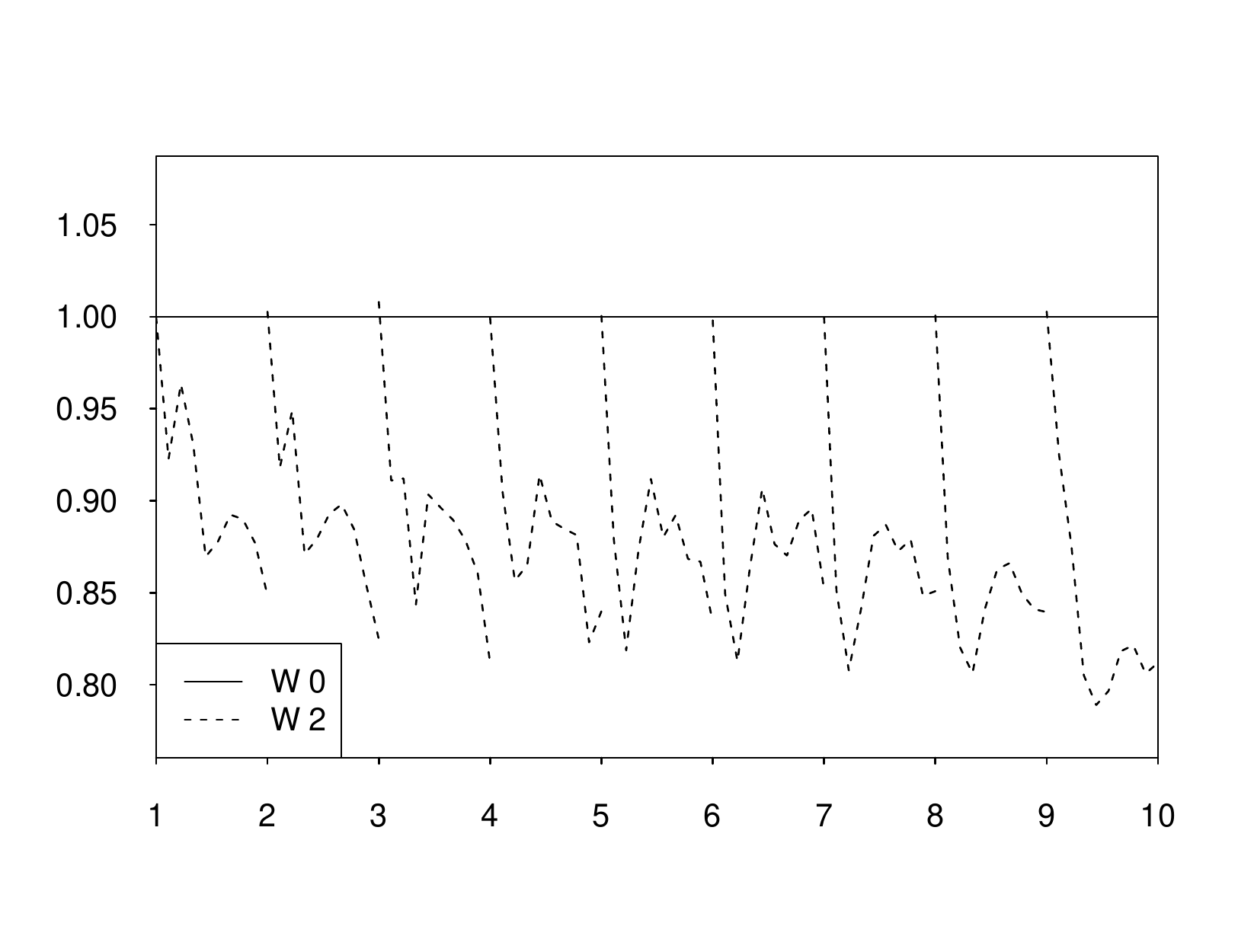}}

\caption{Average train time (left) and corresponding ratio (right) per grid point for different initialization methods 
considering  stopping criteria after with clipped duality gap and \textsf{WSS 2} for the data set \textsc{cal-housing}. 
The graphs comprises for $\tau=0.25$ (top), $\tau=0.50$ (middle) and $\tau=0.75$ (bottom).}
  \label{figure-per-grid time for initialization-cal-housing}
\end{scriptsize}
\end{figure}

\newpage
\vspace{0cm}
\begin{figure}[!ht]
\begin{scriptsize}
\vspace{-2cm}
 \subfloat{\includegraphics[scale=0.49]{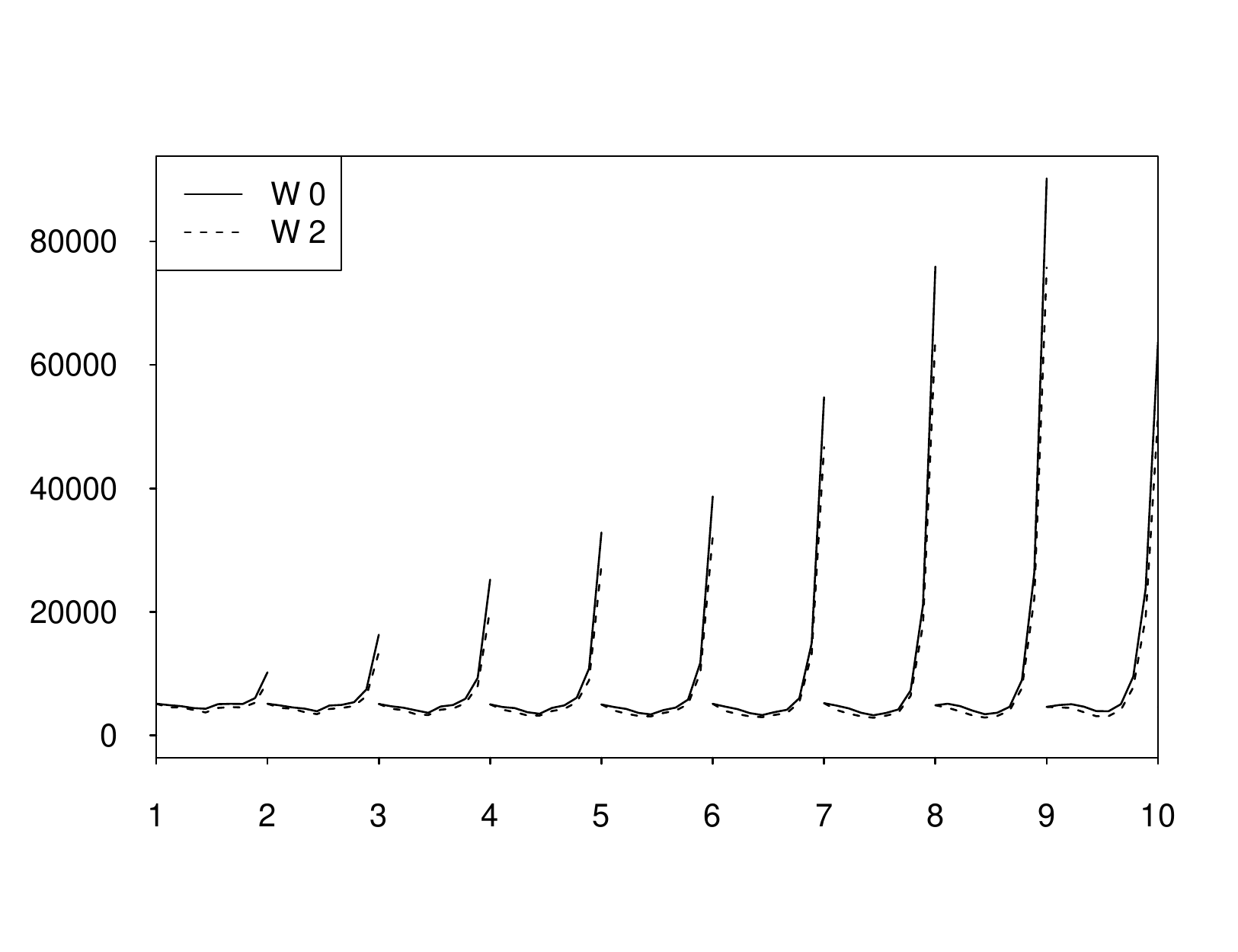}}\hspace{-0.7cm}
 \hfill
 \subfloat{\includegraphics[scale=0.49]{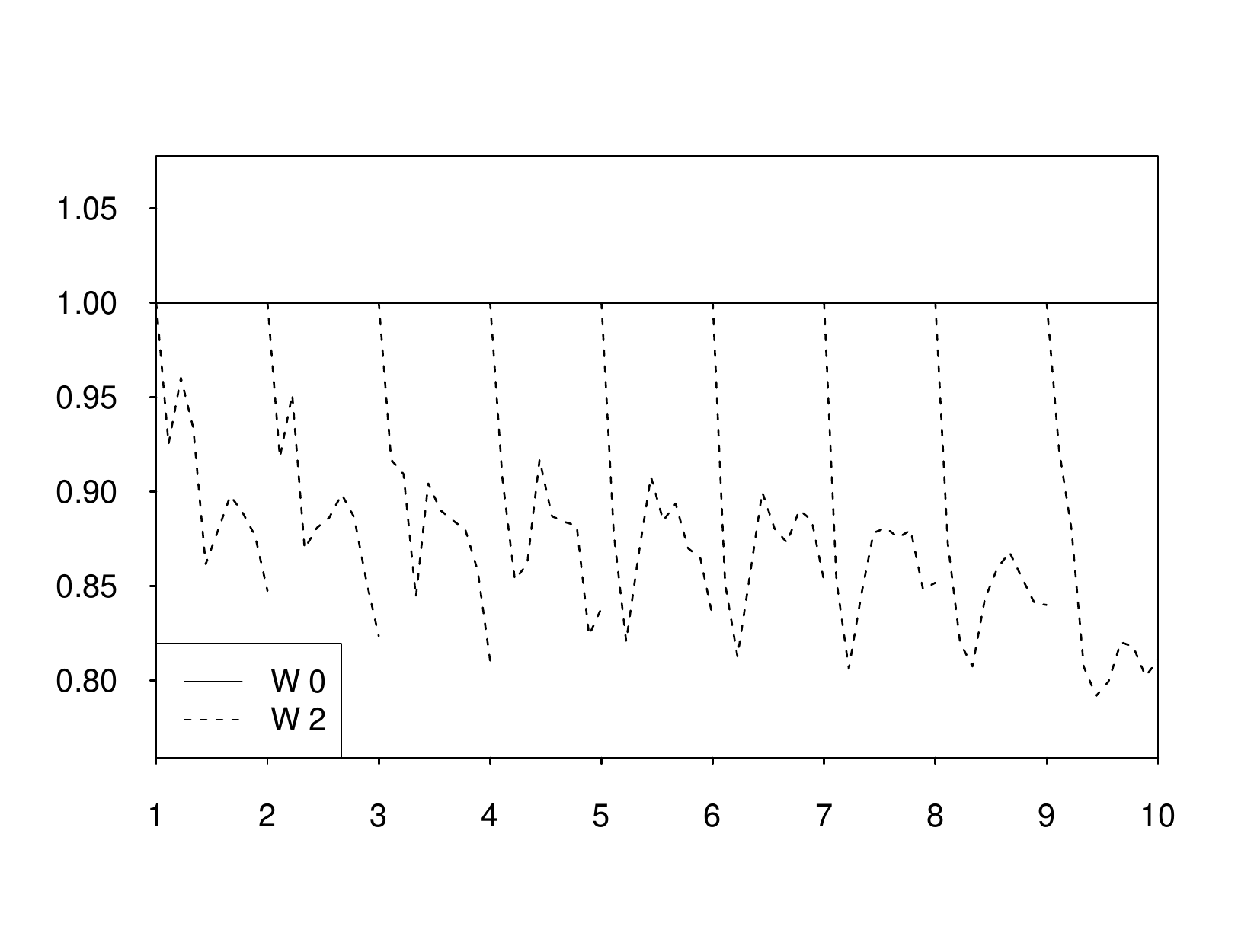}}
 \vspace{-1.2cm}
\subfloat{\includegraphics[scale=0.49]{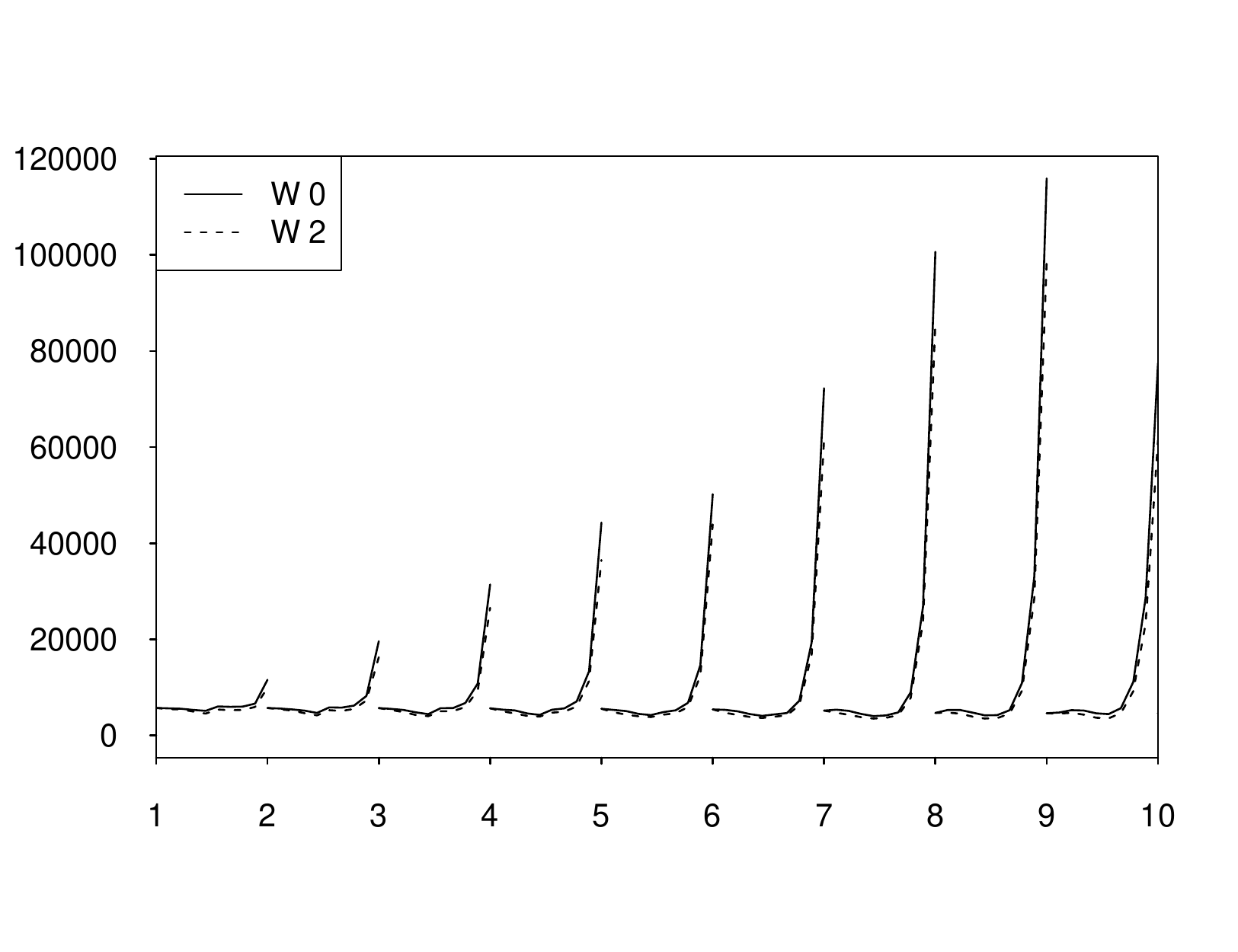}}\hspace{-0.7cm}
\hfill
\subfloat{\includegraphics[scale=0.49]{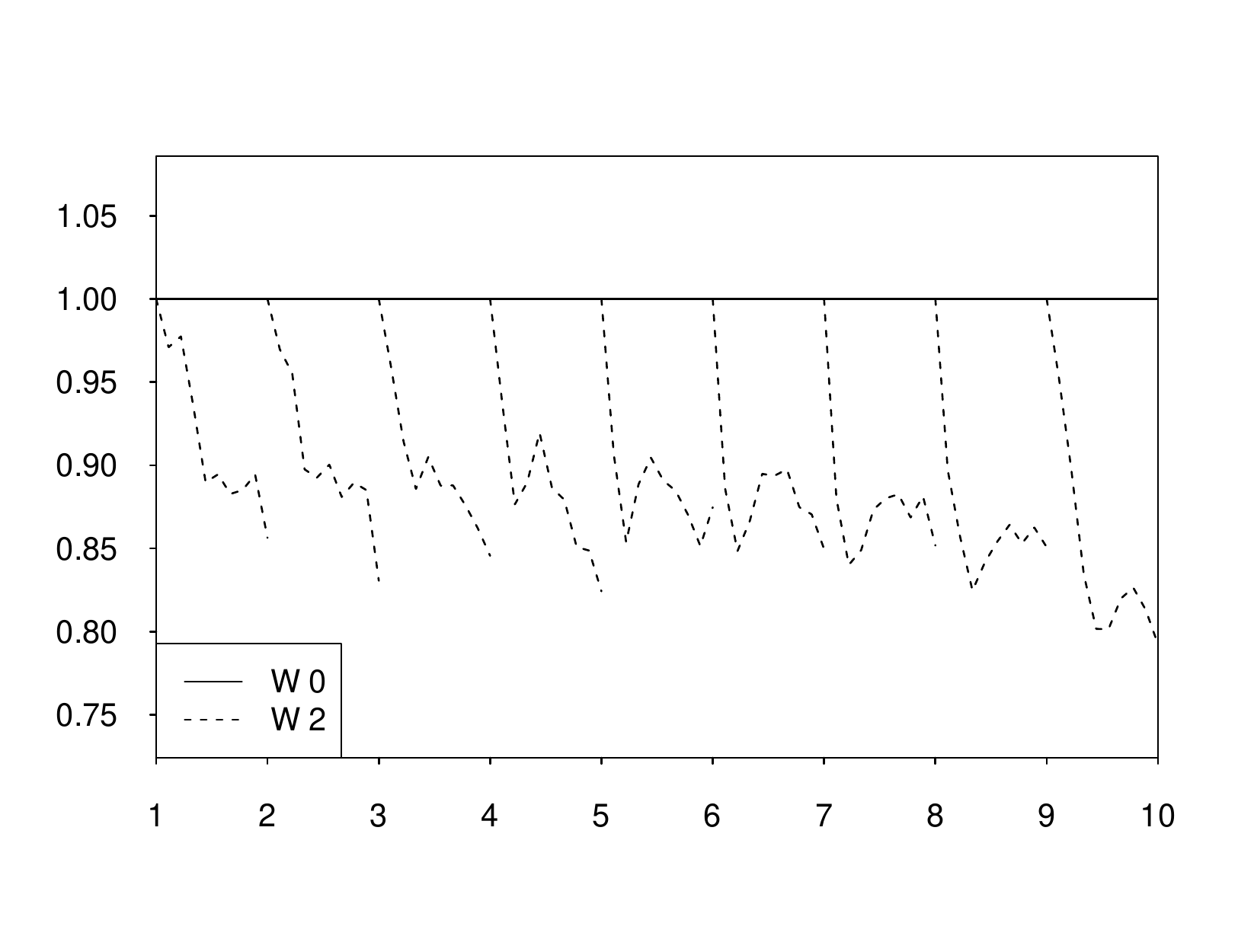}}
\vspace{-1.2cm}
\subfloat{\includegraphics[scale=0.49]{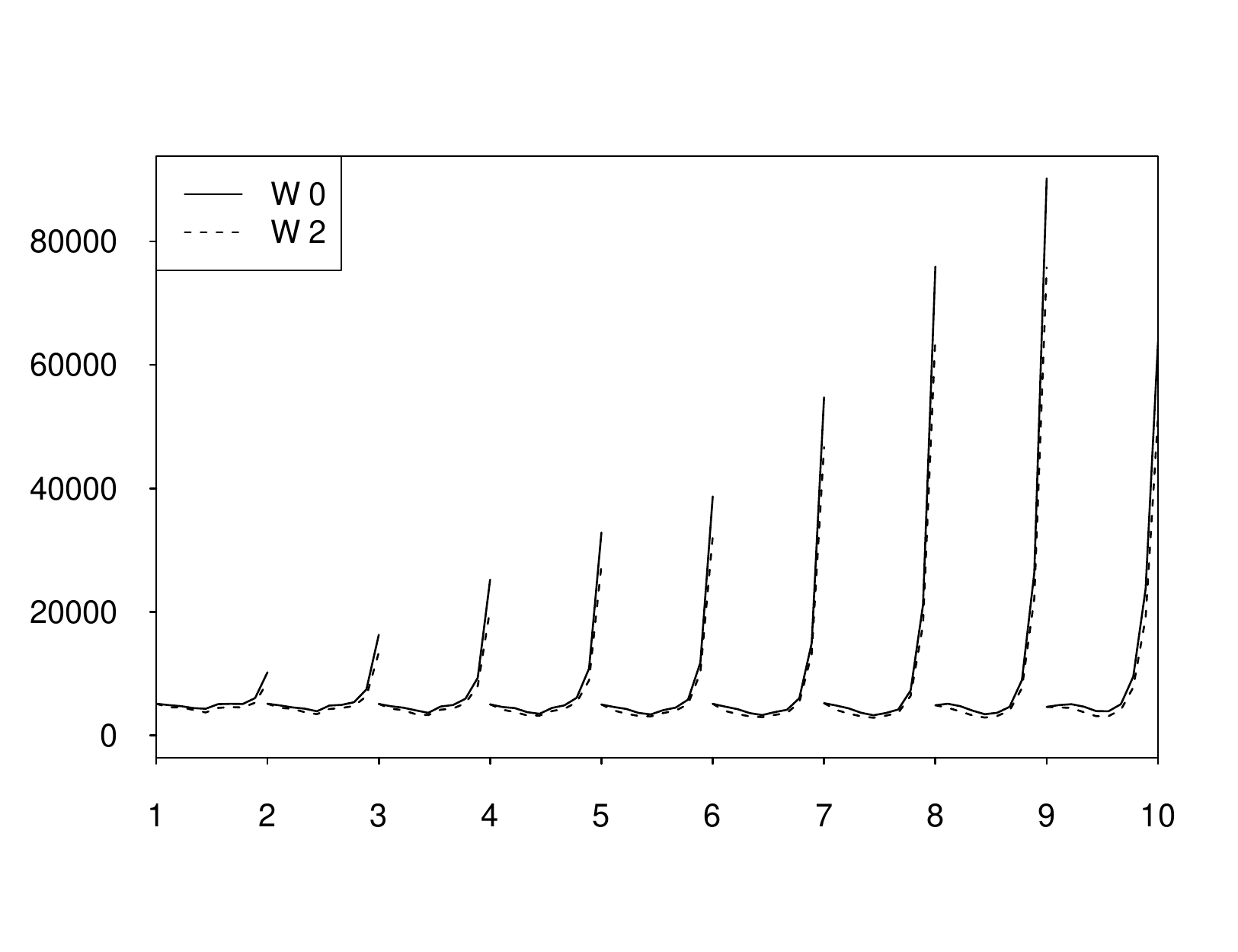}}\hspace{-0.7cm}
\hfill
\subfloat{\includegraphics[scale=0.49]{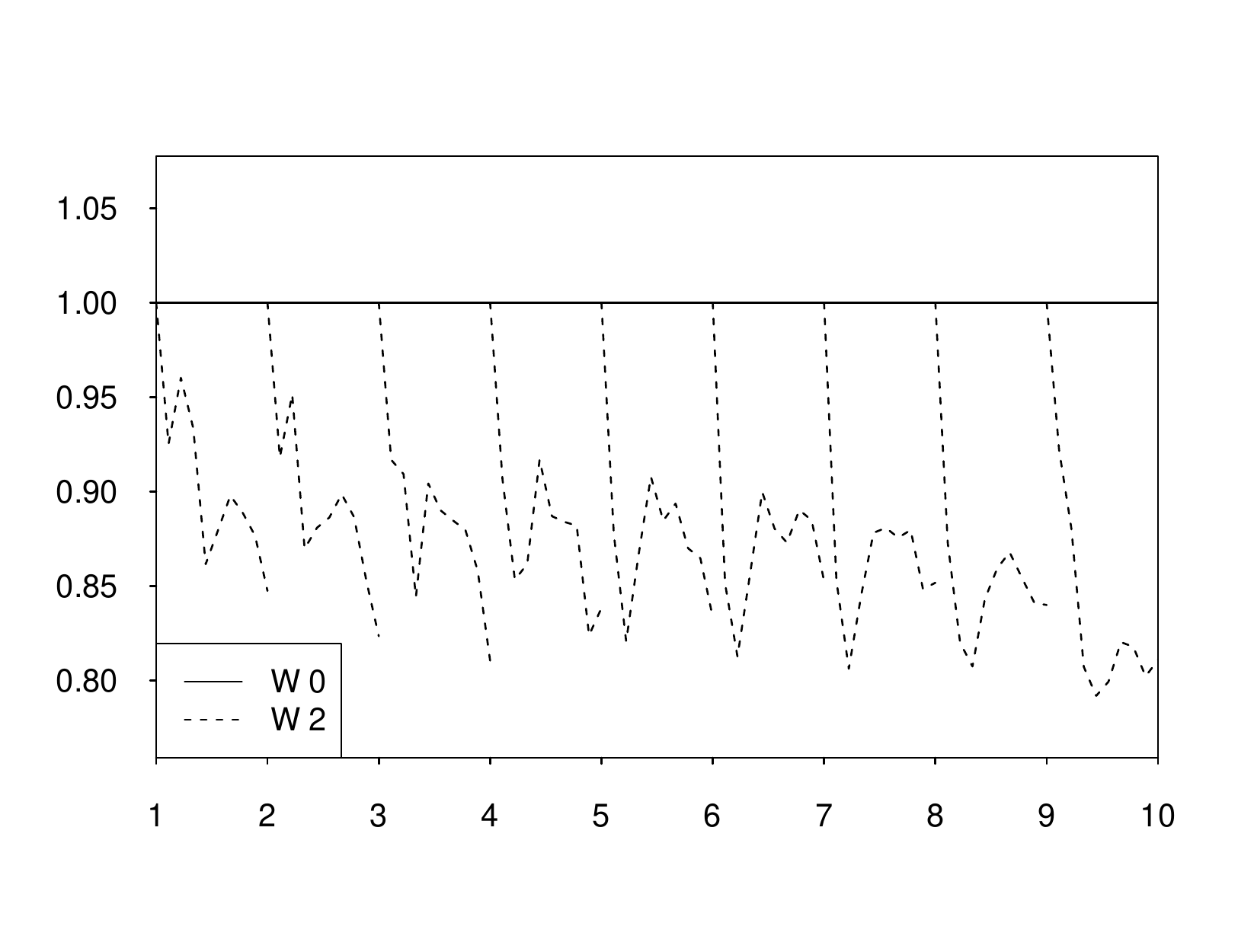}}

\caption{Average train iterations (left) and corresponding ratio (right) per grid point for different initialization methods 
considering  stopping criteria with clipped duality gap and \textsf{WSS 2} for the data set \textsc{cal-housing}. 
The graphs comprises for $\tau=0.25$ (top), $\tau=0.50$ (middle) and $\tau=0.75$ (bottom).}
  \label{figure-per-grid iter for initialization-cal-housing}
\end{scriptsize}
\end{figure}

\newpage
\vspace{-3cm}
\subsection{Results for two Different Stopping Criteria }
\begin{figure}[!ht]
\begin{scriptsize}
 \subfloat{\includegraphics[scale=0.52]{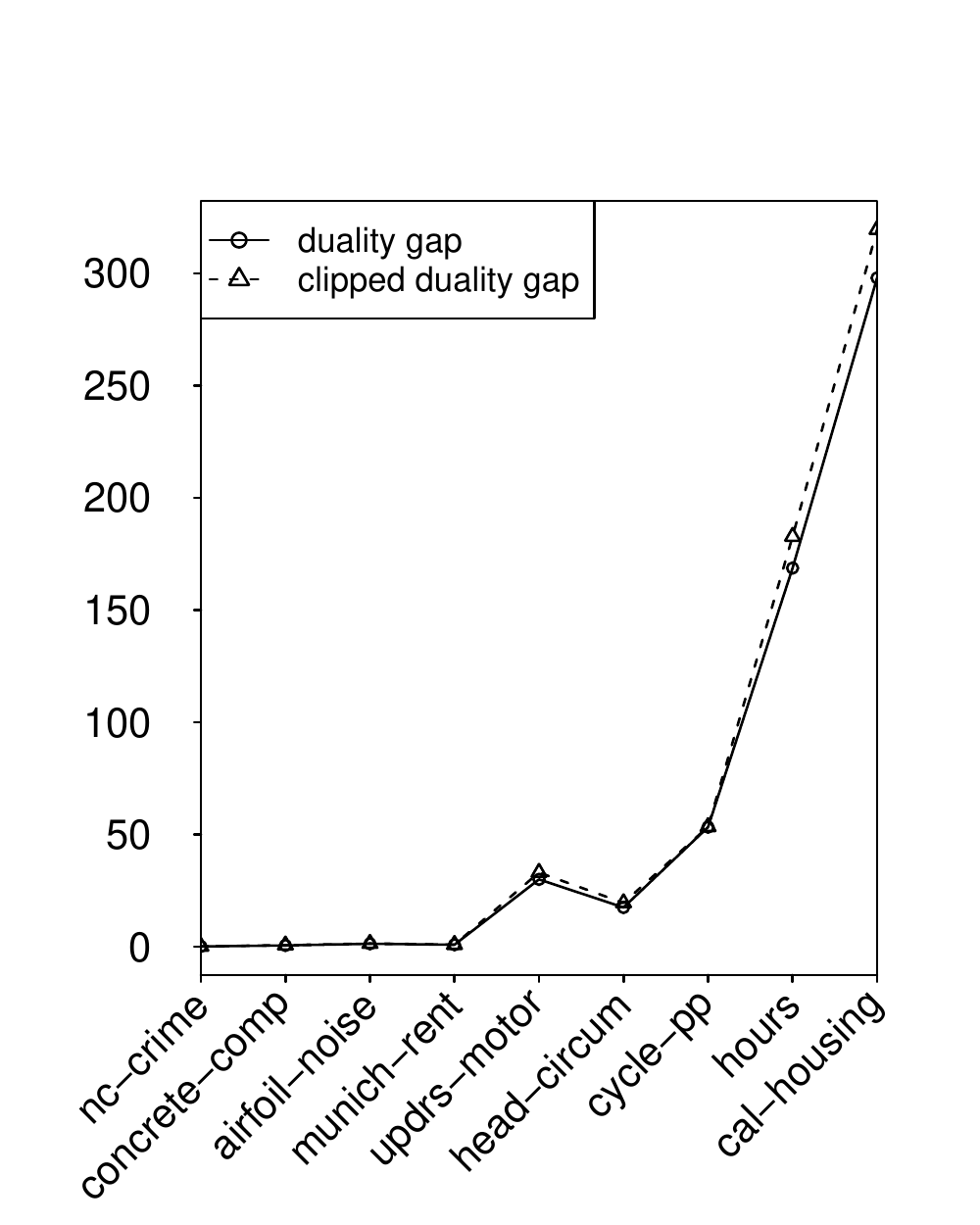}}\hspace{-0.7cm}
 \hfill
 \subfloat{\includegraphics[scale=0.52]{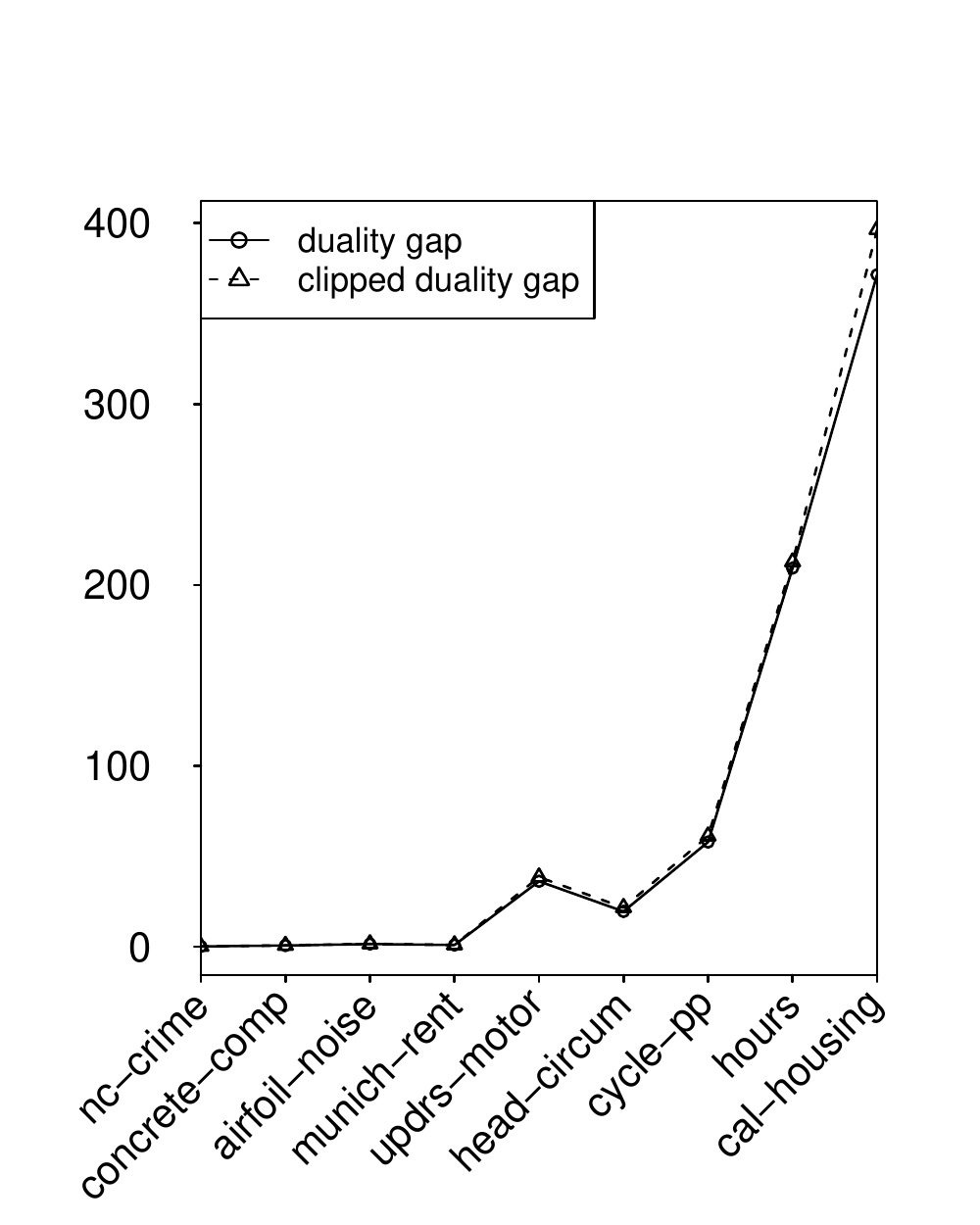}}\hspace{-0.7cm}
\hfill
 \subfloat{\includegraphics[scale=0.52]{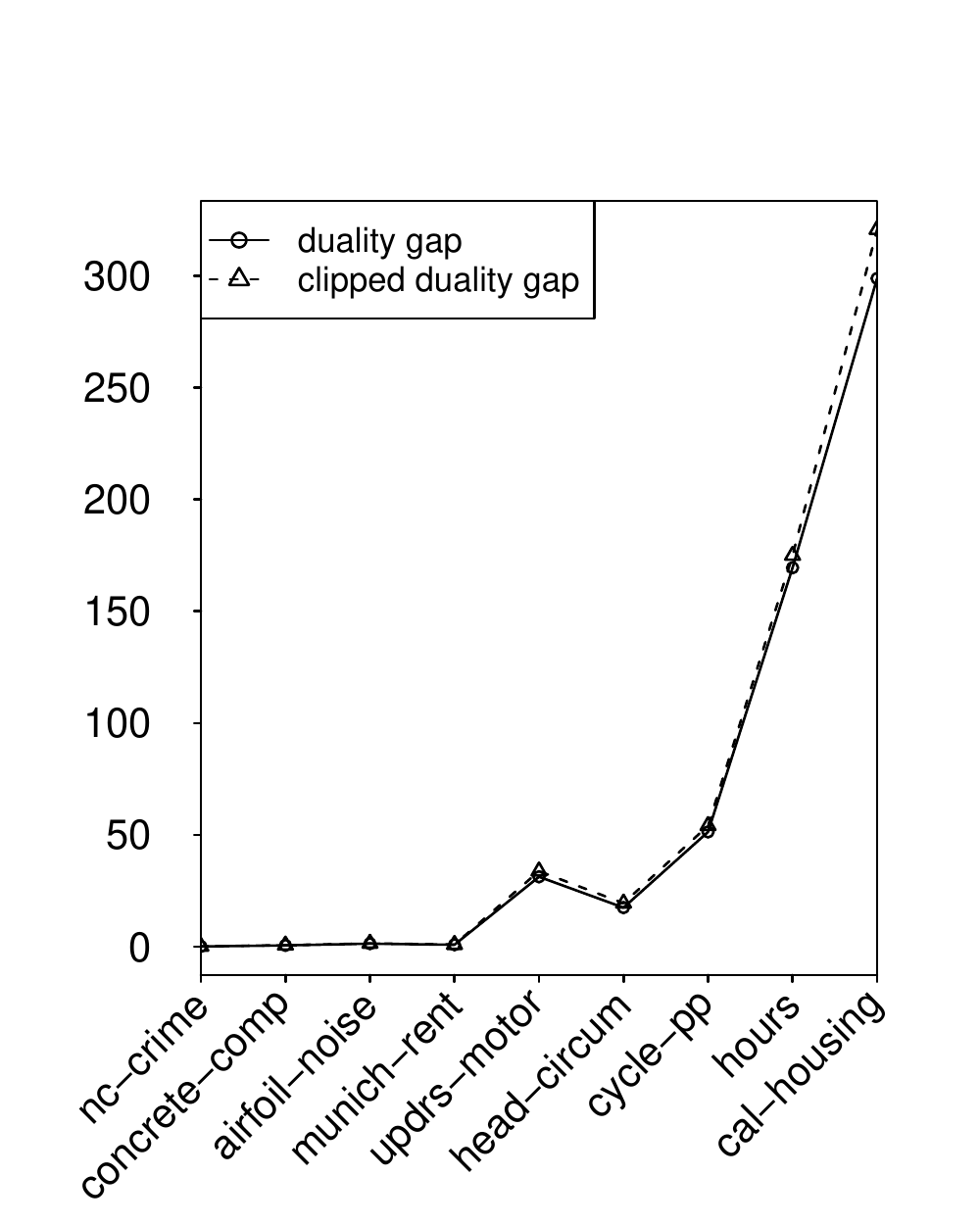}}
 
 \vspace{-1.0cm}
\subfloat{\includegraphics[scale=0.52]{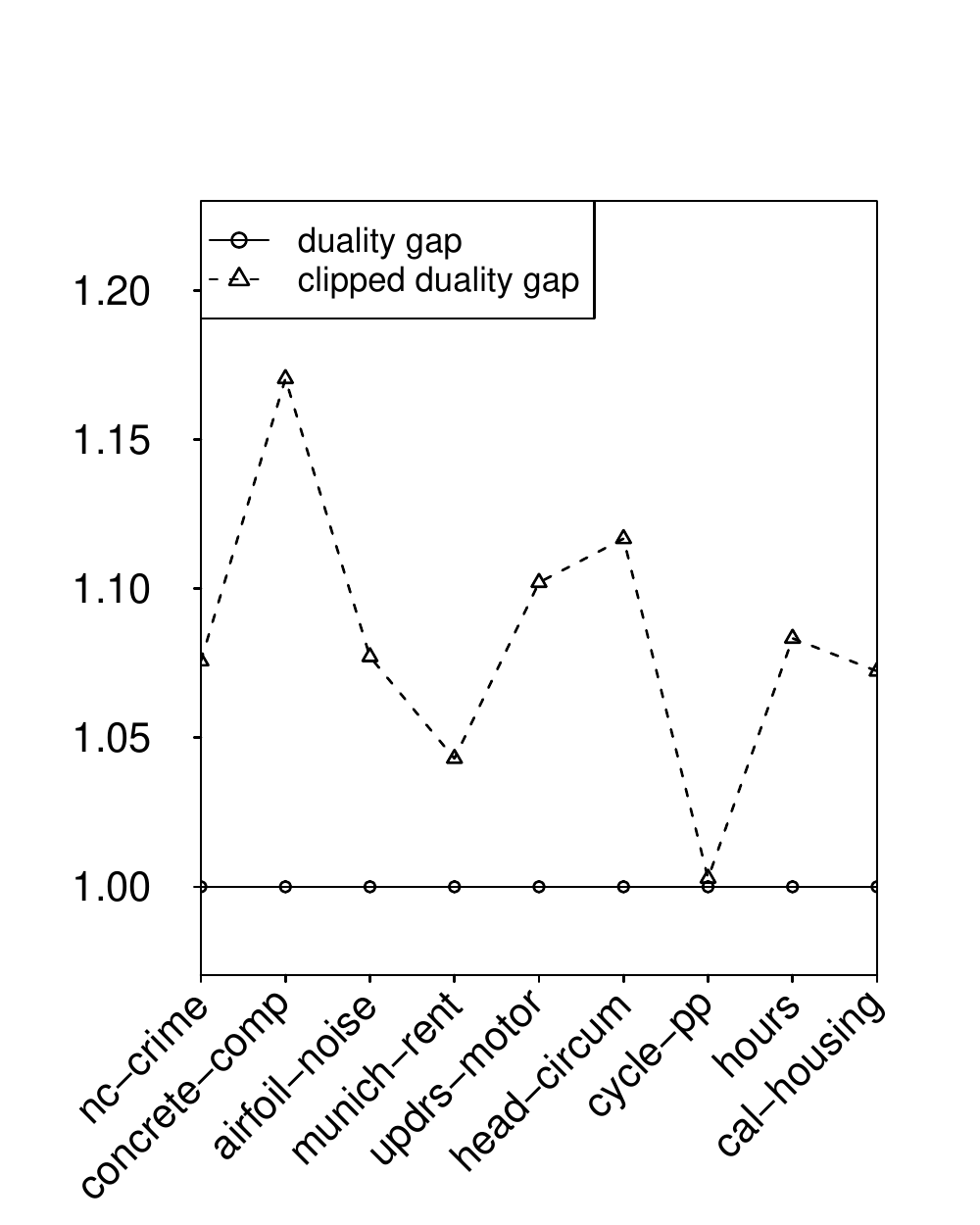}}\hspace{-0.7cm}
 \hfill
 \subfloat{\includegraphics[scale=0.52]{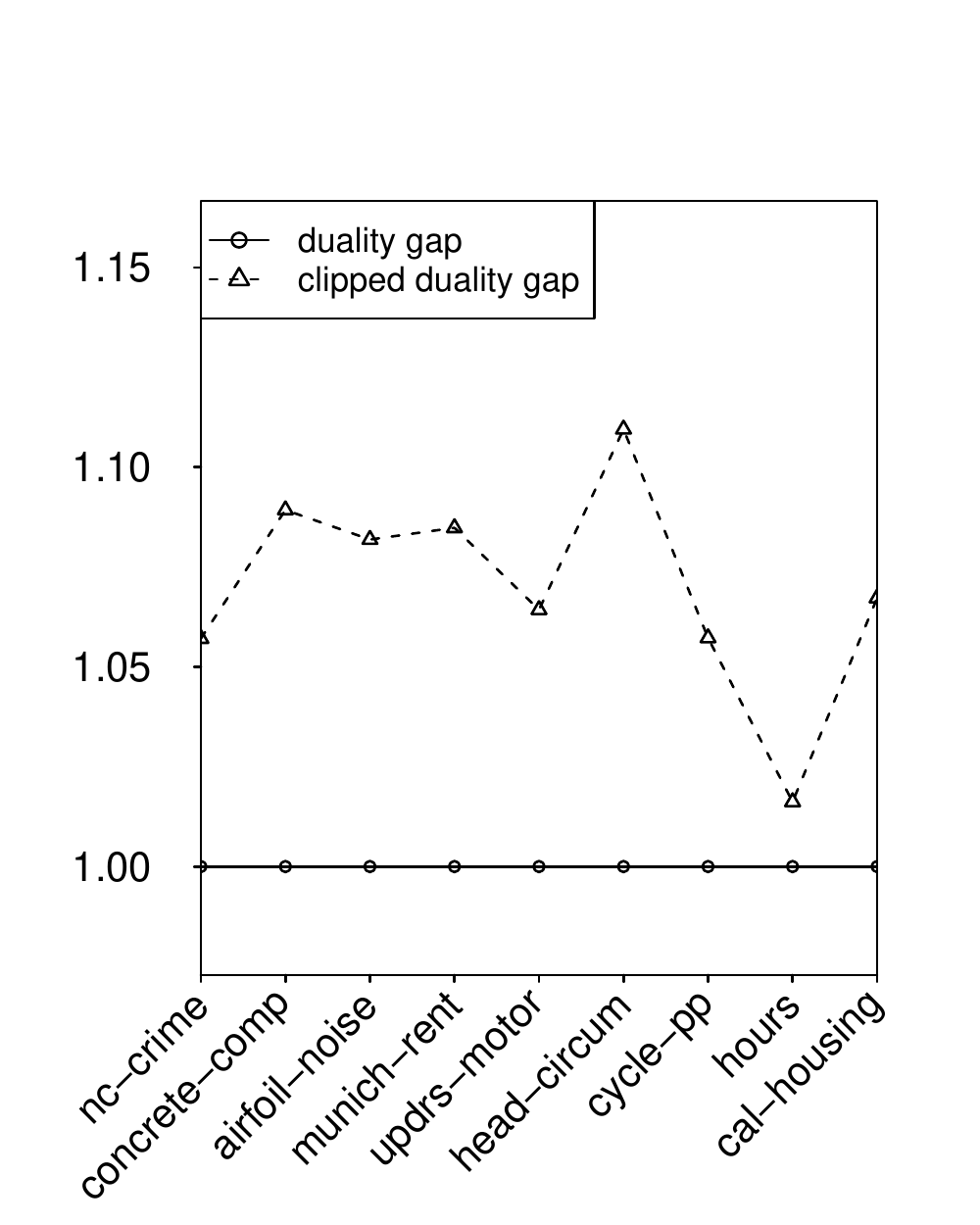}}\hspace{-0.7cm}
\hfill
 \subfloat{\includegraphics[scale=0.52]{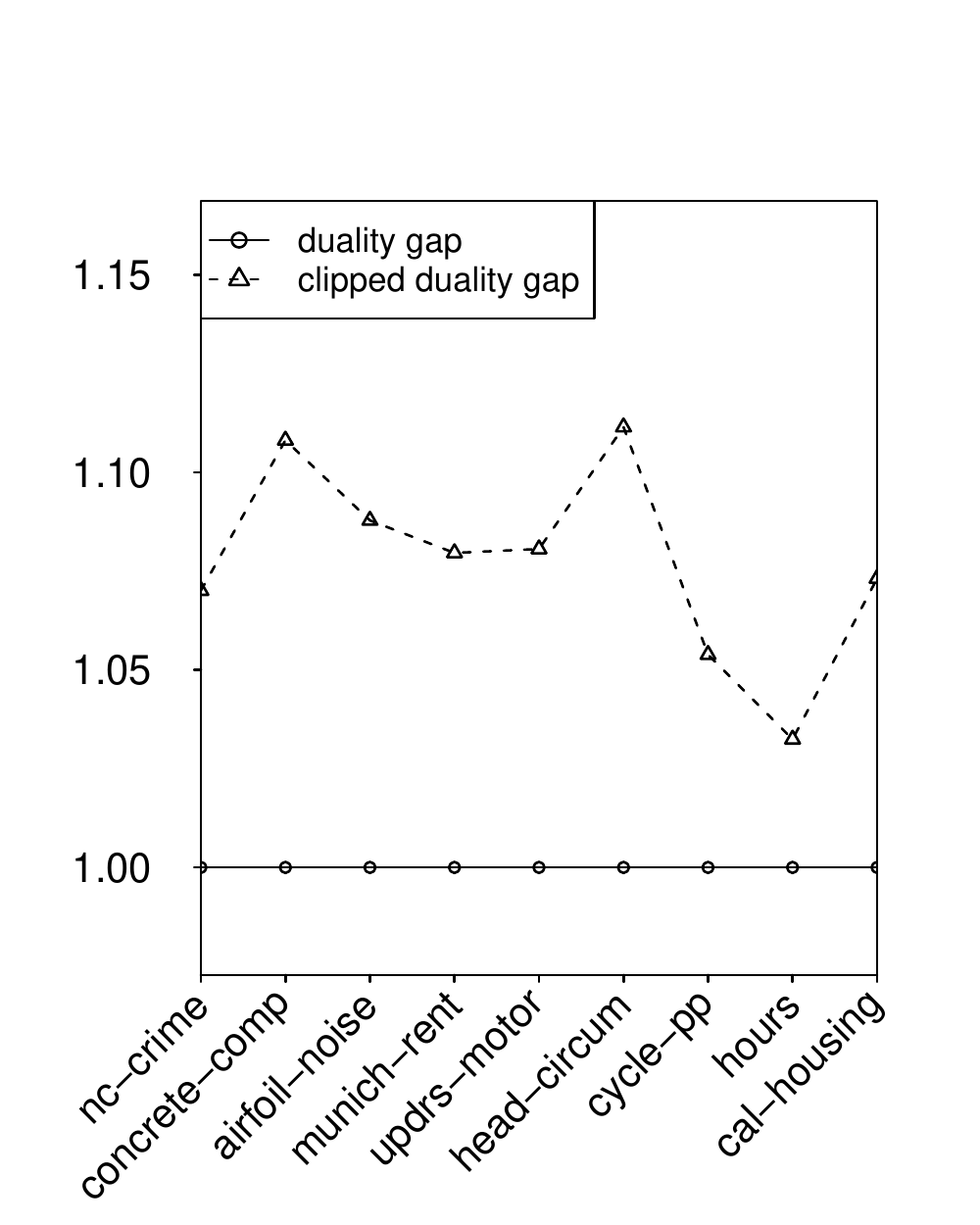}}
 \caption{Train time (top) and corresponding ratio (bottom) of different data sets for different stopping criteria after 
 fixing initialization method as warm start and $NN=15$. The graphs comprises of $\tau=0.25$ (left), $\tau=0.50$ (middle) 
 and $\tau=0.75$ (right).}
  \label{figure-time and ratio-duality gap vs datasets}
\end{scriptsize}
\end{figure}

\newpage
\vspace{-3cm}
\begin{figure}[!ht]
\begin{scriptsize}
 \subfloat{\includegraphics[scale=0.52]{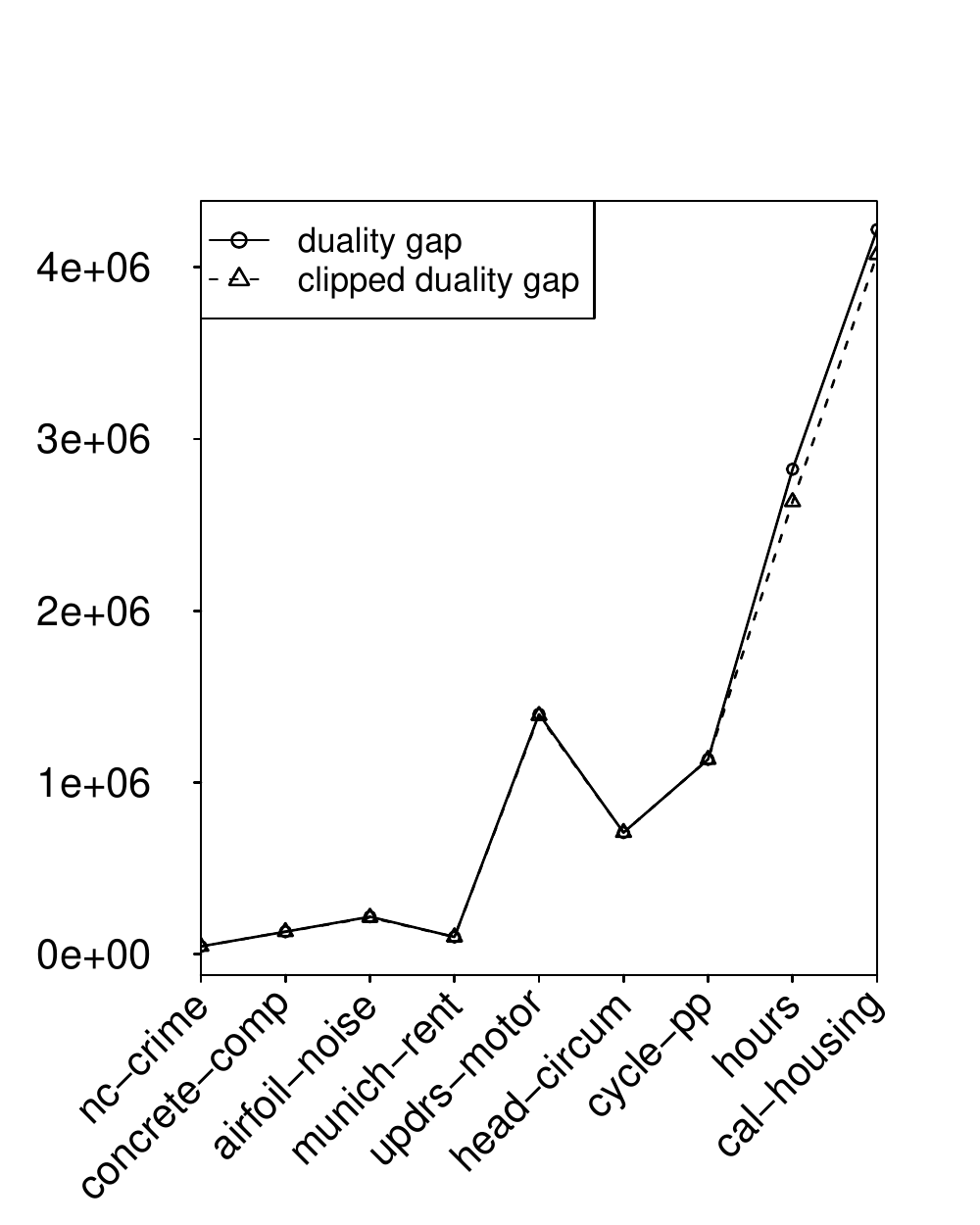}}\hspace{-0.7cm}
 \hfill
 \subfloat{\includegraphics[scale=0.52]{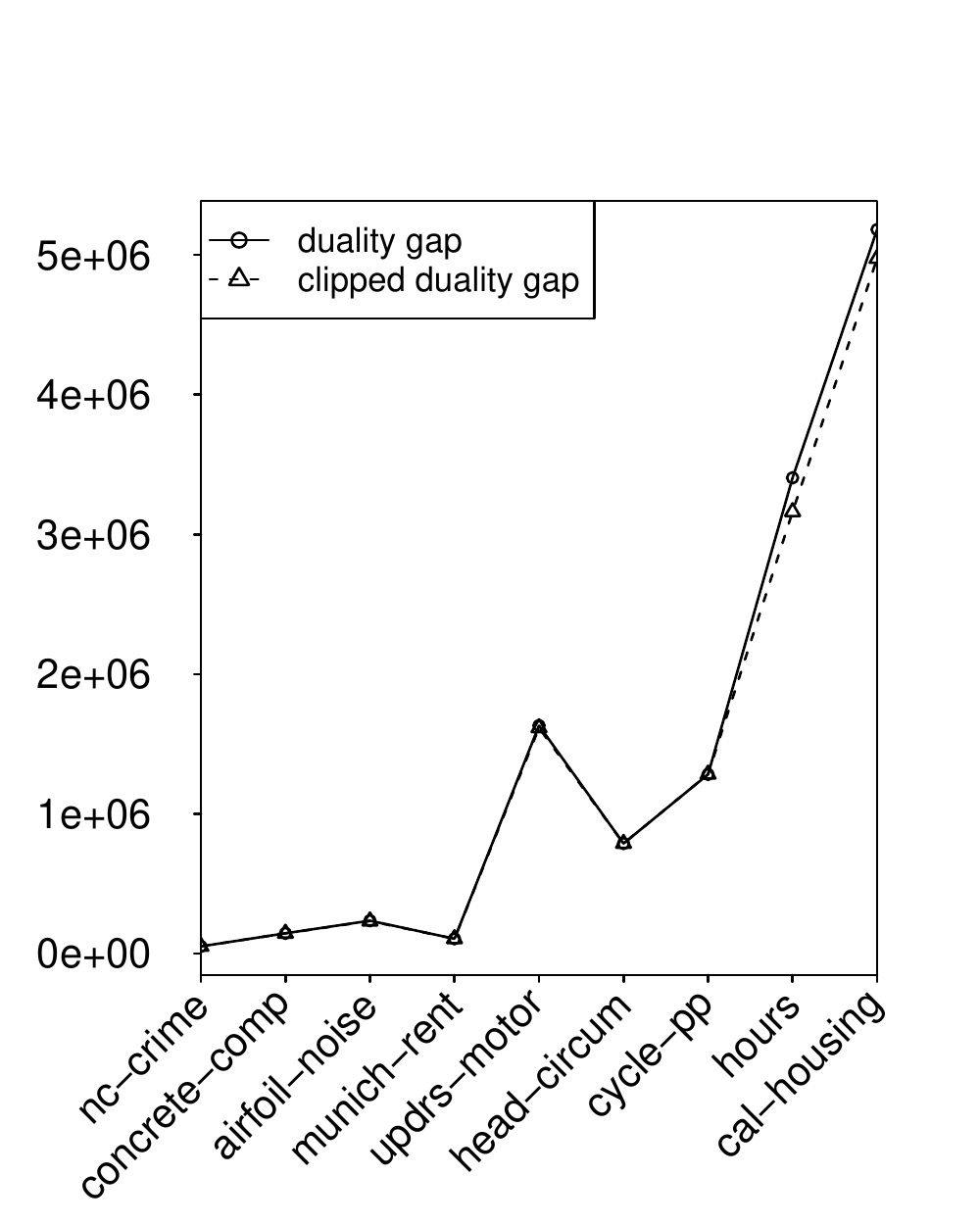}}\hspace{-0.7cm}
\hfill
 \subfloat{\includegraphics[scale=0.52]{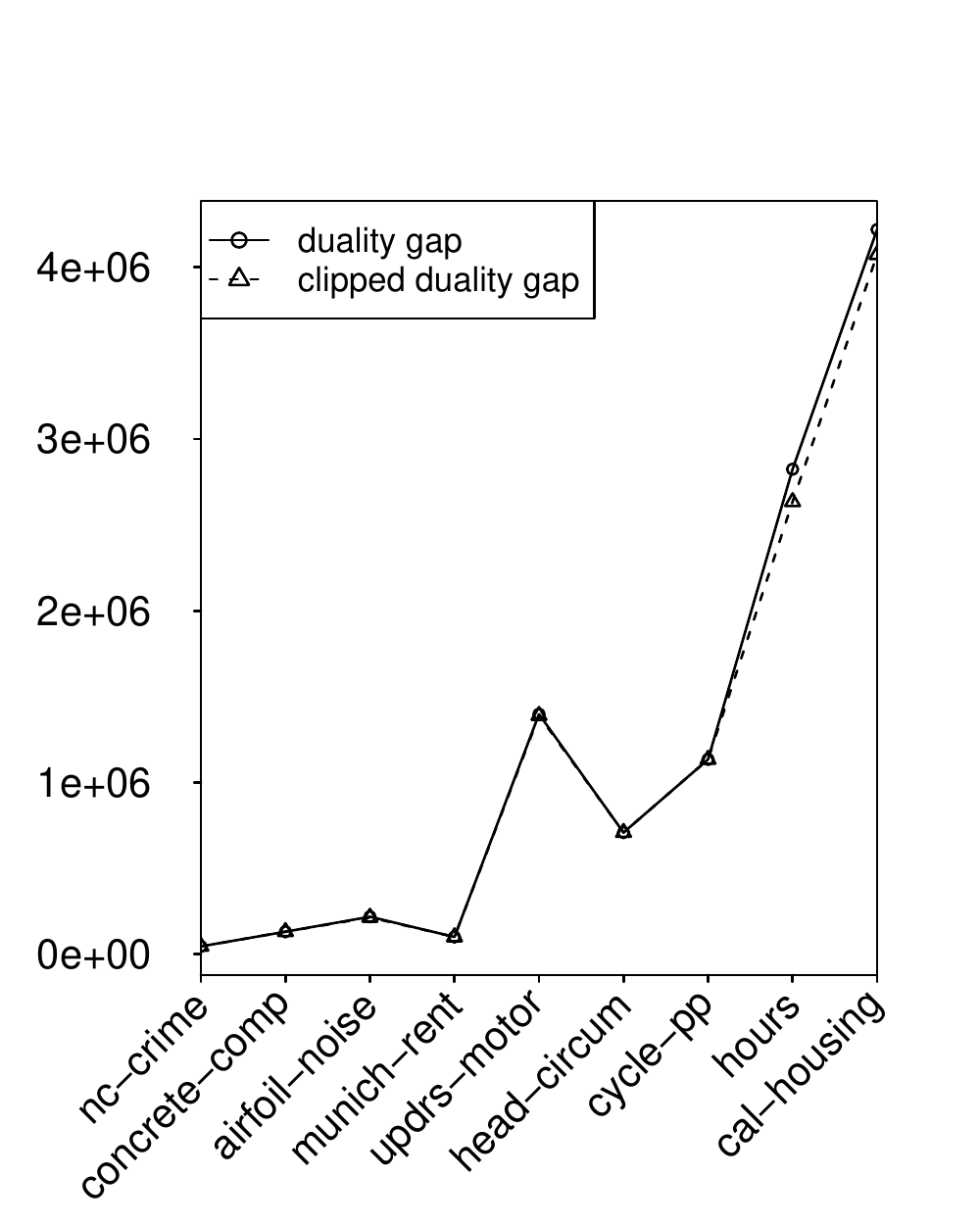}}
 
 \vspace{-1.0cm}
\subfloat{\includegraphics[scale=0.52]{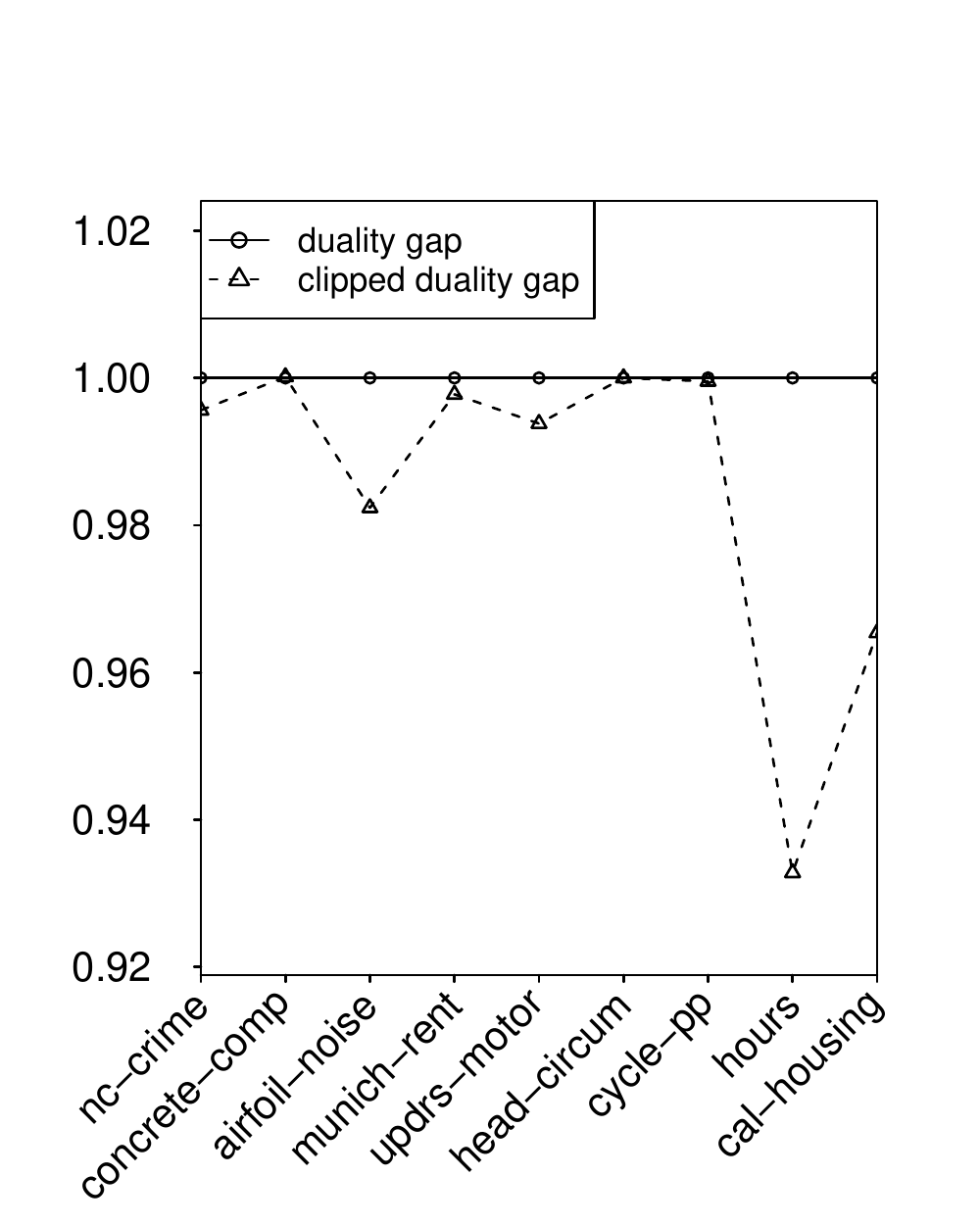}}\hspace{-0.7cm}
 \hfill
 \subfloat{\includegraphics[scale=0.52]{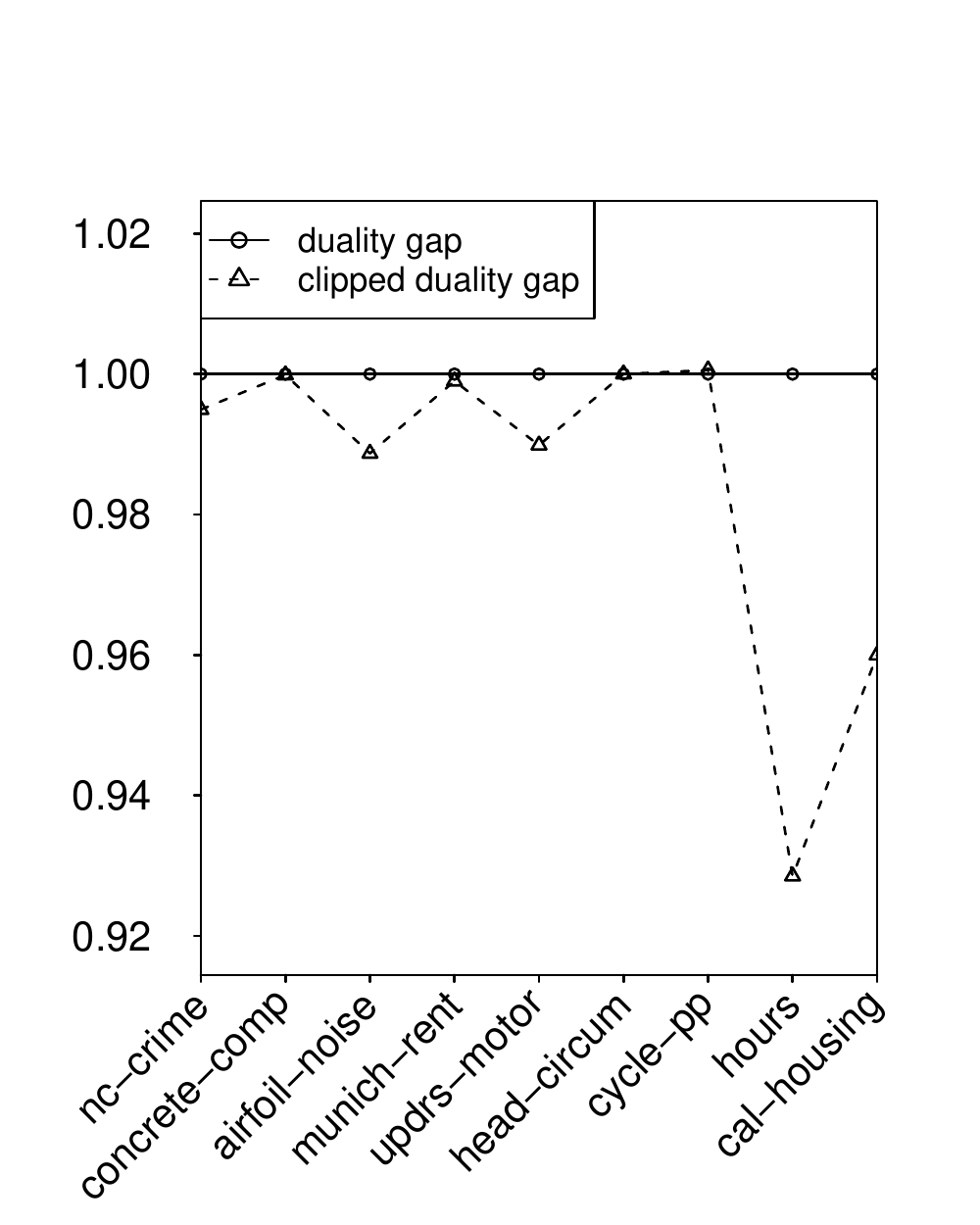}}\hspace{-0.7cm}
\hfill
 \subfloat{\includegraphics[scale=0.52]{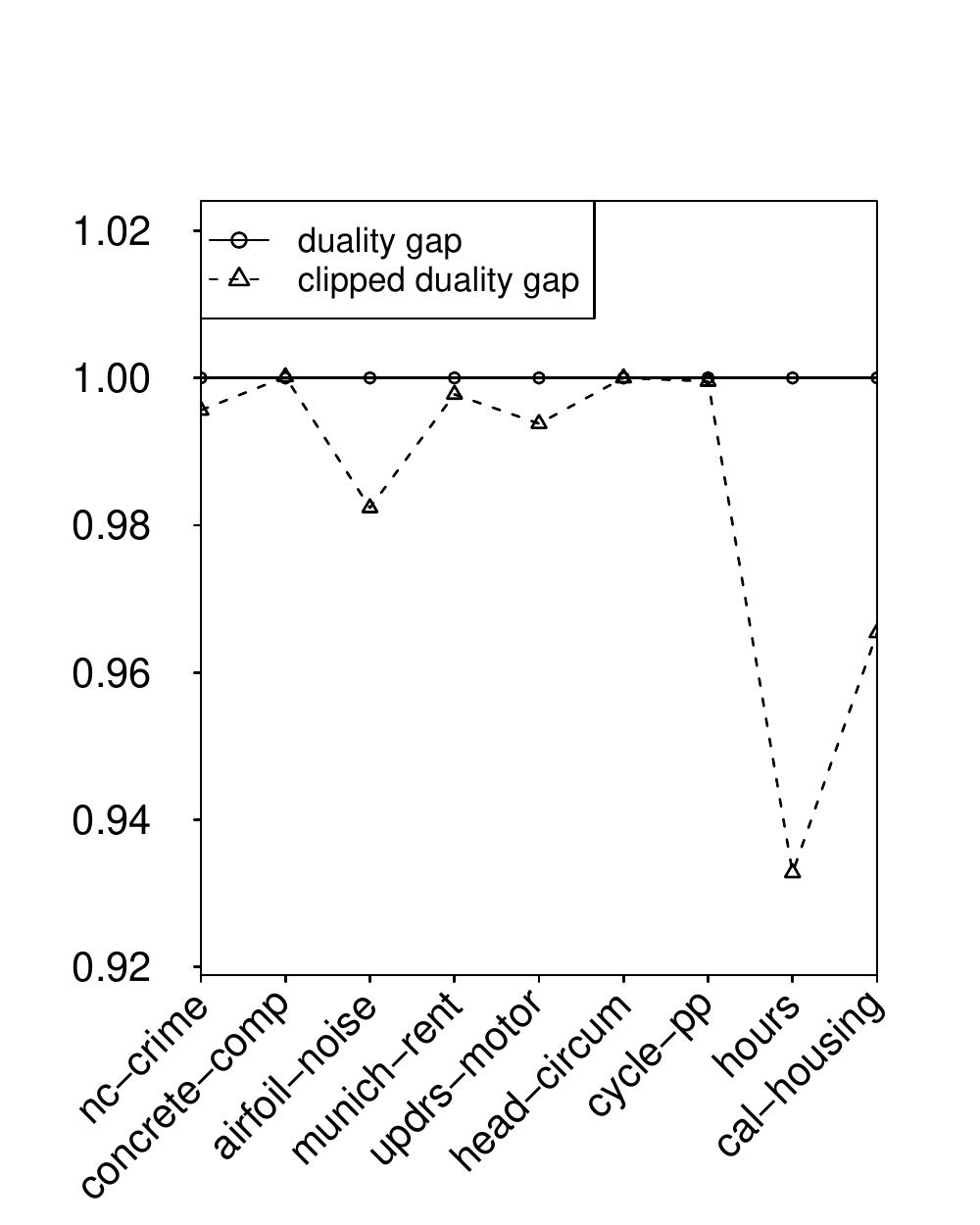}}
 \caption{Train iterations (top) and corresponding ratio (bottom) of different data sets for different stopping criteria 
 after fixing initialization method as warm start and $NN=15$. The graphs comprises of $\tau=0.25$ (left), $\tau=0.50$ 
 (middle) and $\tau=0.75$ (right).}
   \label{figure-iter and ratio-duality gap vs datasets}
\end{scriptsize}
\end{figure}

\newpage

\begin{figure}[!ht]
\begin{scriptsize}
\vspace{-1cm}
 \subfloat{\includegraphics[scale=0.49]{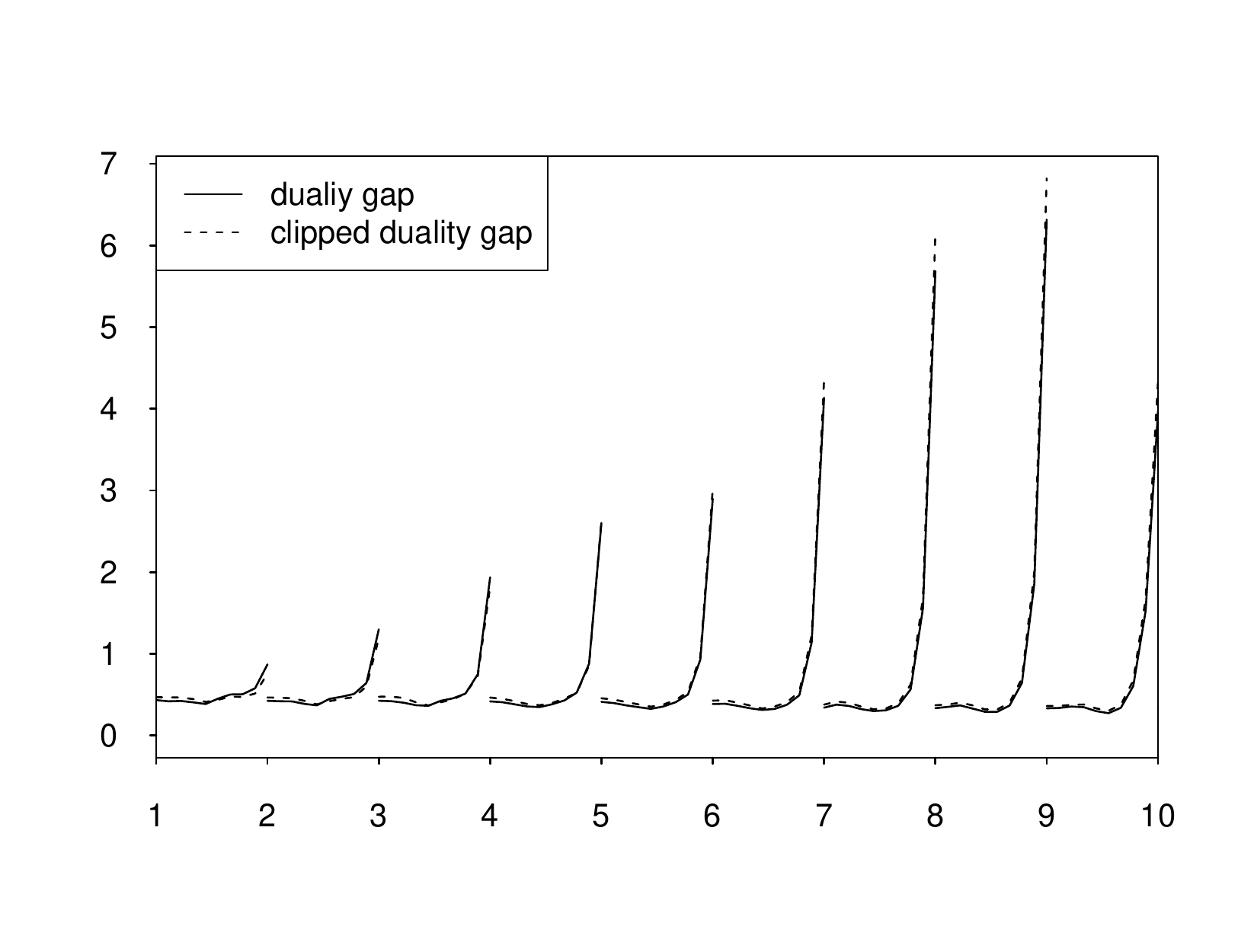}}\hspace{-0.7cm}
 \hfill
 \subfloat{\includegraphics[scale=0.49]{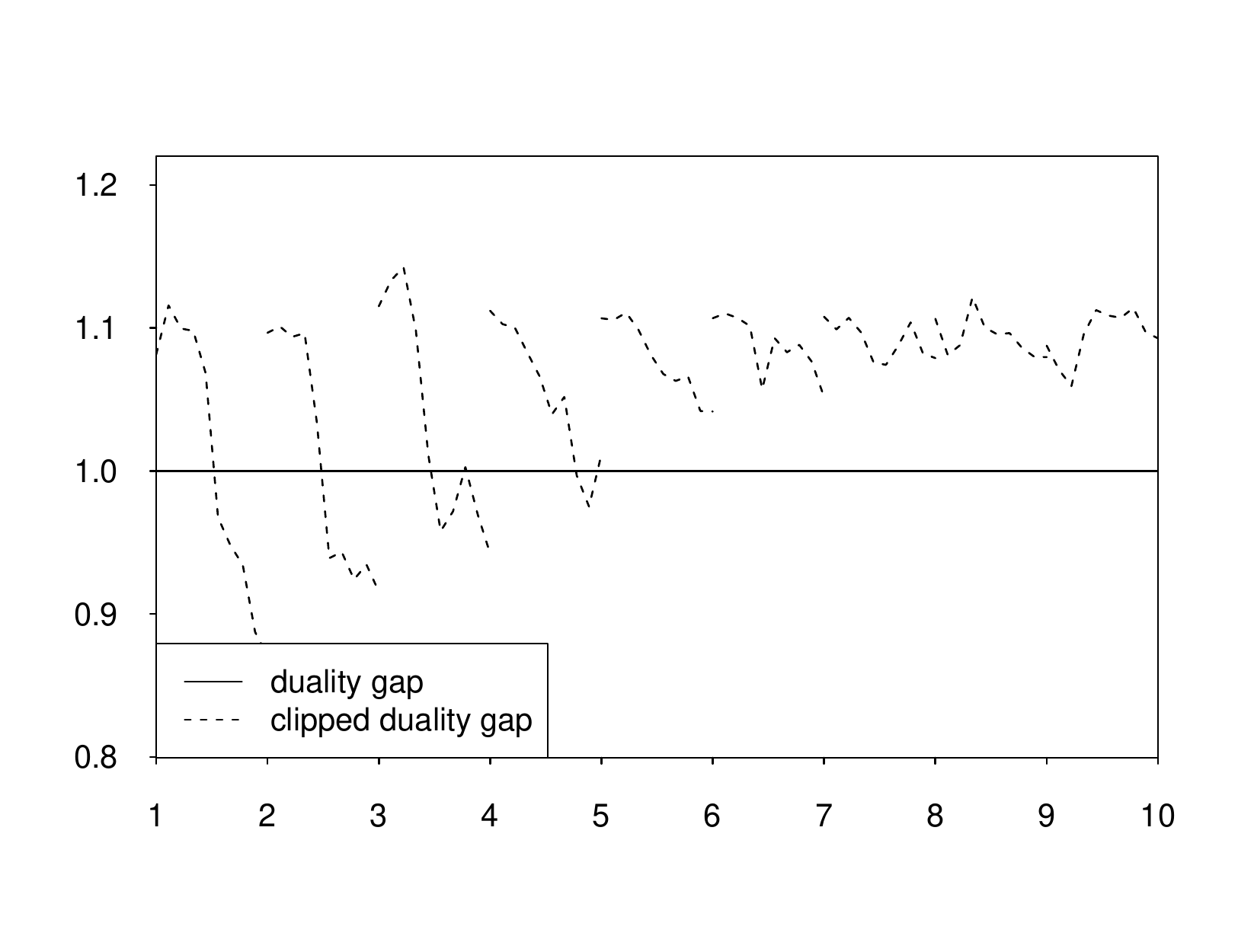}}
 \vspace{-1.9cm}
\subfloat{\includegraphics[scale=0.49]{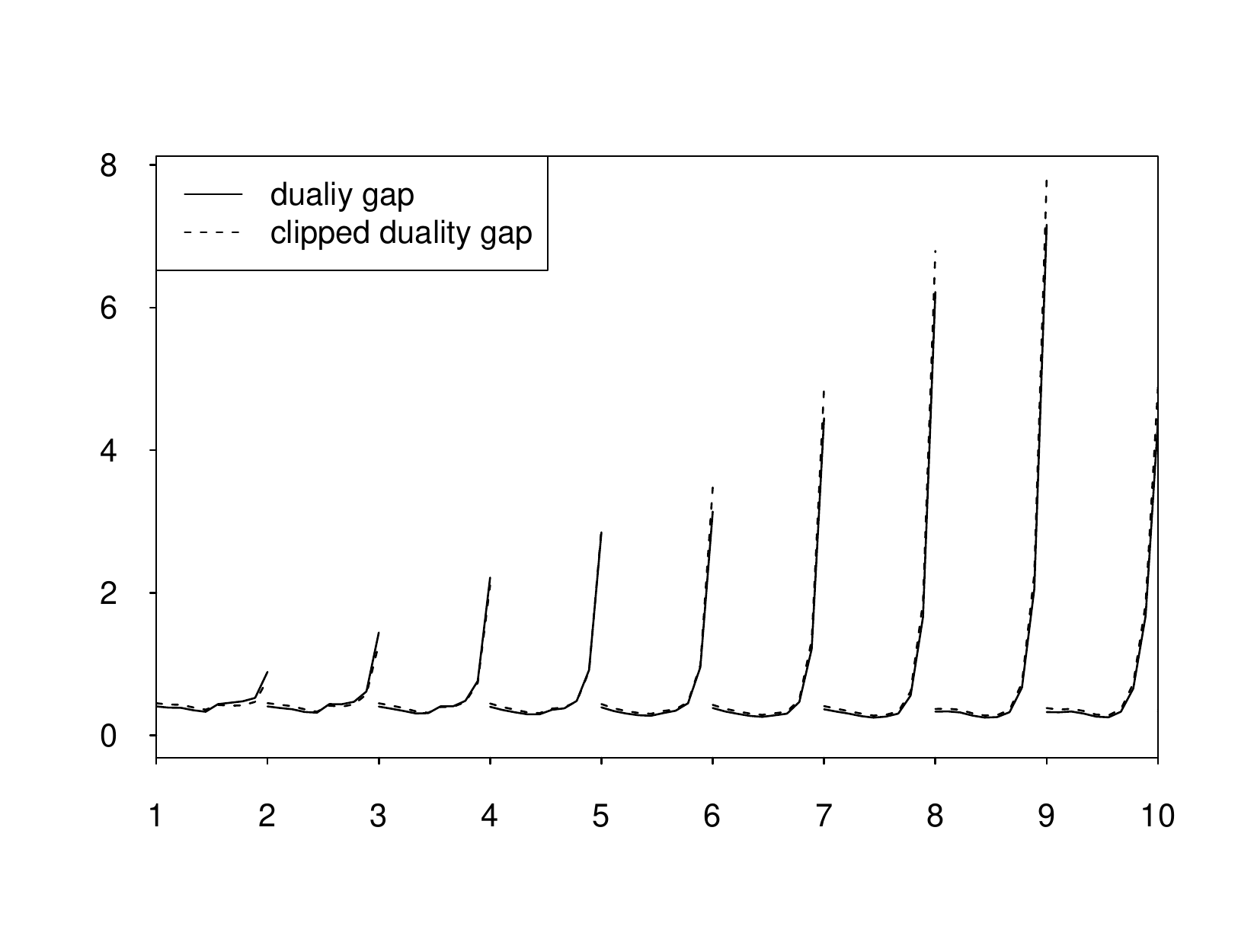}}\hspace{-0.7cm}
\hfill
\subfloat{\includegraphics[scale=0.49]{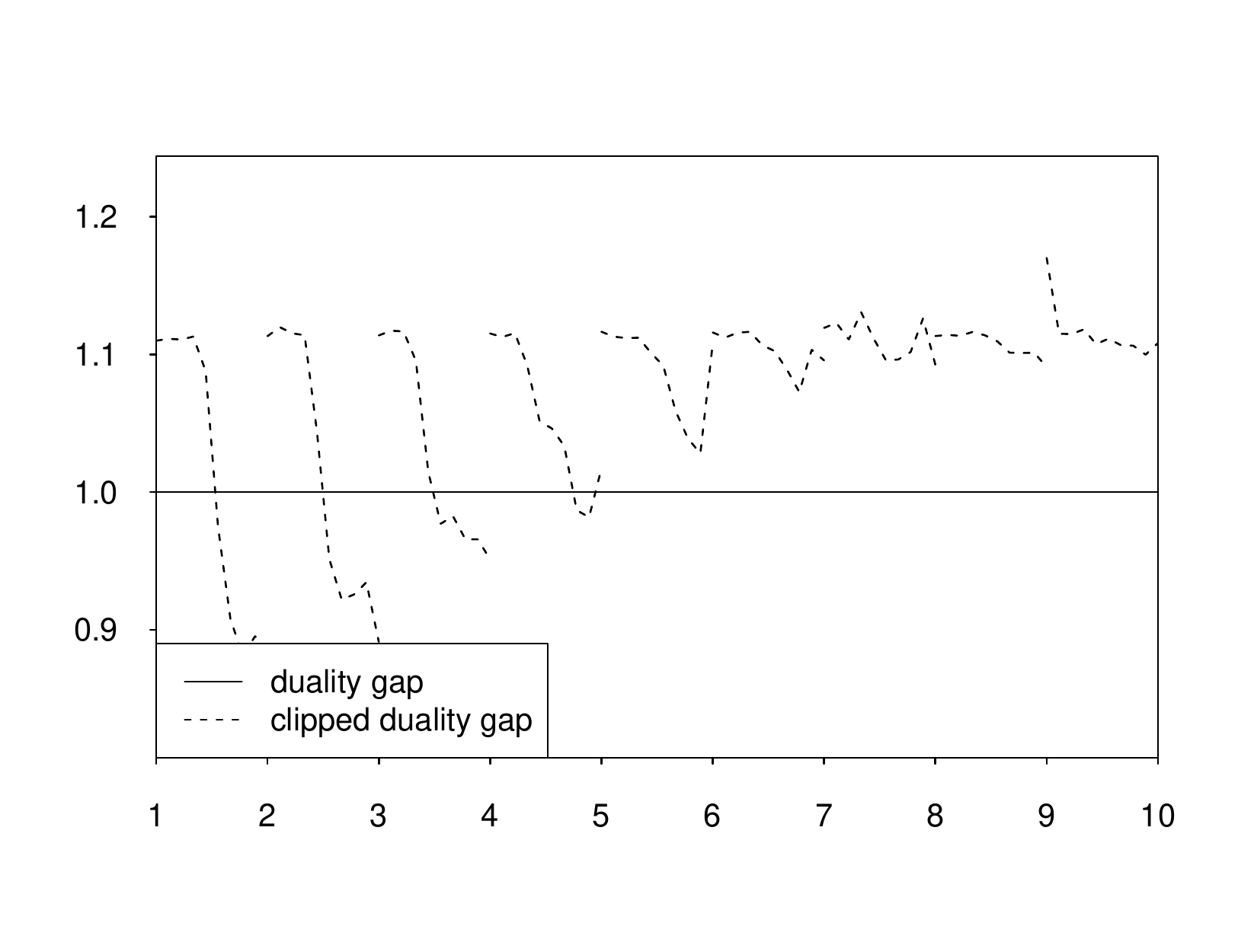}}
\vspace{-1.9cm}
\subfloat{\includegraphics[scale=0.49]{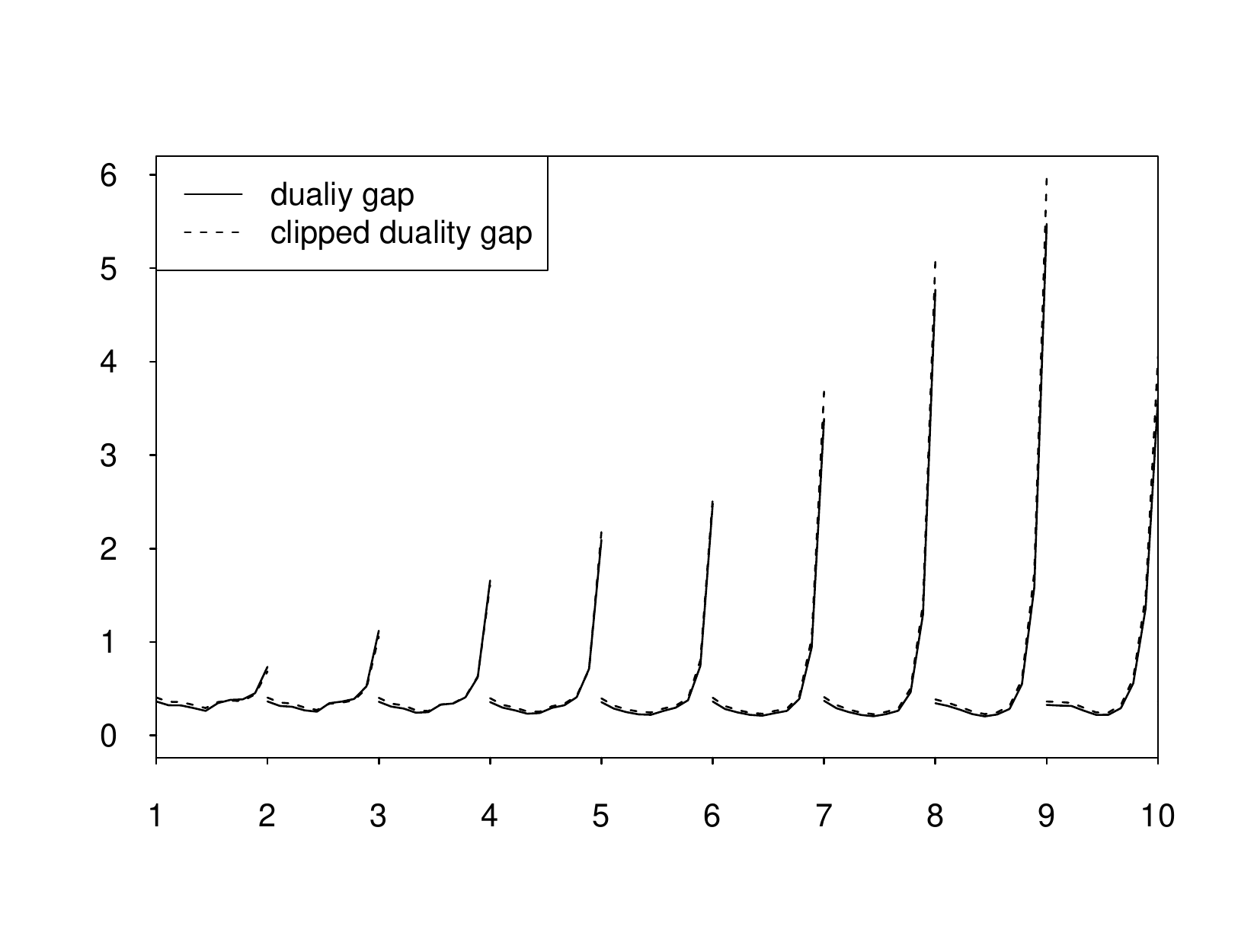}}\hspace{-0.7cm}
\hfill
\subfloat{\includegraphics[scale=0.49]{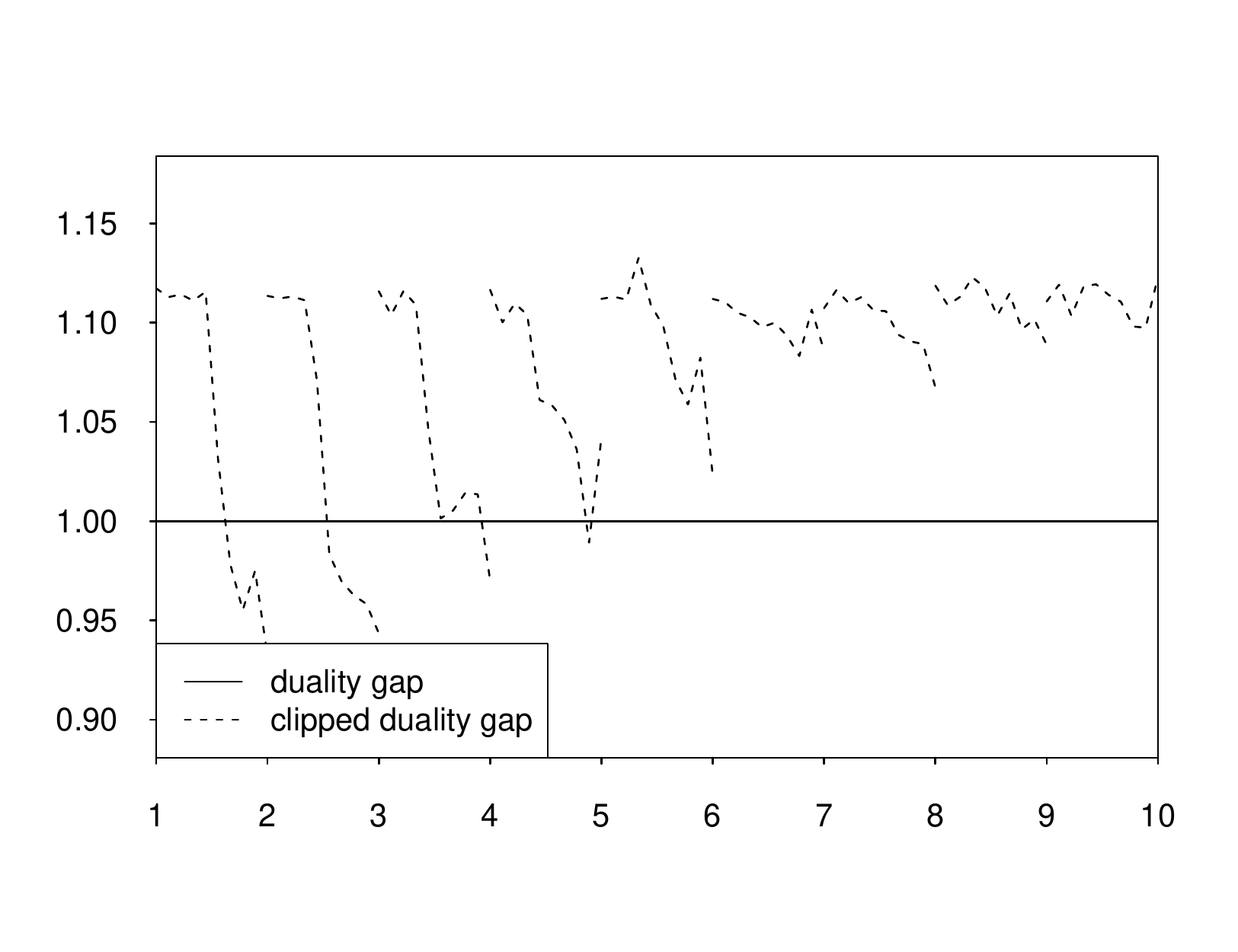}}

\caption{Average train time (left) and corresponding ratio (right) per grid point for different stopping criteria using 15 NN 
and initializing solver with warm start for \textsc{cal-housing}. The graphs comprises for $\tau=0.25$ (top), $\tau=0.50$ (middle)
and $\tau=0.75$ (bottom).}
\label{figure-figure-per-grid time for duality gap-cal-housing}
\end{scriptsize}
\end{figure}
\newpage
\vspace{0cm}
\begin{figure}[!ht]
\begin{scriptsize}
\vspace{-1cm}
 \subfloat{\includegraphics[scale=0.49]{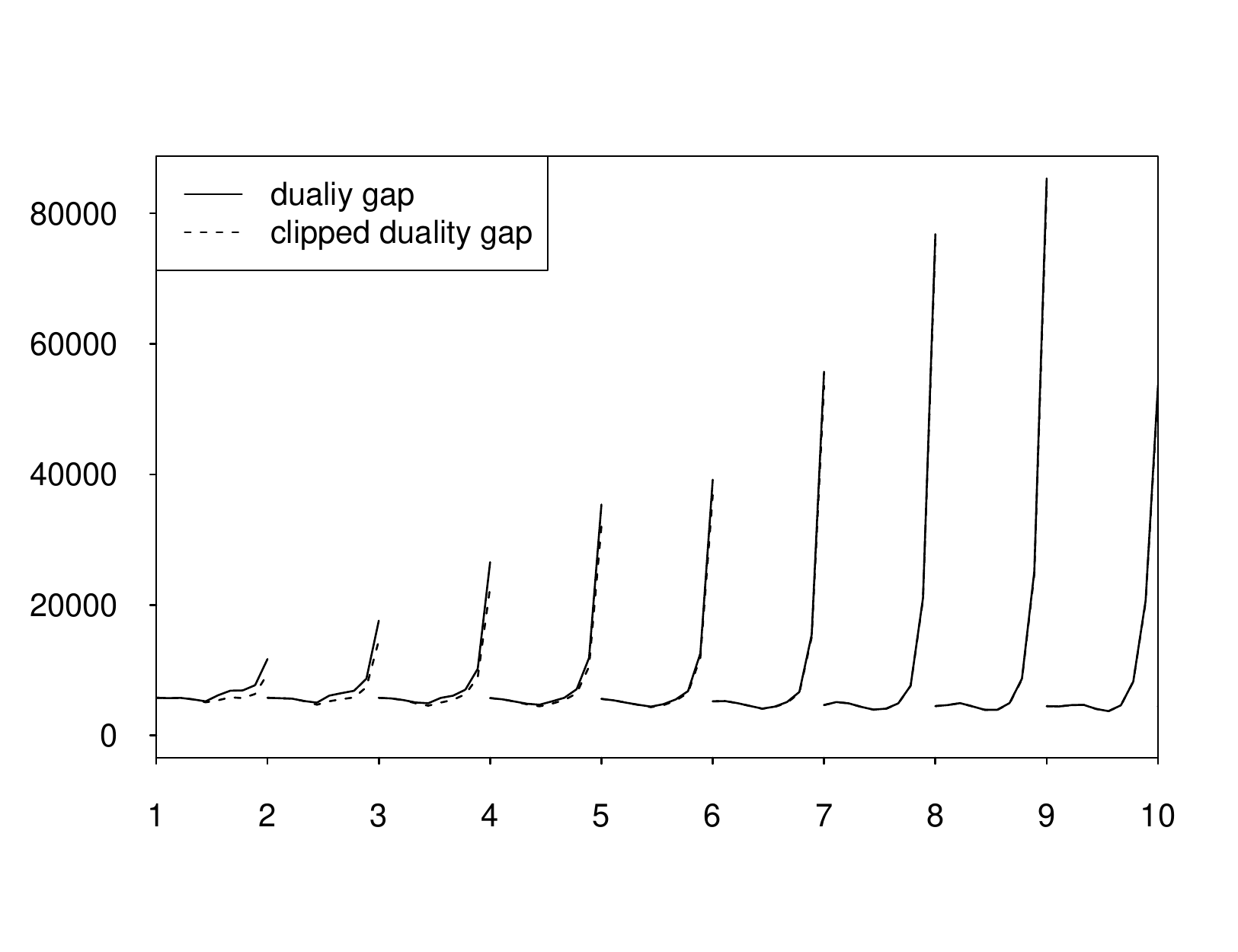}}\hspace{-0.7cm}
 \hfill
 \subfloat{\includegraphics[scale=0.49]{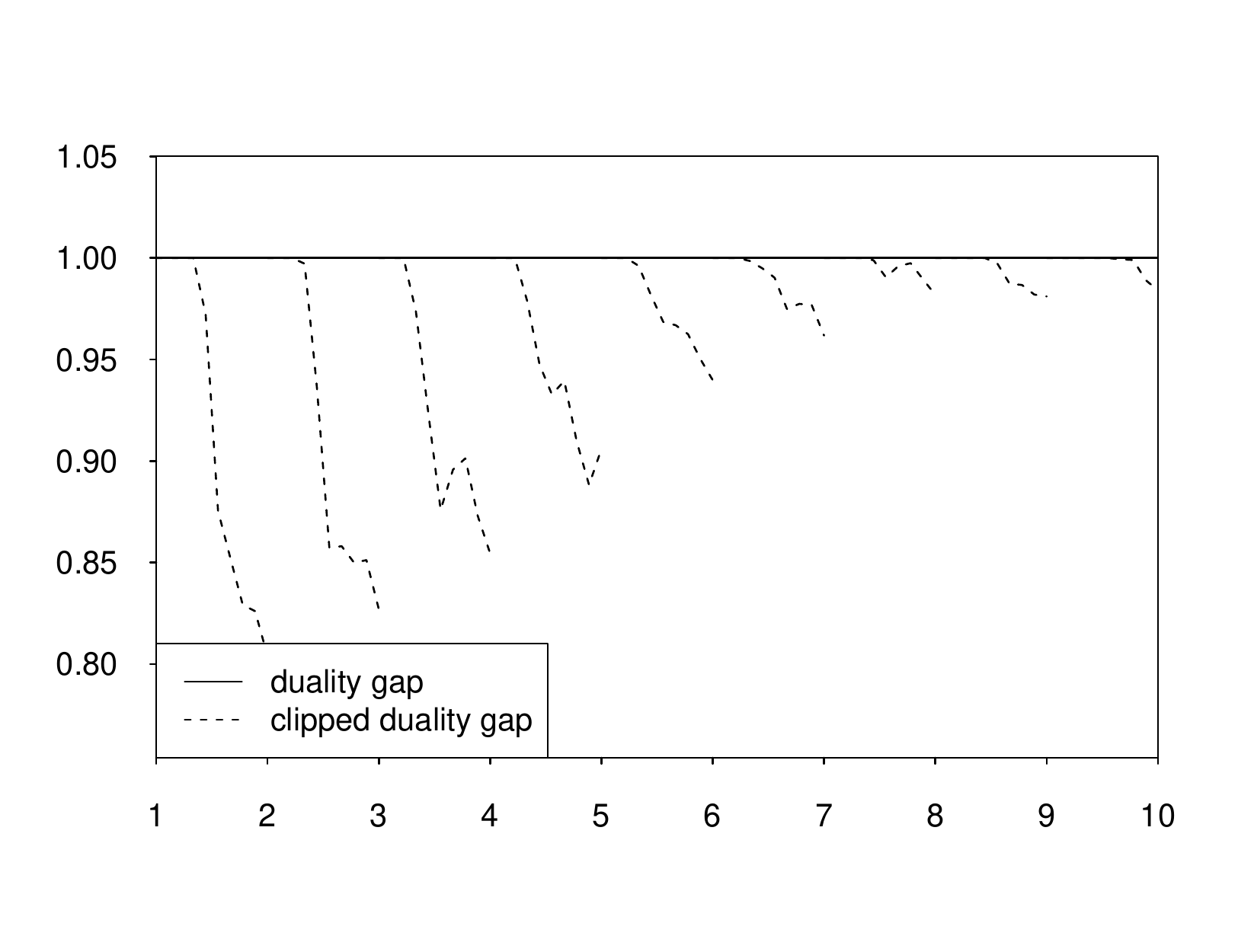}}
 \vspace{-1.9cm}
\subfloat{\includegraphics[scale=0.49]{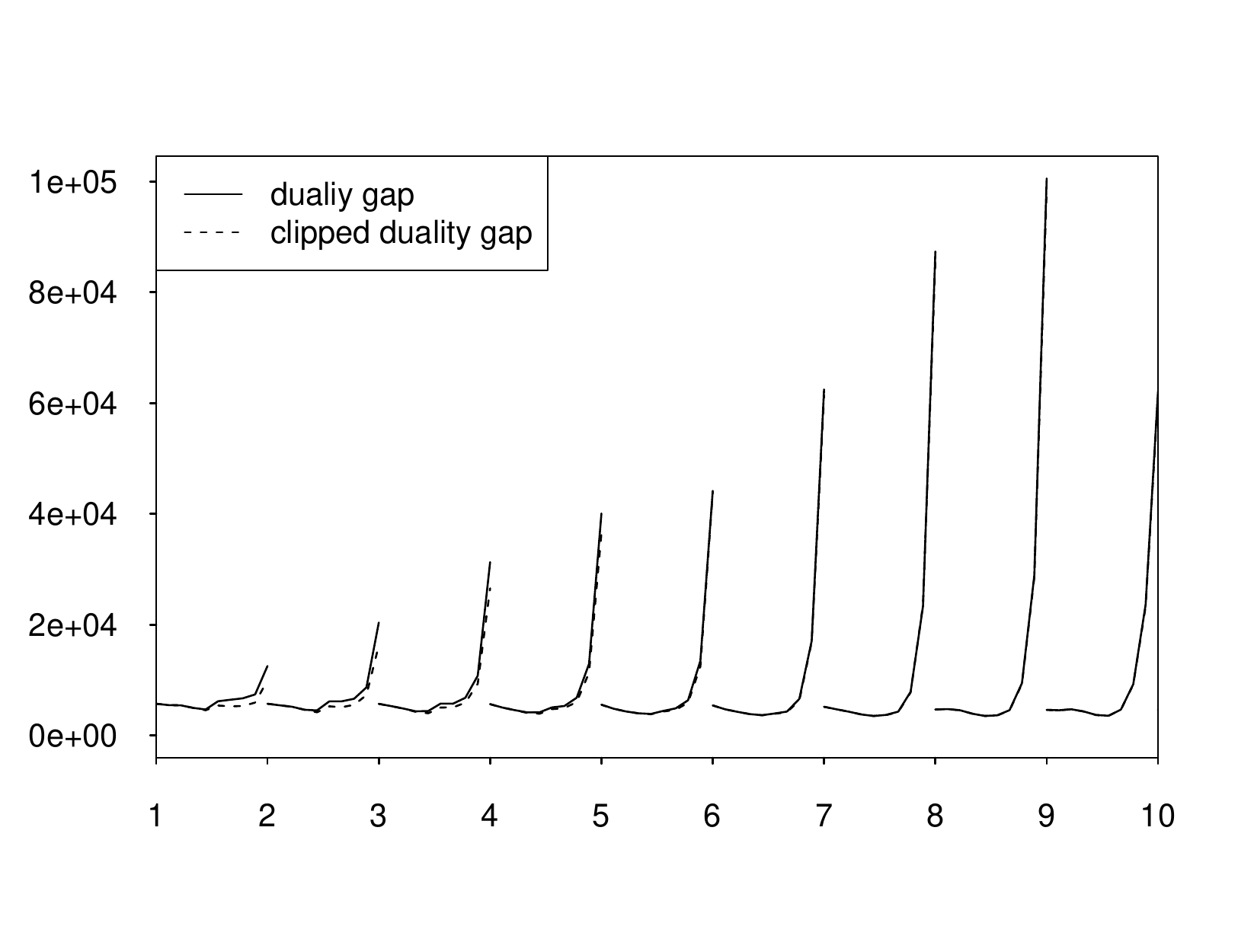}}\hspace{-0.7cm}
\hfill
\subfloat{\includegraphics[scale=0.49]{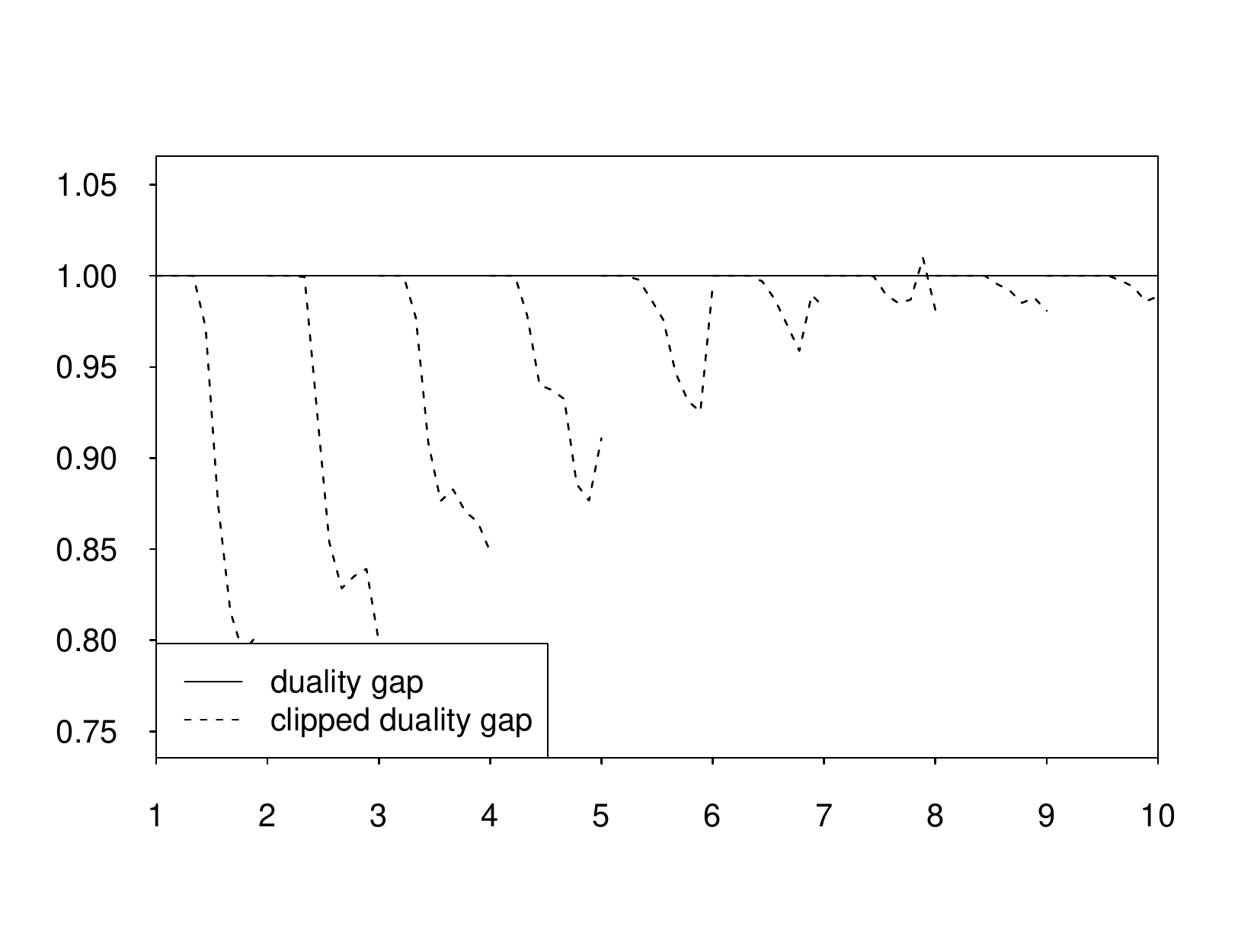}}
\vspace{-1.9cm}
\subfloat{\includegraphics[scale=0.49]{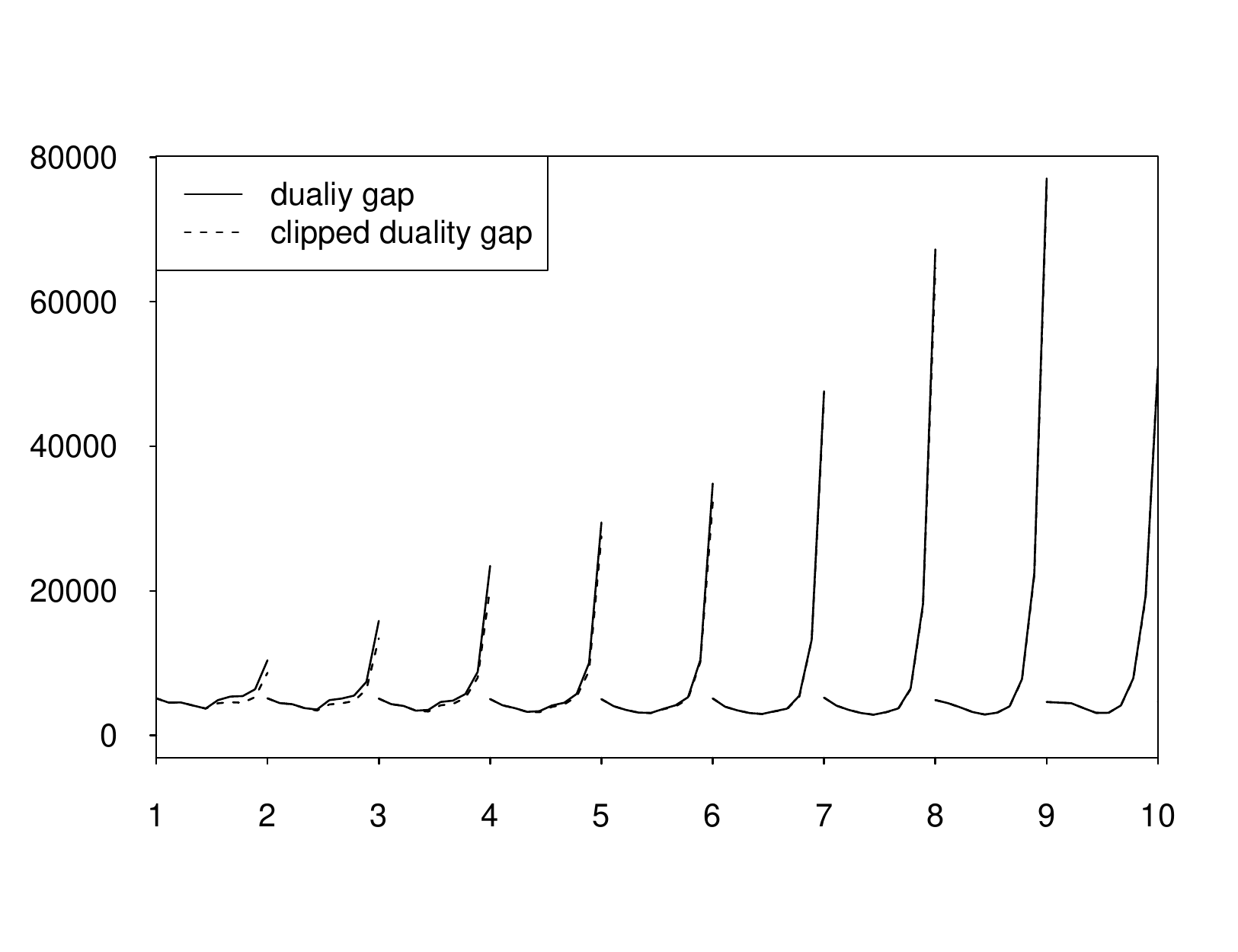}}\hspace{-0.7cm}
\hfill
\subfloat{\includegraphics[scale=0.49]{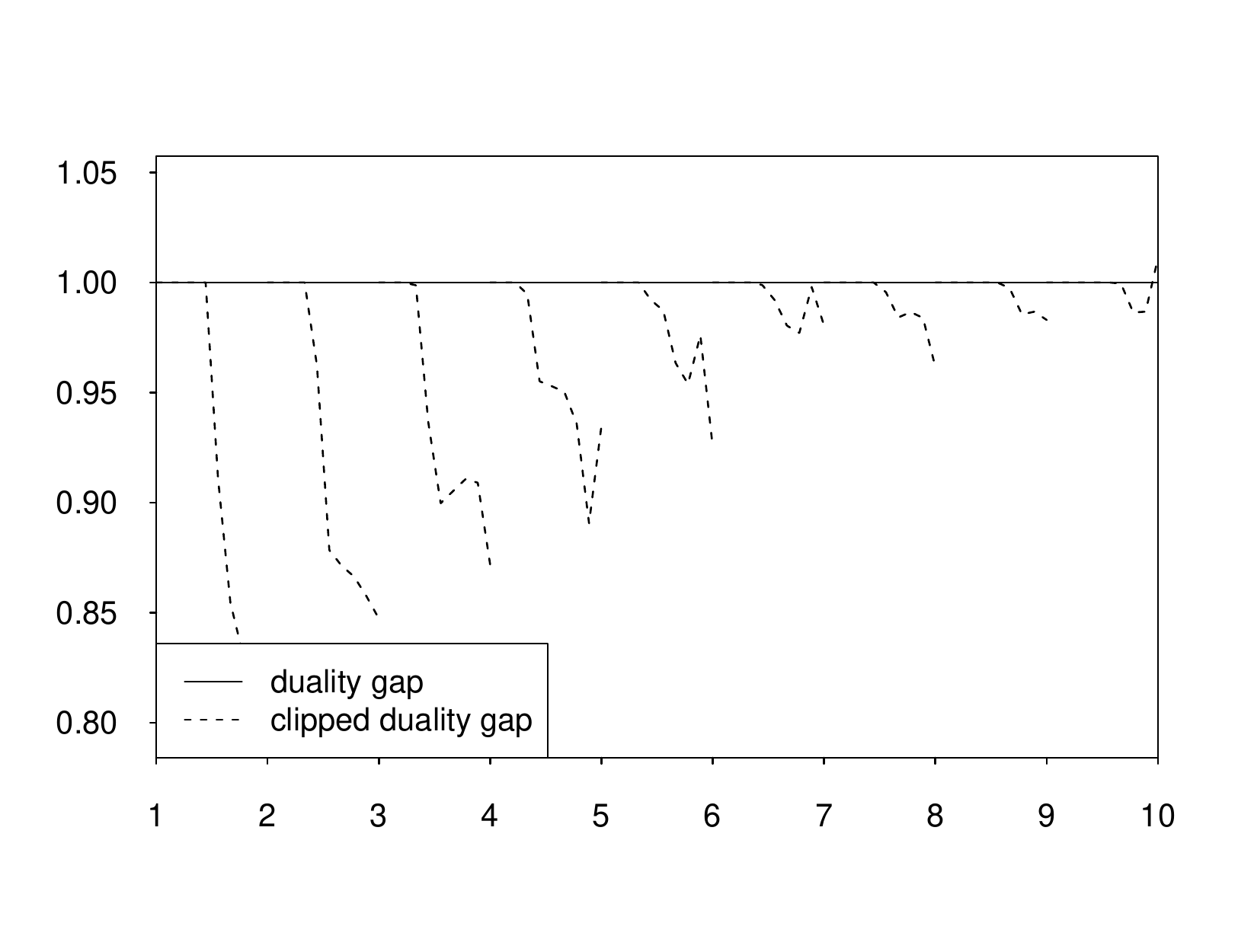}}

\caption{Average number of iterations (left) and corresponding ratio (right) per grid point for different stopping criteria 
using 15 NN and initializing solver with warm start for \textsc{cal-housing}. The graphs comprises for $\tau=0.25$ (top), 
$\tau=0.50$ (middle) and $\tau=0.75$ (bottom).}
\label{figure-figure-per-grid iter for duality gap-cal-housing}
\end{scriptsize}
\end{figure}
 
\end{appendix}

%% file: main.bbl
\small
\begin{thebibliography}{10}
\bibitem{abdous1995relating}
B.~Abdous and B.~Remillard.
\newblock Relating quantiles and expectiles under weighted-symmetry.
\newblock {\em Ann. Inst. Statist. Math.}, 47:371--384, 1995.

\bibitem{anderson2012study}
A.~L. Anderson.
\newblock {\em A Study on Expectiles: Measuring Risk in Finance}.
\newblock PhD thesis, University of Georgia, 2012.

\bibitem{aragon2005conditional}
Y.~Aragon, S.~Casanova, R.~Chambers, and E.~Leconte.
\newblock Conditional ordering using nonparametric expectiles.
\newblock {\em J. Off. Stat.}, 21:617--633, 2005.

\bibitem{aronszajn1950theory}
N.~Aronszajn.
\newblock Theory of reproducing kernels.
\newblock {\em Trans. Amer. Math. Soc.}, 68:337--404, 1950.

\bibitem{BeKlMuGi14a}
F.~Bellini, B.~Klar, A.~M{\"u}ller, and R.~E. Gianin.
\newblock Generalized quantiles as risk measures.
\newblock {\em Insurance Math. Econom.}, 54:41--48, 2014.

\bibitem{berlinet2004reproducing}
A.~Berlinet and C.~Thomas-Agnan.
\newblock {\em Reproducing kernel Hilbert spaces in probability and
  statistics}, volume~3.
\newblock Kluwer Academic Boston, 2004.

\bibitem{breckling1988m}
J.~Breckling and R.~Chambers.
\newblock M-quantiles.
\newblock {\em Biometrika}, 75:761--771, 1988.

\bibitem{chang2011libsvm}
C-C. Chang and C-J. Lin.
\newblock {LIBSVM}: a library for support vector machines.
\newblock {\em ACM Trans. Intell. Syst. Technol.}, 2:27, 2011.

\bibitem{christmann2007svms}
A.~Christmann and I.~Steinwart.
\newblock How {SVM}s can estimate quantiles and the median.
\newblock In {\em Advances in neural information processing systems}, pages
  305--312, 2007.

\bibitem{cristianini2000introduction}
N.~Cristianini and J.~Shawe-Taylor.
\newblock {\em An introduction to support vector machines and other
  kernel-based learning methods}.
\newblock Cambridge University Press, Cambridge, 2000.

\bibitem{eberts2011optimal}
M.~Eberts and I.~Steinwart.
\newblock Optimal learning rates for least squares {SVM}s using gaussian
  kernels.
\newblock In {\em Advances in neural information processing systems}, pages
  1539--1547, 2011.

\bibitem{efron1991regression}
B.~Efron.
\newblock Regression percentiles using asymmetric squared error loss.
\newblock {\em Statist. Sci.}, 1:93--125, 1991.

\bibitem{glasmachers2006maximum}
T.~Glasmachers and C.~Igel.
\newblock Maximum-gain working set selection for {SVM}s.
\newblock {\em J. Mach. Learn. Res.}, 7:1437--1466, 2006.

\bibitem{guler2014mincer}
K.~Guler, P.~T. Ng, and Z.~Xiao.
\newblock Mincer-{Z}arnovitz quantile and expectile regressions for forecast
  evaluations under asymmetric loss functions.
\newblock {\em Northern Arizona University, The WA Franke College of Business.
  Working Paper Series 14-01}, 2014.

\bibitem{hamidi2014dynamic}
B.~Hamidi, B.~Maillet, and J-L. Prigent.
\newblock A dynamic autoregressive expectile for time-invariant portfolio
  protection strategies.
\newblock {\em J. Econom. Dynam. Control}, 46:1--29, 2014.

\bibitem{huang2014asymmetric}
X.~Huang, L.~Shi, and J.~AK Suykens.
\newblock Asymmetric least squares support vector machine classifiers.
\newblock {\em Comput. Statist. Data Anal.}, 70:395--405, 2014.

\bibitem{joachims1999making}
T.~Joachims.
\newblock Making large-scale {SVM} learning practical.
\newblock In {\em Advances in Kernel Methods - Support Vector Learning}. MIT
  Press, Cambridge, MA, USA, 1999.

\bibitem{keerthi2003smo}
S.~S. Keerthi and S.~K. Shevade.
\newblock {SMO} algorithm for least-squares {SVM} formulations.
\newblock {\em Neural comput.}, 15:487--507, 2003.

\bibitem{koenker2005quantile}
R.~Koenker.
\newblock {\em Quantile regression}.
\newblock Cambridge University Press, Cambridge, 2005.

\bibitem{koenker1978regression}
R.~Koenker and G.~Bassett~Jr.
\newblock Regression quantiles.
\newblock {\em Econometrica}, 46:33--50, 1978.

\bibitem{newey1987asymmetric}
W.~K. Newey and J.~L. Powell.
\newblock Asymmetric least squares estimation and testing.
\newblock {\em Econometrica}, 55:819--847, 1987.

\bibitem{platt1999fast}
J.~Platt.
\newblock Fast training of support vector machines using sequential minimal
  optimization.
\newblock In {\em Advances in kernel methods-Support Vector Learning}, pages
  185--208. MIT press, Cambridge, MA., 1999.

\bibitem{schnabel2009analysis}
S.~Schnabel and P.~Eilers.
\newblock An analysis of life expectancy and economic production using
  expectile frontier zones.
\newblock {\em Demographic Res.}, 21:109--134, 2009.

\bibitem{schnabel2009optimal}
S.~K. Schnabel and P.~H. Eilers.
\newblock Optimal expectile smoothing.
\newblock {\em Comput. Statist. Data Anal.}, 53:4168--4177, 2009.

\bibitem{scholkopf2002learning}
B.~Sch{\"o}lkopf and A.J. Smola.
\newblock {\em Learning with kernels: support vector machines, regularization,
  optimization, and beyond}.
\newblock MIT press, Cambridge, MA., 2002.

\bibitem{shim2013expected}
J.~Shim and C.~Hwang.
\newblock Expected shortfall estimation using kernel machines.
\newblock {\em Journal of Korean Data $\&$ Information Science Society},
  24:12--20, 2013.

\bibitem{sobotka2013confidence}
F.~Sobotka, G.~Kauermann, L.~S. Waltrup, and T.~Kneib.
\newblock On confidence intervals for semiparametric expectile regression.
\newblock {\em Stat. Comput.}, 23:135--148, 2013.

\bibitem{sobotka2012geoadditive}
F.~Sobotka and T.~Kneib.
\newblock Geoadditive expectile regression.
\newblock {\em Comput. Statist. Data Anal.}, 56:755--767, 2012.

\bibitem{sobotka2013estimating}
F.~Sobotka, R.~Radice, G.~Marra, and T.~Kneib.
\newblock Estimating the relationship between women's education and fertility
  in {B}otswana by using an instrumental variable approach to semiparametric
  expectile regression.
\newblock {\em J. Roy. Stat. Soc. C- App.}, 62:25--45, 2013.

\bibitem{sobotka2014package}
F.~Sobotka, S.~Schnabel, L.~S. Waltrup, P.~Eilers, T.~Kneib, and G.~Kauermann.
\newblock expectreg: {E}xpectile and qauntile regression. {R} package version
  0.39.
\newblock \url{http://cran.r-project.org/web/packages/expectreg/index.html},
  2014.

\bibitem{stahlschmidt2014expectile}
S.~Stahlschmidt, M.~Eckardt, and W.~K. H{\"a}rdle.
\newblock Expectile treatment effects: An efficient alternative to compute the
  distribution of treatment effects.
\newblock Technical report, Sonderforschungsbereich 649, Humboldt University,
  Berlin, Germany, 2014.

\bibitem{steinwart2008support}
I.~Steinwart and A.~Christmann.
\newblock {\em Support vector machines}.
\newblock Springer, New York, 2008.

\bibitem{steinwart2011estimating}
I.~Steinwart and A.~Christmann.
\newblock Estimating conditional quantiles with the help of the pinball loss.
\newblock {\em Bernoulli}, 17:211--225, 2011.

\bibitem{steinwart2006oracle}
I.~Steinwart, D.~Hush, and C.~Scovel.
\newblock An oracle inequality for clipped regularized risk minimizers.
\newblock In {\em Advances in neural information processing systems}, pages
  1321--1328, 2006.

\bibitem{steinwart2011training}
I.~Steinwart, D.~Hush, and C.~Scovel.
\newblock Training {SVM}s without offset.
\newblock {\em J. Mach. Learn. Res.}, 12:141--202, 2011.

\bibitem{StPaWiZh14a}
I.~Steinwart, C.~Pasin, R.~Williamson, and S.~Zhang.
\newblock Elicitation and identification of properties.
\newblock In M.~F. Balcan and C.~Szepesvari, editors, {\em JMLR Workshop and
  Conference Proceedings Volume 35: Proceedings of the 27th Conference on
  Learning Theory 2014}, pages 482--526, 2014.

\bibitem{takeuchi2006nonparametric}
I.~Takeuchi, Q.~V. Le, T.~D. Sears, and A.~J. S.
\newblock Nonparametric quantile estimation.
\newblock {\em J. Mach. Learn. Res.}, 7:1231--1264, 2006.

\bibitem{taylor2008estimating}
J.~W. Taylor.
\newblock Estimating value at risk and expected shortfall using expectiles.
\newblock {\em J. Financ. Econ.}, 6:231--252, 2008.

\bibitem{vapnik2000nature}
V.~Vapnik.
\newblock {\em The nature of statistical learning theory}.
\newblock Springer-Verlag, New York, 2000.

\bibitem{vogt2002smo}
M.~Vogt.
\newblock {SMO} algorithms for support vector machines without bias term.
\newblock {\em Technische Univ. Darmstadt, Inst. Automat. Contr., Lab. Contr.
  Syst. Process Automat., Darmstadt, Germany}, 2002.

\bibitem{waltrup2014expectile}
L.~S. Waltrup, F.~Sobotka, T.~Kneib, and G.~Kauermann.
\newblock Expectile and quantile regression--{D}avid and {G}oliath?
\newblock {\em Stat. Model.}, page 1471082X14561155, 2014.

\bibitem{wang2011measuring}
Y.~Wang, S.~Wang, and K.~K. Lai.
\newblock Measuring financial risk with generalized asymmetric least squares
  regression.
\newblock {\em Appl. Soft Comput.}, 11(8):5793--5800, 2011.

\bibitem{wright1999numerical}
S.J. Wright and J.~Nocedal.
\newblock {\em Numerical optimization}, volume~2.
\newblock Springer, New York, 1999.

\bibitem{yang2014nonparametric}
Y.~Yang and H.~Zou.
\newblock Nonparametric multiple expectile regression via {ER}-{B}oost.
\newblock {\em J. Stat. Comput. Simulation}, 85:1442--1458, 2015.

\bibitem{yao1996asymmetric}
Q.~Yao and H.~Tong.
\newblock Asymmetric least squares regression estimation: a nonparametric
  approach.
\newblock {\em J. Nonparametr. Statist.}, 6:273--292, 1996.

\end{thebibliography}
